\documentclass[11pt]{article}
\usepackage{jheppub}

\usepackage[usenames,dvipsnames,table]{xcolor}
\usepackage{graphicx,amsmath,amssymb,multirow,array,bm,mathrsfs}
\usepackage{epsf,amsfonts}
\usepackage[numbers,sort&compress]{natbib}

\usepackage{slashed}

\newcommand{\beq}{\begin{equation}}
\newcommand{\eeq}{\end{equation}}
\newcommand{\bea}{\begin{eqnarray}}
\newcommand{\eea}{\end{eqnarray}}

\newcommand{\prn}[1]{\left ( #1 \right )}
\newcommand{\brk}[1]{\left [ #1 \right ]}
\newcommand{\bigbr}[1]{\Bigl\{ #1 \Bigr\} }

\newcommand{\half}{\frac{1}{2}}

\newcommand{\tr}{\mbox{tr}}

\newcommand{\form}[1]{\bm{#1}}

\newcommand{\ic}{\form{i}}
\newcommand{\hodge}{{}^\star}

\newcommand{\hodgeCFT}{ {\hodge}^{\text{\tiny{CFT}}}}

\newcommand{\fu}{\form{u}}

\newcommand{\fomega}{\form{\omega}}

\newcommand{\fA}{\form{A}}

\newcommand{\fF}{\form{F}}

\newcommand{\fGamma}{\form{\Gamma}}

\newcommand{\fR}{\form{R}}

\newcommand{\PhiT}{\Phi_{_T}}

\newcommand{\Sp}{\Sigma}

\newcommand{\fP}{{\form{\mathcal{P}}}}

\newcommand{\ICS}{{\form{I}}_{CS}}

\newcommand{\SpH}{\mathrm{\Sp}_{H}}

\newcommand{\Lag}{\overline{L}}

\newcommand{\fQNoether}{\slashed{\delta}\form{Q}_{_{\text{Noether}}}}

\newcommand{\fKomar}{\form{\mathcal{K}}_\chi}

\newcommand{\fV}{\form{V}}

\newcommand{\N}{\mathrm{N}}
\newcommand{\fN}{\form{\N}}

\newcommand{\GN}{G_{_N}}
\newcommand{\cc}{\Lambda_{_{cc}}}
\newcommand{\gYM}{g_{_{YM}}}

\newcommand{\R}[1]{{}_{_R}\form{#1}}
\newcommand{\F}[1]{{}_{_F}\form{#1}}
\newcommand{\B}[1]{{}_{_{\mathcal{B}}}\form{#1}}
\newcommand{\fchi}{\form{\chi}}

\newcommand{\cR}{\form{{\cal R}}}

\newcommand{\rH}{r_{_H}}
\newcommand{\tV}{\text{(\tiny{V})}}

\newcommand{\fixform}[1]{\texorpdfstring{#1}{~}}
\newcommand{\be}{\begin{equation}}
\newcommand{\ee}{\end{equation}}

\newcommand{\GV}{\mathbb{G}^{\tV }}

\newcommand{\GVCFT}{
\mathbb{G}^{\tV, {\text{\tiny{CFT}} }}
}
\newcommand{\GVCFTmu}{
\mathbb{G}_{\mu}^{{\text{\tiny{CFT}}}}
}
\newcommand{\GVCFTnu}{
\mathbb{G}_{\nu}^{{\text{\tiny{CFT}}}}
}
\newcommand{\HVtemp}{\mathbb{H}^\tV }
\newcommand{\TmnanomCS}{T_{\mu\nu}^{\text{\tiny{anom,CS}}}} 
 
\newcommand{\JmanomCS}{J^{\text{\tiny{anom,CS}}}_{\mu}}
\newcommand{\JnanomCS}{J^{\text{\tiny{anom,CS}}}_{\nu}}

\newcommand{\JSanomCS}{J_S^{\text{\tiny{anom,CS}}}
}

\newcommand{\JSanomCSV}{J_S^{\tV}
}

\newcommand{\TmnCFTCS}{T_{\mu\nu}^{\text{\tiny{CFT,CS}}}} 
 
\newcommand{\TmnCFTEin}{T_{\mu\nu}^{\text{\tiny{CFT,Ein}}}}

\newcommand{\JnCFTCS}{J^{\text{\tiny{CFT,CS}}}_{\nu}}

\newcommand{\JnCFTEin}{J^{\text{\tiny{CFT,Ein}}}_{\nu}}

\newcommand{\WeylW}{\Delta_w }

\newcommand{\fQNoetherEinMax}{(\fQNoether)_{_\text{Ein-Max}} }
\newcommand{\fQNoetherEinMaxCS}{(\fQNoether) }
\newcommand{\fQNoetherHall}{(\fQNoether)_{H} }
\newcommand{\CFT}{{\text{\tiny{CFT}}}}

\newcommand{\athor}{|_{hor}}
\newcommand{\atinfty}{|_{\infty}}
\newcommand{\barA}{\bar{A}}
\newcommand{\barB}{\bar{B}}
\newcommand{\barC}{\bar{C}}
\newcommand{\barD}{\bar{D}}
\newcommand{\barE}{\bar{E}}

\newcommand{\xiH}{\xi_{hor}}
\newcommand{\LambdaH}{\Lambda_{hor}}

\newcommand{\signCA}{s}

\title{\Large
{Anomalies, Chern-Simons Terms and Black Hole Entropy}}

\author[a]{Tatsuo Azeyanagi,}
\author[b]{R. Loganayagam}
\author[c]{and Gim Seng Ng}

\affiliation[a]{
D\'{e}partement de Physique, Ecole Normale Sup\'{e}rieure, CNRS, 24 rue Lhomond,
75005 Paris, France.}
\affiliation[b]{School of Natural Sciences, Institute for Advanced Study, Princeton, NJ 08540, USA.}
\affiliation[c]{Department of Physics, McGill University, Montr\'{e}al, QC H3A 2T8, Canada.}

\emailAdd{tatsuo.azeyanagi@phys.ens.fr}
\emailAdd{nayagam@ias.edu}
\emailAdd{gim.ng@mcgill.ca}
\abstract{
 Recent derivations of Cardy-like formulae in higher dimensional field theories have opened up a way of computing, via AdS/CFT, universal contributions to black hole entropy from gravitational Chern-Simons  terms. Based on the manifestly covariant formulation of the differential Noether charge for Chern-Simons terms  
proposed in \cite{Azeyanagi:2014sna}, we compute the entropy and asymptotic charges for 
the rotating charged AdS black holes in higher dimensions at leading order of the fluid/gravity derivative expansion in the Einstein-Maxwell-Chern-Simons system. 
This  gives a result that exactly matches the field theory predictions from Cardy-like formulae. 
}

\begin{document}
\maketitle

\section{Introduction}

Since the famous demonstration by Bekenstein and Hawking that black hole geometries have entropy, 
the quest for understanding microscopic origins of that entropy has taken us on a long and fascinating adventure connecting 
various fields of physics and mathematics. The advent of string theory and AdS/CFT has injected much 
enthusiasm into this subject, and for a wide class of extremal/near-extremal black holes we can now 
claim to understand  (albeit by somewhat indirect means)  where this entropy comes from \cite{Strominger:1996sh, Callan:1996dv,Cvetic:1996gq,Horowitz:1996ay,Maldacena:1996gb,Johnson:1996ga,Maldacena:1997de}.

Accounting for the entropy of finite-temperature black holes has however been a more difficult endeavor.
The major successes in this regard are often linked to anomalies --- a paradigmatic result in this direction is
the  Cardy formula \cite{Cardy:1986ie} which links the thermodynamic properties of a 2d CFT with the anomaly coefficients 
(the right and left central charges $c_{R,L}$) calculable from the microscopic description. In context of AdS$_3$/CFT$_2$,
this fact  has been repeatedly exploited \cite{Strominger:1997eq} to account for the entropy of finite temperature AdS$_3$ black holes far 
from extremality. It is hence an interesting question to ask whether this success can be extended to higher
dimensions.   

The parity even part of the Cardy formula does not generalize to higher dimensional
field theories.\footnote{For example, consider the famous $3/4$ factor between free energies at strong and weak coupling in four-dimensional $\mathcal{N}=4$ SYM. 
Since the central charge of this theory is coupling independent, the presence of this factor rules out any possibility of  universal formula only involving
central charges. } However, recently the  analogue of the parity odd part of the Cardy formula in higher dimensions
has been conjectured \cite{Landsteiner:2011cp,Landsteiner:2011iq,Loganayagam:2012pz,Loganayagam:2012zg} and proved \cite{Jensen:2012kj, Jensen:2013kka,Jensen:2013rga}. As we will briefly review below, this generalization --- often goes by the name of `replacement rule'--- gives a prescription 
in which one starts with the anomaly polynomial $\fP_{CFT}$ of the field theory under question 
and then, by a series of steps, constructs an expression for the leading parity odd part of the entropy.

It is then natural to enquire whether the replacement rule can be used to account for the leading 
parity odd part of the black hole entropy. The main aim of this work is to show that this is indeed the 
case  and that successes in AdS$_3$/CFT$_2$ can be extended to the 
parity odd sector of any even dimensional CFTs.

The anomaly coefficients we are interested in (along with the Weyl anomaly coefficients to whom they are related to via supersymmetry) 
have been conjectured to determine various terms in different supersymmetric partition functions. Such conjectures
have been investigated actively by various authors including \cite{Spiridonov:2012ww,Assel:2014paa,Buican:2014qla,
Ardehali:2014zba,DiPietro:2014bca,Ardehali:2014esa,Lorenzen:2014pna,Ardehali:2015hya,Assel:2015nca}. Although these 
supersymmetric versions do not yet have a general proof of the type given in \cite{Jensen:2012kj,Jensen:2013rga}, there is
a mounting evidence for their validity. Our arguments in this work about  how anomalies show up in gravity computations 
would hopefully be extended to such supersymmetric versions.

The first step in this direction is to construct a large class of  black hole solutions which 
will play the role of the famous BTZ black hole solution in higher dimensions. In order to have 
a parity odd part to the entropy associated with anomalies, these black holes should 
be solutions of a gravitational theory with Chern-Simons terms. The simplest system of this 
kind is the Einstein-Maxwell-Chern-Simons system with an action 
\begin{equation}\label{eq:action}
\begin{split}
\int d^{d+1} x \,\sqrt{-G} \brk{ \frac{1}{16\pi \GN} \prn{R-2\cc}
-\frac{1}{4\gYM^2}  F_{ab} F^{ab} }+  \int \ICS[\fA,\fF,\fGamma,\fR]\, .
\end{split}
\end{equation}
Here  the Chern-Simons part of the Lagrangian is denoted as $\ICS$ which is a $(d+1)$-form.\footnote{In this paper, we use the same notation and conventions for differential forms as in \cite{Azeyanagi:2013xea, Azeyanagi:2014sna}. We refer the reader to Appendix~\ref{app:notation} for a brief summary of notations.} 
Since Chern-Simons terms are odd forms, this necessarily implies that $d=2n$ with 
a positive integer $n$. 
The cosmological constant $\cc$ is taken to be negative and 
is given by $\cc\equiv  -d(d-1)/2$ such that $G_{ab}$ is an asymptotically AdS$_{d+1}$ metric with unit radius, $F_{ab}$ is 
the Maxwell field strength, $\GN$  and $\gYM$ are the Newton and Maxwell couplings 
respectively. 
For later use, we also define the normalized Maxwell coupling constant $\kappa_q$ by 
$16\pi G_N/\gYM^2 = \kappa_q\, (d-1)/(d-2)$. Large, charged, rotating  black hole solutions of this system were constructed in our recent work \cite{Azeyanagi:2013xea}
using fluid/gravity correspondence. 

The main aim of this work is  to compute the entropy and the asymptotic charges of these solutions. 
In course of our calculations in this somewhat simplified system,
we will exhibit various structural features which we believe would carry over to more complicated examples
in string theory. In particular, the bulk of our Appendices are devoted to proving a kind of `non-renormalization theorem' 
for anomaly-induced entropy which shows how anomaly-induced entropy does not get corrected in 
fluid/gravity expansion. We expect this result (along with the various structural cancellations that lead to it)
to hold in more complicated examples. In fact, the robustness of anomaly induced entropy in field theory 
suggests that such a result should hold even when stringy and quantum gravity corrections are taken into
account ! 

The second step is to develop a coherent  method to compute the entropy of black hole solutions
in the presence of Chern-Simons terms. This involves various subtleties due to the non-covariant nature  
of Chern-Simons terms in the Lagrangian density. In particular, the original Noether procedure 
due to Wald \cite{Lee:1990nz, Wald:1993nt,  Iyer:1994ys} is valid only for the system described by a covariant Lagrangian and thus
fails in the case of Chern-Simons terms. Fortunately, this Wald formalism for constructing differential Noether charges
was  extended to theories with Chern-Simons terms by Tachikawa \cite{Tachikawa:2006sz}. 
As demonstrated by 
Bonora-Cvitan-Prester-Pallua-Smolic \cite{Bonora:2011gz}, however, this extension  suffers from various non-covariance
issues for AdS spacetime with 
dimensions greater than three.  In our recent work \cite{Azeyanagi:2014sna}, we identified  the root cause of  these non-covariance problems to  be the choice of a non-covariant pre-symplectic structure in the Tachikawa
method.  

Further, in that work, we showed that with higher dimensional  Chern-Simons terms, one can instead 
choose a  manifestly covariant  pre-symplectic structure and  implement the Noether procedure 
without any subtleties. One of the main results of that work was a covariant expression for  Chern-Simons contribution $\fQNoetherHall$ 
to the differential Noether charge co-dimension 2 form $\fQNoether$, given as a sum of five terms :
\be
(\fQNoether)_{H}\equiv T_0+T_1 +T_2 +T_3 +T_4\, ,  \label{eq:deltaQcssum1}
\ee where each term on the right hand side is determined by the derivative of the anomaly polynomial 
$ \fP_{CFT} = d \ICS$ ($\ICS$: Chern-Simons terms in the Lagrangian) as
\bea\label{eq:T0toT4intro}
T_0&\equiv &\delta \fA(\Lambda+ \ic_\xi \fA) \frac{\partial^2 \fP_{CFT}}{\partial \fF\partial \fF} 
+\delta \fGamma^a{}_b (\Lambda+ \ic_\xi \fA) \frac{\partial^2 \fP_{CFT}}{\partial \fR^a{}_b \partial \fF}\, , 
\nonumber\\
T_1 &\equiv& \delta \fGamma^a{}_b \nabla_d \xi^c \frac{\partial^2 \fP_{CFT}}{\partial \fR^a{}_b \partial \fR^c{}_d}\, , \nonumber\\
T_2 &\equiv& \delta \fA \nabla_d \xi^c \frac{\partial^2 \fP_{CFT}}{\partial \fF \partial \fR^c{}_d}\, ,\nonumber\\
T_3 &\equiv & -\half \delta G_{ab} (\SpH)^{abc}\, \ic_\xi (\hodge dx_c)\, , \nonumber\\
T_4 &\equiv & -\frac{1}{4} \xi^a \delta\left\{
\left[(\SpH)_a{}^{bc} + (\SpH)^b{}_a{}^c + (\SpH)^{cb}{}_a\right] \hodge (dx_b \wedge dx_c)
\right\}\, . 
\eea  Here the spin Hall current $(\SpH)^{cb}{}_a$ is defined by  $(\SpH)^{cb}{}_a \hodge dx_c\equiv -2( \partial \fP_{CFT}/\partial \fR^a{}_b)$
 and $(\Lambda, \xi^a)$ are the parameters for the $U(1)$ gauge transformation and diffeomorphism, respectively. 
 This $(\fQNoether)_{H}$ is manifestly covariant for any odd dimensional spacetime. 
This expression with its five terms denoted by $T_i$ will play a crucial role in this work. In this paper, we take the third step whereby we  evaluate
this Noether charge  on our black hole solutions  and show that this contribution to the black hole entropy is exactly accounted
for by  the anomaly-induced entropy on the CFT side \cite{Loganayagam:2012pz,Loganayagam:2012zg,Jensen:2012kj,Jensen:2013rga}. In addition, we will also use this covariant 
differential Noether charge to evaluate the asymptotic charges, that is, the energy-momentum tensor and the charge current of the dual CFT. 
 This completes the holographic and systematic derivation of the replacement rules for these quantities initiated in \cite{Azeyanagi:2013xea}. 

Before we proceed to the details of our computation, certain clarifying comments are in order regarding the use of differential Noether charge.
For a time-independent black hole solutions with bifurcate Killing horizon, the differential Noether charge form exhibited above can be integrated over the  bifurcation surface. 
Following Wald, we can then derive an integral expression for the total
Noether charge of stationary solutions\cite{Azeyanagi:2014sna} and this gives the correct modification of Wald  entropy  in the presence of Chern-Simons terms 
as originally conjectured by Tachikawa \cite{Tachikawa:2006sz} :
\begin{equation}\label{eq:WaldTachikawa0}
\begin{split}
 S_{\text{Wald-Tachikawa}} &= \int_{Bif} 2\pi \varepsilon_b{}^a  \frac{\delta\Lag_{cov}}{\delta R^a{}_{bcd}} \varepsilon_{cd}  
 +  \int_{Bif}  \sum_{k=1}^\infty 8\pi k\ \fGamma_N (d\fGamma_N)^{2k-2} \frac{\partial \fP_{CFT}}{\partial\ \text{tr} \fR^{2k} }\\
 &= \int_{S_{\infty}} J_{S,\CFT}^\mu\quad  \hodgeCFT dx_\mu\, , 
 \end{split}
\end{equation} 
where in the first line, the integrals are over the bifurcation surface with $\varepsilon^{ab}$ denoting the binormal at the bifurcation surface. In the second line, we have pulled back these integrals to a spatial slice in the AdS boundary using the ingoing null geodesic prescription for the CFT entropy current $ J_{S,\CFT}^\mu$ following \cite{Bhattacharyya:2008xc}.

In the above entropy formula,  $\Lag_{cov}$ is the covariant part of the gravity Lagrangian which contributes via the famous Wald formula as expected. The CFT anomaly polynomial $\fP_{CFT}= d\ICS$ encodes the information about the Chern-Simons part and we have presented this contribution in terms of the normal bundle connection $ \fGamma_N $ on the bifurcation surface and its curvature $ \fR_N = d\fGamma_N$. They are defined using the binormal as
\begin{equation}
\begin{split}\label{eq:normalbundleGamma}
 \fGamma_N &\equiv \brk{\half \varepsilon_a{}^b \fGamma^a{}_b }_{Bif}\ ,\quad 
 \fR_N \equiv \brk{\half \varepsilon_a{}^b \fR^a{}_b }_{Bif} = d\fGamma_N\, . 
 \end{split}
\end{equation} 

While we will have much to say about the structure of Tachikawa formula \eqref{eq:WaldTachikawa0} especially vis \`a vis the structure of 
the replacement rule,\footnote{In Appendix \ref{appendix:TachikawaPontryagin}, we provide an interesting rewriting of the Tachikawa entropy formula in terms of the Pontryagin classes. 
This expression based on the Pontryagin class would probably be useful to relate the Tachikawa entropy formula to general CFT replacement 
rule in the case when the CFT is on a general curved background. } we will rely directly on $\fQNoetherHall$ for our main results. This is for  a  
computational and a deeper conceptual reason. 

The computational reason is this --- our solutions are naturally written in ingoing Eddington Finkelstein type coordinates 
which are unsuited for examining bifurcation surface geometry  (especially objects like those defined in Eq.\eqref{eq:normalbundleGamma}).
In case of usual Wald formula, this issue does not arise :  according to 
an argument by Jacobson-Kang-Myers(JKM) \cite{Jacobson:1993vj},  for time-independent solutions,
the first term in \eqref{eq:WaldTachikawa0} coming from $\Lag_{cov}$ can be evaluated on an arbitrary spatial slice of the horizon instead
of the bifurcation surface. Unfortunately, we have not been able to formulate such a   JKM type argument for the second term in \eqref{eq:WaldTachikawa0}.\footnote{Note that, in BTZ case, one can explicitly check that such a  Jacobson-Kang-Myers type argument does hold for  many of  the standard coordinate slicings in use. Given the complexity of fluid/gravity solutions in higher dimensions, such an explicit check does not seem feasible.} We will later propose a heuristic expression  for the total Noether charge similar to  Tachikawa formula which, when evaluated in standard fluid/gravity slicings, does reproduce the  answer obtained from the differential Noether charge method
(see \S\S~\ref{ssec:TachiLike}). It would be interesting to come up with a generalization of  \cite{Jacobson:1993vj} to Chern-Simons terms that would justify our
proposal. For these reasons, even in the time-independent case, we will rely on $\fQNoetherHall$ to assign entropy and charges .

The conceptual reason is this --- our fluid/gravity solutions are in general time-dependent and various steps needed for deriving  \eqref{eq:WaldTachikawa0}
are no more valid. We remind the reader that, unlike the discussion in the previous paragraph, the  issue of defining time-dependent entropy current in presence of higher derivative terms is ill-understood even in the absence of CS terms. An implicit assumption here is that the use of differential Noether charge ameliorates these problems, i.e., we advocate that the differential Noether charge is an appropriate way to assign entropy current, energy momentum tensor and charge currents to time dependent black hole solutions at least in the fluid/gravity regime.\footnote{We emphasize that this is indeed an assumption given that Wald-like formalisms do not readily generalize to fluid/gravity regime. More specifically, no such generalization has been proved yet
to give an entropy current which satisfies local version of second law to arbitrary order in derivative expansion.} It would be interesting to see
whether this prescription reproduces  the specifc subleading\footnote{At the leading derivative order we work in this paper, 
there are no time-dependent  corrections to anomalous transport.} time-dependent corrections to anomalous transport predicted by fluid-dynamical considerations
(see Sec.12 of \cite{Haehl:2015pja}).

After this technical aside, let us conclude our introduction by giving the outline of our paper.
We will begin in section \S\ref{sec:review} by briefly reviewing the basic results from previous work that we will need
later on. This review naturally falls into two subsections. In \S\S\ref{ssec:RepRule}, we will state the replacement rule derived from CFT  considerations while in \S\S\ref{ssec:BHSol} we present the black hole solutions of interest  derived in \cite{Azeyanagi:2013xea}.
This is followed by \S\S\ref{ssec:TachiForm} which contains a description of  the manifestly covariant differential Noether charge constructed in \cite{Azeyanagi:2014sna}. 
This section ends with \S\S\ref{ssec:summary} which is a summary of new results from this paper for the convenience of the readers.

In the next section \S\ref{sec:RepEntropy}, we show how the replacement rule 
for the anomaly-induced entropy current is reproduced from 
the differential Noether charge for the Chern-Simons terms evaluated at the horizon of our rotating charged AdS black hole solution. 
This is preceded by a heuristic derivation of the same result using a  Tachikawa-like formula on the horizon.    

Moving on to  section \S\ref{sec:RepTJ}, our differential Noether charge is used to derive the CFT stress tensor and current which reproduce field theory expectations. We conclude in section \S\ref{sec:Disc} with discussions on future directions.

We relegate various simple examples and many technical details to our Appendices.
First, we provide a series of Appendices containing a list of notation (Appendix  \ref{app:notation}) and a collection of useful formulas (Appendix \ref{app:useForm} and \ref{app:useForm2}) we use throughout this paper. 
Rewriting of the Tachikawa entropy formula in terms of the Pontryagin classes is explained in Appendix \ref{appendix:TachikawaPontryagin}. 
Using our differential Noether charge,
we review in  Appendix \ref{app:BTZ}  the well-known AdS$_3$ derivation of the Cardy formula in the presence of gravitational anomalies. The simple case of Abelian Chern-Simons
terms are dealt with in Appendix \ref{app:AbelianCS}. The following Appendix works out in detail the AdS$_5$ case which shows the essential structures necessary for the computations in arbitrary dimensions. We then describe in Appendix \ref{app:replacementT3T4} some structural results regarding the $T_3$ and $T_4$ terms appearing in 
our differential Noether charge Eq.~\eqref{eq:deltaQcssum} on the rotating charged AdS black hole background. Appendice \ref{sec:T0T1T2hor0th1st} and \ref{sec:entropyhighercontribution}  are devoted to the evaluation of 
$T_0$, $T_1$ and $T_2$ terms in the differential Noether charge Eq.~\eqref{eq:deltaQcssum}
at the horizon. In Appendix \ref{sec:roughrefinedestimate} and \ref{sec:chargeshighercontribution}  we compute the asymptotic 
charges for our black hole solution. 
Finally in Appendix \ref{app:entropyEM}, we compute the Einstein-Maxwell contribution to the entropy. 

\section{Review of previous works and summary}\label{sec:review} 
In this section, we will review a few relevant recent results which will be useful in the computations of 
the stress tensor/current and entropy. We start with the recent field-theoretical results on the replacement rule of stress tensor/current 
and entropy. After this, we move to the dual gravity side and briefly review some important results from our previous papers :
the rotating charge-AdS black hole solution dual to charged rotating fluid \cite{Azeyanagi:2013xea} as well as
the manifestly covariant differential Noether charge for Chern-Simons terms and the Tachikawa entropy formula derived from it \cite{Azeyanagi:2014sna}. 
In the final part of this section, we summarize our main results 
in the current paper and compare them with predictions coming from the CFT replacement rule.

\subsection{Entropy current and stress tensor/current from CFT replacement rule}\label{ssec:RepRule}

Through the recent studies on the hydrodynamic description of systems with global anomalies, 
it has been shown that the leading order anomaly-induced transports are completely captured by  the `the replacement rule' \cite{Loganayagam:2012pz,Loganayagam:2012zg,Jensen:2012kj,Jensen:2013rga}. 
The statement of the replacement rule is as follows : 
let us consider an even-dimensional quantum field theory with global anomalies characterized by an anomaly polynomial $\fP_{CFT}$.
We define the pseudo-vector $V_\mu$ as $\hodgeCFT \form{V} = \fu \wedge (d\fu)^{n-1} \equiv \fu \wedge (2\fomega)^{n-1}$ 
where $\fu\equiv u_{\mu}dx^{\mu}$  is the fluid velocity one-form and $\fomega \equiv  (1/2)\omega_{\mu\nu}dx^{\mu}\wedge dx^{\nu}$ is the vorticity 2-form.
Then,  the leading anomaly-induced contribution to the Gibbs free energy current $\GVCFTmu$ is along $V_\mu$. If we write
 $\GVCFTmu= \GVCFT V_\mu+\ldots$ where $\ldots$ denotes the non-anomalous (and the sub-leading anomalous) contributions, 
the replacement rule claims that $\GVCFT$ is completely determined by the anomaly polynomial : 
\begin{equation}\label{eq:repwB0}
\begin{split}
\GVCFT= \fP_{CFT}\left[
\fF\rightarrow \mu;\quad
\tr[ \fR^{2k}] \rightarrow 2(2\pi T)^{2k}\right] \,. 
\end{split}
\end{equation}  
Here, $\mu$ is the chemical potential for the $U(1)$ charge while $T$ is the temperature.
We note that $\tr[ \fR^{2k}]\rightarrow 2(2\pi T)^{2k}$ means the replacement of each $\tr[ \fR^{2k}]$
($k$: positive integer) appearing in the anomaly polynomial by $2(2\pi T)^{2k}$. Following the standard thermodynamic relations
\begin{equation}
\mathcal{M}=G+\mu \,\mathcal{Q} + T \mathcal{S}=G-\mu \prn{\frac{\partial G}{\partial \mu}}_{T}-T \prn{\frac{\partial G}{\partial T}}_{\mu}
\, ,
\quad  
\mathcal{Q} = -\prn{\frac{\partial G}{\partial \mu}}_{T} 
\, ,
\quad  
\mathcal{S}=- \prn{\frac{\partial G}{\partial T}}_{\mu}\, ,  \,
\end{equation} (where $G$ : Gibbs free energy, $\mathcal{M}$ : energy, $\mathcal{Q}$ : $U(1)$ charge, $\mathcal{S}$ : entropy) we also obtain the replacement rule for 
 the anomaly-induced contribution to the stress tensor $T^{\text{\tiny{anom}}}_{\mu\nu}$, $U(1)$ current $J^{\text{\tiny{anom}}}_{\nu} $ 
 and entropy current $(J^{\text{\tiny{anom}}}_S)_{\nu}$ as 
\begin{eqnarray}
\label{eq:Tmnanom0}
T^{\text{\tiny{anom}}}_{\mu\nu} &=&
\prn{
\GVCFTmu
 -\mu\   \frac{\partial \GVCFTmu }{\partial \mu}
-T\  \frac{\partial \GVCFTmu
 }{\partial T} } u_\nu 
 +u_\mu \prn{\GVCFTnu
 -\mu\   \frac{\partial \GVCFTnu }{\partial \mu}
-T\  \frac{\partial \GVCFTnu }{\partial T} } \, , \nonumber \\
J^{\text{\tiny{anom}}}_{\nu} &=& - \prn{ \frac{\partial \GVCFTnu }{\partial \mu} }_T \, , \qquad 
\label{eq:Sentanom0}
(J^{\text{\tiny{anom}}}_S)_{\nu}=  - \prn{ \frac{\partial \GVCFTnu }{\partial T} }_{\mu}\,.
\end{eqnarray}

We note that this replacement rule was first conjectured in \cite{Loganayagam:2012pz,Loganayagam:2012zg} based on  observations in free theories 
and was then proved via formal Euclidean methods in \cite{Jensen:2012kj,Jensen:2013rga}.

\subsection{Rotating charged AdS black hole solution}\label{ssec:BHSol}
Our main interest in the current paper is to consider Einstein-Maxwell-Chern-Simons theory with 
a negative cosmological constant in $(d+1)$ dimensions ($d=2n$ with a positive integer $n$)
and evaluate the Hall contribution to the differential Noether charge 
constructed in Ref.~\cite{Azeyanagi:2014sna} both at the boundary and horizon 
of the rotating charged AdS black hole solutions in five and higher dimensions.  
These black hole solutions are derived via the fluid/gravity derivative expansion 
in Ref.~\cite{Azeyanagi:2013xea}.
For later use, we summarize some key results on these black hole solutions from Ref.~\cite{Azeyanagi:2013xea} 
and on the differential Noether charge from Ref.~\cite{Azeyanagi:2014sna}. 
The action of the Einstein-Maxwell-Chern-Simons theory with a negative cosmological 
constant is given in Eq.~\eqref{eq:action}. 
The rotating charged AdS black hole solution on which we are going to 
evaluate the Chern-Simons contribution to the differential Noether charge \eqref{eq:deltaQcssum} 
takes the following form \cite{Azeyanagi:2013xea} : 
\begin{equation}\label{eq:flugravsoln}
\begin{split}
ds^2 &= -2 u_\mu dx^\mu \otimes_{sym} dr 
+r^2\brk{- f(r,m,q)\ u_\mu u_\nu + P_{\mu\nu} } dx^\mu \otimes_{sym} dx^\nu +\ldots  \\
&\qquad\qquad +
2 g_{_V}(r,m,q)\, u_\mu V_\nu\ dx^\mu \otimes_{sym} dx^\nu + \ldots , \\
\fA &=  \Phi(r,q)\ u_\mu\ dx^\mu + \ldots \\
&\qquad\qquad + a_{_V}(r,m,q)   V_\mu\ dx^\mu +\ldots \, . 
\end{split}
\end{equation}
Here $P_{\mu\nu} \equiv g_{\mu\nu} + u_\mu u_\nu$ is the projection operator and
\begin{equation}
\begin{split}
f(r,m,q) &\equiv 1- \frac{m}{r^d} + \half \kappa_q \frac{q^2}{r^{2(d-1)}} \ ,\qquad
\Phi(r,q) \equiv \frac{q}{r^{d-2}}\, ,\\
\PhiT(r,m,q) &\equiv \half r^2 \frac{df}{dr} 
= \frac{1}{2r^{d-1}} \brk{ m d- \kappa_q(d-1)\frac{q^2}{r^{d-2}} }\, . \\
\end{split}
\end{equation}
We denote the location of the horizon by $r=r_H$ (which satisfies $f(r_H, m, q)=0$).  
The parameters $m$ and $q$ determine the mass and electric charge of this black hole solution.   
We also define $\Psi(r) \equiv r^2 f(r, m, q)/2$ for later use. 
Throughout this paper, we set the boundary metric to be flat, $g_{\mu\nu}=\eta_{\mu\nu}$ 
and set the velocity vector $u^\mu$ (normalized such that $u_{\mu}u^{\mu}=-1$) to 
pure rotation, i.e. $\partial_{(\mu} u_{\nu)}=0$ and $u^\mu \omega_{\mu\nu}=0$ 
for the vorticity $\omega_{\mu\nu} =\partial_{[\mu} u_{\nu]}$.  
We note that, at the horizon $r=r_H$, 
we have  $\Phi(r=r_H) = \mu$ and $\Phi_T(r=r_H) = 2\pi T$ 
where $\mu$ and $T$ are the $U(1)$ chemical potential and the Hawking temperature, respectively. 
We will refer the reader to  Appendix~\ref{sec:app_useful_previouspaper_0th1st} 
and our previous paper \cite{Azeyanagi:2013xea} for a  more detailed analysis
of these black hole solutions.

The first lines of the metric and gauge field in Eq.~\eqref{eq:flugravsoln} 
give the AdS Reissner-Nordstrom solution
boosted by a velocity $u^\mu$. 
In particular, we note that, when we take $u^\mu$ to be a uniformly rotating configuration on a sphere, 
this first line gives the leading order terms of AdS Kerr-Newman solutions 
expanded in the fluid/gravity expansion.  
The second line of the metric and gauge field in Eq.~\eqref{eq:flugravsoln} 
is proportional to a pseudo-vector 
\begin{equation} \label{eq:pseudoV}
\begin{split} 
V^\mu &\equiv \varepsilon^{\mu\nu\lambda_1 \sigma_1 \lambda_2 \sigma_2 \ldots \lambda_{n-1} \sigma_{n-1}}
u_\nu (\nabla_{\lambda_1} u_{\sigma_1}) (\nabla_{\lambda_2} u_{\sigma_2}) \ldots (\nabla_{\lambda_{n-1}} u_{\sigma_{n-1}})\, , 
\end{split}
\end{equation}
and describes how the AdS Kerr-Newman black hole is dressed by the Chern-Simons contributions.

\subsection{Manifestly covariant differential Noether charge for Chern-Simons terms}\label{ssec:TachiForm}
In Ref.~\cite{Azeyanagi:2014sna}, we constructed a manifestly covariant 
differential Noether charge for the Einstein-Maxwell-Chern-Simons system.  
The charge is split into two contributions
\be\label{eq:deltaQsplit}
\fQNoetherEinMaxCS=\fQNoetherEinMax+\fQNoetherHall\,, 
\ee
where the first term on the right hand side comes from the Einstein-Maxwell part of the Lagrangian
while the second one arises from the Chern-Simons terms. 

The explicit form of the Einstein-Maxwell contribution $\fQNoetherEinMax$ is given by 
\begin{eqnarray} \label{eq:einmaxnoether}
\fQNoetherEinMax 
&=& (\nabla_a \xi^b)\, \delta\left[\frac{\hodge (dx^a\wedge dx_b)}{16\pi G_N} \right] 
+ \delta \fGamma^b{}_a \wedge  \ic_\xi \left[\frac{\hodge (dx^a\wedge dx_b)}{16\pi G_N} \right]  \nonumber \\ 
&&\quad  + 
(\Lambda +  \ic_\xi \fA) \wedge \delta \left[\frac{\hodge \fF}{\gYM^2}\right] 
+ \delta \fA \wedge \ic_\xi \left[\frac{\hodge \fF}{\gYM^2}\right] \, . 
\end{eqnarray}
We will call the first term in each line as the Komar contribution,  while the second term is referred to as the non-Komar part.  The Chern-Simons contribution $\fQNoetherHall$ on the other hand is given by 
\be
(\fQNoether)_{H}\equiv T_0+T_1 +T_2 +T_3 +T_4\, ,  \label{eq:deltaQcssum}
\ee where each term on the right hand side is determined by the derivative of the anomaly polynomial 
$ \fP_{CFT}[\fF, \fR] = d \ICS$ ($\ICS[\fA, \fF, \fGamma, \fR]$ : Chern-Simons terms in the Lagrangian) as
\bea\label{eq:T0toT4sec2}
T_0&\equiv &\delta \fA(\Lambda+ \ic_\xi \fA) \frac{\partial^2 \fP_{CFT}}{\partial \fF\partial \fF} 
+\delta \fGamma^a{}_b (\Lambda+ \ic_\xi \fA) \frac{\partial^2 \fP_{CFT}}{\partial \fR^a{}_b \partial \fF}\, , 
\nonumber\\
T_1 &\equiv& \delta \fGamma^a{}_b \nabla_d \xi^c \frac{\partial^2 \fP_{CFT}}{\partial \fR^a{}_b \partial \fR^c{}_d}\, , \nonumber\\
T_2 &\equiv& \delta \fA \nabla_d \xi^c \frac{\partial^2 \fP_{CFT}}{\partial \fF \partial \fR^c{}_d}\, ,\nonumber\\
T_3 &\equiv & -\half \delta G_{ab} (\SpH)^{abc}\, \ic_\xi (\hodge dx_c)\, , \nonumber\\
T_4 &\equiv & -\frac{1}{4} \xi^a \delta\left\{
\left[(\SpH)_a{}^{bc} + (\SpH)^b{}_a{}^c + (\SpH)^{cb}{}_a\right] \hodge (dx_b \wedge dx_c)
\right\}\, . 
\eea  Here the spin Hall current $(\SpH)^{cb}{}_a$ is defined by 
 $(\SpH)^{cb}{}_a \hodge dx_c\equiv -2( \partial \fP_{CFT}/\partial \fR^a{}_b)$. 
 In this paper, we consider the anomaly polynomials of the form\footnote{
 We use the notation  $c_{_M}$ (as well as $c_{_A}$ and  $c_{_g}$)  to denote the anomaly coefficients,viz., the 
 numerical coefficients in  the anomaly polynomial.} 
\begin{eqnarray}  \label{eq:expressiongeneralanompoly}
\fP_{CFT}=c_{_M}\,\fF^l\wedge \tr[\fR^{2k_1}]\wedge \tr[\fR^{2k_2}]\wedge\ldots \wedge\tr[\fR^{2k_p}]\, , 
\end{eqnarray} 
(here $n=2k_{tot}+l-1$ with $k_{tot}\equiv \sum_{i=1}^{p} k_i$, and $n\ge2$) or, more generally, a linear combination of it.

\subsection{Main results}\label{ssec:summary}
In this subsection, we summarize the main results of this paper :  the Chern-Simons contribution to the 
 differential Noether charge evaluated at the boundary and horizon of the rotating charged AdS black hole solution \eqref{eq:flugravsoln}.

\subsubsection{Anomaly-induced currents}  
Through AdS/CFT correspondence, we can write the differential Noether charge evaluated at the boundary 
in terms of the stress tensor and current of the CFT living on the boundary~:  
\be\label{eq:deltaQandTmnCFT}
 \fQNoetherEinMaxCS \atinfty
 =  -\brk{\xi^\mu|_{\infty} (\delta T^{\text{\tiny{CFT}}}_{\mu\nu})\ 
 +(\Lambda+\ic_{\xi} \fA)|_{\infty}\, (\delta J_\nu^{\text{\tiny{CFT}}})
 } \hodgeCFT dx^\nu\,. 
 \ee 
 Here, we have used the notation $(\ldots)\atinfty$ or simply $(\ldots)_\infty$ to denote a quantity $(\ldots)$ evaluated 
at the boundary $r\to\infty$.  

In the above equation, the differential Noether charge has been evaluated over a diffeomorphism $\xi^a$ 
and $U(1)$ gauge transformation $\Lambda$ which, near AdS boundary, asymptote to 
an arbitrary diffeomorphism and a flavor transformation of the CFT, that is,
at the boundary of AdS,  we will choose $\xi^a$ and $\Lambda$ such that 
\be\label{eq:infsub}
\xi^a {\atinfty} \rightarrow {\cal O}(r^0)\,,\qquad
\Lambda {\atinfty} \rightarrow {\cal O}(r^0)\,.
\ee 
For simplicity,  we choose the vector $\xi^a$ to be a boundary vector satisfying $\xi^r=0$ and $ \partial_r \xi^a=0$ 
(for example, the Killing vectors corresponding to translations and rotations). We will also take $\Lambda$
to be independent of $r$ : $ \partial_r \Lambda=0$. We note that these $\{\xi^a,\Lambda\}$ used for computing the stress tensor and
currents are \emph{state-independent} with $\{\delta\xi^a=0,\delta\Lambda=0\}$.

 Our main interest is the anomaly-induced part of the CFT stress tensor, current or charges.  
These quantities are 
proportional to $\hodgeCFT \fV = \hodgeCFT (V_\mu dx^\mu)= \hodgeCFT (\fu\wedge (2\fomega)^{n-1})$
and thus are at $\omega^{n-1}$ order in the derivative expansion. 
Throughout this paper, we in general add a superscript `anom' to all expressions to denote such types of contribution. For example, 
the anomaly-induced parts of the stress-energy tensor and current 
(we simply call them as anomaly-induced currents) are
 \bea
 T^{\text{\tiny{CFT}}}_{\mu\nu}&=&
 T^{\text{\tiny{anom}}}_{\mu\nu}+\ldots\,, \qquad \quad
 J_\mu^{\text{\tiny{CFT}}}=J_\mu^{\text{\tiny{anom}}}+\ldots\, . 
 \eea  
 To simplify the notation for the differential Noether charge, we will drop the superscript `anom'  
 and denote the anomaly-induced part by $(\fQNoether)$ in the following part of the paper. 

 Furthermore, due to the splitting in Eq.~\eqref{eq:deltaQsplit}, we can also define the two contributions 
 to  the anomaly-part of the stress tensor and current :
  \bea
 T^{\text{\tiny{anom}}}_{\mu\nu}&=&\ T^{\text{\tiny{anom,Ein-Max}}}_{\mu\nu}+\TmnanomCS \,,\qquad 
  J^{\text{\tiny{anom}}}_{\mu}= J^{\text{\tiny{anom,Ein-Max}}}_{\mu}+\JmanomCS\,.
  \label{eq:currentssplit}
 \eea
 The first terms come from the Einstein-Maxwell part of the Lagrangian, while 
 the second ones are from the Chern-Simons terms. 
 In Ref.~\cite{Azeyanagi:2013xea},  
 the anomaly-induced currents are obtained 
 from the gravity side as\footnote{Strictly speaking, Eqs.~\eqref{eq:Tmnanom1} and \eqref{eq:TJanomCS0} are only true for AdS$_{2n+1}$ for $n\ge2$. In AdS$_3$, 
 the anomaly part of the CFT stress tensor is reproduced by the Chern-Simons part of the bulk differential Noether charge only. 
 We refer the readers to Appendix \ref{app:BTZ} for detailed discussions (in particular, see Eqs.~\eqref{eq:TmnCFTCS3D} and \eqref{eq:TmnCFTEM3D}).
 }
\begin{eqnarray}\label{eq:Tmnanom1}
T^{\text{\tiny{anom,Ein-Max}}}_{\mu\nu} &=&
\prn{\mathbb{G}_\mu
 -\Phi\   \frac{\partial \mathbb{G}_\mu }{\partial \Phi}
-\PhiT\  \frac{\partial \mathbb{G}_\mu }{\partial \PhiT} }_{hor} u_\nu 
 +u_\mu \prn{\mathbb{G}_\nu
 -\Phi\   \frac{\partial \mathbb{G}_\nu }{\partial \Phi}
-\PhiT\  \frac{\partial \mathbb{G}_\nu }{\partial \PhiT} }_{hor}  \, , \nonumber \\
J^{\text{\tiny{anom,Ein-Max}}}_{\mu} &=& - \prn{ \frac{\partial \mathbb{G}_\mu }{\partial \Phi} }_{hor} \, . 
\label{eq:TabCFTlogan}
\end{eqnarray}
Here $\mathbb{G}_{\mu}$ is defined by 
$\mathbb{G}_{\mu} = \sum_{\{V\}}  \mathbb{G}^\tV V_\mu$ where $\mathbb{G}^\tV$ 
is obtained from the anomaly polynomial via the bulk replacement rule \cite{Azeyanagi:2013xea} :
\begin{equation}\label{eq:repwB}
\begin{split}
\mathbb{G}^\tV 
\equiv \fP_{CFT}\left[
\fF\rightarrow \Phi;\quad
\tr[ \fR^{2k}] \rightarrow 2\PhiT^{2k}\right].\
\end{split}
\end{equation}
We note that at the horizon, this reduces to the boundary CFT replacement rule in Eq.~(\ref{eq:Tmnanom0}) by recalling $\Phi(r=r_H) = \mu$ and $\Phi_T(r=r_H) = 2\pi T$.  
As was done in Ref.~\cite{Azeyanagi:2013xea}, these results \eqref{eq:TabCFTlogan} match with the ones
derived from the CFT Gibbs current by following the discussions in Refs.~\cite{Loganayagam:2011mu,Loganayagam:2012pz,Banerjee:2012cr} (see \S\S\ref{ssec:RepRule} above), under the assumption
\bea
 \TmnanomCS=0\, , \qquad \JmanomCS=0\,.  \label{eq:TJanomCS0}
\eea

In this paper, we have confirmed the correctness of this assumption \eqref{eq:TJanomCS0} 
by directly computing these quantities from the differential Noether charge on the gravity side on the rotating charged AdS black hole background.

\subsubsection{Entropy current }  
In order to evaluate the entropy current of our solution, we proceed as follows \cite{Wald:1993nt,Iyer:1994ys}.
We consider the differential Noether charge associated with  a specific \emph{state-dependent}  diffeomorphism
and $U(1)$ transformation on our black hole solution. As before, we choose the diffeomorphism
$\xi^a$ to be a boundary vector satisfying $\xi^r=0$ and $ \partial_r \xi^a=0$ and the $U(1)$ transformation
$\Lambda$ to be independent of $r$ : $ \partial_r \Lambda=0$. However, the boundary components are 
now chosen to depend on the particular fluid state under question : we take 
\be\label{eq:horsub}
\xi^\mu = \xiH^\mu \equiv u^\mu/ T  +{\cal O}(\omega^2)\,,\quad
\Lambda =\LambdaH \equiv -\ic_{\xi} \fA |_{r=r_H}=\mu/T +{\cal O}(\omega^2)\,.
\ee 
For convenience, we will use the notation  $(\ldots)\athor$ or simply $(\ldots)_{hor}$  to denote a quantity $(\ldots)$ evaluated 
at the horizon with the substitutions given in Eq.~\eqref{eq:horsub} and then pulled back to the boundary using ingoing
null geodesics \cite{Bhattacharyya:2008xc} .

As discussed in \cite{Wald:1993nt,Iyer:1994ys}, 
the differential Noether charge associated with \eqref{eq:horsub} evaluated on the horizon corresponds 
to the entropy of the black hole solution.  We can write this entropy 
by introducing an entropy current as 
$\prn{J_S^{\text{\tiny{CFT,anom}}}}_\mu$ as 
\be
 \fQNoetherEinMaxCS \athor=\delta\prn{J_S^{\text{\tiny{CFT,anom}}}}_\mu\ \hodgeCFT dx^\mu\, . 
\ee 
This should be thought of as the local version of the formalism developed in \cite{Wald:1993nt,Iyer:1994ys}
using the pull-back prescription of  \cite{Bhattacharyya:2008xc}.

Here, as in \eqref{eq:currentssplit}, we can split the entropy current into the contribution 
coming from the Einstein-Maxwell part of the Lagrangian and the one from the Chern-Simons terms : 
\be \label{eq:entropycurrentsplit}
({J_S^{\text{\tiny{CFT,anom}}}})_{\mu}
=({J_S^{\text{\tiny{anom,Ein-Max}}}})_{\mu}
+ (\JSanomCS)_\mu\,, 
\ee 
where the first term vanishes as we have shown in Appendix \ref{app:entropyEM}
\be
({J_S^{\text{\tiny{anom,Ein-Max}}}})_{\mu}=0\,.
\ee 
For later convenience, it is also useful to define $(\JSanomCSV)_l$ by expanding 
the entropy current with respect to $U(1)$ chemical potential :
\be \label{eq:entropycurrentexpansion}
\JSanomCSV\equiv \sum_{l} (\JSanomCSV)_l \, (\mu^l)\, .
\ee

In this paper, we derive the following expression for the entropy current
from the gravity side by computing the entropy of the rotating charged AdS black hole \eqref{eq:flugravsoln} : 
  \be
 (\JSanomCS)_\mu =    \JSanomCSV V_\mu\,
\qquad{\rm with} \qquad
\JSanomCSV\equiv -2\pi  \left(\frac{\partial \GV}{\partial \Phi_T} \right)_{hor}\,.
\label{eq:defJSanomCSV}
\ee 
This result is consistent with the CFT prescription for the anomaly-induced entropy current given in Eq.~(\ref{eq:Sentanom0}).

\subsubsection{Examples}
For the reader's convenience, in Table~\ref{table:results1}  we present the explicit expressions of $\GV$ and 
the anomaly-induced entropy current $\JSanomCSV$ for various anomaly polynomials in AdS$_3$, AdS$_5$ and AdS$_7$.
The results for the more general cases are given in the next section and in the Appendices.
\begin{table}[ht]
\centering 
\begin{tabular}{| c | c | c | c| c|} 
\hline 
$n$ &  $({\text V}) $& $\fP_{CFT} $ &$\GV$   &$\JSanomCSV$ 
\\ [0.5ex]
\hline
 1 & $\fu(2\fomega)^0$ & $ c_{_A} \fF^2$ &$c_{_A} \Phi^2$&0\\   \cline{3-5}
   & &  $c_{_g} \tr[\fR^2]$ &$ c_{_g} (2\PhiT^2)$&
 $-2\pi c_{_g}\times 2 \times 2 \times (2\pi T)  $
  \\  \hline
 2 & $\fu(2\fomega)^1$ & $ c_{_A} \fF^3$ &$c_{_A} \Phi^3$&0\\   \cline{3-5}
   & &  $c_{_M} \fF\wedge \tr[\fR^2]$ &$ c_{_M}\Phi (2\PhiT^2)$&
 $-2\pi c_{_M}\times 2 \times 2 \times (2\pi T)\times \mu $
  \\  \hline
3 & $\fu(2\fomega)^2$ &  $c_{_A} \fF^4$& $c_{_A} \Phi^4$ &0 \\
    \cline{3-5}
 &  &  $c_{_M} \fF^2\wedge \tr[\fR^2]$& $  c_{_M} \Phi^2 (2\PhiT^2)$
 &$-2\pi c_{_M} \times 2 \times 2 \times  (2\pi T) \times  \mu^2 $ \\
     \cline{3-5}
 &  &  $c_{_g}  \tr[\fR^2]\wedge \tr[\fR^2]$& $c_{_g} (2\PhiT^2)^2$ &$-2\pi c_{_g} \times 2^2 \times 4 \times (2\pi T)^3 $ \\
     \cline{3-5}
 &  &  $c_{_g}  \tr[\fR^4]$& $c_{_g} (2\PhiT^4)$ &$-2\pi c_{_g}\times 2\times 4 \times (2\pi T)^3 $ \\
 \hline 
\end{tabular}
\caption{$\GV$ and $\JSanomCSV$ for AdS$_{3}$, AdS$_{5}$ and AdS$_7$.}
\label{table:results1}
\end{table}


\section{CFT replacement rule for entropy current from gravity}
\label{sec:RepEntropy}
In this section, we will show that the CFT entropy current in Eq.~(\ref{eq:defJSanomCSV}) is reproduced by the black hole entropy coming from the Chern-Simons terms.
The parity odd part of the Einstein-Maxwell contribution to the entropy turns out to be zero. The details of the computations are rather standard and straightforward 
and thus will be presented in Appendix \ref{app:entropyEM} instead. This section will be devoted to some explicit computations of the Chern-Simons contribution to the entropy. 

As discussed in the previous section (as well as Introduction), our main derivation here is  based on the differential Noether charge at the horizon. 
Given the length of this computation, however, let us begin  instead with a heuristic derivation inspired by Tachikawa entropy formula \eqref{eq:WaldTachikawa0}.
The aim here is to make various assumptions that would directly get us to the heart of how the replacement rule appears in holography. In the next subsection, 
we will present the results from evaluating the differential Noether charge \eqref{eq:deltaQcssum}
on the rotating charged AdS black hole background \eqref{eq:flugravsoln}.

\subsection{Entropy I :  Tachikawa-like formula at arbitrary horizon slice}\label{ssec:TachiLike}
 Let us begin by recalling from Ref.~\cite{Azeyanagi:2014sna} that for stationary solutions, at the bifurcation surface, the Chern-Simons terms contribute to black hole entropy through the Tachikawa entropy formula (that is, the second term of \eqref{eq:WaldTachikawa0})
\begin{equation}
\label{eq:TachikawaFormula}
 S_{\text{Tachikawa}} = \int_{Bif}  \sum_{k=1}^\infty 8\pi k\ \fGamma_N \fR_N^{2k-2} \frac{\partial \fP_{CFT}}{\partial\ \text{tr} \fR^{2k} } \, , \\
\end{equation} for a general anomaly polynomial $\fP_{CFT}$.
We notice that the rotating charged AdS black hole solution that we constructed in \cite{Azeyanagi:2013xea} is obtained in the ingoing Eddington-Finkelstein coordinates. This makes the evaluation on the bifurcation surface difficult (since it is at the boundary of such coordinates). 

As we described in Introduction, in the usual Wald formula (that is, the first term of \eqref{eq:WaldTachikawa0}), these difficulties can be
tackled using the arguments of Jacobson-Kang-Myers (JKM)~\cite{Jacobson:1993vj}. This JKM type argument ensures that, for stationary solutions,
the first term of \eqref{eq:WaldTachikawa0} evaluated on an arbitrary horizon slice gives the same answer as when evaluate over the 
bifurcation surface. In contrast, it is unclear how to convert the Tachikawa formula \eqref{eq:TachikawaFormula} into an expression that 
can be evaluated over an arbitrary horizon slice. It is especially unclear how to interpret objects like $ \fGamma_N$ over an arbitrary slice.
This is an indication of broader slice-dependence  issues when Wald-like formulae for time-dependent solutions are considered.\footnote{We
would like to thank Sayantani Bhattacharyya, Shiraz Minwalla and Mukund Rangamani for enlightening discussions regarding related issues.}
 
We will not solve this important issue here. However,  we will now make a simple proposal which seems to give the right answers which are consistent 
with the direct computation based on the differential Noether charge and also  the expectations from the CFT side. 
Let us define  
\be\label{eq:GammaNHor}
\tilde{\fGamma}_N \equiv \frac{1}{4\pi} \nabla_b \xi^a \fGamma^b{}_a \athor\, ,\quad
\tilde{\fR}_N \equiv d \tilde{\fGamma}_N 
  \,.
\ee
We will then assume that with these definitions, the entropy is given by a formula similar to \eqref{eq:TachikawaFormula}  where 
$\fGamma_N$ is replaced by $\tilde{\fGamma}_N$ and  $\fR_N$ is replaced by $\tilde{\fR}_N$.
Thus, our goal now is to explicitly evaluate 
\begin{equation}
\label{eq:TachikawaFormulaHor}
 \tilde{S}_{\text{Tachikawa}} = \int_{\text{arbitrary slice}}  \sum_{k=1}^\infty 8\pi k\ \tilde{\fGamma}_N \tilde{\fR}_N^{2k-2} \frac{\partial \fP_{CFT}}{\partial\ \text{tr} \fR^{2k} }\, 
\end{equation} 
to obtain the replacement rule for anomaly-induced entropy current. The integrand can then be taken (after pull-back) to be the CFT entropy-current form.

Using Eqs.~\eqref{eq:trgradxiGamma}, \eqref{eq:Fhor} and \eqref{eq:horgradxiR}, we find that $\tilde{\fGamma}_N$ starts at order $\omega^0$, 
while $\fF \athor$ and $\tilde{\fR}_N$ start at $\omega^1$. In particular, the leading order expressions are given by
\be
\fF \athor= \mu (2 \fomega )+\ldots,\quad
\tilde{\fGamma}_N  = -(2\pi T)\fu+\ldots\, ,\quad
\tilde{\fR}_N =  -(2\pi T) (2\fomega)+\ldots\, .
\ee  Finally, using  Eq.~\eqref{eq:trR2k}, we obtain the leading contribution to $ \tilde{S}_{\text{Tachikawa}}$ for the rotating charged AdS black hole (which is at order $\omega^{n-1}$) :
\begin{equation}
\label{eq:TachikawaFormulaHorExplicitly}
 \tilde{S}_{\text{Tachikawa}} =\int_{\text{arbitrary slice}}   
 \prn{J_S^{\text{\tiny{CFT,anom}}}}_\mu\ \hodgeCFT dx^\mu \, , 
 \\
\end{equation} 
where
\be
  \prn{J_S^{\text{\tiny{CFT,anom}}}}_\mu =
   -2\pi  \left(\frac{\partial \GV}{\partial \Phi_T} \right)_{hor} V_\mu\, , 
\ee
for any anomaly polynomial $\fP_{CFT}$ in AdS$_{2n+1}$.  Remarkably, this reproduces the CFT replacement rule for the anomaly-induced entropy current, agreeing with the result based on the differential Noether charge without any assumptions in the next subsection. 
This crude computation inspired by the Tachikawa entropy formula simplifies the treatment of 2nd and higher order terms drastically. 
That is, since the above derivation uses each building block at its lowest order, obviously if any of them is at 2nd or higher order, the contribution to the entropy will be higher than $\omega^{n-1}$. This computation also shows that the replacement rule follows from the general structure of black hole solutions 
and indicates how it might be a robust statement holding beyond the simple model under consideration.

\subsection{Entropy II : Evaluation of differential Noether charge at horizon}
We will now turn to a more honest (but more lengthy) derivation of the replacement rule via the differential Noether charge with no ad-hoc assumptions. 
To derive the CFT replacement rule for the anomaly-induced entropy current, we start with the manifestly covariant differential 
Noether charge summarized in Eqs.~\eqref{eq:deltaQcssum} and  \eqref{eq:T0toT4sec2}. 
The goal of this subsection is to briefly explain the evaluation of $T_0, T_1, T_2, T_3$ and $T_4$ terms in \eqref{eq:T0toT4sec2} 
on the horizon of the rotating charged AdS black hole solution \eqref{eq:flugravsoln}. 
We consider the general anomaly polynomial of the form \eqref{eq:expressiongeneralanompoly}.
The detail of the computation is provided in
the Appendices. For AdS$_3$ and AdS$_5$, it is given in Appendices \ref{app:BTZ} and \ref{sec:deltaQcsads5eq0}.
For the general cases, the evaluation of $T_3$ and $T_4$ (for an arbitrary fixed $r$) is given in Appendix \ref{app:replacementT3T4}, while  
the rest of the terms, $T_0$, $T_1$ and $T_2$ on the horizon are calculated in Appendices \ref{sec:T0T1T2hor0th1st} and \ref{sec:entropyhighercontribution}.

Here is one remark : In the evaluation of the differential Noether charge at the horizon, there are two potential prescriptions depending on whether one does the variation or the evaluation first.
In the first prescription, one first sets the radial coordinate to be $r=r_H$ and then does the variation with respect to the parameters of the black hole solution, 
while in the second prescription one first does the variation and set $r=r_H$ afterward. In Appendix \ref{app:BTZ}, for the rotating BTZ black holes, we have explicitly evaluated the Chern-Simons contribution to the differential Noether charge at the horizon  by using these two prescriptions and then obtained the same result. 
Furthermore, in the case of entropy coming from the Einstein-Maxwell terms, we have also explicitly checked in Appendix \ref{app:entropyEM} that both prescriptions give the same answer.
Although the distinction has not often been discussed in the literature (even in the case of a covariant Lagrangian), it is not known to us if these two prescriptions should also yield the same answer.\footnote{Essentially, the physical difference of these two prescriptions for the evaluation at the horizon traces back to whether one evaluates at the horizon (before the variation) or the new horizon (after the variation). 
An argument as to why both prescriptions gives the same answer in our computations is that if we include the {\it total} contribution to the entropy (including all the terms appearing in the Lagrangian), then on-shell $\partial_r(\fQNoether)=0$ because of $d(\fQNoether)=0$.  This implies that $\fQNoether$ does not depend on $r$ and thus it is the same whether it is evaluated on the original or the new horizon (after the variation).} We take the case of BTZ black hole entropy as well as the Einstein-Maxwell part of the entropy as a hint that it is reasonable to expect that these prescriptions agree in general, and for the rest of this paper we will use the first prescription (due to simplifications in the computations). 

Another important comment is that in the evaluation of the differential Noether charge \eqref{eq:deltaQcssum} 
with $T_0$, $T_1$, $T_2$, $T_3$ and $T_4$ in \eqref{eq:T0toT4sec2}, since we evaluate the charge at a given fixed $r$-surface, 
the terms proportional to $dr$ do not contribute. 
Therefore, throughout this paper, we neglect these terms unless otherwise mentioned 
(i.e. `=' in the evaluation of the differential Noether charge and $T_i$'s is valid up to these terms). 
In addition to this, we also set $\delta r=0$.

\subsubsection{Terms $T_3$ and $T_4$ (from Appendix~\ref{app:replacementT3T4})}
\label{sec:T3T4horizon}
As a result of the detailed computation in Appendix~\ref{app:replacementT3T4}, 
the terms $T_3$ and $T_4$ for a general anomaly polynomial \eqref{eq:expressiongeneralanompoly}
at an arbitrary fixed $r$ surface of the black hole solution \eqref{eq:flugravsoln} are given by 
\bea \label{eq:appendixT3T4summary}
T_3 
&=&
\frac{1}{2r} 
(2\Psi- r^2 )
 \frac{d}{dr}\left[
r \frac{\partial \GV}{\partial \Phi_T} 
\right] \ic_{\xi}\left[
(\delta \fu)\wedge \fu \wedge (2\fomega)^{n-1}\right]\, , \\
T_4 &= & 
\xi^\mu \delta\Biggl\{
-\frac{r^2}{2} 
\frac{d}{dr}\left[
 \frac{\partial \GV}{\partial \Phi_T} 
\right]  V_\mu \hodgeCFT \fu
-  u_\mu \frac{\Psi}{r^2}\frac{d}{dr}\left[r^2
 \frac{\partial \GV}{\partial \Phi_T} 
\right]\fu\wedge(2\fomega)^{n-1}
\nonumber\\
&&\qquad\qquad
+4r^{-2}\Psi 
\HVtemp
\left[
r^2V_\mu \hodgeCFT\fu
-2 \Psi u_\mu  \fu\wedge (2\fomega)^{n-1}
\right]  \Biggr\}\, ,
\eea 
where $\GV$ and $\HVtemp$ are determined by the corresponding anomaly polynomial 
$\fP_{CFT}[\fF, \fR]$ via the bulk replacement rule
\bea \label{eq:gvhv}
 \GV&\equiv &\fP_{CFT}\left[
 \fF \rightarrow \Phi; \quad
\tr[\fR^{2k}] \rightarrow 2\Phi_T^{2k}
 \right]\, ,  \nonumber\\
  \HVtemp&\equiv & 
   \frac{\partial  \fP_{CFT}}{\partial( \tr[\fR^2])}   [\fF \rightarrow \Phi
 ; \quad
\tr[\fR^{2k}] \rightarrow 2\Phi_T^{2k}]\, .
  \eea 

By evaluating these expression for $T_3$ and $T_4$ at the horizon $r=r_H$, we obtain 
\begin{eqnarray}
 (T_3 + T_4) \athor &=& \frac{\rH}{4\pi T}\,  J_S^{\tV} \delta \fu \wedge (2\fomega)^{n-1}
 \qquad{\rm with} \qquad
\JSanomCSV\equiv -2\pi  \left(\frac{\partial \GV}{\partial \Phi_T} \right)_{hor}\,
 \nonumber\, . 
\end{eqnarray}

We note that the above result can be obtained by assuming that 2nd and higher order terms 
in the  building blocks (that is, $\fR$, $\fF$ etc.) do not contribute to $T_3$ and $T_4$ at the leading order of the 
derivative expansion. We can directly show that the 2nd and higher order terms does not 
generate the same or lower order contribution by using essentially the 
same argument for the Einstein source in Appendix D.6 of Ref.~\cite{Azeyanagi:2013xea}.
For the readers' convenience, this argument is briefly reviewed in Appendix~\ref{sec:2ndorderT3andT4}. 
For more details of the computations and argument related to $T_3$ and $T_4$, please refer 
to Appendix~\ref{app:replacementT3T4}.

\subsubsection{Terms $T_0$, $T_1$, $T_2$ at horizon (from Appendices~\ref{sec:T0T1T2hor0th1st} and \ref{sec:entropyhighercontribution})}
In Appendix \ref{sec:T0T1T2hor0th1st}, we have carried out the evaluation of $T_0, T_1$ and $T_2$ at the horizon, 
by taking into account the zeroth and 1st order terms in the building blocks (that is, $\fR$, $\fF$ etc.) only. 
The result for each term at the leading order of the derivative expansion (which turns out to be of order $\omega^{n-1}$) is given by 
\begin{equation}\begin{split}
T_0 \athor& =0\, , \\
T_1 \athor& =  \sum_l \delta \brk{ (J_S^\tV)_l \fu \wedge (2\fomega)^{n-l-1} } \wedge (2\fomega \mu)^l - \rH /(4\pi T)  J_S^\tV \delta \fu \wedge (2\fomega)^{n-1}\, , \\
T_2 \athor& =   \sum_l (J_S^\tV)_l \fu \wedge (2\fomega)^{n-l-1} \wedge \delta \brk{ (2\fomega \mu)^l }\, , 
\end{split}\end{equation}
where we have used $(J_S^\tV)_l$ defined as in Eq.~\eqref{eq:entropycurrentexpansion}. 
As in the case of $T_3$ and $T_4$, we can obtain this result by assuming that 2nd and higher order terms 
in the  building blocks do not contribute. 
In Appendix \ref{sec:entropyhighercontribution}, we have also taken into account the 2nd and higher order terms in the building blocks 
and confirmed that these terms do not generate any contribution to $T_0, T_1$ and $ T_2$ at the $\omega^{n-1}$ order or lower.

\subsubsection{Anomaly-induced contribution to entropy} 
As summarized in Appendix \ref{app:entropyEM}, the Einstein-Maxwell part of the differential Noether charge gives no 
parity odd contribution to black hole entropy. Combined this fact with the results on the Chern-Simons part summarized 
in the above two subsections, we finally obtain
the parity odd contribution to the black hole entropy in the leading order of the derivative expansion as 
\begin{equation}\begin{split}
 \fQNoetherEinMaxCS \athor=\delta\prn{J_S^{\text{\tiny{CFT,anom}}}}_\mu\ \hodgeCFT dx^\mu\,  
 \qquad{\rm with} \qquad
\JSanomCSV\equiv -2\pi  \left(\frac{\partial \GV}{\partial \Phi_T} \right)_{hor}\,. 
\end{split}\end{equation}
This indeed reproduces the CFT replacement rule for the anomaly-induced entropy current.


\section{CFT current and stress tensor from differential Noether charge}
\label{sec:RepTJ}

In this section, we will explicitly compute the differential Noether charge evaluated at the boundary 
$ \fQNoetherEinMaxCS \atinfty$ of the rotating charged AdS black hole background \eqref{eq:flugravsoln} in the leading order of the fluid/gravity derivative expansion.
We recall that the differential Noether charge (evaluated at the boundary)
splits into the Einstein-Maxwell part and the Chern-Simons part 
\be
\fQNoetherEinMaxCS\atinfty=\fQNoetherEinMax\atinfty+\fQNoetherHall\atinfty\,. 
\label{eq:decomposeNoetheratinfty}
\ee 
In the following part of this section, we will briefly explain the evaluation 
of these two terms on the right hand side separately, while the details are 
provided in the Appendices. 

In \S\S\ref{subsec:asympchargeEM},  
we will confirm Eqs.~\eqref{eq:Tmnanom1} and \eqref{eq:repwB} 
from the evaluation of the first term $\fQNoetherEinMax\atinfty$. 
We note that the Einstein-Maxwell terms of the Lagrangian are parity even
and thus, in the evaluation of the anomaly-induced contribution \eqref{eq:Tmnanom1} 
from the Einstein-Maxwell part at the leading order of the derivative expansion, 
we drop all parity even contributions
with derivatives along the boundary coordinates in \S\S\ref{subsec:asympchargeEM}. 
In \S\S\ref{sec:38summary}, we provide a summary for the evaluation of 
the second term in Eq.~\eqref{eq:decomposeNoetheratinfty}, which results in 
\be
\fQNoetherHall\atinfty=0\, , 
\ee
and thus justifies the assumptions of Eq.~\eqref{eq:TJanomCS0} (which are used in Ref.~\cite{Azeyanagi:2013xea}). 
Since the detail of the evaluation of this part involves various technical points and is lengthy, interested readers are kindly referred to Appendix \ref{sec:roughrefinedestimate} and \ref{sec:chargeshighercontribution}.

\subsection{Asymptotic charges I : Einstein-Maxwell contribution}\label{subsec:asympchargeEM}

In this subsection, we compute the Einstein-Maxwell contribution to the differential Noether charge,  
Eq.~\eqref{eq:einmaxnoether}, at the boundary of the rotating charged AdS black hole background 
\eqref{eq:flugravsoln}. As in Eq.~\eqref{eq:einmaxnoether}, we separate this contribution into the Komar part and the non-Komar part, 
and evaluate them separately.
We recall that the Komar part (from the first line in Eq.~\eqref{eq:einmaxnoether}) is given by 
\be \label{eq:komarEM}
(\fKomar)_{_\text{Ein-Max}} 
\equiv\nabla_a\xi^b \frac{\hodge (dx^a\wedge dx_b)}{16\pi\GN} +(\Lambda+\ic_{\xi} \fA) \cdot \frac{\hodge \fF}{\gYM^2}
\, , \\
\ee and the non-Komar part (the second line in Eq.~\eqref{eq:einmaxnoether}) is  
\be \label{eq:nonkomarEM}
 \delta \fGamma^b{}_a \wedge  \ic_\xi \left[\frac{\hodge (dx^a\wedge dx_b)}{16\pi G_N} \right]  + \delta \fA \wedge \ic_\xi \left[\frac{\hodge \fF}{\gYM^2}\right] \,.
\ee

\subsubsection{Komar charge}
Straightforward computation with the help of Eqs.~\eqref{eq:hodgeFconstr} and \eqref{eq:gradxidxdx} 
gives the following expression for  the Komar charge for the Einstein-Maxwell part on the rotating charged AdS black hole 
background \eqref{eq:flugravsoln} :\footnote{We refer the readers to Ref.~\cite{Azeyanagi:2013xea} 
for the definitions and details of the functions $\mathcal{Q}^\tV(r)$ and $ \mathcal{M}^\tV (r)$ we use here. We note that they satisfy Eq.~(\ref{eq:MQCFTloganCSentropy}) at infinity.
}
 
\begin{equation} \label{eq:komarEMexpanded}
\begin{split} 
(\fKomar)_{_\text{Ein-Max}} 
&= \xi^\mu  \frac{r^{d-1}}{16\pi\GN}
\brk{2rf \eta_{\mu\nu} -r^2\frac{df}{dr} u_\mu u_\nu}\ \hodgeCFT dx^\nu +\ldots\\
&\qquad\quad  +\xi^\mu 
\brk{\frac{r^{d+1}}{16\pi\GN} \frac{d}{dr}\prn{\frac{g_{_V}}{r^2}}V_\mu u_\nu
+ \frac{r^{d+1}f^2}{16\pi\GN}\frac{d}{dr}\prn{\frac{g_{_V}}{r^2f}}u_\mu V_\nu }
\ \hodgeCFT dx^\nu +\ldots\, \\
&\quad 
-\frac{1}{\gYM^2}(\Lambda+\ic_{\xi}\fA)
  \brk{(d-2)q\ u_\mu  } \  \hodgeCFT dx^\mu+\ldots \, \\
&\quad\qquad
-\frac{1}{\gYM}(\Lambda+\ic_{\xi}\fA)
  \brk{ \gYM^2 \mathcal{Q}^\tV\  V_\mu  } \  \hodgeCFT dx^\mu+\ldots \, .
\end{split}
\end{equation}
We note that the first and third lines respectively are the parity even contribution
from the first and second terms of \eqref{eq:komarEM} in the leading order of the derivative expansion. 
On the other hand, the second and fourth lines in Eq.~\eqref{eq:komarEMexpanded} respectively
are the parity odd contribution from the first and second terms of \eqref{eq:komarEM} 
in the leading order of the derivative expansion. 

Now we evaluate this Komar part at the boundary, $r\to \infty$. We look at the contribution from the 
Einstein part (the first and second lines in Eq.~\eqref{eq:komarEMexpanded}) 
and Maxwell part (the third and fourth lines in Eq.~\eqref{eq:komarEMexpanded} separately. 
We first take the limit $r\to \infty$ of the Einstein part after subtracting the empty AdS contribution 
(the second term on the left hand side in the following expression) :
\begin{equation}\label{eq:KomarInf}
\begin{split}
&\brk{ \nabla_a\xi^b \frac{\hodge (dx^a\wedge dx_b)}{16\pi\GN}
- \xi^\mu  \frac{r^d \eta_{\mu\nu}}{8\pi \GN} \hodgeCFT dx^\nu } \Biggl |_{\infty}\\
&\quad =-\xi^\mu\bigbr{\frac{m}{16\pi\GN} \prn{2\eta_{\mu\nu} + d\ u_\mu u_\nu } 
+  \mathcal{M}^\tV (r=\infty)\ \prn{ V_\mu u_\nu+\ u_\mu V_\nu }\ } 
\ \hodgeCFT dx^\nu \, , 
\end{split}
\end{equation}
where we have used 
\begin{equation}
\begin{split}
\brk{\frac{r^{d+1}}{16\pi\GN} \frac{d}{dr} \prn{\frac{g_{_V}}{r^2}} } \Biggl |_{\infty}=
\brk{\frac{r^{d+1}f^2}{16\pi\GN} \frac{d}{dr} \prn{\frac{g_{_V}}{r^2f}} } \Bigg|_{\infty}
=- \mathcal{M}^\tV (r=\infty) \, . 
\end{split}
\end{equation}
On the other hand the Maxwell part of the Komar charge is evaluated at the boundary as 
\begin{equation}\label{eq:MaxwellKomarCharge}
\left. \left[\prn{\Lambda+\ic_\xi \fA}  \cdot  \frac{\hodge \fF}{\gYM^2} 
\right] \right|_{\infty}
=  -\prn{\Lambda+\ic_\xi \fA}_{\infty}\brk{\frac{(d-2)q}{\gYM^2}\ u_\mu + \ldots 
+ \mathcal{Q}^\tV(r=\infty)\  V_\mu +\ldots } \  \hodgeCFT dx^\mu\, .
\end{equation}
The two $\ldots$ in the above expression denote higher derivative contribution 
to the parity even and parity odd part, respectively.

\subsubsection{Non-Komar variation}
We next proceed to the evaluation of the non-Komar part \eqref{eq:nonkomarEM}.
We separately 
evaluate the Einstein part and Maxwell part (the first and the second term in \eqref{eq:nonkomarEM}, respectively). 

First, we start with the evaluation of the Einstein part. We note that, by adding Eqs.~\eqref{eq:non-Komaruse1}-\eqref{eq:non-Komaruse4}, we obtain
\begin{equation}\label{eq:NonKomar}
\begin{split}
\delta&\fGamma^b{}_a\wedge \frac{\hodge(dx^a \wedge dx_b)}{16\pi\GN}\\
&= \bigbr{\frac{1}{16\pi\GN}\frac{d}{dr}(r^{d+1}\delta f)
+\brk{\frac{r^{d+1}f^2}{16\pi\GN}  \frac{d}{dr}\prn{\frac{g_{_V}}{r^2f}}
-\frac{r^{d+1}}{16\pi\GN} \frac{d}{dr}\prn{\frac{g_{_V}}{r^2}} }
V^\mu\delta u_\mu }
\ \hodgeCFT 1 \\
&\quad +\bigbr{\ldots}_\mu\ dr\wedge\hodgeCFT dx^\mu \, . 
\end{split}
\end{equation}
We pull-back this on a radial slice, contract it with $\xi^\mu$ using 
$ \ic_\xi \hodgeCFT 1 = \eta_{\mu\nu} \xi^\mu\ \hodgeCFT dx^\nu$ and then take $r\to\infty$ limit. In the end, we have the following expression : 
\begin{equation}
\label{eq:nonKomarEH}
\ic_\xi\brk{\delta\fGamma^b{}_a\wedge \frac{\hodge(dx^a \wedge dx_b)}{16\pi\GN}} \Bigg|_{\infty}
= - \delta\brk{\xi^\mu \frac{ m}{16\pi\GN}\eta_{\mu\nu} \hodgeCFT dx^\nu } \, . 
\end{equation}

As a next step we evaluate the non-Komar part of the Maxwell contribution at the boundary. 
We note that this term evaluated for the empty AdS is trivially zero. Near the boundary, since
$\delta\fA |_{\infty} \rightarrow \delta \fA_{\infty} +{\cal O}\left(r^{-(d-2)}\right)$ and the gauge field fall-offs as in Eq.~\eqref{eq:usefuluseful1}, we conclude that 
\begin{equation}\label{eq:nonKomarMW}
\left.\left[
-\ic_\xi   \left( \delta \fA\cdot \frac{\hodge \fF}{\gYM^2}\right)
\right]\right|_{\infty} \rightarrow \,\,0\,. 
\end{equation} 
Therefore the non-Komar part of the Maxwell contribution vanishes at the boundary.

\subsubsection{Einstein-Maxwell contribution to asymptotic charges}
Now we combine all the results above to compute the Einstein-Maxwell part of 
the differential Noether charge \eqref{eq:einmaxnoether} at the boundary. 
Subtracting all the non-Komar contributions (see Eq.~\eqref{eq:nonKomarEH} and Eq.~\eqref{eq:nonKomarMW}) from the variation of Komar contribution in Eq.~\eqref{eq:KomarInf} and Eq.~\eqref{eq:MaxwellKomarCharge}, we finally 
obtain 
\begin{equation}
\begin{split}
\fQNoetherEinMax\atinfty
&= -\delta\brk{ T^{\text{\tiny{CFT}}}_{\mu\nu} \xi^\mu+\prn{\Lambda+\ic_\xi \fA}_{\infty} J^{\text{\tiny{CFT}}}_{\nu}}\ \hodgeCFT dx^\nu\,  , \\
\end{split}
\end{equation}
with
\begin{equation}
\begin{split}\label{eq:CFTfluidST}
&T^{\text{\tiny{CFT}}}_{\mu\nu} = \frac{m}{16\pi\GN} \prn{\eta_{\mu\nu} + d\ u_\mu u_\nu } +\ldots\\
&\qquad\qquad\qquad +  \prn{\mathbb{G}_\mu
 -\Phi\   \frac{\partial \mathbb{G}_\mu }{\partial \Phi}
-\PhiT\  \frac{\partial \mathbb{G}_\mu }{\partial \PhiT} }_{hor} \ \prn{ V_\mu u_\nu+\ u_\mu V_\nu } +\ldots \, , \\
&
J^{\text{\tiny{CFT}}}_{\mu} =
\frac{(d-2)q}{\gYM^2}\ u_\mu + \ldots  \\
&\qquad\qquad\qquad
 - \prn{ \frac{\partial \mathbb{G}_\mu }{\partial \Phi} }_{hor}\  V_\mu +\ldots   \,. 
\end{split}
\end{equation}
where we have used
\bea
{\mathcal M}^\tV(r=\infty) &=& \prn{\mathbb{G}^\tV
 -\Phi\   \frac{\partial \mathbb{G}^\tV }{\partial \Phi}
-\PhiT\  \frac{\partial \mathbb{G}^\tV }{\partial \PhiT} }_{r=\rH}\, ,  \nonumber\\
{\mathcal Q}^\tV (r=\infty)&=&- \prn{ \frac{\partial \mathbb{G}^\tV }{\partial \Phi} }_{r=\rH} \, ,
\label{eq:MQCFTloganCSentropy}
\eea
which are shown in Eq.~(5.9) of \cite{Azeyanagi:2013xea}.
{Here again the $\ldots$ in the first and third (second and fourth) lines of Eq.~(\ref{eq:CFTfluidST}) denote the higher order terms in the parity even (odd) contribution.}
The expressions in the first lines (i.e. non-$V_\mu$ parts) in $T^{\text{\tiny{CFT}}}_{\mu\nu}$ and $J^{\text{\tiny{CFT}}}_{\mu} $ above are exactly the perfect-fluid ones with pressure $p=m/(16\pi \GN)$. The anomaly-induced parts (i.e. terms proportional to $V_\mu$) match Eq.~\eqref{eq:Tmnanom1}  exactly  as claimed.

Of course, when the Lagrangian density contains the  Chern-Simons terms,  
there is potential extra contribution to the CFT stress tensor and current 
from the Hall part to the differential Noether charge $(\fQNoether)_{H}$ at the boundary. 
The goal of the next subsection is to confirm that there is no such contribution, that is,  
to prove Eq.~\eqref{eq:TJanomCS0}.


\subsection{Asymptotic charges II : Chern-Simons contribution}
\label{sec:38summary}
This subsection is devoted to a summary of the results in the evaluation of
$(\fQNoether)_{H}$  with the anomaly polynomial  \eqref{eq:expressiongeneralanompoly} 
at the boundary of the rotating charged AdS black hole background. 
The detail of the computation contains many technical points
and thus is provided in 
Appendix \ref{app:replacementT3T4}-\ref{sec:chargeshighercontribution}. 
In the following, we summarize the key results we obtained for the general anomaly 
polynomial \eqref{eq:expressiongeneralanompoly} in AdS$_{2n+1}$ ($n\ge1$).
As in Appendix~\ref{app:replacementT3T4}, 
for $T_3$ and $T_4$ in \eqref{eq:T0toT4sec2}, we can obtain their exact expression valid for any fixed $r$ 
in a relatively simple way. On the other hand, for $T_0$, $T_1$ and $T_2$ 
in AdS$_{7}$ and higher, we compute them only at the horizon and boundary in 
Appendices~\ref{sec:T0T1T2hor0th1st}-\ref{sec:entropyhighercontribution} 
and \ref{sec:roughrefinedestimate}-\ref{sec:chargeshighercontribution}, respectively. 
As a comparison, we also briefly comment on the AdS$_5$ case (see Appendix \ref{sec:deltaQcsads5eq0} for detail).

We again stress that, in the evaluation of $T_0$, $T_1$, $T_2$, $T_3$ and $T_4$, 
we neglect terms proportional to $dr$, since these terms do not contribute to $(\fQNoether)_{H}$ 
at any fixed $r$.

\subsubsection{Terms $T_3$ and $T_4$ at boundary (from Appendix~\ref{app:replacementT3T4})}
As summarized at the beginning of \S\S\S \ref{sec:T3T4horizon}, 
the terms $T_3$ and $T_4$ for a general anomaly polynomial at arbitrary fixed $r$ surface are given by \eqref{eq:appendixT3T4summary} with \eqref{eq:gvhv}. 
At the boundary $r\to \infty$, we note that $(T_3+T_4)\atinfty$ itself vanishes for AdS$_7$ and higher. We note that,
for AdS$_5$, as can be seen in Appendix \ref{sec:deltaQcsads5eq0}, this sum is nonzero but cancels with the rest terms $(T_0+T_1+T_2)\atinfty$.

\subsubsection{Terms $T_0$, $T_1$, $T_2$ at boundary (from Appendices~\ref{sec:roughrefinedestimate} 
and \ref{sec:chargeshighercontribution})}
At the boundary, for AdS$_{2n+1}$ ($n\ge 3$), all the terms $T_0$, $T_1$ and $T_2$
at $r\to \infty$ vanish up to the leading order of the derivative expansion. Therefore, 
for the sum of these three terms, we also have 
\begin{eqnarray}
(T_0+T_1+T_2)\atinfty=0\,.
\end{eqnarray}
In Appendix~\ref{sec:roughrefinedestimate}, we confirmed the above statement 
by taking into account zeroth and first order terms in the building blocks, 
while in Appendix~\ref{sec:chargeshighercontribution}, we consider 
the 2nd and higher order terms and then proved that the statement still holds. 

We note again that for AdS$_5$, the sum $(T_0+T_1+T_2)\atinfty$ is nonzero but cancels with $(T_3+T_4)\atinfty$ 
as can be seen in Appendix \ref{sec:deltaQcsads5eq0}.

\subsubsection{Chern-Simons contribution to asymptotic charges}
By combining with the results of $(T_0+T_1+T_2)\atinfty$ and $(T_3+T_4)\atinfty$ obtained in the previous subsections, 
we finally confirm that $(\fQNoether)_{H}$ at the boundary vanishes for AdS$_5$ and higher. Therefore, we have
\be
 (\fQNoether)_{H}\atinfty=- \delta\left[
\xi^\mu \TmnanomCS+(\Lambda+\ic_{\xi} \fA)|_{ \infty}\JnanomCS \right]\hodgeCFT dx^\nu\, , 
 \ee
 where 
 \be
 \TmnanomCS=\JmanomCS=0\,.\quad
\ee This verifies one of the main results we claimed in Eq.~\eqref{eq:TJanomCS0}.

\section{Discussions and conclusions}
\label{sec:Disc}

The first main result of this paper is that we have reproduced the CFT replacement rule for anomaly-induced entropy current from the dual gravity side. 
We started with the manifestly covariant differential Noether charge derived in \cite{Azeyanagi:2014sna} and evaluated it at the horizon of the 
rotating charged AdS black hole solution constructed by using the fluid/gravity derivative expansion in \cite{Azeyanagi:2013xea}. 
In \S\S\ref{ssec:TachiLike}, we also showed that the same result can be also obtained by a heuristic Tachikawa-like entropy formula at the horizon. 
As mentioned in \S\ref{sec:RepEntropy}, this simpler derivation is based on somewhat ad-hoc proposal about how to lift the 
bifurcation surface normal bundle connection $\fGamma_N$ onto an arbitrary slice (Eqn.\eqref{eq:GammaNHor}). It would be interesting to give a direct 
derivation of this proposal.

The second main result of this paper is to show that at the boundary  the Chern-Simons part of our differential Noether charge vanishes in five dimensions and higher (while there is some nontrivial contribution in the case of three dimensions).   
This completes the holographic and systematic derivation of the replacement rule for the anomaly-induced contribution to the CFT stress tensor and current initiated in \cite{Azeyanagi:2013xea}. 

For future works, there are various exciting possibilities  generalizing our computations. First of all, it will be useful 
(in particular, in setups embedded in string theories) to 
include covariant higher derivative terms in the Lagrangian in the presence of the Chern-Simons terms.\footnote{Alternately, one can add scalars 
coupling to the gauge field via non-minimal Maxwell terms. Such a system was considered in \cite{Gursoy:2014ela,Gursoy:2014boa}.}
 Since the field theoretical results do not get corrected, we expect that the extra higher derivative covariant terms would correct the fluid/gravity metric in such a way that the final results for the anomaly-induced currents and stress tensor still agree with the replacement rule.  It would present yet another non-trivial check of the replacement rule.

Looking towards a different direction, we recall that for simplicity we have set the magnetic field to be zero and the metric at the boundary to be flat in the current paper. It would be interesting to turn on some nontrivial profiles for them. These generalization will hopefully lead us to a deeper 
understanding of the replacement rule and, in particular, clarify the role of higher  Pontryagin classes. In a similar vein, the various structures of products of Riemann curvatures that we observed (and relied heavily on in our computations) are still somewhat mysterious. 
It will be meaningful to understand what kind of physical insights or geometric properties of these black hole are encoded in them. Similar objects were also used in \cite{Chapman:2012my} in studying the proposal of a local entropy current related to the Wald's construction of black hole entropy as a Noether charge. This enables \cite{Chapman:2012my}  to study the validity of the second law (i.e. the non-negativity of the divergence of the entropy current). It will be interesting to relate our work to their study as well as extending the analysis of the second law to our setup.

In light of the recent excitement in the area of holographic entanglement entropy \cite{Ryu:2006bv,Lewkowycz:2013nqa,Dong:2013qoa,Camps:2013zua} (see also \cite{Bhattacharyya:2013jma,Bhattacharyya:2013gra,Bhattacharyya:2014yga}) which potentially sheds light on the emergence of geometry in the context of gauge/gravity dualities \cite{Lashkari:2013koa,Faulkner:2013ica,Banerjee:2014oaa,Faulkner:2014jva}, a generalization to this line of investigation to include Chern-Simons terms is of great interest. The case of AdS$_3$ with the gravitational Chern-Simons term
was studied in \cite{Castro:2014tta} whereby an interesting ribbon-like structure in the bulk encodes the anomaly-induced contribution to the entanglement entropy in CFT$_2$ with gravitational anomaly. It begs the question of what then generalizes this structure in the higher dimensional holographic entanglement entropy due to Chern-Simons terms. This may well be a concrete arena where we can sharply investigate how 
the entanglement at the boundary CFT (at least for the chiral degrees of freedom of the field theories) manifests itself geometrically in the bulk. 
We will report on this aspect in the near future \cite{Azeyanagi:2015apr}.


\section*{Acknowledgements}
The authors especially thank M.~Rodriguez for  collaboration  and  useful discussions in the early stages of this work.
They would also like to thank G.~Comp\`{e}re, S.~Detournay, N.~Iqbal, K.~Jensen, M.~Rangamani, Y.~Tachikawa and A.~Yarom for valuable discussions.  
T.~A. is grateful to Okinawa Institute for Science and Technology, Galileo Galilei Institute for Theoretical Physics and in particular to Harvard University for hospitality.  
T.~A. was in part supported by INFN during his stay in the Galileo Galilei Institute for the workshop ``Holographic Methods for Strongly Coupled System." 
T.~A. would like to thank the participants of the YITP workshop ``Holographic vistas on Gravity and Strings" and ``Strings and Fields". 
T.~A. and G.~N. are grateful to the participants and organizers of the Solvay Workshop on ``Holography for Black Holes and Cosmology".  
T.~A. was financially supported by the LabEx ENS-ICFP: ANR-10-LABX-0010/ANR-10-IDEX-0001-02 PSL*. 
R.~L. was supported by Institute for Advanced Study, Princeton. G.~N. was supported by DOE grant DE-FG02-91ER40654 and the Fundamental Laws Initiative at Harvard.  G.~N. was also supported by an NSERC Discovery Grant.

\appendix

\section{Notation}
\label{app:notation}
Regarding conventions and notations used throughout this paper, we follows our previous papers, 
Ref. \cite{Azeyanagi:2013xea,Azeyanagi:2014sna}.
Here we summarize some notations that we frequently use and also define some new useful notation 
for the purpose of the current paper. 

\subsection{Summary of notations}
\begin{itemize}
\item
Sometimes, for convenience or to avoid cluttering of indices, we suppress all matrix indices. 
It is always assumed that objects next to each others are multiplied as matrix multiplications. In particular, we think of $(\nabla_b \xi^a)$ as a matrix $(\nabla \xi)^a{}_b$. For example,
\bea
\left(\delta \fGamma \nabla \xi \fR\right)\equiv
(\delta \fGamma^a{}_b) \nabla_c \xi^b \fR^c{}_d\, ,\quad
\tr[\nabla \xi \fR]
\equiv  \nabla_b \xi^a \fR^b{}_a\,.
\eea

Another important terminology we will frequently use throughout this paper is the phrase {\it  `building block'} 
which is defined as one of the following objects :
\be\label{eq:defbuildingblocks}
\fA,\quad \delta \fA,\quad \fF,\quad
G_{ab},\quad\fGamma^a{}_b,\quad \delta \fGamma^a{}_b,\quad \fR^a{}_b,\quad 
\nabla_a\xi^b.
\ee These are the basic objects whose products (and appropriate index-contractions)  form the basis  
in the evaluation of  $T_0,T_1,T_2,T_3$ and $T_4$.

\item 
When carrying out the fluid/gravity derivative expansion, we need to write down 
some differential forms at a particular order in the derivative expansion.  For this purpose, it is convenient to introduce the following notation :
the $m$-th order terms in the derivative expansion of a form $\mathcal{B}$ is denoted as  $(\B{m})^{a}{}_{b}$ while the product of $k$ matrix-valued 2-forms, $(\B{m_1}),  \cdots$ and  $(\B{m_k})$ is written as $(\B{m_1}\ldots \B{m_{k}})$, so that the matrix-valued 2-forms inside the brackets are always multiplied through matrix multiplication. The $k$-th power of $(\B{m})$ is denoted $(\B{m}^k)$. 
For example, for the field strength $\fF$, we have
\be
(\F{0}) = \Phi' dr \wedge \fu \qquad
(\F{1}) =  (2\Phi\fomega) \,.
\ee 

\item
In discussing products of curvature two-forms, the cases containing purely $(\R{0})$'s and $(\R{1})$'s are of particular significance throughout the computations of 
the differential Noether charge. Therefore we define 
$\fchi_m$ ($m=0, 1, 2, \cdots$) to denote products of $(\R{0})$'s and $(\R{1})$'s with exact $m$-number of $(\R{0})$'s wedged with an arbitrary number of $(\R{1})$'s.
For example, $\fchi_0$ only includes product of $(\R{1})$'s, i.e. $(\R{1}\R{1}\ldots \R{1}\R{1})$. Another example is $\fchi_1$, which for example contains some of the following possibilities :
\be
(\R{0}\R{1} \R{1})\,,\qquad (\R{1}\R{0} \R{1})\,,\qquad
(\R{1}\R{1}\R{0})\, .
\ee  The classification of $\fchi_m$ has been carried out in Appendix B.2 of Ref.~\cite{Azeyanagi:2013xea} and 
will be reviewed in Appendix \ref{sec:app_useful_previouspaper}.
We also note that we sometimes use $\fchi_m$ to simply denote an element in $\fchi_m$.

\item
We define a convenient symbol $\form{\upsilon}$ is defined to represent a string made of 2nd or higher order curvature two-form 
 $(\R{m})$'s with $m\ge2$ (for example, $(\R{2}^2\R{5}\R{3})$).

\item
We also define $
(\cR^q_{(p)})
$ to denote all possible structures (including those consist of zeroth and first order 
building blocks only) that can contribute to $(\fR^q)$ at $\omega^p$ order.
For example, the non-trivial possible structures in $(\cR^3_{(2)})$ are
\be
(\R{0}\R{1}\R{1})\,,\quad
(\R{1}\R{0}\R{1})\,,\quad
(\R{1}\R{1}\R{0})\,,\quad
(\R{0}\R{2}\R{0})\,.
\ee A complete classification of $(\cR^q_{(p)})$ for $0 \le p \le q+1$ is given in Appendix \ref{sec:classificationRpq}.

\end{itemize}

\subsection{Orientation convention in holography} 
\label{sec:Orientation convention in Holography}

Let $\varepsilon_{\mu_1\mu_2\ldots \mu_d}$ be the $\varepsilon$-tensor of the
spacetime in which the field theory CFT$_d$ lives. Then, there are two possible 
conventions for the orientation of the dual AdS$_{d+1}$. 

Let $r$ be the radial coordinate such that $r\to\infty$ is the conformal boundary
of the dual AdS$_{d+1}$. It is usual in gravity to fix the orientation such that
$\varepsilon_{r\mu_1\mu_2\ldots \mu_d}$ has an opposite sign as compared to 
$\varepsilon_{\mu_1\mu_2\ldots \mu_d}$. For example, if $\brk{txyz}$ formed
a right-handed coordinate basis in CFT$_d$ then, in this convention $\brk{trxyz}$
forms a right-handed coordinate basis in AdS$_{d+1}$.

We use here the opposite convention whereby $\varepsilon_{r\mu_1\mu_2\ldots \mu_d}$
has the same sign as $\varepsilon_{\mu_1\mu_2\ldots \mu_d}$ which is more natural
from the viewpoint of pullback. Our expressions can easily be adopted to the reverse bulk orientation 
(we will keep the CFT orientation unchanged)
by replacing
$\hodge$ with $-\hodge$ and $\varepsilon_{abcd\ldots}$ with  $-\varepsilon_{abcd\ldots}$.
Note that we keep $\hodgeCFT$ unchanged. In the new convention, the relation between
the barred and the unbarred forms becomes $\form{V}=\hodge\overline{\form{V}}$.



\section{Useful relations I}
\label{app:useForm}
This Appendix collects some results needed for the computations in this paper. 

\subsection{Results from our previous paper}
\label{sec:app_useful_previouspaper}
Here we summarize some results from Ref.~\cite{Azeyanagi:2013xea} which are valid for any $r$. 

\subsubsection{0th and 1st order terms}
\label{sec:app_useful_previouspaper01only}

\noindent
\underline{\bf Bulk metric and gauge field} \\
The bulk metric  $G_{ab}$ (in the coordinate $x^a=\{r,x^{\mu}\}$) and its inverse for the rotating charged AdS 
black hole in $d+1$ dimensions ($d=2n$ with a positive integer $n$) is given 
up to the first order of the derivative expansion as 
\bea \label{eq:metricandinverse}
G_{rr} & = & 0\,,\quad
G_{r\mu} = - u_\mu\, ,\quad
G_{\mu\nu} = -2 \Psi(r)u_{\mu} u_{\nu}
+r^2 P_{\mu\nu}\,, \nonumber\\
G^{rr}  & = & 2\Psi(r)\,  ,   \quad
G^{r\mu}   =  u^\mu\,,   \quad
G^{\mu\nu}  = r^{-2} P^{\mu\nu}\,,
\eea while the gauge field $A_a$ up to this order is 
\begin{eqnarray}
\fA = A_a dx^a = \Phi \fu \, . 
\end{eqnarray}
Here $f(r)$, $\Psi(r)$ and $\Phi$ ($\Phi_T$ that we will use later) are defined as
\bea
f(r,m,q) &\equiv &1- \frac{m}{r^d} + \half \kappa_q \frac{q^2}{r^{2(d-1)}} \,,\\
\Phi(r,q) &\equiv& \frac{q}{r^{d-2}}\,,\\
\Psi(r,m,q) &\equiv& \half r^2 f = \frac{r^2}{2}\left[
1- \frac{m}{r^d} + \half \kappa_q \frac{q^2}{r^{2(d-1)}} \right] \,,\\
\PhiT(r,m,q) &\equiv& \half r^2 \frac{df}{dr} 
= \frac{1}{2r^{d-1}} \brk{ m d- \kappa_q(d-1)\frac{q^2}{r^{d-2}} }. \label{eq:Phi_Tsol}
\eea 
The horizon of the black hole is located at $r_H$ satisfying $f(r_H, m, q)=0$. 
The parameters $m$ and $q$ are related to the mass and electric charge of the black hole solution 
and $\kappa_q$ is a normalized Maxwell coupling constant defined as 
$16\pi G_N/\gYM^2 = \kappa_q(d-1)/(d-2)$ where $G_N$ and $\gYM$ are 
the Newton constant and Maxwell coupling constant, respectively.  

For the boundary fields, the boundary metric $g_{\mu\nu}$ is set to be flat, i.e. $g_{\mu\nu}=\eta_{\mu\nu}$
throughout this paper. The fluid velocity field $u^\mu$ (normalized as $u^2\equiv g^{\mu\nu} u_{\mu} u_{\nu}=-1$) is chosen so that it corresponds to a pure rotation configuration, which implies $u^\mu \omega_{\mu\nu}=0$ and $\partial_{(\mu}u_{\nu)}=0$. 
The vorticity field $\omega_{\mu\nu}$ is defined as $\omega_{\mu\nu}\equiv (\partial_{\mu} u_{\nu}-\partial_{\nu} u_{\mu})/2$ while the projection operator $P_{\mu\nu} \equiv g_{\mu\nu}+u_{\mu}u_{\nu}$
satisfies $P^{\mu\nu} u_\nu=0$ and  $P_{\mu}^{\,\,\, \rho} P_{\rho\nu} =P_{\mu\nu}$.
We also frequently use the velocity one-form and vorticity two form defined by 
$\fu = u_\mu dx^\mu$ and $\fomega =(1/2) d\fu = (1/2)\omega_{\mu\nu}dx^\mu\wedge dx^\nu$, respectively. 
\vspace{0.5cm}\\
\noindent
\underline{\bf Christoffel connection} \\
The components of the Christoffel connection for the bulk metric $G_{ab}$ are given by
\begin{eqnarray}\label{eq:christoffel1}
\Gamma^a_{\ rr}  & = & 0,\quad
\Gamma^r_{\ r\nu} = u_\nu \Psi' ,\quad
\Gamma^r_{\ \rho\nu} =
 2\Psi\left[ \Psi' u_{\rho}u_{\nu} -r P_{\rho\nu} \right]\, ,\quad
\Gamma^\mu_{\ r \nu} = r^{-2} \left[ r P^\mu{}_\nu +\omega^\mu{}_{\nu} \right]\,, \nonumber\\
\Gamma^\mu_{\ \rho\nu} &=& u^{\mu} \left[ u_\rho u_\nu \Psi' -r P_{\rho \nu} \right]
-2 r^{-2}   (2\Psi-r^2) u_{(\nu}\omega_{\rho)}{}^{\mu}\,,
 \end{eqnarray} 
or in terms of the connection 1-form $\fGamma^a{}_b\equiv \Gamma^a{}_{bc} dx^c$, by 
\bea
\fGamma^r{}_r&=&  \Psi' \fu \,,\quad
\fGamma^r{}_\rho= u_\rho \Psi' dr + 2\Psi\left( \Psi'u_\rho \fu - r P_{\rho\nu} dx^\nu\right)\,, \quad
\fGamma^\mu{}_r=r^{-2}\left(
r P^\mu{}_\nu+\omega^{\mu}{}_\nu
\right)dx^\nu\,, \nonumber\\
\fGamma^\mu{}_\rho&=&
r^{-2}\left(r P^\mu{}_\rho+\omega^\mu{}_\rho\right)dr
+\left[u^\mu u_\rho \Psi' + r^{-2}(2\Psi-r^2)\,\omega^\mu{}_\rho \right]\fu
\nonumber\\
& &
-r u^\mu P_{\rho\nu}dx^\nu
+r^{-2}\left(2 \Psi-r^2\right) u_\rho\, \omega^\mu{}_\nu dx^\nu\,. 
\eea 

\noindent
\underline{\bf $\nabla_a\xi^b$}\\
For $\nabla_a\xi^b$, since we are considering $\xi^a$ with $\xi^r=0$ and $\partial_r \xi^a=0$,
we obtain (at any fixed $r$)
\bea \label{eq:nablaxigeneralr}
\nabla_r \xi^r &=& \Psi' u_\beta \xi^\beta
,\quad
\nabla_r \xi^\mu = \left[
r^{-1}P^\mu{}_\beta
+r^{-2} \omega^\mu{}_\beta
\right]\xi^\beta,
\quad
\nabla_\mu \xi^r =
2\Psi\left(
\Psi' u_\mu u_\beta
-r P_{\mu\beta}
\right)
\xi^\beta\, , \nonumber\\
\nabla_\mu \xi^\nu&=&
\partial_\mu \xi^\nu
+u^\nu\left(
u_\mu u_\beta \Psi'
-r P_{\mu\beta}
\right)
\xi^\beta
-2r^{-2}(2\Psi-r^2)u_{(\mu}\omega_{\beta)}{}^\nu \xi^\beta\, . 
\eea

\noindent
\underline{\bf $U(1)$ field strength} \\
By using the notation in Appendix~\ref{app:notation}, 
the zeroth and first order terms of  the field strength $\form{F}$ are given by 
\be\label{eq:F0F1F2}
(\F{0}) = \Phi' dr \wedge \fu\, , \qquad 
(\F{1}) =  (2\Phi\fomega) \,.
\ee 

\noindent
\underline{\bf Curvature two-form} \\
The curvature 2-form, defined from the Riemann tensor $R^a{}_{bcd}$ 
as $\fR^a{}_b\equiv (1/2) R^a{}_{bcd} \,dx^c \wedge dx^d$, at zeroth order is given by 
\bea
(\R{0})^r{}_r
&=&\Psi''dr\wedge \fu\,,  \nonumber\\
(\R{0})^{\mu}{}_{r}
&=& r^{-1} \Psi'  dx^\mu \wedge \fu\,,
 \nonumber\\
 (\R{0})^{r}{}_{\rho}
 &=&
 r \Psi' P_{\rho\nu}dx^\nu\wedge dr
-2\Psi \Psi'' u_\rho  \fu\wedge dr\,,
\\
(\R{0})^\mu{}_{\rho}
&=&
-r^{-1} \partial_r \left(r \Phi_T\right) u^\mu u_\rho
\fu\wedge dr
+r^{-1} \Psi' u_\rho dx^\mu \wedge dr
\nonumber\\
& &
-r \Phi_T u^\mu
P_{\rho\nu} dx^\nu \wedge \fu
+2 r^{-1} \Psi dx^\mu \wedge\left[
\Psi' u_\rho \fu
-r P_{\rho\nu} dx^\nu
\right]\,, \nonumber
\eea
while the first order terms are 
\bea
(\R{1})^r{}_{r} &=&(2 \Phi_T \fomega)\,, \nonumber\\
(\R{1})^\mu{}_{r} &=&
-r^{-2}   \fu
\wedge (\Phi_T \omega^\mu{}_{\nu} dx^\nu)\,, \\
(\R{1})^r{}_{\rho} &=&
  dr\wedge(\Phi_T \omega_{\rho\nu} dx^\nu)
+2 \Psi \left[
u_\rho (2\Phi_T\fomega)
+  \fu\wedge (\Phi_T\,\omega_{\rho\nu}\, dx^\nu)
\right]\,, \nonumber\\
(\R{1})^\mu{}_{\rho} &=&
-2 r^{-2} \Phi_T\, \omega^\mu{}_{\rho}\,\fu \wedge dr
+r^{-2}\,  u_\rho\, dr\wedge  (\Phi_T\,\omega^\mu{}_{\nu}\, dx^\nu)\nonumber\\
&&
+ u^\mu\left[
  u_\rho (2\Phi_T \fomega)
+ \fu \wedge (\Phi_T\,\omega_{\rho\nu}\, dx^\nu )\right]\,.\nonumber 
\eea

\noindent
\underline{\bf Products of curvature two-form} \\
Let us now consider wedge products of $(\R{0})$'s and $(\R{1})$'s.
We first consider the products of two curvature two-forms: 
When two $(\R{0})$'s are multiplied, we have
\bea
&&(\R{0}\R{0}) = 0\,.  
\eea
Therefore, the products of two or more $(\R{0})$'s vanish identically. 
For the products of one $(\R{0})$ and one $(\R{1})$, there are following two possibilities : 
\bea
\left(\R{0}\R{1}\right)^r{}_r &=&
 r^{-1} \partial_r(r \Phi_T) dr\wedge \fu \wedge (2\Phi_T \fomega)\,, \nonumber \\
\left(\R{0}\R{1}\right)^\mu{}_r &=& r^{-1}\Phi_T dx^\mu \wedge \fu \wedge (2\Phi_T \fomega)\,,  \nonumber\\
\left(\R{0}\R{1}\right)^r{}_\rho &=&0\,, \\
\left(\R{0}\R{1}\right)^\mu{}_\rho&=&
- \Phi_T' u^\mu u_\rho dr\wedge \fu \wedge (2\Phi_T \fomega)
- r^{-1} \Phi_T\, u_\rho\, dx^\mu \wedge dr\wedge (2\Phi_T \fomega)
\nonumber\\
&& -2r^{-1} \Phi_T dr \wedge \fu \wedge dx^\mu \wedge (\Phi_T \,\omega_{\rho\nu}\, dx^\nu)\,, \nonumber
\eea
and
\bea
\left(\R{1}\R{0}\right)^r{}_r &=& 
 r^{-1} \partial_r(r \Phi_T) dr\wedge \fu \wedge (2\Phi_T\fomega)\, , \nonumber\\
 \left(\R{1}\R{0}\right)^\mu{}_r &=&0\, , \\
\left(\R{1}\R{0}\right)^r{}_\rho &=&- r\Phi_T dr\wedge (P_{\rho\nu}\, dx^\nu)\wedge (2\Phi_T\fomega)
+2 r \Psi \Phi_T (P_{\rho\nu }\,dx^\nu)\wedge \fu\wedge (2\Phi_T\fomega)\, , \nonumber \\
\left(\R{1}\R{0}\right)^\mu{}_\rho &=&
2r^{-1}\Phi_T   (P_{\rho\nu}\, dx^\nu)  \wedge \fu\wedge dr \wedge (\Phi_T \omega^\mu{}_\sigma dx^\sigma)
+ r \Phi_T  u^\mu P_{\rho \nu}\, dx^\nu \wedge \fu \wedge (2\Phi_T \fomega)
\nonumber\\
& &
+  r^{-1}\partial_r(r \Phi_T)  u^\mu u_\rho \fu\wedge dr \wedge (2 \Phi_T\fomega)\, .  
\eea
Finally, the following is the products of two $(\R{1})$'s :
\bea
\left(\R{1}\R{1}\right)^r{}_r &=& (2\Phi_T \fomega)^2\, , \nonumber \\
\left(\R{1}\R{1}\right)^\mu{}_r &=&
-r^{-2} \fu\wedge(\Phi_T \,\omega^\mu{}_\nu\, dx^\nu )\wedge (2\Phi_T \fomega)\, ,
 \nonumber\\
\left(\R{1}\R{1}\right)^r{}_\rho &=&
  dr\wedge (\Phi_T\,\omega_{\rho\nu}\, dx^\nu)\wedge (2\Phi_T \fomega)\, , \\
\left(\R{1}\R{1}\right)^\mu{}_\rho &=&
- u^\mu u_\rho (2 \Phi_T\fomega)^2 - 2r^{-2} \Psi  u_\rho  \fu \wedge (\Phi_T\,\omega^\mu{}_\nu\, dx^\nu)\wedge (2\Phi_T\fomega) \nonumber\\
 & &-  r^{-2}  u_\rho dr\wedge (\Phi_T\,\omega^\mu{}_\nu\, dx^\nu)  \wedge (2\Phi_T \fomega)
  -u^\mu \fu\wedge (\Phi_T\,\omega_{\rho\nu}\,dx^\nu) \wedge (2\Phi_T \fomega)\, . \nonumber
\eea

Next we consider the products of three curvature two-forms.  
As a result of $(\R{0}\R{0})=0$, the following products trivially vanish :
\begin{eqnarray}
(\R{0}\R{0}\R{0})=(\R{0}\R{0}\R{1})=(\R{1}\R{0}\R{0})=0\, . 
\end{eqnarray}
Therefore, there is only one nontrivial case with two $(\R{0})$'s in the product : 
\bea
\left(\R{0}\R{1}\R{0}\right)^r{}_r &=&\left(\R{0}\R{1}\R{0}\right)^r{}_\rho=\left(\R{0}\R{1}\R{0}\right)^\mu{}_r=0\,,
\nonumber \\
\left(\R{0}\R{1}\R{0}\right)^\mu{}_\rho&=&
2\,\Phi_T^2\, dr\wedge \fu \wedge dx^\mu \wedge (P_{\rho\nu} dx^\nu )\wedge (2\Phi_T \fomega)\,.
\eea
For the products with only one $(\R{0})$, there three possibilities depending on where 
$(\R{0})$ is located. The first case is 
\bea
\left(\R{0}\R{1}\R{1}\right)^r{}_r &=&
 r^{-1}\partial_r(r \Phi_T)dr\wedge \fu \wedge (2\Phi_T\fomega)^2\, , \nonumber\\
\left(\R{0}\R{1}\R{1}\right)^\mu{}_r &=&
 r^{-1} \Phi_T  \,dx^\mu \wedge \fu \wedge(2\Phi_T \fomega)^2\, , \nonumber\\
\left(\R{0}\R{1}\R{1}\right)^r{}_\rho &=&
2\Psi  r^{-1}\partial_r \left(r  \Phi_T \right) u_\rho dr\wedge \fu \wedge (2\Phi_T\fomega)^2\, , \\
\left(\R{0}\R{1}\R{1}\right)^\mu{}_\rho &=&
 \Phi_T' \,u^\mu u_\rho\, dr\wedge \fu \wedge (2\Phi_T\fomega)^2
+ r^{-1} \Phi_T\, u_\rho dx^\mu\wedge dr\wedge (2\Phi_T\fomega)^2\nonumber\\
&& +2 r^{-1} \Psi \Phi_T u_\rho dx^\mu \wedge \fu \wedge (2\Phi_T\fomega)^2\, . \nonumber
\eea
The second case is 
\bea\label{eq:R101}
\left(\R{1}\R{0}\R{1}\right)^r{}_r & = &
 \Phi_T' \,dr\wedge \fu\wedge (2\Phi_T\fomega)^2\, , \nonumber\\
\left(\R{1}\R{0}\R{1}\right)^\mu{}_r& = &0\,, \nonumber \\
\left(\R{1}\R{0}\R{1}\right)^r{}_\rho & = & 2\Psi \Phi_T' u_\rho \,dr\wedge \fu\wedge (2\Phi_T\fomega)^2\, ,\\
\left(\R{1}\R{0}\R{1}\right)^\mu{}_\rho & = & \Phi_T'\, u^\mu u_\rho\, dr\wedge \fu \wedge (2\Phi_T\fomega)^2\, , \nonumber
\eea
and the final case is 
\bea
\left(\R{1}\R{1}\R{0}\right)^r{}_r & = &
 r^{-1} \partial_r (r \Phi_T)dr\wedge \fu \wedge(2\Phi_T \fomega)^2\, , \nonumber \\
\left(\R{1}\R{1}\R{0}\right)^\mu{}_r & = & 0\, , \\
\left(\R{1}\R{1}\R{0}\right)^r{}_\rho & = &
 2\Psi r^{-1} \partial_r (r \Phi_T ) u_\rho \,dr\wedge \fu \wedge(2\Phi_T \fomega)^2
+ r \,\Phi_T P_{\rho\nu} dx^\nu \wedge dr \wedge (2\Phi_T \fomega)^2\, ,
\nonumber\\
\left(\R{1}\R{1}\R{0}\right)^\mu{}_\rho&=&
 r^{-1} \partial_r(r \Phi_T) u^\mu u_\rho dr\wedge \fu \wedge(2\Phi_T \fomega)^2
+r \Phi_T u^\mu \fu \wedge P_{\rho\nu} dx^\nu\wedge (2\Phi_T\fomega)^2\,,  \nonumber
\eea
We also have the case with no $(\R{0})$'s in the product : 
\bea
\left(\R{1}\R{1}\R{1}\right)^r{}_r &=&(2 \Phi_T \fomega)^3\, ,  \nonumber\\
\left(\R{1}\R{1}\R{1}\right)^\mu{}_r &=&
- r^{-2} \fu \wedge (\Phi_T \, \omega^\mu{}_\nu\, dx^\nu )\wedge (2\Phi_T\fomega)^2\, , \\
\left(\R{1}\R{1}\R{1}\right)^r{}_\rho &=&
 2\Psi  u_\rho (2\Phi_T\fomega)^3+2 \Psi  \fu \wedge(\Phi_T\, \omega_{\rho\nu}\, dx^\nu) \wedge (2\Phi_T\fomega)^2
 \nonumber\\
 & &+ dr \wedge (\Phi_T\, \omega_{\rho\nu}\, dx^\nu) \wedge  (2\Phi_T \fomega)^2\, ,  \nonumber\\
\left(\R{1}\R{1}\R{1}\right)^\mu{}_\rho &=&
 r^{-2}  u_\rho dr \wedge(\Phi_T\,\omega^\mu{}_\nu\, dx^\nu) \wedge  (2\Phi_T\fomega)^2\nonumber\\
 && -2r^{-2} dr \wedge \fu \wedge(\Phi_T\, \omega^\mu{}_\sigma\, dx^\sigma) \wedge 
 (\Phi_T\, \omega_{\rho\nu}\,dx^\nu)\wedge  (2\Phi_T\fomega)\nonumber\\
 &&+ u^\mu u_\rho (2\Phi_T\fomega)^3
  +u^\mu \fu \wedge( \Phi_T\,\omega_{\rho\nu}\,dx^\nu) \wedge  (2\Phi_T\fomega)^2\, . \nonumber
\eea

We sometimes find it useful to reduce products of $(\R{1})$'s using
\be
dr\wedge \fu
\wedge(\R{1}\R{1}\R{1})^a{}_b=
dr\wedge \fu
\wedge(\R{1})^a{}_b \wedge (2\Phi_T \fomega)^2\, , \quad
(\R{1}\R{1}\R{1}\R{1})=(2\Phi_T\fomega)^2\wedge  (\R{1}\R{1})\,.
\ee

\noindent
\underline{\bf Classification of the products of the curvature two-forms} \\
As we have shown in Appendix B.2 of Ref.~\cite{Azeyanagi:2013xea},  
the building block of the wedge product of $\fR$'s 
made of  $(\R{0})$ and $(\R{1})$ only reduces to the following possibilities : 
\begin{eqnarray}
\label{eq:fchin}
\fchi_0 & : & \{(\R{1})\,, (\R{1}\R{1})\,, (\R{1}\R{1}\R{1})\}\, , \nonumber\\
\fchi_1 & : & \{(\R{0}),\,(\R{0}\R{1})\, ,\, (\R{1}\R{0})\, ,\, (\R{0}\R{1}\R{1})\, ,\, (\R{1}\R{1}\R{0})\, ,\,(\R{1}\R{0}\R{1})\, ,\,(\R{1}\R{0}\R{1}\R{1})\, ,\,(\R{1}\R{1}\R{0}\R{1})\}\, ,\nonumber\\
\fchi_2 & : & \{(\R{0}\R{1}\R{0})\}\, ,
\end {eqnarray}
Here by `reduce' we mean
the products of four or more $\fR$'s (with $(\R{0})$ and $(\R{1})$ only) 
are zero or written as the one of the above elements wedged by an appropriate power of 
$(2\Phi_T \fomega)$.  
In particular, wedge products containing three or more $(\R{0})$'s 
(with the rest of $\fR$'s equal to $(\R{1})$'s) are zero, i.e. $\fchi_m=0$ for $m\ge3$.
For more detail, we refer the readers to Appendix B.2 of Ref.~\cite{Azeyanagi:2013xea}.
Another useful notation for later purpose is
$\tilde{\fchi}_1$, which is defined as all elements in ${\fchi}_1$ excluding $(\R{0})$.
 \vspace{0.5cm} \\
 \noindent
\underline{\bf Trace of the products of the curvature two-form} \\
Finally, the traces of the wedge products of $(\R{0})$'s and $(\R{1})$'s are rather simple :
\bea\label{eq:trR2k}
\tr\left[(\R{0}\R{1}\left(\R{1}\R{1}\right)^{k}) \right]&=&
2   \Phi'_T dr\wedge \fu\wedge (2\Phi_T\fomega)^{2k+1}\, , 
\nonumber\\
 \tr\left[ \left(\R{1}\R{1}\right)^{k+1} \right]&=& 2(2  \Phi_T\fomega)^{2k+2}\, , 
\eea
for $k\ge 0$.

\subsubsection{2nd order terms}
\label{sec:app_useful_previouspaper_0th1st}
Here we summarize some results at the second order of the fluid/gravity 
derivative expansion which are useful for the purpose of this paper. 

In Ref.~\cite{Azeyanagi:2013xea}, the rotating charged AdS black hole 
solutions in $(d+1)$ dimensions is constructed up to the second order
in the fluid/gravity derivative expansion (assuming stationary fluid configurations). 
It is given by 
\bea
\label{eq:2ndOrderMetric}
ds^2 &=&
-2 u_{\mu} dx^{\mu} dr -r^2 f(r,m,q) \, u_{\mu} u_{\nu}  \, dx^{\mu} dx^{\nu}
+ r^2 P_{\mu\nu} dx^{\mu} dx^{\nu}
\nonumber\\
& &
-{\omega_{\mu}}^{\alpha}\omega_{\alpha\nu} \, dx^{\mu} dx^{\nu}
+g(r,m,q)\,\omega_{\alpha\beta}\omega^{\alpha\beta} \,u_{\mu}u_{\nu}\,dx^{\mu}dx^{\nu}
\nonumber\\
&&
+h(r,m,q)\left[{\omega_{\mu}}^{\alpha}\,\omega_{\alpha\nu}
+\frac{1}{d-1} \omega_{\alpha\beta}\omega^{\alpha\beta} P_{\mu\nu}\right]dx^{\mu}dx^{\nu}\,,\nonumber\\
\fA &=& \Phi(r,q) \brk{1- \frac{1}{2r^2} \omega_{\alpha\beta}\omega^{\alpha\beta} }\, u_{\mu} dx^{\mu} ,
\eea
where $g(r,m,q)$ and $h(r,m,q)$ are given by 
\begin{eqnarray}
\label{eq:2ndOrderMetric1}
g(r, m, q)&=& -\frac{m}{2 r^d} + \frac{\kappa_q}{2}\frac{q^2}{r^{2(d-1)}} \left[1-\frac{1}{(d-1)(d-2)}\right]\, ,  \\
\label{eq:2ndOrderMetric2}
h(r, m , q)&=&- \frac{d}{d-2}\ \kappa_q\ \frac{ r^2 q^2}{r_{_H}^{2d}}
\int_{r/r_{_H}}^{\infty}\frac{\zeta^d-1}{\zeta^{2d+1} f(\zeta\ r_{_H},m,q)}d\zeta\, .
\end{eqnarray}
For the $U(1)$ field strength, the 2nd order contribution is calculated 
from the solution as 
\begin{eqnarray}
(\F{2})\propto dr \wedge \fu \times \omega^{\alpha\beta}\omega_{\alpha\beta}\,.
\end{eqnarray}

Furthermore, from Appendix B.1 of Ref.~\cite{Azeyanagi:2013xea}, the second order curvature 2-form has 
two types of non-trivial contributions. The one coming from the second order metric and 
the one coming purely from the zeroth order metric (and its derivatives).
To distinguish from the whole 2nd order curvature 2-form $(\R{2})$, we denote the 
latter contribution by $(\R{2}')$. They are given by
\bea\label{eq:R2prime}
(\R{2}')^r{}_{r} &=&0\,, \nonumber\\
(\R{2}')^\mu{}_{r} &=&
 -
r^{-4} \left(\omega^\mu{}_\nu \omega^\nu{}_\sigma dx^\sigma\right)\wedge dr
-r^{-4} \left(2\Psi-r^2\right) \left(\omega^\mu{}_\nu \omega^\nu{}_\sigma dx^\sigma\right)\wedge \fu\,,
 \nonumber\\
 (\R{2}')^r{}_{\rho} &=&0\,,
 \nonumber\\
(\R{2}')^\mu{}_{\rho} &=&
r^{-2} \partial_\nu \omega^{\mu}{}_\rho dx^\nu \wedge dr \label{eq:2ndcurvaturepart}\\
& &
+r^{-2}(2\Psi-r^2)
\left[
\partial_\nu \omega^{\mu}{}_\rho dx^\nu \wedge \fu
+2\omega^{\mu}{}_\rho\fomega
-(\omega_{\rho\sigma} dx^\sigma)\wedge(\omega^\mu{}_\nu dx^\nu)
\right]\nonumber\\
 & &
-r^{-4}\left(2\Psi-r^2\right) \left(
u_\rho \omega^\mu{}_\sigma \omega^\sigma{}_\nu  dx^\nu \right)\wedge dr
-r^{-4}\left(2\Psi-r^2\right)^2 \left(
u_\rho \omega^\mu{}_\sigma \omega^\sigma{}_\nu  dx^\nu \right)\wedge \fu\,. \nonumber
\eea

\subsection{Structures of (\texorpdfstring{$\fR^{k}$}{~}) at arbitrary \texorpdfstring{$r$}{r}}
\label{eq:singleproductstructure}
In this Appendix, we study the general structures of $(\fR^{k})$ at any fixed $r$.
All equalities are evaluated at a fixed $r$ and terms proportional to $dr$ are ignored. 

\subsubsection{Some relations}
\label{sec:somerelations}
First it is useful to recall from Eq.~(\ref{eq:nablaxigeneralr}) (which is valid at any $r$) that
\bea\label{eq:useful4}
({}^{(0)} \nabla_r \xi^r) &=& \Psi' u_\nu \xi^\nu
,\quad
({}^{(0)} \nabla_r \xi^\mu) = 
r^{-1}P^\mu{}_\nu \xi^\nu, 
\nonumber\\
({}^{(0)} \nabla_\rho \xi^a)&=&
G^{ar}\left(
\Psi' u_\rho u_\nu 
-r P_{\rho\nu}
\right)
\xi^\nu,
\eea where $G^{rr}=2\Psi$ and $G^{r\alpha} =u^\alpha$
and we have used that in our setup $\partial_r \xi^a=0$ and $\xi^r=0$.
We also have\be
({}^{(0)} \nabla_b \xi^a)= \ic_{\xi}({}^{(0)} \fGamma^a{}_b)\,.
\ee
On the other hand,
\be
^{(0)} \fGamma^r{}_r = \Psi' \fu\,,\quad
^{(0)} \fGamma^\mu{}_r=r^{-1} P^\mu{}_\nu dx^\nu\, ,  \quad
^{(0)} \fGamma^a{}_\rho = G^{ar} (\Psi' u_\rho \fu -r P_{\rho\nu}dx^\nu)\, , 
\ee where we have ignored $dr$ terms. Thus, for the variation of the Christoffel connection, we have
\bea
^{(0)} \delta \fGamma^r{}_r &=& (\delta\Psi') \fu+\Psi' (\delta \fu)\,,\quad
^{(0)} \delta\fGamma^\mu{}_r=r^{-1} (u^\mu \delta \fu + \fu\delta u^\mu)\, ,  \nonumber\\
^{(0)}\delta \fGamma^a{}_\rho &=&
( \delta G^{ar}) (\Psi' u_\rho \fu -r P_{\rho\nu}dx^\nu)
+ G^{ar}  \left[(\delta\Psi' ) u_\rho \fu
+ (\Psi' -r) (u_\rho \delta \fu +\fu \delta u_\rho) \right].\nonumber\\
\eea 
We note that
\be\label{eq:useful2}
^{(0)}\delta \fGamma^a{}_\rho P^\rho{}_\nu dx^\nu \wedge \fu=0\, ,\quad
^{(0)}\delta \fGamma^a{}_\rho u^\rho  \fu \propto G^{ar}( \delta \fu) \wedge \fu\, . 
\ee
Furthermore, we recall that 
\bea
\label{eq:useful3}
&&(\R{0}\R{2})^a{}_r
=(\R{0}\R{2})^r{}_b
=(\R{2}\R{0})^a{}_r
=(\R{2}\R{0})^r{}_b=0\, ,\nonumber\\
&&
(\R{0}\R{2})^\mu{}_\rho \propto (\R{2}\R{0})^\mu{}_\rho \propto \fu \wedge (P^{\mu}{}_\nu dx^\nu) \wedge (P_{\rho\sigma} dx^\sigma)\,.
\eea  
Moreover, at constant $r$ we have the following relations :
\be\label{eq:useful5}
P_{\beta \nu} dx^\nu \wedge (\R{2})^{\beta r}=
\fu  \wedge u_\beta (\R{2})^{\beta r}=
0\,.
\ee 
Using the above identities, we can show that
\begin{eqnarray}
\label{eq:useful101}
&&
({}^{(0)} \delta \fGamma^a{}_b)( {}^{(0)}\nabla_c \xi^b)(\R{0}\R{2})^c{}_a  =
({}^{(0)} \delta \fGamma^a{}_b)( {}^{(0)}\nabla_c \xi^b)(\R{2}\R{0})^c{}_a   =0\,,
\nonumber\\
 && 
({}^{(0)} \delta \fGamma^a{}_b)(\R{0}\R{2})^b{}_c( {}^{(0)}\nabla_a \xi^c)=
({}^{(0)} \delta \fGamma^a{}_b)(\R{2}\R{0})^b{}_c( {}^{(0)}\nabla_a \xi^c)=0 \,,
\nonumber\\
&&
 ({}^{(0)}\delta \fGamma^a{}_b) (\R{0})^b{}_c  ({}^{(0)}\nabla_d \xi^c ) (\R{2})^d{}_a = ({}^{(0)}\delta \fGamma^a{}_b) (\R{2})^b{}_c  ({}^{(0)}\nabla_d \xi^c ) (\R{0})^d{}_a=0\, .
 \end{eqnarray}

\subsubsection{Contributions to (\fixform{$\fR^k$}) up to $\omega^{k+1}$ order}
\label{sec:classificationRpq}
Consider a product of $\fR$ of the form $(\fR^k)$ at constant $r$ (i.e. $dr=0$). We will classify all the structures appearing at order $\omega^{k-1}, \omega^k$ and $\omega^{k+1}$. The reason why $\omega^{k-1}$ is the lowest non-trivial order is because $\fchi_2\propto dr$.

Before we start the classifications, we remind the readers of the possible structures in a single product of curvature two-forms denoted as Case $\barA, \barB, \barC, \barD$ or $\barE$ (as we did in the Einstein sources in Appendix D.6 of Ref.~\cite{Azeyanagi:2013xea}) :
 \bea\label{eq:casestructures}
\mbox{Case }\barA&:& \left(\form{\upsilon}\,\fchi \, \form{\upsilon}\,\fchi
\ldots \form{\upsilon}\,\fchi \,\form{\upsilon}\right)\,, \nonumber \\
\mbox{Case }\barB&:& \left(\form{\upsilon}\,\fchi\form{\upsilon}\,\fchi
\ldots \form{\upsilon}\,\fchi\right)\,,\quad
 \left(\fchi\form{\upsilon}\,\fchi\,\form{\upsilon}
\ldots \fchi\,\form{\upsilon}\right)\, , 
\nonumber\\
\mbox{Case }\barC&:&  \left(\fchi\,\form{\upsilon}\,\fchi\,\form{\upsilon}
\ldots \fchi\,\form{\upsilon}\,\fchi\right)\,,   \nonumber \\
\mbox{Case }\barD&:&  \left(\form{\upsilon}\right)\,, \nonumber \\
\mbox{Case }\barE&:&  \left(\fchi\right)\, .
\eea We remind the readers of the definition of $\fchi_m$ (see Eq.~\eqref{eq:fchin}) which denotes a product of $(\fR)$'s consisting of $m$ number of $(\R{0})$'s with the remaining $(\fR)$'s being $(\R{1})$'s.
The symbol $\form{\upsilon}$ is defined to represent a string made of 2nd or higher order terms 
 $(\R{m})$ with $m\ge2$.
Now, let us start analyzing Case $\barA$ to Case $\barE$ one by one.
We further note that whenever we encounter two $\fchi_1$'s somewhere in the wedge product, 
the only non-zero cases are
\be\label{eq:fchi1fchi1}
(\fchi_1\ldots \fchi_1)=\left\{(\R{0}\ldots \R{0}),
\quad
(\tilde{\fchi}_1\ldots \R{0}),\quad
(\R{0}\ldots \tilde{\fchi}_1)\right\},
\ee  where we define $\tilde{\fchi_1}$ to be the $\fchi_1$ containing more than one $\fR$'s, i.e.
\be\label{eq:fchitilde}
\tilde{\fchi_1} \equiv \left\{ (\R{0}\R{1}^{k-1});\quad (\R{1}^{k-1}\R{0}) \right\}\, . 
\ee  We note that since $(\R{1}\R{0}\R{1}) \propto dr$ (see Eq.~\eqref{eq:R101}), the above are the only two possibilities in $\tilde{\fchi_1}$. 
Furthermore, it is useful to know that $\tilde{\fchi_1}\propto \fu$.
Finally, another helpful fact is that $(\R{2}\fchi_1)$ and $(\fchi_1 \R{2})$ 
reduce to $(\R{2}\R{0})$ and $(\R{0}\R{2})$ respectively.\\
\vspace{0.01cm} \\
\noindent
{\bf \underline{Case $\barA$ :}} \\
\noindent
Since each  $\form{\upsilon}$ is at least of order $\omega^2$
and all $\fchi$'s need to be either $\fchi_0$ or $\fchi_1$, 
the product is in total of order  $\omega^{k+1}$ or higher. 
At $\omega^{k+1}$-th order, 
the only nontrivial case is when all $\form{\upsilon}$'s are $(\R{2})$'s 
and all $\fchi$'s are $\fchi_1$'s, i.e. $(\fR^k)=(\R{2}\,\fchi_1\,\R{2}\,\fchi_1\ldots \R{2} \fchi_1\, \R{2})$. 
However, because of $(\R{2}\,\fchi_1)\propto (\R{2}\R{0})\propto \fu$ and $(\fchi_1 \R{2})\propto (\R{0}\R{2})\propto \fu$, we have 
\be\label{eq:useful51}
(\fchi_1\, \R{2} \ldots \fchi_1 \,\R{2})=(\R{2}\, \fchi_1  \ldots \fchi_1 \,\R{2})=(\R{2} \,\fchi_1  \ldots \R{2}\,\fchi_1 )=0\,.
\ee 
 Thus, we conclude that case $\barA$ only has one non-trivial structure
\be
{\cal O}( \omega^{k+1}):\quad (\fR^k)=(\R{2}\R{0}\R{2})\,.
\ee
\noindent
{\bf \underline{Case $\barB$ :}} \\
\noindent
For case $\barB$, the lowest order starts at $\omega^{k}$ for which all $\form{\upsilon}$'s 
are $(\R{2})$'s and all $\fchi$'s are $\fchi_1$'s, i.e. $(\fR^k)=(\R{2}\fchi_1\R{2}\fchi_1\ldots \R{2} \fchi_1)$ 
or $(\fchi_1\R{2}\fchi_1\R{2}\fchi_1\ldots \R{2} )$.  
As in the arguments in Case $\barA$, this type of the product is non-zero only
 when $(\fR^k)=(\R{2}\fchi_1)$ or $(\fR^k)=(\fchi_1 \R{2})$.

For the order $\omega^{k+1}$ contributions, there is a  few possible structures 
listed below :
\bea
&&(\R{3} \fchi_1)\,,\quad( \fchi_1 \R{3})\, , \quad
(\R{2}\fchi_0 )\, , \quad
(\R{2}\R{2}\fchi_1 )\, , \quad
(\fchi_0 \R{2})\, , \quad
(\fchi_1\R{2}\R{2} )\, , \quad
\eea  where we have used Eq.~(\ref{eq:useful51}) and
\be\label{eq:useful52}
(\fchi_1 \R{2} \fchi_1)
=(\fchi_1\R{2}\fchi_0)=
(\fchi_0 \R{2} \fchi_1)
=
(\R{2}\fchi_1 \ldots \R{m} \fchi_1)
=
(\fchi_1\R{2} \ldots \R{m} \fchi_1)
=
0\, ,
\ee for any $m\ge 0$.
 Thus, in summary we have
\bea
 {\cal O}( \omega^{k}) :\quad (\fR^k)&=&\left\{ (\R{2}\R{0}),\quad
(\R{0}\R{2})\right\} \, , \nonumber\\
 {\cal O}( \omega^{k+1}) :\quad (\fR^k)&=&
 \left\{(\R{3} \fchi_1)\,,  \quad( \fchi_1 \R{3})\, , \quad
(\R{2}\fchi_0 )\, , \quad
(\R{2}\R{2}\R{0} )\, , \quad
\right.
\nonumber\\
&&\left.
(\fchi_0 \R{2})\, , \quad
(\R{0}\R{2}\R{2} )\right\}.
\eea
\noindent
{\bf \underline{Case $\barC$ :}} \\
\noindent
For case $\barC$, the lowest order starts at order $\omega^{k-1}$ for which all $\form{\upsilon}$'s
 are $(\R{2})$'s and all $\fchi$'s are $\fchi_1$'s, i.e. $(\fR^q)=(\fchi_1\R{2}\fchi_1\R{2}\fchi_1\ldots \R{2} \fchi_1)$. However, this product is zero since $(\fchi_1 \R{2}\fchi_1)=0$. 

The order $\omega^{k}$ terms consist of
\be
 (\R{0} \R{3} \R{0})\,,
\ee where we have used Eq.~(\ref{eq:fchi1fchi1}), Eq.~(\ref{eq:useful51}), Eq.~(\ref{eq:useful52}) and
\be
(\tilde{\fchi}_1  \R{m} \R{0})=
(   \R{0}\R{m}\tilde{\fchi}_1)=
0\, ,
\ee for all $m\ge0$.

Similarly, the order $\omega^{k+1}$ terms are
\bea
&&
(\R{0} \R{4}\R{0}) \, , \quad
 (\fchi_0 \R{3}\fchi_1)\, ,  \quad
(\fchi_1 \R{3}\fchi_0)\, , 
\nonumber\\
&&
 (\fchi_0 \R{2}\fchi_0) \, ,\quad
   (\fchi_0 \R{2}\R{2}\fchi_1) \,,\quad
  (\fchi_1 \R{2}\R{2}\fchi_0) \,,\quad
\eea where we have used Eq.~(\ref{eq:useful51}) and Eq.~(\ref{eq:useful52}).
To summarize, we have
\bea
{\cal O}( \omega^{k}):\quad (\fR^k)&=&
\left\{
 (\R{0} \R{3} \R{0})
\right\} \nonumber\\
 {\cal O}( \omega^{k+1}) :\quad (\fR^k)&=&
\left\{
  (\R{0} \R{4}\R{0}) \, ,\quad
 (\fchi_0 \R{3}\fchi_1)\,, \quad
(\fchi_1 \R{3}\fchi_0)\, , \right.\nonumber\\
&&\left.
 (\fchi_0 \R{2}\fchi_0) \, ,\quad
   (\fchi_0 \R{2}\R{2}\R{0}) \,,\quad
  (\R{0} \R{2}\R{2}\fchi_0) \right\}.
\eea
\noindent
{\bf \underline{Case $\barD$ :}} 

We note that in this case,  each  $\form{\upsilon}$ is at least of order $\omega^2$ and hence the total order is at least $\omega^{2 k}$. For $k\ge1$, this means that it could only contribute to order $\omega^{k+1}$ for $k=1$ (i.e. $(\fR^k)=\fR$)  where $\form{\upsilon} =(\R{2})$. Thus we conclude that Case $\barD$ contains only one structure
\be
{\cal O}( \omega^{k+1}) :\quad (\fR^k)=(\R{2})\,.
\ee

\noindent
{\bf \underline{Case $\barE$ :}} \\
\noindent
For case $\barE$, we can have the following two structures
\bea
{\cal O}( \omega^{k-1})& :&\quad (\fR^k)=(\fchi_1)=\left\{
(\R{0})\, , \quad \tilde{\fchi_1}
\right\} \,,\nonumber\\
 {\cal O}( \omega^{k})&:&\quad (\fR^k)=(\fchi_0)\,,
\eea

At this point, we would like to summarize the results from the above analysis.
Before doing so, we first introduce a useful symbol
\be
(\cR^q_{(p)})
\ee to denote all possible structures (including those consist of zeroth and first order 
building blocks only) that can contribute to $(\fR^q)$ at $\omega^p$ order. Thus, the summary of the results from above could be stated as :
\bea\label{eq:classification0}
(\cR^q_{(q-1)})&\equiv&
 (\fchi_1)\,, \nonumber\\
(\cR^q_{(q)})&\equiv&\left\{ 
(\fchi_0)\, ,\quad 
 (\R{2} \R{0})\,,\quad
 (\R{0} \R{2} )\,,\quad
 (\R{0} \R{3} \R{0})\right\} \,, \nonumber\\
(\cR^q_{(q+1)})&\equiv&
\left\{
 (\R{2}) \,,\quad
  (\R{2}\R{0}\R{2}) \,,\right.
   \nonumber\\
       &&
 ( \R{3}\fchi_1) \,,\quad
(\fchi_1 \R{3}) \,,\quad
\nonumber\\
&&
 (\R{2}\fchi_0) \,,\quad
      (\R{2}\R{2}\R{0}) \,,\quad
            (\R{0}\R{2}\R{2}) \,,\quad
  (\fchi_0 \R{2}) \,,\quad
\nonumber\\
&&
 (\R{0} \R{4}\R{0}) \,,
\nonumber\\
&&
 (\fchi_0 \R{3}\fchi_1)\,, \quad
(\fchi_1 \R{3}\fchi_0)\,,\nonumber\\
&&\left.
 (\fchi_0 \R{2}\fchi_0) \,,\quad
   (\fchi_0 \R{2}\R{2}\R{0}) \,,\quad
  (\R{0} \R{2}\R{2}\fchi_0) \right\}.
\eea 
We remind the readers that in the Einstein source computations in Appendix D.6 of Ref.~\cite{Azeyanagi:2013xea}, it was proved that 2nd and higher order terms 
in $\fR$ do not contribute to $(\fR^q)$ up to order $\omega^{q-1}$. 
This is why 2nd and higher order terms in $\fR$ do not appear in $(\cR^q_{(q-1)})$.  
Furthermore, up to  $\omega^{q-2}$,  all contributions (including zeroth and first order building blocks) to 
$\fR^q$ vanish.

\subsubsection{Structures of \fixform{$\tr[\nabla \xi \fR^{2k-1}]$} and \fixform{$\tr[\delta \fGamma \fR^{2k-1}]$}}
\label{sec:trgradxiR}
Let us consider the traces $\tr[\nabla \xi \fR^{2k-1}]$ and $\tr[\delta \fGamma \fR^{2k-1}]$ at constant $r$ (i.e. $dr=0$). We will classify the various structures appearing up to order $\omega^{2k-1}$.

 First, due to $\fchi_2 \propto dr$ and
\bea
&&\tr[({}^{(0)} \delta \fGamma)( \R{0}\R{1}^{2k-2})]
=\tr[({}^{(0)} \delta \fGamma)(\R{1}^{2k} \R{0})]=0\, , \nonumber\\
&&\tr[({}^{(0)} \nabla \xi) (\R{0}\R{1}^{2k-2})]
=\tr[({}^{(0)}  \nabla \xi)(\R{1}^{2k} \R{0})]
=0\, ,
\eea 
(for $k\ge1$), the lowest-order contributions to these traces are
\bea
 {\cal O}( \omega^{2k-1}) :
 \quad \tr[(\delta \fGamma)  \fR^{2k-1}]
 &=&\{ \quad
 \tr[({}^{(0)} \delta \fGamma)(\cR^{2k-1}_{(2k-1)})]\, , \quad
 \tr[({}^{(1)} \delta \fGamma)(\fchi_1)
 \quad\}\, ,  \nonumber\\
  \tr[(\nabla \xi) \fR^{2k-1}]&=&
  \{ \quad
 \tr[({}^{(0)}\nabla \xi)(\cR^{2k-1}_{(2k-1)})]\, , \quad
 \tr[({}^{(1)} \nabla \xi)(\fchi_1)]\quad
 \}\,.
\eea
Once one substitutes the possible structures of $(\cR^{2k-1}_{(2k-1)})$ from Eq.~(\ref{eq:classification0}), 
using Eq.~(\ref{eq:R0gradxiR0}) and
\be
\tr[
({}^{(0)} \nabla \xi) (\R{0}\R{2})
]=
\tr[
({}^{(0)} \nabla \xi) (\R{2}\R{0})
]=0\,,
\ee
one finds that 
\bea
 {\cal O}( \omega^{2k-1}) :
 \quad \tr[(\delta \fGamma)  \fR^{2k-1}]
 &=&\{ 
 \tr[({}^{(0)} \delta \fGamma)(\fchi_0)]\, , \quad
 \tr[({}^{(1)} \delta \fGamma)(\fchi_1)]
 \}\, ,  \nonumber\\
  \tr[(\nabla \xi) \fR^{2k-1}]&=&
  \{ 
 \tr[({}^{(0)}\nabla \xi)(\fchi_0]\, , \quad
 \tr[({}^{(1)} \nabla \xi)(\fchi_1)]
 \}\,.
\eea  
Thus, we conclude that 
$\tr[\nabla \xi \fR^{2k-1}]$ and $\tr[\delta \fGamma \fR^{2k-1}]$ start at $\omega^{2k-1}$ order 
and that second or higher order building blocks do not contribute to
 $\tr[\nabla \xi \fR^{2k-1}]$ and $\tr[\delta \fGamma \fR^{2k-1}]$ at order $\omega^{2k-1}$.

\subsection{Behavior at horizon}

In this part, we summarize various quantities evaluated at the horizon and some relation valid at horizon.
In the rest of this Appendix, we assume that all equalities are evaluated at the horizon and ignore terms proportional to $dr$. 

\subsubsection{0th and 1st order terms}
In particular, the metric at the horizon and its inverse simplify as
\bea\label{eq:metrichor}
&&G_{rr} = 0\,,\quad
G_{r\mu} = - u_\mu \,,\quad
G_{\mu\nu}= r_H^2 P_{\mu\nu}\, , \quad \nonumber \\
&&G^{rr}= 0 \,,\quad
G^{r\mu}= u^\mu\,,\quad
G^{\mu\nu}=r_H^{-2} P^{\mu\nu}\,.\nonumber
\eea The gauge field and the connection 1-form at the 0th and 1st orders are evaluated at the horizon as
\bea\label{eq:Gammahor}
\fA&=& \mu \fu\, , \nonumber\\
\fGamma^r{}_r&=& (2\pi T) \fu\,,\quad
\fGamma^r{}_\rho = 0
\,,\quad
\fGamma^\mu{}_r=r_H^{-2}\left(
r_H P^\mu{}_\nu+\omega^{\mu}{}_\nu
\right)dx^\nu\,, \nonumber\\
\fGamma^\mu{}_\rho&=&
\left[u^\mu u_\rho (2\pi T) -\omega^\mu{}_\rho \right]\fu
-r_H u^\mu P_{\rho\nu}dx^\nu
- u_\rho \omega^\mu{}_\nu dx^\nu\,,
\eea while its variation is
\begin{equation}\begin{split}
{}^{(0)} \delta \fA&= \delta(\mu \fu) \, ,\\
{}^{(0)}\delta \fGamma^r{}_r&=
\delta[ (2\pi T)\fu]\,,\quad
{}^{(0)}\delta \fGamma^r{}_\rho
=0\,,\\
{}^{(0)}\delta \fGamma^\mu{}_r&=
r_H^{-2}\left(
r_H u^\mu \delta \fu
+r_H  \delta u^\mu \fu
\right)-r_H^{-2}
(\delta r_H) (P^\mu{}_\nu dx^\nu)
\,, \\
{}^{(0)}\delta \fGamma^\mu{}_\rho&=
\left[ (u_\rho  \delta u^\mu+u^\mu \delta u_\rho )(2\pi T)
+u^\mu u_\rho  \delta (2\pi T)
\right]\fu
+u^\mu u_\rho (2\pi T)\delta \fu
\\
& 
-\delta r_H u^\mu P_{\rho\nu}dx^\nu
-r_H \delta u^\mu P_{\rho\nu}dx^\nu
-r_H u^\mu \delta u_\rho \fu
-r_H u^\mu u_\rho \delta \fu
\,,
\end{split}\end{equation}
for the 0th order part and
\bea
{}^{(1)} \delta \fA&=& 0\, ,\nonumber\\
{}^{(1)}\delta \fGamma^r{}_r&=&
{}^{(1)}\delta \fGamma^r{}_\rho=0
\,,
\nonumber\\
{}^{(1)}\delta \fGamma^\mu{}_r&=&
r_H^{-2}\left(
\delta \omega^\mu{}_\nu dx^\nu
\right)-2r_H^{-3}
(\delta r_H) ( \omega^\mu{}_\nu dx^\nu)
\,, \nonumber\\
{}^{(1)}\delta \fGamma^\mu{}_\rho&=&
-\omega^\mu{}_\rho \delta \fu
-\delta u_\rho \omega^\mu{}_\nu dx^\nu
-u_\rho \delta \omega^\mu{}_\nu dx^\nu
-\delta \omega^\mu{}_\rho \fu
\,,
\eea
for the 1st order part. 

As for the covariant derivative of $\xi^a$ at the horizon,
we have at the lowest order of derivative expansion :
\be\label{eq:gradxi}
({}^{(0)}\xi^a)\partial_a = \frac{1}{T}u^{\mu}\partial_{\mu}\,,~~
({}^{(0)} \nabla_r \xi^r)=-2\pi\,,~~
({}^{(0)}\nabla_\mu \xi^r)=({}^{(0)} \nabla_r \xi^\rho)=0\,,~~
({}^{(0)}\nabla_\mu \xi^\nu)=- 2\pi u_\mu u^\nu\,.
\ee
We note that there is no 1st order term in the above equation and
higher order terms of the derivative expansion start with the 2nd order.
Therefore, we will drop the `$(0)$' superscript on $\nabla \xi$ in the discussions in this subsection.

Let us consider the products of $\fF$'s or $\fR$'s evaluated at the horizon. 
Since $(\F{0})\propto dr\wedge \fu$, we have that
\be\label{eq:Fhor}
(\fF^k) =(\F{1}^k)= (2\mu \fomega)^k.
\ee For the product of $\fR$'s, all products containing more than one $(\R{0})$ vanish or are proportional to $dr$, i.e. $\fchi_m=0$ or proportional to $dr$ for $m\ge 2$.
We next consider the products of the curvature 2-form containing only one $(\R{0})$, i. e. $\fchi_1$ terms.
At the horizon, nontrivial components of this kind of the products are (for $k\ge 0$)
\bea
\label{eq:chi1horizon}
(\R{0}\R{1}^k)^{\mu}{}_r &=& (r_H^{-1} (2\pi T))dx^{\mu}\wedge \fu \wedge (2(2\pi T)\fomega)^k\,, \nonumber\\
(\R{1}^k\R{0})^{\mu}{}_{\rho} &=& 
(-1)^{(k+1)}r_H(2\pi T) (2(2\pi T)\fomega)^{k} \wedge (u^{\mu}P_{\rho\nu}dx^{\nu})\wedge \fu
\,. 
\eea
We also note that the existence of more than one $\fchi_1$ in a given term implies that such term vanishes, since all the nonzero elements in $\fchi_1$ are proportional to $\fu$ at horizon. 

For the products with $(\R{1})$'s only,  i.e. $\fchi_0$ terms, we have
\bea
(\R{1}^m)^r{}_{r} &=& (2(2\pi T) \fomega)^m \, , \nonumber \\
(\R{1}^m)^\mu{}_{r} &=&
(\ldots)\wedge \omega^\mu{}_\sigma dx^\sigma \wedge \fu\,, \nonumber \\
  (\R{1}^m)^r{}_{\rho} &=&0
  \,, \\
(\R{1}^m)^{\ \mu}{}_\rho &=&
(-1)^{m+1} (2(2\pi T) \fomega)^m u^\mu u_\rho
+(\ldots)\wedge u^\mu \omega_{\rho\nu}dx^\nu \wedge \fu
\, . \nonumber 
\eea 
The explicit form of $(\ldots)$'s appearing in the above expressions
are not need for our calculation.

The following relations on traces are also useful
 \bea\label{eq:superuseful}
\tr[ \delta \fGamma \nabla \xi]
&=&
-(2\pi ) \left[
2\delta (2\pi T \fu)
 - r_H \delta \fu
+ u_\mu \delta \omega^\mu{}_\nu dx^\nu
\right] \, ,   \label{eq:hordeltaGgradxi}\\
\tr[\nabla \xi \fR] 
&=&-(4\pi)\times  (2\pi T)(2\fomega)\, ,\label{eq:horgradxiR} \\
\tr[\delta \fGamma\fR]
&=&
    r_H \fu \wedge (2(2\pi T)\delta \fomega)
+(2\delta (2\pi T \fu)-r_H \delta \fu)\wedge(2(2\pi T)\fomega),\label{eq:hordeltaGR}
\\
\tr[ \nabla \xi (\R{1}^{2k})] &=& 0 \label{eq:R1hor}\, ,\\
\tr[\nabla \xi (\R{1}^{2k+1})] &=&-(2\pi)\times 2(2(2\pi T)\fomega)^{2k+1}\,,\\
\tr[({}^{(0)} \delta \fGamma) (\R{1}^{2k+1})] &=&
(2 \delta(2\pi T\fu) -r_H \delta \fu)
\wedge(2(2\pi T)\fomega)^{2k+1}\, \\
\tr[({}^{(0)} \delta \fGamma)( \R{0}\R{1}^{2k-2})]
&=&\tr[({}^{(0)} \delta \fGamma)(\R{1}^{2k} \R{0})]=0\, , \label{eq:usefulrelation105}\\
\tr[({}^{(0)} \nabla \xi) (\R{0}\R{1}^{2k-2})]
&=&\tr[({}^{(0)}  \nabla \xi)(\R{1}^{2k} \R{0})]
=0\,  \label{eq:usefulrelation106}.
\eea 
Here  we stress that only 0th and 1st order terms are considered 
for $\fR$, $\delta\fGamma$ and $\nabla\xi$, while all terms proportional to $dr$ have been dropped.
Furthermore, at the horizon, we can prove the following relations :
\bea
&&
\delta \fGamma^r{}_\rho =0\,,
\quad
u_\mu u^\rho \delta \fGamma^\mu{}_\rho=
\delta (2\pi T \fu)
-r_H \delta \fu
+u_\mu (\delta \omega^\mu{}_\nu dx^\nu)\,,
\nonumber\\
&&
u_\mu (\R{0}\R{1}^k)^\mu{}_r =
u^\rho (\R{1}^k\R{0})^\mu{}_\rho=0\,,\nonumber\\
&&
u_\mu (\R{1}^m)^\mu{}_r = 0\,,\quad
u^\rho (\R{1}^m)^\mu{}_\rho =(-1)^m u^\mu (2(2\pi T)\fomega)\,,\nonumber\\
&&
({}^{(0)} \delta \fGamma^\mu{}_\rho) u^\rho \omega_{\mu\nu} dx^\nu \wedge \fu=0\, , 
\eea
\bea\label{eq:someeq1}
&&
\delta\fGamma^c{}_d(\fchi_1)^d{}_a (\nabla_b\xi^a)=
( \nabla_b\xi^a)(\R{0}\R{1}^k)^b{}_c
=( \nabla_b\xi^a)(\R{1}^{k}\R{0})^b{}_r
=( \nabla_b\xi^r)(\R{1}^{k}\R{0})^b{}_c
=0\,,\nonumber\\
&&( \nabla_b\xi^\alpha)(\R{1}^{k}\R{0})^b{}_\gamma ({}^{(0)}\delta\fGamma^\gamma{}_\delta)
\propto 
u^\alpha P_{\delta \nu}dx^\nu \wedge \fu\,, 
\eea
and in particular it follows that 
\be\label{eq:someeq111}
\tr[(\nabla\xi)\fchi_1]=
\tr[( \nabla \xi) \fchi_1 {}^{(0)}\delta \fGamma]=(\nabla_b\xi^a)(\R{1}^{k}\R{0})^b{}_c {}^{(0)}\delta\fGamma^c{}_d (\fchi_0)^d{}_a
=0\,.
\ee  
Another useful identity is
 \be\label{eq:anotheruseful1}
 (\delta \fGamma)^c{}_d(\fchi_1)^d{}_a(\nabla_b\xi^a)=0\,.
\ee

For computing the Chern-Simons contribution to the entropy using Tachikawa-like formula, we also need
\be\label{eq:trgradxiGamma}
\tr[(\nabla \xi)\fGamma]=-(4\pi)(2\pi T)\fu.
\ee

\subsubsection{2nd and higher order terms}
Now we summarize some useful relations relevant to 2nd and higher order terms evaluated 
at the horizon.  
First, since $(\F{2})\propto dr$, we do not need to deal with it. 
For product of curvature two-forms, one of the most important relations for our our purpose is 
$\tr[\R{2}\fchi_1]=0$ which is valid for any $r$. 
Moreover, the non-zero components of $(\R{0}\R{2})$ and $(\R{2}\R{0})$ are
\be
(\R{0}\R{2})^\mu{}_\rho\,,\, (\R{2}\R{0})^\mu{}_\rho 
\propto \fu  \wedge (P^{\mu}{}_{\sigma}dx^\sigma) \wedge(P_{\rho\nu}dx^\nu)\,, 
\ee which lead to
\be
(\fchi_0 \R{0}\R{2})=
(\fchi_0 \R{2}\R{0})
=( \R{0}\R{2} \fchi_0)
=( \R{2}\R{0} \fchi_0)
=0\,,
\ee and
\be\label{eq:deltaGamma02}
\tr[({}^{(0)}\delta \fGamma)(\R{0}\R{2})]
=\tr[({}^{(0)}\delta \fGamma)(\R{2}\R{0})]
=0\,.
\ee
 
Furthermore, in $(\R{2}\fchi_1)$ and $(\fchi_1 \R{2})$, the only non-zero objects are (for $k\ge0$)
\be
(\R{2}\R{0}\R{1}^k)\, ,\quad
(\R{2}\R{1}^k \R{0})\sim (\R{2}\R{0})\,,\quad
(\R{1}^k \R{0} \R{2})\,, \quad
 (\R{0}\R{1}^k \R{2})\sim (\R{0}\R{2})\,,
 \ee leading to
 \be\label{eq:fchi02fchi1}
(\fchi_0 \R{2} \fchi_1)=(\fchi_1 \R{2} \fchi_0)=0\,.
\ee 
We note that the non-zero possibilities in $(\R{2}\fchi_1)$ and $(\fchi_1\R{2})$ are reducible to
$ (\R{2}\R{0})$ and $(\R{0}\R{2})$ (wedged by an appropriate power of $(2\Phi_T \fomega)$).

The following relations are also useful :
\bea\label{eq:anotheruseful2}
&&({}^{(0)} \nabla_b\xi^a)(\R{0}\R{2})^b{}_c
=(\R{0}\R{2})^a{}_b({}^{(0)} \nabla_c\xi^b)
=({}^{(0)} \nabla_b\xi^a)(\R{2}\R{0})^b{}_c
=(\R{2}\R{0})^a{}_b({}^{(0)} \nabla_c\xi^b)
=0\,,  \nonumber \\
&&\fu \wedge (P_{\mu\nu} dx^\nu) \wedge (\R{2})^{\mu}{}_\rho u^\rho=0\,.
\label{eq:special2prop}
\eea

\subsection{Asymptotic fall-offs at boundary}
\label{sec:infinitybehavior}
We summarize the asymptotic behavior of some quantities at $r=\infty$. 
We note here again that here we ignore terms proportional to $dr$. 

\subsubsection{0th and 1st order terms}
In this subsection, all building blocks are considered up to first order and in particular we drop their superscript that we usually use to denote their derivative orders.
For $\Phi$, $\Phi_T$ and $\Psi$, we have 
\bea \label{eq:fields_limitinfty}
\Phi\atinfty &=& \frac{q}{r^{d-2}},\quad 
\Psi\atinfty\rightarrow \frac{r^2}{2}
-\frac{m}{2 r^{d-2}}+{\cal O}\left(\frac{1}{r^{2(d-2)}}\right),\quad \Phi_T\atinfty \rightarrow \frac{ m d}{2 r^{d-1}}+{\cal O}\left(\frac{1}{r^{2d-3}}\right)\,,\nonumber\\
\delta\Phi\atinfty& =& \frac{\delta q}{r^{d-2}}\,,\quad 
\delta \Psi\atinfty\rightarrow 
-\frac{\delta m}{2 r^{d-2}}+{\cal O}\left(\frac{1}{r^{2(d-2)}}\right)
,\quad \delta \Phi_T\atinfty \rightarrow \frac{ (\delta m)d }{2 r^{d-1}}+{\cal O}\left(\frac{1}{r^{2d-3}}\right).\nonumber\\
\eea  
For $\nabla_a\xi^b$, from \eqref{eq:nablaxigeneralr} we have 
\bea
\nabla_r \xi^r \atinfty&=& r u_\beta \xi^\beta+{\cal O}(r^{-d+1})\, , \qquad 
\nabla_r \xi^\mu \atinfty = 
r^{-1}P^\mu{}_\beta \xi^\beta
+{\cal O}(r^{-2})\, , \\
\nabla_\mu \xi^r \atinfty&=&
-r^3\eta_{\mu\beta}\xi^\beta
+
{\cal O}(r^{-d+3}) \, , \qquad 
\nabla_\mu \xi^\nu\atinfty=
-r u^\nu \eta_{\mu\beta}
\xi^\beta
+\partial_\mu \xi^\nu
+ {\cal O}(r^{-d+1})\, , \nonumber 
\eea and the gauge field and the connection one-form are
\bea\label{eq:falloffGamma0}
\fA \atinfty& =&  r^{-d+2}(q\fu)\, ,\nonumber\\
\fGamma^r{}_r \atinfty&=& r \fu +{\cal O}(r^{-d+1})\, , \qquad
\fGamma^r{}_\rho \atinfty=-r^3
 \eta_{\rho\nu} 
dx^\nu
+O(r^{-d+3}) \, , 
\nonumber\\ 
\fGamma^\mu{}_r \atinfty&=& r^{-1} P^\mu{}_\nu + r^{-2} \fomega^\mu{}_\nu dx^\nu\, , \qquad
\fGamma^\mu{}_\rho \atinfty=
-r u^\mu \eta_{\rho \nu}dx^\nu
 + {\cal O}(r^{-d+1})\, .
\eea Now assuming $\delta r=0$, we have the variation of the gauge field and the connection one-form 
as follows:
\bea\label{eq:falloffconnection}
\delta \fA\atinfty& =& r^{-d+2}\delta (q \fu) \, ,\nonumber\\
\delta \fGamma^r{}_r \atinfty&=& r (\delta \fu) +{\cal O}(r^{-d+1})\, , \qquad 
\delta \fGamma^r{}_\rho\atinfty =
O(r^{-d+3}) \, , 
\\ 
\delta\fGamma^\mu{}_r \atinfty&=& r^{-1} (u^\mu \delta \fu+ \delta u^\mu \fu) + r^{-2} (\delta \fomega^\mu{}_\nu) dx^\nu\, , \nonumber\\
\delta \fGamma^\mu{}_\rho \atinfty&=&
-r (\delta u^\mu) \eta_{\rho \nu}dx^\nu
 + {\cal O}(r^{-d+1})\, . \nonumber 
\eea For the 0th and 1st order terms in the field strength and curvature two-form, from the explicit form of these terms, 
we have  (note that $(\F{0})\propto dr\wedge \fu$)
\bea\label{eq:usefuluseful1}
(\F{1})\atinfty&=& r^{-d+2}q(2  \fomega)\, ,\nonumber\\
(\R{0})^r{}_r \atinfty& =& 0 \, , \qquad 
(\R{0})^\mu{}_r\atinfty  = dx^\mu \wedge \fu+{\cal O}(r^{-d}) \, ,  \nonumber\\
(\R{0})^r{}_\rho \atinfty &=& 0 \, , \qquad 
(\R{0})^\mu{}_\rho \atinfty =
-r^2
dx^\mu\wedge(
 \eta_{\rho \nu}dx^\nu)
+{\cal O}( r^{-d+2}) \, , \\ 
(\R{1})^r{}_r \atinfty& =& {\cal O}(r^{-d+1}) \, , \qquad 
(\R{1})^\mu{}_r \atinfty={\cal O}(r^{-d-1}) \, , \nonumber\\
(\R{1})^r{}_\rho\atinfty &=&
{\cal O}(r^{-d+3})\, , \qquad 
(\R{1})^\mu{}_\rho \atinfty={\cal O}(r^{-d+1})\, . 
\eea

\subsubsection{Fall-off of 2nd order term at boundary from direct computations}
\label{sec:falloffR2}
For the field strength, since $(\F{2})\propto dr$, we do not need to take this into account 
when we evaluate the fall-off behaviors of $T_1$ and $T_2$ in Appendix.~\ref{sec:T0T1T2hor0th1st}. 

For the the second order terms in the curvature two-form $(\R{2})$, 
we can find their fall-off behaviors from the explicit metric (up to second order) in Eq.~\eqref{eq:2ndOrderMetric}.  
The contributions to $(\R{2})$ purely from the zeroth order metric can be found in Eq.~\eqref{eq:R2prime} and hence we can just take the $r\rightarrow \infty$ limit to see the falloffs :\footnote{To simplify the 
notation, we often neglect the subscript `$\infty$', 
even when a given quantity is evaluated at the boundary. }
\be
(\R{2})^r{}_r |_{G_{(0)}}  =  (\R{2})^r{}_\rho |_{G_{(0)}}  = 0\,,\quad 
(\R{2})^\mu{}_r |_{G_{(0)}} ={\mathcal O}(r^{-d-2})\,,\quad
(\R{2})^\mu{}_\rho |_{G_{(0)}} ={\mathcal O}(r^{-d})\,.
\ee
Here, for a general 2nd order building block $(\B{2})$ (or, ${}^{(2)}\form{\mathcal B}$)
made of the metric 
(and its derivatives), we introduced the notation $(\B{2})|_{G_{(0)}}$
(or, ${}^{(2)}\form{\mathcal B}|_{G_{(0)}}$) 
to denote the contribution containing the zeroth order metrics only. 
The remaining part (which contains a second order metric) is defined 
as  $(\B{2})|_{G_{(2)}}= (\B{2})-(\B{2})|_{G_{(0)}}$
(or in general ${}^{(2)}\form{\mathcal B}|_{G_{(2)}} = {}^{(2)}\form{\mathcal B}-{}^{(2)}\form{\mathcal B}|_{G_{(0)}}$).  

 On the other hand, the rest of the contribution to the second order curvature two-form (in which 
 there is a second order metric and the derivatives inside of the definition of 
 the curvature two-form are all $\partial_r$) at the boundary is evaluated as follows:
 We recall the definition of the curvature 2-form 
$(\fR)^a{}_b= d
\fGamma^a{}_b +\fGamma^a{}_c \wedge \fGamma^c{}_b$. 
The second order metric can not contribute to 
$(\R{2})^a{}_b$ through the $d\fGamma^a{}_b$ term (at order $\omega^2$) since $ d[ ({}^{(2)}  \fGamma)^a{}_b |_{G_{(2)}} ]$ 
is proportional to $dr$ or of order $\omega^3$.  
Therefore, we only need to compute the 2nd order metric contribution to the 
term $(\fGamma^a{}_c \wedge \fGamma^c{}_b)$, which is given by
\be
{}^{(2)}(\fGamma^a{}_c \wedge \fGamma^c{}_b) |_{G_{(2)}}
= 
 ({}^{(2)}  \fGamma^a{}_c ) |_{G_{(2)}} 
\wedge
({}^{(0)} \fGamma^c{}_b) 
+
({}^{(0)} \fGamma^a{}_c) \wedge
 ({}^{(2)}  \fGamma^c{}_b ) |_{G_{(2)}}\, .
\ee 
Now, from the explicit second order metric, we deduce the fall-offs of the connection one-form from the second order metric :
\bea
 ({}^{(2)}  \fGamma)^r{}_r |_{G_{(2)}} & = &({}^{(2)}  \fGamma)^\mu{}_\rho |_{G_{(2)}}  ={\mathcal O}(r^{-1})\, ,
\nonumber\\
 ({}^{(2)}  \fGamma)^\mu{}_r |_{G_{(2)}} &=&  {\mathcal O}(r^{-3}) ,\quad
 ({}^{(2)}  \fGamma) ^r{}_\rho |_{G_{(2)}} = {\mathcal O}(r)\, .
\eea Combining above and the fall-offs of $({}^{(0)} \fGamma^a{}_b)$ from Eq.~(\ref{eq:falloffGamma0}), we obtain that
\be
{}^{(2)}(\fGamma \wedge \fGamma)^r{}_r |_{G_{(2)}} = 
{}^{(2)}(\fGamma \wedge \fGamma)^\mu{}_r |_{G_{(2)}} =  
{}^{(2)}(\fGamma \wedge \fGamma)^\mu{}_\nu |_{G_{(2)}} = {\mathcal O}(r^0)\,,\quad
{}^{(2)}(\fGamma \wedge \fGamma)^r{}_\nu |_{G_{(2)}} ={\mathcal O}(r^2)\,,\quad
\ee and thus
\be
(\R{2})^r{}_r |_{G_{(2)}}=(\R{2})^\mu{}_r |_{G_{(2)}} =(\R{2})^\mu{}_\rho |_{G_{(2)}}  = {\mathcal O}(r^0)\,,\quad
(\R{2})^r{}_\rho |_{G_{(2)}} = {\mathcal O}(r^2)\, .
\ee Therefore, combining the contributions from $G_{(0)}$ and $G_{(2)}$, we find
\be\label{eq:falloffR2}
(\R{2})^r{}_r = (\R{2})^\mu{}_r  =  (\R{2})^\mu{}_\rho = {\mathcal O}(r^0)\,,\quad
(\R{2})^r{}_\rho  ={\mathcal O} (r^2)\,, 
\ee
at infinity. In particular, this indicates that $(\R{2})$ under the rough estimates is 
\be\label{eq:R2falloff}
(\R{2}) \sim r^2\, . 
\ee

We note that in the above derivations, we rely heavily on the explicit form of the second order metric solutions. In general, due to Weyl covariance of the CFT$_{2n}$ and its extension to the bulk, the fall-off in Eq.~\eqref{eq:R2falloff} can be derived without making use of any information about the explicit second order metric solutions. The proof of Eq.~\eqref{eq:R2falloff} using Weyl scaling is presented in Appendix \ref{sec:WeylWeight}.
Therefore, the estimate in Eq.~\eqref{eq:R2falloff}  is in fact valid in more general setups.


\subsection{Explicit structures in \texorpdfstring{AdS$_5$}{AdS5} }
In Appendix~\ref{sec:deltaQcsads5eq0}, we need to deal with traces of products of 
$\delta{\fGamma}^a{}_b$, $\nabla_a\xi^b$ and $\fR$. Here we summarize some 
useful results for the computations of these traces.
We note that we will ignore all order $\omega^2$ and higher order building blocks in this Appendix.

Let us first consider a product of $\delta{\fGamma}^a{}_b$ and $\fR$. 
We have the following results related to this :
\bea\label{eq:deltaGammaR}
\delta \fGamma^r{}_r (\R{0})^r{}_r &=& \delta \fGamma^\rho{}_r (\R{0})^r{}_\rho =
\delta \fGamma^r{}_\mu (\R{0})^\mu{}_r =0\,,\nonumber \\
\delta \fGamma^\mu{}_\rho (\R{0})^\rho{}_\mu &=&
-2r^{-1}   \fu \wedge
\left[
(\delta \Psi) (2\Phi_T\fomega)
+\left(\Psi-\half r^2\right) (2\Phi_T\delta \fomega)
\right] \,,\nonumber\\
\delta \fGamma^r{}_r (\R{1})^r{}_r &=&
(\Psi' \delta \fu +\fu \delta \Psi')\wedge(2\Phi_T\fomega) \,,\nonumber\\
\delta \fGamma^\rho{}_r (\R{1})^r{}_\rho &=&
 -2r^{-1} (\delta \fu) \Psi \wedge (2\Phi_T \fomega )\,,
 \nonumber\\
\delta \fGamma^r{}_\rho (\R{1})^\rho{}_r  &=&
-2r^{-1} (\delta \Psi)  \fu  \wedge(2\Phi_T \fomega)\,,
\nonumber\\
 \delta \fGamma^\mu{}_\rho (\R{1})^\rho{}_\mu &=&
\left[
 (\Psi'\delta \fu+ \fu\delta\Psi' )
 -r \delta \fu
 \right]\wedge(2\Phi_T \fomega)\, .
 \eea   
 For a product of $\nabla_a\xi^b$ and $\fR$, we have 
 \begin{eqnarray}\label{eq:gradxiR}
&& \nabla_r \xi^r (\R{0})^r{}_r=
\nabla_r \xi^\mu (\R{0})^r{}_\mu
= 0\, , \nonumber \\
&&\nabla_\mu \xi^r (\R{0})^\mu{}_r=  
-2    \Psi \Psi' \form{\xi} \wedge \fu\, , 
\nonumber\\
&&\nabla_\mu \xi^\rho (\R{0})^\mu{}_\rho= 
\partial_\mu \xi^\rho
\left[
-r\Phi_T u^\mu \eta_{\rho\beta}dx^\beta \wedge \fu
+2 r^{-1} \Psi  \Psi' u_\rho dx^\mu \wedge \fu
-2 \Psi dx^\mu \wedge P_{\rho\nu} dx^\nu
\right]
\nonumber\\
&&\qquad\qquad\qquad\qquad
+2 
 \Psi  \Psi'  \form{\xi} \wedge \fu
+r^{-2}(2\Psi-r^2)\omega_{\beta\nu} \xi^\beta\left[
r\Phi_T  
+2 \Psi
\right]\fu\wedge dx^\nu \nonumber\\
&&\qquad\qquad\qquad\qquad
+2r^{-2}(2\Psi-r^2)u_{\beta} \xi^\beta
\Psi  (2\fomega)\, , \nonumber \\
&&\nabla_r \xi^r (\R{1})^r{}_r=  \Psi'(u_\beta \xi^\beta) (2\Phi_T \fomega)\,,
 \nonumber\\
&&\nabla_r \xi^\rho (\R{1})^r{}_\rho= r^{-1}   (2\Psi)\fu\wedge(\Phi_T \xi^\beta\omega_{\beta\nu}dx^\nu)\,, \nonumber\\
&&\nabla_\mu \xi^r (\R{1})^\mu{}_r=  r^{-1}
(2\Psi )\fu\wedge(\Phi_T \xi^\beta\omega_{\beta\nu} dx^\nu)\,,
 \nonumber\\
&&\nabla_\mu \xi^\rho(\R{1})^\mu{}_\rho= 
\Psi'( u_\beta  \xi^\beta)
 (2\Phi_T \fomega)\, ,
 \end{eqnarray}

For a product of $\delta\fGamma^a{}_b$ and $\nabla_c\xi^d$, we obtain
\begin{eqnarray} \label{eq:deltaGammagradxi}
\delta \fGamma^r{}_r \nabla_r \xi^r
&=& \Psi' u_\beta \xi^\beta\left[
(\delta \Psi') \fu
+\Psi' \delta \fu\right]+\cal O (\omega)\, , 
\nonumber\\
\delta \fGamma^r{}_\rho \nabla_r \xi^\rho
&=& 
\xi^\beta\left[
-(2\delta \Psi)( P_{\beta\nu} dx^\nu)
+(2\Psi)(r^{-1}\Psi'-1) (\delta u_\beta) \fu
\right]+\cal O (\omega)\, , 
\nonumber\\
\delta \fGamma^\mu{}_r \nabla_\mu \xi^r
&=&
-2\Psi \xi^\beta\left[(\delta u_\beta)
  \fu
+r^{-1} 
 \Psi'  u_\beta
\delta \fu\right]+\cal O (\omega)\, , 
\nonumber\\
\delta \fGamma^\mu{}_\rho \nabla_\mu \xi^\rho
&=&
 \Psi' u_\beta  \xi^\beta 
\left[
   (\delta \Psi') \fu
 +  (\Psi'-r) \delta \fu
\right] +  r \Psi'
 (\xi^\beta\delta u_\beta)  \fu+\cal O (\omega)\, . 
 \eea

\section{Useful relations II}
\label{app:useForm2}
This Appendix summarizes some useful relations that are relevant to the computation of the 
the Einstein-Maxwell part of asymptotic charges and entropy.  

\subsection{Gauge field, metric and Hodge duals}
Here we explicitly compute the Hodge-duals of some quantities on the rotating charged AdS black hole background \eqref{eq:flugravsoln}. 
The results summarized here will be useful for the evaluation of the differential Noether charges. 

For convenience, we start with a summary of the bulk metric and gauge field for the rotating charged AdS black hole solution \eqref{eq:flugravsoln} :
\begin{equation}
\begin{split}
G_{rr} &= 0\ ,\ G_{r\mu} =  - u_\mu \ ,\\ 
G_{\mu\nu} &= -r^2 f u_\mu u_\nu + r^2 P_{\mu\nu}+\ldots + g_{_V} \brk{ u_\mu V_\nu + u_\nu V_\mu }+\ldots \ ,\\
G^{rr} &= r^2 f+\ldots \ ,\qquad
G^{\mu\nu}= \frac{1}{r^2}P^{\mu\nu} +\ldots\ , \\
G^{r\mu} &=   u^\mu + \ldots + \frac{ g_{_V}}{r^2}V^\mu +\ldots\, \\
\fA &= \fA_{\infty}+ \Phi(r,q)\ u_\mu\ dx^\mu + \ldots + a_{_V}(r,m,q)   V_\mu\ dx^\mu +\ldots \, , \\
\end{split}
\end{equation}
where we have retained only the leading order contributions to the parity even and parity odd part in the derivative expansion.
The gauge field strength is given by 
\begin{equation}
\begin{split}
\fF &=  dr\wedge \brk{ \frac{d\Phi}{dr} u_\mu + \ldots 
+ \frac{da_{_V}}{dr}  V_\mu +\ldots } \ dx^\mu + \ldots \, , 
\end{split}
\end{equation} 

It is useful to summarize the Hodge-duals of the following zero forms and one forms :
\begin{equation}
\begin{split}
\hodge 1 
&= \sqrt{-G}\  dr\wedge\hodgeCFT 1= r^{d-1}dr\wedge\hodgeCFT 1+ \ldots\, , \\
\hodge dr 
&=\sqrt{-G}\ G^{rr}\ \hodgeCFT 1+\sqrt{-G}\ G^{r\mu}\eta_{\mu\nu}\ dr\wedge \hodgeCFT dx^\nu \\
&=r^{d+1}f\ \hodgeCFT 1+ r^{d-1}\brk{u_\mu +  \frac{ g_{_V}}{r^2}V_\mu}\ dr\wedge \hodgeCFT dx^\mu
+\ldots \, , \\ 
\hodge dx^\mu 
&=\sqrt{-G}\ G^{\mu r}\ \hodgeCFT 1
- \sqrt{-G}\ G^{\mu \nu}\eta_{\nu\lambda}\ dr\wedge\hodgeCFT dx^\lambda\\ 
&=r^{d-1}(u^\mu+\frac{g_{_V}}{r^2} V^\mu) \ \hodgeCFT 1
- r^{d-3}P^\mu_\nu \ dr\wedge\hodgeCFT dx^\nu + \ldots \,  ,
\end{split}
\end{equation}
where we have dropped again the sub-leading contributions in the fluid/gravity derivative expansion.
Here we also summarize the formulae for Hodge-duals of 2-forms :
\begin{equation}
\begin{split}
\hodge(dr\wedge dx^\mu) 
&=\sqrt{-G}\ \brk{G^{rr}G^{\mu\nu}-G^{r\nu}G^{\mu r}}\eta_{\nu\lambda}
\ \hodgeCFT dx^\lambda\\
&\quad + \sqrt{-G}\ G^{r\alpha}G^{\mu\beta}\eta_{\lambda\alpha}\eta_{\sigma\beta}
\ dr\wedge \hodgeCFT (dx^\lambda \wedge  dx^\sigma)\\
&=
r^{d-1}\brk{fP^\mu_\nu-(u^\mu+\frac{g_{_V}}{r^2} V^\mu)(u_\nu+\frac{g_{_V}}{r^2} V_\nu)}
\ \hodgeCFT dx^\nu \\
&\quad + r^{d-3} \brk{u_\lambda+\frac{g_{_V}}{r^2} V_\lambda}P^\mu_\sigma
\ dr\wedge \hodgeCFT (dx^\lambda \wedge  dx^\sigma)+ \ldots\, ,
\end{split}
\end{equation}
and
\begin{equation}
\begin{split}
\hodge(dx^\mu\wedge dx^\nu) 
&=\sqrt{-G}\ \brk{G^{\mu r}G^{\nu\lambda}-G^{\mu\lambda}G^{\nu r}}\eta_{\lambda\alpha}
\ \hodgeCFT dx^\alpha\\
&\ + \sqrt{-G}\ G^{\mu \alpha}G^{\nu\beta}\eta_{\lambda\alpha}\eta_{\sigma\beta}
\ dr\wedge \hodgeCFT (dx^\lambda \wedge  dx^\sigma)\\ 
&=
r^{d-3}\brk{(u^\mu+\frac{g_{_V}}{r^2} V^\mu)P^\nu_\lambda-(u^\nu+\frac{g_{_V}}{r^2} V^\nu)P^\mu_\lambda }
\ \hodgeCFT dx^\lambda\\
&\ +r^{d-5}\ P^\mu_\alpha P^\nu_\beta
\ dr\wedge \hodgeCFT (dx^\alpha\wedge dx^\beta)+ \ldots\, . 
\end{split}
\end{equation}
From this expression, we can obtain 
\begin{equation}\label{eq:2formsUpDown}
\begin{split}
\hodge(dx^r\wedge dx_r) 
&=-r^{d-1}\prn{u_\mu+\frac{g_{_V}}{r^2} V_\mu}
\ \hodgeCFT dx^\mu + dr \wedge (\ldots) + \ldots \, , \\
\hodge(dx^r\wedge dx_\mu) 
&=r^{d+1}f\ \eta_{\mu\nu}
\ \hodgeCFT dx^\nu + dr \wedge (\ldots)+ \ldots \, , \\
\hodge(dx^\mu \wedge dx_r) 
&=- r^{d-3}P^\mu_\nu
\ \hodgeCFT dx^\nu + dr \wedge (\ldots) + \ldots \, , \\
\hodge(dx^\mu \wedge dx_\nu) 
&=  r^{d-1}\prn{ u^\mu +\frac{g_{_V}}{r^2} V^\mu  }
\eta_{\nu\lambda}\ \hodgeCFT dx^\lambda
+ dr \wedge (\ldots)+ \ldots \, . 
\end{split}
\end{equation} By using this, we can also compute the Hodge-dual of the gauge field strength two-form as 
\begin{equation}\label{eq:hodgeFconstr}
\begin{split}
\hodge\fF 
&=  \brk{ r^{d-1}\frac{d\Phi}{dr} u_\mu + \ldots 
+ r^{d-1}\prn{f\frac{da_{_V}}{dr}+\frac{g_{_V}}{r^2}\frac{d\Phi}{dr} }  V_\mu +\ldots } \  \hodgeCFT dx^\mu  
\\
&
\quad+ r^{d-3} \frac{da_{_V}}{dr}  
\ dr\wedge  (2\fomega)^{n-1}+ \ldots\,  \\
&=  -\brk{(d-2)q\ u_\mu + \ldots 
+ \gYM^2 \mathcal{Q}^\tV\  V_\mu +\ldots } \  \hodgeCFT dx^\mu + dr\wedge (\ldots)\, . 
\end{split}
\end{equation}
The definition and detail of the function $\mathcal{Q}^\tV\ $ is provided in 
Ref.~\cite{Azeyanagi:2013xea}.

For the evaluation of the differential Noether charges at the horizon, it is useful to 
explicitly write down the Hodge-duals of the two-forms evaluated at the horizon 
\begin{equation}\label{eq:2formsUpDownhor}
\begin{split}
\hodge(dx^r\wedge dx_r) \athor
&=-r_H^{d-1}
\ \hodgeCFT \fu + \ldots \, , \\
\hodge(dx^r\wedge dx_\mu)  \athor
&= \ldots \, , \\
\hodge(dx^\mu \wedge dx_r) \athor
&=- r_H^{d-3}P^\mu_\nu
\ \hodgeCFT dx^\nu  + \ldots \, , \\
\hodge(dx^\mu \wedge dx_\nu) \athor
&=  r_H^{d-1}u^\mu
\eta_{\nu\lambda}\ \hodgeCFT dx^\lambda
+  \ldots \, . 
\end{split}
\end{equation} which also lead to  
\begin{equation}\label{eq:2formsUpDownhorixi}
\begin{split}
\ic_{\xi}\hodge(dx^r\wedge dx_r)  \athor
&=
\ic_{\xi} \hodge(dx^r\wedge dx_\mu) \athor
=   \ldots \, , \\
\ic_{\xi}\hodge(dx^\mu \wedge dx_r) \athor
&=- T^{-1} r_H^{d-3}P^\mu_\nu
\  \hodgeCFT (dx^\nu \wedge \fu) + \ldots \, , \\
\ic_{\xi} \hodge(dx^\mu \wedge dx_\nu) \athor
&=  T^{-1} r_H^{d-1}u^\mu
\eta_{\nu\lambda}\hodgeCFT (dx^\lambda\wedge \fu)
+ \ldots \, . 
\end{split}
\end{equation} Here we have used $\xi^\mu \athor=u^\mu/T$.

Another useful identity about Hodge duals is the following : Let $\fN$ be a $p$-form
with legs only along the boundary direction and is completely transverse to the velocity
$u_\mu$. Then, it satisfies
\begin{equation}
\begin{split}
(\hodgeCFT \fN)\wedge dr = \frac{(-1)^d}{r^{d-1-2p}} \hodge \fN \, . 
\end{split}
\end{equation}

\subsection{Christoffel symbols}
In the course of the computation of the differential Noether charge, we will also frequently encounter the evaluation 
of the Christoffel symbols on the rotating charged AdS black hole background. Here we summarize the explicit form of 
the Christoffel symbols with all the indices lowered on this background :
\begin{equation}
\begin{split}
\Gamma_{rrr} &= \Gamma_{rr\mu} = \Gamma_{r\mu r} = \Gamma_{\mu rr} = \Gamma_{\mu\nu\lambda} = 0+\ldots\, ,  \\
\Gamma_{\mu\nu r} &= -\Gamma_{r\mu\nu} = \half \partial_r G_{\mu\nu} +\ldots \, , 
\end{split}
\end{equation}
where we have retained only the leading order contributions in the fluid/gravity derivative expansion. This leads to
\begin{equation}\label{eq:flugravGamma}
\begin{split}
\Gamma^r{}_{rr} &=\Gamma^\mu{}_{rr}= 0 +\ldots \, , \\
\Gamma^r{}_{r\mu} 
&=\half \frac{d(r^2f)}{dr} u_\mu - \half r^2 \frac{d}{dr}\prn{\frac{g_{_V}}{r^2}}V_\mu +\ldots\, , \\
\Gamma^r{}_{\mu\nu} 
&=\half r^2f \frac{d(r^2f)}{dr} u_\mu u_\nu
-r^3 f P_{\mu\nu}
- \half r^2 f\frac{dg_{_V}}{dr}(V_\mu u_\nu+V_\nu u_\mu) +\ldots \, , \\
\Gamma^\mu{}_{\nu r} 
&=\frac{1}{r} P^\mu_\nu +\frac{1}{2r^2} \frac{dg_{_V}}{dr} V^\mu u_\nu  +\ldots \, , \\
\Gamma^\mu{}_{\nu \lambda} 
&=\half\frac{d(r^2f)}{dr} \prn{u^\mu  + \frac{ g_{_V}}{r^2}V^\mu }u_\nu u_\lambda
-r\prn{u^\mu  + \frac{ g_{_V}}{r^2}V^\mu }P_{\nu\lambda}\\
&\qquad - \half \frac{dg_{_V}}{dr}u^\mu \brk{ u_\nu V_\lambda + u_\lambda V_\nu }+\ldots  \, . 
\end{split}
\end{equation}

In particular, at the horizon, the Christoffel symbols simplify to 
\begin{equation}\label{eq:flugravGammahor}
\begin{split}
\Gamma^r{}_{rr} \athor &=\Gamma^\mu{}_{rr} \athor= \Gamma^r{}_{\mu\nu} \athor
=0 +\ldots \, , \\
\Gamma^r{}_{r\mu}  \athor
&=(2\pi T) u_\mu - \half  \left.\left[\frac{dg_{_V}}{dr}\right]\right\athor V_\mu +\ldots\, , \\
\Gamma^\mu{}_{\nu r}  \athor
&=r_H^{-1} P^\mu_\nu +\frac{1}{2r_H^2} \left.\left[\frac{dg_{_V}}{dr}\right] \right\athor V^\mu u_\nu  +\ldots \, , \\
\Gamma^\mu{}_{\nu \lambda} \athor
&=(2\pi T) u^\mu   u_\nu u_\lambda
-r_Hu^\mu  P_{\nu\lambda} - \half\left. \left[\frac{dg_{_V}}{dr} \right]\right\athor u^\mu \brk{ u_\nu V_\lambda + u_\lambda V_\nu }+\ldots  \, . 
\end{split}
\end{equation} where we have used $(1/2)r_H^2 f'(r_H)=\Phi_T(r_H)=(2\pi T)$. 
We can also write down the connection one-form $\fGamma^a_{b}=\Gamma^a_{bc}dx^c$ at the horizon
by using this result :
\begin{equation}\label{eq:flugravGammahorform}
\begin{split}
\fGamma^r{}_{r} \athor &= (2\pi T) \fu
- \half 
\left.\left[\frac{dg_{_V}}{dr}\right]\right\athor \fV+\ldots\, , 
\\
\fGamma^\mu{}_{r} \athor &=
r_H^{-1} P^\mu_\nu dx^\nu +\frac{1}{2r_H^2} \left.\left[\frac{dg_{_V}}{dr}\right] \right\athor V^\mu \fu 
+\ldots\, , 
\\
\fGamma^r{}_{\mu} \athor &= 
\ldots\, , \\
\fGamma^\nu{}_{\mu} \athor&=
(2\pi T) u^\nu    u_\mu \fu 
-r_Hu^\nu  P_{\mu \lambda}dx^\lambda
 - \half\left. \left[\frac{dg_{_V}}{dr} \right]\right\athor u^\nu \prn{ u_\mu \fV +  V_\mu \fu}
+
+\ldots\, . 
\end{split}
\end{equation}

\subsection{\texorpdfstring{$\nabla \xi$}{grad xi}}
Let us consider a vector $\xi^\mu$ along the boundary directions which  is only dependent on the boundary
coordinates. Then, we have
\begin{equation}\label{eq:gradXi}
\begin{split}
\nabla_r\xi^r 
&=\xi^\mu \brk{\half \frac{d(r^2f)}{dr} u_\mu - \half r^2 \frac{d}{dr}\prn{\frac{g_{_V}}{r^2}}V_\mu
+\ldots}\, , \\
\nabla_r \xi^\mu
&=\xi^\nu\brk{\frac{1}{r} P^\mu_\nu +\frac{1}{2r^2} \frac{dg_{_V}}{dr} V^\mu u_\nu  +\ldots}\, , \\
\nabla_\mu \xi^r
&=\xi^\nu\brk{\half r^2f \frac{d(r^2f)}{dr} u_\mu u_\nu
-r^3 f P_{\mu\nu}
- \half r^2 f\frac{dg_{_V}}{dr}(V_\mu u_\nu+V_\nu u_\mu) +\ldots}\, , \\
\nabla_\mu \xi^\nu
&=\xi^\lambda\brk{\half\frac{d(r^2f)}{dr} \prn{u^\nu  + \frac{ g_{_V}}{r^2}V^\nu }u_\mu u_\lambda
-r\prn{u^\nu  + \frac{ g_{_V}}{r^2}V^\nu }P_{\mu\lambda}}\\
&\qquad - \half \xi^\lambda \frac{dg_{_V}}{dr}u^\nu \brk{ u_\mu V_\lambda + u_\lambda V_\mu }+\ldots \, . 
\end{split}
\end{equation}
We note that at the horizon, each components of $\nabla_a \xi^b$ simplifies to 
\begin{equation}\begin{split}\label{eq:gradxihorV}
\nabla_r\xi^r  \athor
&= -2\pi
+\ldots\, , \\
\nabla_r \xi^\mu \athor
&=-T^{-1}\frac{1}{2r_H^2}\left.\left[ \frac{dg_{_V}}{dr}\right]\right\athor V^\mu   +\ldots\, , \\
\nabla_\mu \xi^r \athor
&= \ldots\,, \\
\nabla_\mu \xi^\nu \athor
&=-2\pi u^\nu u_\mu  +\half T^{-1} \left. \left[ \frac{dg_{_V}}{dr}\right]\right\athor u^\nu  V_\mu +\ldots \, , 
\end{split}
\end{equation}
where we have used $\xi^\mu \athor = u^\mu/T$.

\subsection{Some useful relations for Einstein-Maxwell charges}
In the end of this Appendix, we summarize some results useful for the evaluation of the Einstein-Maxwell part of the 
differential Noether charge on rotating charged AdS black hole back ground. 

The first useful result is related to $\nabla_a \xi^b$. 
Contracting these with the 2-forms in Eq.~\eqref{eq:2formsUpDown}, we obtain\footnote{ 
This answer can also be derived  by directly evaluating
\begin{equation}
\begin{split}
 \nabla_a\xi^b &\hodge (dx^a\wedge dx_b)\\ 
&= \hodge d\form{\xi} = \xi^\mu \partial_r G_{\mu\nu} \hodge (dr\wedge dx^\nu)
= \xi^\lambda r^{d-1} (\partial_r G_{\lambda\mu}) 
\brk{fP^\mu_\nu-(u^\mu+\frac{g_{_V}}{r^2} V^\mu)(u_\nu+\frac{g_{_V}}{r^2} V_\nu)}
\ \hodgeCFT dx^\nu\, . 
\end{split}
\end{equation}
}  
\begin{equation}
\label{eq:gradxidxdx}
\begin{split}
&\nabla_a \xi^b\hodge(dx^a \wedge dx_b )\\
&\ = r^{d-1}\xi^\mu 
\brk{2rf \eta_{\mu\nu} -r^2\frac{df}{dr} u_\mu u_\nu}\ \hodgeCFT dx^\nu +\ldots\\
&\quad +\xi^\mu 
\brk{r^{d+1} \frac{d}{dr}\prn{\frac{g_{_V}}{r^2}}V_\mu u_\nu
+ r^{d+1} f^2\frac{d}{dr}\prn{\frac{g_{_V}}{r^2f}}u_\mu V_\nu }
\ \hodgeCFT dx^\nu +\ldots\, . 
\end{split}
\end{equation}

Another set of results useful in the computation of the non-Komar part of the Einstein-Maxwell differential Noether charge is
\begin{equation}
\begin{split}\label{eq:non-Komaruse1}
\delta&\fGamma^r{}_r\wedge \hodge(dx^r \wedge dx_r)\\
&= \bigbr{\half r^{d-1}\frac{d(r^2\delta f)}{dr}
- \half r^{d-1}\brk{r^2 \frac{d}{dr}\prn{\frac{g_{_V}}{r^2}}+\frac{g_{_V}}{r^2} 
\frac{d(r^2 f)}{dr} } V^\mu\delta u_\mu }
\ \hodgeCFT 1 \, , 
\end{split}
\end{equation}

\begin{equation}
\begin{split}\label{eq:non-Komaruse2}
\delta\fGamma^\mu{}_r\wedge \hodge(dx^r \wedge dx_\mu)
&=0 +\ldots \, , 
\end{split}
\end{equation}
\begin{equation}
\begin{split}\label{eq:non-Komaruse3}
\delta\fGamma^r{}_\mu &\wedge \hodge(dx^\mu \wedge dx_r)\\
&= \bigbr{r^2\delta f \frac{d}{dr} r^{d-1}  
+ r^{d-1} f \frac{dg_{_V}}{dr}V^\mu\delta u_\mu }
\ \hodgeCFT 1 \\
&\quad +\bigbr{\ldots}_\mu\ dr\wedge\hodgeCFT dx^\mu \, , 
\end{split}
\end{equation}
and
\begin{equation}
\begin{split}\label{eq:non-Komaruse4}
\delta&\fGamma^\nu{}_\mu\wedge \hodge(dx^\mu \wedge dx_\nu)\\
&=  \bigbr{
\half r^{d-1}\frac{d(r^2\delta f)}{dr}
- \half r^{d-1}\brk{ r^2\frac{d}{dr}\prn{\frac{g_{_V}}{r^2}}+\frac{g_{_V}}{r^2} 
\frac{d(r^2 f)}{dr} } V^\mu\delta u_\mu }\ \hodgeCFT 1 \\ 
&\quad +\bigbr{\ldots}_\mu\ dr\wedge\hodgeCFT dx^\mu \, . 
\end{split}
\end{equation}



\section{Tachikawa formula in terms of Pontryagin classes}
\label{appendix:TachikawaPontryagin}
The aim of this Appendix is to rewrite Tachikawa entropy formula in terms of
Pontryagin classes of the curvature. This gives a useful expression for the entropy formula, since the Pontryagin classes
evaluated on our solution often take a simpler form as compared to traces of the wedge products of the curvature two-form.
For mathematical details of the Pontryagin classes, please refer to \cite{kobayashi1996foundations,nakahara2003geometry}.

We start with the Tachikawa entropy formula (the second term of \eqref{eq:WaldTachikawa0}) written 
in terms of the trace of the wedge products of the curvature two-form. The goal of this Appendix is to rewrite this in terms of Pontryagin classes $p_{_k}(\fR)$ which are defined through 
\begin{equation}
\label{eq:PontryaginDef}
 \text{det}\brk{1+\frac{t\fR}{2\pi} }
=   \sum_{k=0}^\infty t^{2k} p_{_k}(\fR)\, , 
\end{equation}
where $t$ can be thought of formally as a $(-2)$ form. It follows that $p_{_k}(\fR)$ is a $4k$ form.
The first few Pontryagin classes are given by 
\begin{equation}
\begin{split}
p_{_0}(\fR) &= 1 \, ,   \qquad 
p_{_1}(\fR)  = -\frac{\text{tr}\fR^2}{2(2\pi)^2}  \, , \qquad 
p_{_2}(\fR) = -\frac{1}{2} \brk{\frac{\text{tr}\fR^4}{2(2\pi)^4}-
\prn{\frac{\text{tr}\fR^2}{2(2\pi)^2} }^2\,
}\,, \\
p_{_3}(\fR) &= - \brk{\frac{1}{3}\ \frac{\text{tr}\fR^6}{2(2\pi)^6}-\frac{1}{2}\ \frac{\text{tr}\fR^2}{2(2\pi)^2}  \wedge \frac{\text{tr}\fR^4}{2(2\pi)^4}+\frac{1}{6}\prn{\frac{\text{tr}\fR^2}{2(2\pi)^2} }^3} \, . 
\end{split}
\end{equation}
These can be inverted to obtain
\begin{equation}
\begin{split}
\frac{\text{tr}\fR^2}{2(2\pi)^2} & = -p_{_1}(\fR) \, , \qquad 
\frac{\text{tr}\fR^4}{2(2\pi)^4} =-2p_{_2}(\fR)+p^2_{_1}(\fR) \, , \\
\frac{\text{tr}\fR^6}{2(2\pi)^6} &=-3p_{_3}(\fR) + 3 p_{_1}(\fR)p_{_2}(\fR)-p^3_{_1}(\fR)\, .   \\
\end{split}
\end{equation}

In order to rewrite the Tachikawa entropy formula in terms of the Pontryagin classes, we will need to use the following
property  of the Pontryagin classes :
\begin{equation}
\label{eq:PontDeriv}
\begin{split}
p_{_{j-k}}(\fR)= -2k(2\pi)^{2k}\frac{\partial\ p_{_j}(\fR) }{\partial\ \text{tr} \fR^{2k} }\, , 
\end{split}
\end{equation}
which follows by differentiating Eq.\eqref{eq:PontryaginDef} with respect to $\text{tr} \fR^{2k}$ :
\begin{equation}
\begin{split}
 \frac{\partial\  }{\partial\ \text{tr} \fR^{2k} }  \text{det}\brk{1+\frac{t\fR}{2\pi} }
=  -\frac{t^{2k}}{2k(2\pi)^{2k}}\text{det}\brk{1+\frac{t\fR}{2\pi} }\, . 
\end{split}
\end{equation}
Using Eq.\eqref{eq:PontDeriv}, we can then use  the chain rule to rewrite the Tachikawa entropy formula as
\begin{equation}\label{eq:PontDoubleSum}
\begin{split}
 S_{\text{Tachikawa}} 
&= -\int_{Bif} 2\ \frac{\fGamma_N}{2\pi} \sum_{k=1}^\infty \sum_{j=k}^\infty 
\  p_{_{j-k}}(\fR) \prn{\frac{\fR_N}{2\pi}}^{2k-2} 
\frac{\partial \fP_{CFT}}{\partial\ p_{_j}(\fR) }\\
&= -\int_{Bif} 2\ \frac{\fGamma_N}{2\pi} \sum_{j=1}^\infty \sum_{k=1}^j
\  p_{_{j-k}}(\fR) \prn{\frac{\fR_N}{2\pi}}^{2k-2} 
\frac{\partial \fP_{CFT}}{\partial\ p_{_j}(\fR) }\, , 
\end{split}
\end{equation}
The sum above can be simplified by using 
\begin{equation}\label{eq:PontSumBif}
\begin{split}
 \brk{\sum_{k=1}^j
\  p_{_{j-k}}(\fR) \prn{\frac{\fR_N}{2\pi}}^{2k-2} }_{Bif}
&= p_{_{j-1}}(\fR+\fR_N \varepsilon) \Big|_{Bif}\, , 
\end{split}
\end{equation}
or in terms of the defining polynomial of the Pontryagin classes 
\begin{equation}
\begin{split}
\text{det}\brk{1+\frac{t \fR}{2\pi} }
 \text{det}^{-1}\brk{1-\frac{t \fR_N \varepsilon}{2\pi} } = \text{det}\brk{1+\frac{t (\fR+\fR_N\varepsilon)}{2\pi} } \,.
\end{split}
\end{equation}
This in turn follows from the matrix identity 
\begin{equation}
\begin{split}
\brk{1-\frac{t \fR_N \varepsilon}{2\pi} } \brk{1+\frac{t (\fR+\fR_N\varepsilon)}{2\pi} } = 1+\frac{t \fR}{2\pi}\, .
\end{split}
\end{equation}
where we have used $\varepsilon(\fR+\fR_N\varepsilon) = - \varepsilon^2 \fR_N +\varepsilon^2 \fR_N =0$
on the bifurcation surface. Thus, using Eq.~\eqref{eq:PontSumBif} in Eq.~\eqref{eq:PontDoubleSum}, we finally obtain
\begin{equation}\label{eq:TachikawaPont}
\begin{split}
 S_{\text{Tachikawa}} 
&= -\int_{Bif} 2\ \frac{\fGamma_N}{2\pi} \wedge \sum_{j=1}^\infty p_{_{j-1}}(\fR+\fR_N \varepsilon)
\frac{\partial \fP_{CFT}}{\partial\ p_{_j}(\fR) }\, , 
\end{split}
\end{equation}
which is the Tachikawa entropy formula written in terms of the Pontryagin classes. 


\section{Example I : Chern-Simons terms on BTZ black hole}
\label{app:BTZ}
Here we consider $U(1)$ and gravitational Chern-Simons terms in 
three dimensions and evaluate $(\fQNoether)_{H}$ on the BTZ black hole background. 
We can obtain the metric, Christoffel symbol and $\nabla \xi$
 by setting $d=2$, $\kappa_q=0$ and $q=0$ in Eqs.~\eqref{eq:metricandinverse}-\eqref{eq:Phi_Tsol}, 
 \eqref{eq:christoffel1} and \eqref{eq:nablaxigeneralr}. We also note that 
 $g_{V}=0$ for the BTZ black hole since the presence of the Chern-Simons terms does not correct 
 the solution.  
We note that for the covariant part of the Lagrangian, we only consider the Einstein term without the Maxwell part.
 
In the following, we start with the Hall contribution to the differential Noether charge for an arbitrary fixed $r$ surface 
of the BTZ black hole solution. 
We then evaluate it at the boundary and the horizon to obtain the CFT stress tensor, current and entropy. 
At the horizon, there are two possible prescriptions for the evaluation of the differential Noether charge. One way is to 
first do the variation with respect to the parameters of the BTZ black hole solution and then set the radial coordinate to be $r=r_H$,  
while the another way is to first set $r=r_H$ and then do the variation. 
We will comment more on this point in the middle of this Appendix and confirm that we can obtain the same result from these two prescriptions. 
At the end of this Appendix, we also compare our result with Ref.~\cite{Tachikawa:2006sz}.  
 
\subsection{Differential Noether charge} 
The anomaly polynomial for the three-dimensional Chern-Simons terms is given by 
\begin{eqnarray}
\fP_{CFT}=c_{_g} \tr[\fR^2  ]+c_{_A} \fF^2\, . 
\end{eqnarray} For later use, we define $\signCA\equiv \mbox{sign}(c_A)$.

We note that the BTZ black hole is locally AdS$_3$ everywhere and thus the Riemann tensor 
is given by $R^a{}_{bcd} = -2\delta^a_{[c}G_{d]b}$ and 
then the spin Hall current in this case becomes
$(\SpH)^{abc}=  -4 c_{_g}(1/2!)\, \varepsilon^{cef} R^{ab}{}_{ef} = 
4 c_{_g} \varepsilon^{abc} $. By using this,
the Hall contribution to the differential Noether charge $(\fQNoether)_{H}$ is written as 
\begin{equation}
\begin{split}
(\fQNoether)_{H} 
&=   2 c_{_g} \nabla_a \xi^b\delta \fGamma^a{}_b
+2c_{_g}(\delta G_{ab})\ \xi^a dx^b  + 2 c_{_A} (\delta \fA)\cdot (\Lambda+\ic_{\xi} \fA)
\, . 
\end{split}
\end{equation}

Let us now explicitly evaluate this $(\fQNoether)_{H}$ on the BTZ black hole solution. The Christoffel symbols in \eqref{eq:flugravGamma}
simplify by using two facts : in $d=2$, $\kappa_q =0$ and hence $(1/2)(d(r^2 f)/dr)=r$. 
Further, we have $g_{_V}=0$ since the BTZ black hole solution does not 
get corrected in the presence of the Chern-Simons terms. Using these, we have
\[ G_{\mu\nu}=r^2 \eta_{\mu\nu} + m u_\mu u_\nu\, ,  \]
and\footnote{
We note that for a  $U(1)$ gauge field in AdS$_3$/CFT$_2$, the boundary conditions depend on the sign of $c_{_A}$. See section.~3.1 of \cite{Kraus:2006wn}.
}
\begin{eqnarray}
&&\fA= 
\fA_\infty+\mu (\fu-s\fV)\, , \nonumber
\\
&&\Gamma^r{}_{rr} =\Gamma^\mu{}_{rr}= 0\, ,  \qquad 
\Gamma^r{}_{r\mu}  =r u_\mu \, ,  \\
&& \Gamma^r{}_{\mu\nu} 
= -r^3 f \eta_{\mu\nu}\, , \qquad 
\Gamma^\mu{}_{\nu r} 
=\frac{1}{r} P^\mu_\nu \, , \qquad \Gamma^\mu{}_{\nu \lambda} 
=-ru^\mu\eta_{\nu\lambda}\, .  \nonumber
\end{eqnarray}
We also note that Eq.~\eqref{eq:gradXi} simplifies to
\begin{eqnarray}
&&\nabla_r\xi^r 
=r u_\mu \xi^\mu\, ,  \qquad
\nabla_r \xi^\mu
=\frac{1}{r} P^\mu_\nu \xi^\nu\, , \nonumber \\
&&\nabla_\mu \xi^r
=-\xi^\nu r^3f\eta_{\mu\nu}\, ,  \qquad 
\nabla_\mu \xi^\nu
=-\xi^\lambda ru^\nu \eta_{\mu\lambda}\, . 
\end{eqnarray}

Using the above results and neglecting the terms proportional to $dr$,  we obtain the following expression for the differential Noether charge $(\fQNoether)_{H}$ on a fixed $r$ slice :
\begin{equation} \label{eq:chargeBTZgeneralr}
\begin{split} 
(\fQNoether)_{H} 
&=   \delta\brk{ 2 c_{_g} m \prn{u_\mu V_\nu+V_\mu u_\nu} \xi^\mu \ 
-s c_{_A}  \mu^2( u_\nu -s  V_\nu )(u_\mu-s V_\mu)  
\xi^\mu} \hodgeCFT dx^\nu 
\\
&\qquad
-  \delta\brk{ 
2 s c_{_A} (\mu u_\nu-s \mu V_\nu ) (\Lambda+\ic_{\xi} \fA_\infty)
}\hodgeCFT dx^\nu \,.
 \end{split}
\end{equation}
Here we have used the relation $\varepsilon^{\nu\lambda} = u^\nu V^\lambda -V^\nu u^\lambda$
which follows from the fact that the two orthogonal vectors $u^\mu$ and $V^\mu=\varepsilon^{\mu\nu}u_\nu$ form a complete basis.  
We have also used the fact that in $d=2$, $\eta_{\nu\alpha}= V_\nu V_\alpha- u_\nu u _\alpha$ and thus
\be \label{eq:idV1}
(u_\nu - s V_\nu) (u_\alpha - s V_\alpha)
=\eta_{\nu\alpha}+2 u_{\nu} u_{\alpha} -s (u_\nu V_\alpha+ V_\nu u_\alpha)\,.
\ee 

For later use, here we provide another useful result related to $V^\mu$ : 
\be \label{eq:idV2}
\delta(u_\nu - s V_\nu) (u_\alpha - s V_\alpha)
=s(V^\beta \delta u_\beta) (u_\nu - s V_\nu) (u_\alpha - s V_\alpha)
=(u_\nu - s V_\nu) \delta(u_\alpha - s V_\alpha),
\ee where we have used the fact that any change in $V^\mu$ is orthogonal $V^\mu$ (and thus parallel to $u^\mu$) since $V_\mu V^\mu= 1$, 
and we also have $\delta u^\mu \propto V^\mu$ from a similar argument. Then, from Eqs.~\eqref{eq:idV1} and  \eqref{eq:idV2}, we obtain 
\be\label{eq:idV3}
\delta(u_\nu - s V_\nu) (u_\alpha - s V_\alpha)
=\half \delta \left[\eta_{\nu\alpha}+2 u_{\nu} u_{\alpha} -s (u_\nu V_\alpha+ V_\nu u_\alpha)\right]\,.
\ee 

\subsection{Asymptotic charge}
Now we evaluate the the Hall contribution to the differential Noether charge $(\fQNoether)_{H}$ at the boundary. 
By using Eqs.~\eqref{eq:chargeBTZgeneralr} and \eqref{eq:idV3}, 
we have the following expression at the boundary $r\to\infty$ :
\begin{equation}
\begin{split}
\brk{(\fQNoether)_{H}}_{\infty}
&=  - \delta\brk{ 
\TmnCFTCS\ \xi^\mu\
+
\JnCFTCS\cdot  (\Lambda+\ic_{\xi} \fA)_\infty 
\, } \hodgeCFT dx^\nu \, ,\end{split}
\end{equation}
with 
\begin{eqnarray}
\label{eq:TmnCFTCS3D}
\TmnCFTCS&= &
s c_{_A} \mu^2 \left[
\eta_{\mu\nu} + 2u_\mu u_\nu
\right]
-(2 c_{_g} (2\pi T)^2 +c_{_A}\mu^2)\prn{u_\mu V_\nu+V_\mu u_\nu}
 \, , \nonumber\\
\JnCFTCS&=&
2 s c_{_A}  \mu  u_\nu
-2 c_{_A}  \mu V_\nu
 \, .
\end{eqnarray} 
On the other hand, the Einstein contribution from Eq.~\eqref{eq:CFTfluidST} to the stress tensor $\TmnCFTEin$ and current $\JnCFTEin$ in this case reduces to
\be\label{eq:TmnCFTEM3D}
\TmnCFTEin =
  (2\pi T^2)\times  \frac{3 }{ 24 \GN}
(\eta_{\mu\nu}+2 u_\mu u_\nu),\quad
\JnCFTEin =0 \,.
\ee

Combining the Einstein and  Chern-Simons contributions, we can write the total CFT stress tensor and current as
\bea
T_{\mu\nu}^{\text{\tiny{CFT}}}
&=&T_{\mu\nu}^{\text{\tiny{CFT, perfect fluid}}} +T_{\mu\nu}^{\text{\tiny{CFT,anom}}} \, ,\nonumber\\
J_{\nu}^{\text{\tiny{CFT}}}
&=&J_{\nu}^{\text{\tiny{CFT, perfect fluid}}} +J_{\nu}^{\text{\tiny{CFT,anom}}} \, ,
\eea where the perfect fluid part with the pressure $p=(2\pi T^2) (3 /24\GN)+s c_{_A} \mu^2$ is given by
\be
T_{\mu\nu}^{\text{\tiny{CFT, perfect fluid}}} = p (\eta_{\mu\nu}+2 u_\mu u_\nu)\, ,\qquad
J_{\nu}^{\text{\tiny{CFT, perfect fluid}}} =\frac{\partial p}{\partial \mu}u_\nu\, , 
\ee
while the anomaly-induced contribution to the stress tensor and current  are
\bea
T_{\mu\nu}^{\text{\tiny{CFT,anom}}}&=&
\left.\prn{\mathbb{G}_\mu
 -\Phi\   \frac{\partial \mathbb{G}_\mu }{\partial \Phi}
-\PhiT\  \frac{\partial \mathbb{G}_\mu }{\partial \PhiT} }\right\athor u_\nu+u_\mu
\left. \prn{\mathbb{G}_\nu
 -\Phi\   \frac{\partial \mathbb{G}_\nu }{\partial \Phi}
-\PhiT\  \frac{\partial \mathbb{G}_\nu }{\partial \PhiT} } \right \athor  \, , \\
J_{\nu}^{\text{\tiny{CFT,anom}}}&=& -\Phi\  \left( \frac{\partial \mathbb{G}_\nu }{\partial \Phi}\right) \Biggr{|}_{hor}\, .
\eea Here $\mathbb{G}_{\mu} =  \mathbb{G}^\tV V_\mu$ with $\GV=c_{_g} \times 2\Phi_T^2+c_{_A} \Phi^2$.
This agrees with the result coming from the replacement rule.

For more details on the construction of the commutator relations and the central extensions such that the charges generate a 
$U(1)$ Kac-Moody-Virasoro algebra, see \cite{Kraus:2006wn}.

\subsection{Entropy}
\label{app:entropyBTZwhole}
In this part, we compute the differential Noether charge at the horizon to obtain the entropy current. 
We will carry out this computation by using two different prescriptions which end up with the same final result. 
We will also comment on the consistency with the original computation on the Chern-Simons contribution to 
the black hole entropy by Ref.~\cite{Tachikawa:2006sz}. 

\subsubsection{Entropy: prescription I}

As a next step, we evaluate the differential Noether charge at the horizon of the BTZ black hole. 
Here we carry out the variation (with $\delta r=0$) first and then set $r=r_H$. 
By expanding out the variation as 
\begin{equation}
\begin{split}
(\fQNoether)_{H} 
&= 2 c_{_g} (\delta m) \xi^\nu \brk{ u_\nu V_\mu+V_\nu u_\mu} \ \hodgeCFT dx^\mu\\
&\quad +2 c_{_g} m \xi^\nu \brk{u_\nu (\delta V_\mu)+(\delta V_\nu) u_\mu} \ \hodgeCFT dx^\mu\\
&\quad + 2 c_{_g} m \xi^\nu\brk{(\delta u_\nu) V_\mu+V_\nu (\delta u_\mu)} \ \hodgeCFT dx^\mu \, \\
& 
\quad + 2 c_{_A}  (\Lambda+\ic_{\xi} \fA)\delta (\mu u_\nu )   dx^\nu
\, . 
\end{split}
\end{equation}
Then by setting $r=\rH$ (where $\xi^\mu |_{hor} = \xi^\mu_h= u^\mu/T$ and $(\Lambda+\ic_{\xi} \fA)\athor=0$) we obtain 
\begin{equation} \label{eq:deltaQhorBTZ}
\begin{split}
 (\fQNoether)_{H}\athor 
&= - 2 c_{_g}(2\pi)^2\bigbr{2(\delta T) V_\mu 
+T \brk{2(\delta V_\mu)-P^\nu_\mu \delta V_\nu} }\ \hodgeCFT dx^\mu\,  \\
 &= \delta\brk{ \prn{J_S^{\text{\tiny{CFT,anom}}}}_\mu\ \hodgeCFT dx^\mu}\, , 
\end{split}
\end{equation}
where the Hall contribution to the entropy current 
\begin{equation}
\begin{split}
\prn{J_S^{\text{\tiny{CFT,anom}}}}_\mu = -4c_{_g} (2\pi)^2 T V_\mu
=  -\frac{\partial}{\partial T}\brk{2c_{_g} (2\pi T)^2 V_\mu}\, .
\end{split}
\end{equation}
In the above derivation, we have used the fact that any change in $V^\mu$ is orthogonal $V^\mu$ and thus parallel to $u^\mu$ since $V_\mu V^\mu= 1$.
In particular it follows that $P^\nu_\mu \delta V_\nu=0$. 
This result for the entropy current agrees with the CFT replacement rule.

\subsubsection{Entropy: prescription II}
In the previous section, the computation of $(\fQNoether)_{H}$ was carried out by first taking the variation and then set $r=r_H$ of all quantities. 
Her we consider the second prescription in which
we first set $r=r_H$ in the expression \eqref{eq:chargeBTZgeneralr} 
and then carry out the variation.
Then $(\fQNoether)_{H}$ at the horizon is evaluated as 
\bea
(\fQNoether)_{H}  \athor
&=&   -2(2\pi )^2  c_{_g}  \left[
2\delta ( T \fu)
 -  T \delta \fu
\right] 
+2(2\pi )^2 c_{_g} T(  \delta \fu)  \nonumber\\
&=& \delta\brk{ \prn{J_S^{\text{\tiny{CFT,anom}}}}_\mu\ \hodgeCFT dx^\mu}\, , 
\eea which agrees with the result from prescription I in Eq.~\eqref{eq:deltaQhorBTZ}.

\subsubsection{Comparision with Ref.~\cite{Tachikawa:2006sz}}
We now compare our covariant computation with the prescription provided in
Ref.~\cite{Tachikawa:2006sz}.
According to Ref.~\cite{Tachikawa:2006sz}, 
for the gravitational Chern-Simons term in three dimensions, one can write down 
a non-covariant Komar charge which is twice of what would be 
expected by the usual Wald formula :
\begin{equation}
\begin{split}
(\fKomar)_{H}^{\text{\tiny{Tachikawa}}} &= 2\nabla_a \xi^b \frac{\partial \ICS}{\partial \fR^b{}_a} 
=2 c_{_g} \nabla_a \xi^b \fGamma^a{}_b \\
&= 4c_{_g} \xi^\mu\brk{-r^2\eta_{\mu\nu}+ m (\eta_{\mu\nu}+u_\mu u_\nu)}
\varepsilon^{\lambda\nu}\ \hodgeCFT dx_\lambda\, , 
\end{split}
\end{equation}
where in the last line we have evaluated the Komar charge in our coordinates. Taking
$r=\rH$ with $\xi^\mu = u^\mu /T$, this gives the same result as our covariant computation :
\begin{equation}
\begin{split}
 (\fKomar)_{H}^{\text{\tiny{Tachikawa}}} |_{hor}
= \prn{J_S^{\text{\tiny{CFT,anom}}}}_\mu\ \hodgeCFT dx^\mu\, .  
\end{split}
\end{equation}

\section{Example II : Abelian Chern-Simons terms}
\label{app:AbelianCS}

For the Abelian Chern-Simons term in $(2n+1)$ dimensions, the anomaly polynomial is given by 
\be
\fP_{CFT}= c_{_A}\, \fF^{n+1} \, , 
\ee
where $c_{_A}$ is a constant.
Here we consider the case with $n\ge 2$.  
The Hall contribution to the differential Noether charge $(\fQNoether)_{H}$ 
in this case (see \cite{Compere:2009dp}) comes only from the first term of $T_0$ in Eqs.~\eqref{eq:deltaQcssum}
and \eqref{eq:T0toT4sec2} :
\be
(\fQNoether)_{H} =c_{_A} (n+1) n\, (\Lambda+ \ic_\xi \fA)\, \delta \fA\wedge \fF^{n-1}\, .
\ee

The evaluation of $(\fQNoether)_{H}$ on the background \eqref{eq:flugravsoln} goes as follows 
at the leading order of the derivative expansion. At any fixed $r$, 
the leading order contribution to $\fF^{n-1}$ is $(\F{1}^{n-1})=(2\Phi\fomega)^{n-1}$ 
since $(\F{0})\propto dr\wedge \fu$. The leading order contribution
to $(\fQNoether)_{H}$ therefore is 
\begin{eqnarray} \label{eq:leadingabeliannoether}
(\fQNoether)_{H}  =  c_{_A} (n+1) n\, (\Lambda+ \Phi \xi^\mu u_\mu)\, \delta(\Phi \fu) 
\wedge (2\Phi\fomega)^{n-1}\, , 
\end{eqnarray}
which is of order $\omega^{n-1}$. 
We note that the 2nd and higher order terms in $\fA$ and $\fF$ etc. do not contribute to this order or lower, 
since this type of contribution is always accompanied by at least one $(\F{0})$. 

Now we evaluate $(\fQNoether)_{H}$ in particular at the horizon ($r =  r_H$) and 
at the boundary ($r\to\infty$). At the horizon, we substitute
$\xi^a \athor=\xiH^a$, $\Lambda\athor = \LambdaH$ (see Eq.~\eqref{eq:horsub}) and $\Phi(r=r_H) = \mu$
into \eqref{eq:leadingabeliannoether}, 
while, at the boundary $r\to\infty$, we use the fall-off behaviors $\xi^a\atinfty \to {\cal O}(r^0)$ and $\Phi\atinfty \to  {\cal O}(r^{-(2n-2)})$ (see Eq.~\eqref{eq:infsub}). 
Then we finally have 
\begin{eqnarray}
(\fQNoether)_{H} \athor  = 0\, ,  \qquad 
(\fQNoether)_{H} \atinfty
\rightarrow  {\cal O} \left(\frac{1}{r^{2n(n-1)}} \right)\, . 
\end{eqnarray}
Therefore the Hall contribution to the differential Noether charge $(\fQNoether)_{H}$ vanishes both at the horizon and boundary. 
In terms of the currents defined in Eqs.~\eqref{eq:currentssplit} and \eqref{eq:entropycurrentsplit} , 
this result is rewritten as   
 \be
  \TmnanomCS=\JmanomCS=\JSanomCSV=0\, . 
 \ee 
We note that this is consistent with holographic renormalisation type computation in \cite{Sahoo:2010sp,Kharzeev:2011ds}.
 

\section{Example III: Hall contribution in \texorpdfstring{AdS$_5$}{AdS5}}
\label{sec:deltaQcsads5eq0}
Before discussing the general anomaly polynomials in general odd dimensional AdS, 
it is instructive to explain some detail computation of  the Hall contribution to 
the differential Noether charge $(\fQNoether)_{H}$ for AdS$_5$ with the anomaly polynomial 
\be \label{eq:ads5mixedanom}
\fP_{CFT}=  c_{_M}\, \fF \wedge \tr[{\fR^2}]\, .
\ee 
In this Appendix, we analyze this example in detail. 

We note that, for the anomaly polynomial given in Eq.~\ref{eq:ads5mixedanom}, 
each term in the expression Eqs.~\eqref{eq:deltaQcssum} and \eqref{eq:T0toT4sec2} 
in this case is given by 
\bea
\label{eq:T0T1T2forAdS5}
T_0
&=&
2 c_{_M}(\Lambda+ i_\xi \fA) \wedge \tr[\delta \fGamma  \fR]
\, , 
\nonumber\\
T_1 &=& 
2c_{_M}\fF\wedge\tr[  \nabla \xi \delta \fGamma] \, , 
\nonumber\\
T_2 &=& 2c_{_M} (\delta \fA) \wedge\tr[\nabla \xi \fR]\, ,
\nonumber\\
T_3 &= & -\half \delta G_{ab} (\SpH)^{abc} \ic_\xi (\hodge dx_c)\, , \nonumber\\
T_4 &= & -\frac{1}{4} \xi^a \delta\left\{
\left[(\SpH)_a{}^{bc} + (\SpH)^b{}_a{}^c + (\SpH)^{cb}{}_a\right] \hodge (dx_b \wedge dx_c)
\right\}\,,
\eea  where the spin Hall current is given by 
$(\SpH)^{cb}{}_a \hodge dx_c= -4c_{_M} \fF \wedge \fR^b{}_a$
or in components
$(\SpH)^{cb}{}_a = -c_{_M}\epsilon^{c p_1 p_2 p_3 p_4} F_{p_1 p_2} R^b{}_{ap_3p_4}$\,.

Before starting the computations of each term above,  we massage the expression of $T_3$ and $T_4$ into the following forms for later use:
\bea
T_3 &=& -\half (\delta G_{a\beta}) (\SpH)^{a\beta r}
r^{3} \, (\hodgeCFT \form{\xi})
\, , \qquad
T_4 \equiv \xi^\mu \delta( \tilde{T}_4)_\mu\, , \label{eq:t3t4simplifiedads5}
\eea
where $\form{\xi} \equiv \eta_{\mu\nu}\xi^{\mu}dx^\nu$ and 
\bea
(\SpH)^{a\beta r} &=& (\SpH)^{a\beta }{}_r (2\Psi) + (\SpH)^{a\beta}{}_\gamma u^\gamma\, , 
\label{eq:sigmaalphabetarads5}
 \\
(\tilde{T}_4)_\mu
&=&
\left[
G_{\mu a}(\SpH)^{(a \gamma)r}
+\frac{1}{2}
(\SpH)^{ r \gamma}{}_\mu
\right]
\,  r^{3} (\hodgeCFT \eta_{\gamma\rho}dx^\rho)
\equiv (\tilde{T}_4^{(a)} )_\mu+ (\tilde{T}_4^{(b)})_\mu
\, .  \label{eq:T4dividedads5}
\eea
To derive this, we have used 
$\xi^r=0$ and the identities 
$\hodge dx_r=r^{3} (\hodgeCFT 1)$ and  
$\ic_{\xi} \hodgeCFT \form{C} = \hodgeCFT(\form{C} \wedge \form{\xi})$
valid for any boundary form $\form{C}$
as well as the anti-symmetric property of the spin Hall current, 
$(\SpH)^{abc}=-(\SpH)^{acb}$.

When we evaluate the leading order contribution to $(\fQNoether)_{H}$, 
the treatment of 2nd and higher order building blocks is trivial in AdS$_5$ case, 
since we are concerned about contributions up to order $\omega^1$.
Thus, in the computation of $T_0,T_1,T_2, T_3$ and $T_4$, we do not need to take into account
2nd and higher order building blocks.  In the rest of this Appendix, we therefore 
deal with 0th and 1st order terms in the building blocks only. 
We will confirm soon that the leading order contributions to $(\fQNoether)_{H}$ 
for AdS$_5$ indeed start with $\omega^1$. 

Now we compute each term in \eqref{eq:T0T1T2forAdS5} from the gravity side one by one 
to evaluate the differential Noether charge at the horizon and boundary.

\subsection{Term \texorpdfstring{$T_0$}{T0}}
For the evaluation of the term $T_0$ one can compute $\tr[\delta\fGamma \fR]$ 
by using Eq.~\eqref{eq:deltaGammaR} and in the end we have the following expression 
valid at any fixed $r$ :
\bea
T_0&=& 
2c_{_M}(\Lambda+\Phi\xi^\mu u_\mu)\times\nonumber\\
&& \left[
-2r^{-1}    \left(\Psi-\half r^2 \right) \fu \wedge (2\Phi_T\delta \fomega)
+(\Phi_T-r+\Psi')(\delta \fu)\wedge(2\Phi_T\fomega)
+2(\delta \Phi_T)\fu\wedge(2\Phi_T\fomega)
\right].\nonumber\\
\eea

Let us next evaluate this expression at the horizon and boundary. 
By substituting $\xi^\mu\athor=\xiH^\mu = u^{\mu}/T$, $\Lambda\athor = \LambdaH$ as in 
Eq.~\eqref{eq:horsub} at the horizon and the fall-off behaviors of the parameters and fields at infinity 
summarized in Eqs.~\eqref{eq:infsub}  and \eqref{eq:fields_limitinfty}, 
we can evaluate $T_0$ there as 
\begin{eqnarray} \label{eq:adsT0hor}
T_0 \athor=0\, , \qquad T_0 \atinfty \rightarrow {\cal O}\left(\frac{1}{r^{6}}\right)\,. 
\end{eqnarray} 
Therefore we conclude that $T_0$ vanishes both at the horizon and boundary 
up to $\omega^1$ order. 

\subsection{Term \texorpdfstring{$T_1$}{T1}}
As in the case of the Abelian Chern-Simons terms, we can replace $\fF$ in $T_1$ 
by $(\F{1})=(2\Phi\fomega)$ since $(\F{0})\propto dr\wedge \fu$\,. By directly computing 
the rest part of $T_1$ by using Eq.~\eqref{eq:deltaGammagradxi}, 
we obtain the leading order contribution to this term as follows : 
\bea
T_1
&=&2c_{_M}(2\Phi \fomega)\times\nonumber\\
&&\left\{
-(2\delta \Psi)(\xi^\beta P_{\beta\nu} dx^\nu)
+\left[(2\Psi)(r^{-1}\Psi'-1)
+  r \Phi_T \right]
 (\xi^\beta\delta u_\beta)  \fu
+4 r^{-1} \Psi (\xi^\beta u_\beta )(\delta \Psi') \fu
\right.
\nonumber\\
&&\qquad+2r^{-1}  (\Psi'-r)     
  \Psi  (\xi^\beta u_\beta)\delta \fu
+4r^{-1}\Psi       \Phi_T    (\xi^\beta u_\beta)\delta \fu
+4r^{-1}\Phi_T (\xi^\beta u_\beta )(\delta \Psi) \fu
\nonumber\\
&&\left.
\qquad+2\Phi_T (\xi^\beta u_\beta )(\delta \Phi_T) \fu
+\Phi_T\left[     
- r
+2  \Phi_T  \right]  (\xi^\beta u_\beta)\delta \fu
\right\}\, . 
\eea
It  is of order $\omega^1$. 
 
By substituting $\xi^\mu\athor=\xiH^\mu = u^{\mu}/T$, $\Lambda\athor = \LambdaH$ as in 
Eq.~\eqref{eq:horsub} at the horizon and the fall-off behaviors of the parameters and fields at infinity 
summarized in Eqs.~\eqref{eq:infsub}  and \eqref{eq:fields_limitinfty}, 
we can evaluate $T_1$ there as
\beq\label{eq:ads5T1hor}
T_1 \athor=
\delta\left[
  (\JSanomCSV)_{l=1}\,\fu
\right]\wedge (2\mu \fomega)
-\left(\frac{r_H}{4\pi T}\right)(\JSanomCSV)_{l=1}
\left[ (\delta \fu)
 \wedge (2\mu \fomega)\right]\, , 
 \qquad 
 T_1\atinfty \rightarrow  {\cal O}\left(\frac{1}{r^{4}}\right)\, .
 \eeq
 Here $(\JSanomCSV)_{l=1}$ is defined from the entropy current 
 \be\label{eq:GvJsV}
\JSanomCSV=
 -2\pi  \left(\frac{\partial \GV}{\partial \Phi_T} \right)_{hor}
=-2\pi\times c_{_M}  \times 4 \times (2\pi T)\times \mu\, , 
\ee
through the expansion in  \eqref{eq:entropycurrentexpansion} as 
$(\JSanomCSV)_{l=1}=-2\pi \times c_{_M}  \times 4 \times (2\pi T)$. 
 
The above result \eqref{eq:ads5T1hor} will be naturally generalized to higher dimensions, as summarized in \S\S~\ref{sec:38summary} 
(for the detail of the computation, see Appendices~\ref{app:replacementT3T4}-\ref{sec:entropyhighercontribution}).

\subsection{Term \texorpdfstring{$T_2$}{T2}}
By using Eq.~\eqref{eq:gradxiR}
we can compute $T_2$ at the leading order of the derivative expansion as  
\bea
 T_2&=&
2c_{_M} \delta(\Phi \fu)\left\{
\partial_\mu \xi^\rho\left[
-r\Phi_T u^\mu \eta_{\rho\beta}dx^\beta \wedge \fu
+2  \Psi  (r^{-1}\Psi'-1) u_\rho dx^\mu \wedge \fu
-2 \Psi dx^\mu \wedge \eta_{\rho\nu} dx^\nu
\right]\right.
\nonumber\\
&&\qquad\qquad\qquad
+\left[
2\Psi\left(
2r^{-2}\Psi-1
 +3r^{-1}
 \Phi_T \right)
-r\Phi_T 
\right]\fu\wedge( \xi^\beta\omega_{\beta\nu} dx^\nu)
\nonumber\\
&&\left.\qquad\qquad\qquad
+\left[2 \Psi\left(
2r^{-2}\Psi
- 1\right)  
+2\Psi'\Phi_T\right](u_{\beta} \xi^\beta)\wedge
 (2 \fomega)\right\}.
\eea
We note that this leading order contribution is of order $\omega^1$. 

The second term $T_2$ at the horizon and the boundary is then calculated as follows. 
By substituting $\xi^a \athor=\xiH^a$, $\Lambda\athor = \LambdaH$ as in Eq.~\eqref{eq:horsub} 
at the horizon and using the fall-off behaviors Eqs.~\eqref{eq:infsub} and \eqref{eq:fields_limitinfty}
at the boundary, we have
\bea \label{eq:ads5T2horv2}
T_2 \athor
 &=& -2 c_{_M} (2\pi) 
 \delta(2\mu \fu)
\wedge (2(2\pi T) \fomega)
 =
\left[ (\JSanomCSV)_l\,\fu\right]
 \wedge  \delta(2\mu \fomega)\, ,  \\
 T_2 \atinfty &\rightarrow&
-2 c_{_M}(d\form{\xi}) \wedge\delta(\fu q)
=-2c_{_M}\form{\xi} \wedge\delta(2\fomega q)\,,  
\eea 
where in the second equality for the evaluation at the horizon, we have used 
 \be
 \label{eq:integrationbypart} g(r)(\delta \fu) \wedge \fomega^{m+1}
=g(r)\fu\wedge (\delta \fomega) \wedge \fomega^{m}
+d(\ldots)+(\ldots)dr\,, 
 \ee 
 for any function $g(r)$ of $r$ and a integer $m\ge0$. 
We also note that we have dropped the terms proportional to $dr$ as well as the total derivative terms 
$d(\ldots)$ in the above expression since they do not contribute the Noether charge after the integration. 
 
 \subsection{Term \texorpdfstring{$T_3$}{T3}}
\label{sec:T3ads5}
The term $T_3$ depends on the spin Hall current, but we note that,  from the expression of \eqref{eq:t3t4simplifiedads5} and  \eqref{eq:sigmaalphabetarads5},
we only need to know the specific components of the spin Hall current $(\SpH)^{a\beta r}$ to evaluate $T_3$.
When only zeroth and first order terms in the building blocks are considered, 
there are three types of terms in $(\SpH)^{ab}{}_c$ up to $\omega^1$ order : 
\bea \label{eq:ads5SpH}
{\mathcal O}(\fomega^0)\, : \, 
(\SpH)'^{ab}{}_c&=&-c_{_M} \epsilon^{a p_1 p_2 p_3 p_4} (\F{0})_{p_1 p_2} (\R{0})^b{}_{cp_3p_4}\,, \nonumber\\
{\mathcal O}(\fomega^1)\, : \, 
(\SpH^{(1)})^{ab}{}_c&=&-c_{_M} \epsilon^{a p_1 p_2 p_3 p_4} (\F{0})_{p_1 p_2} (\R{1})^b{}_{cp_3p_4}\,, \nonumber\\
(\SpH^{(2)})^{ab}{}_c&=& -c_{_M}\epsilon^{a p_1 p_2 p_3 p_4} (\F{1})_{p_1 p_2} (\R{0})^b{}_{cp_3p_4}\,.
\eea
From this expression, we can have $(\SpH)^{r\beta r}=0$ up to $\omega^1$ order, since 
$(\SpH)'$ and $(\SpH^{(1)})$ contains $(\F{0})\propto dr\wedge \fu$ which leads to 
$(\SpH)'^{rb}{}_c=(\SpH^{(1)})^{rb}{}_c=0$ and $(\R{0})^{\beta r}$ in $(\SpH^{(2)})$ is proportional to $dr$. As a result, 
we can simplify the expression of $T_3$ in \eqref{eq:t3t4simplifiedads5} as 
\begin{eqnarray} \label{eq:rewriteT3}
T_3 &=&
  -\half (\delta G_{\alpha\beta}) (\SpH)^{(\alpha\beta) r} 
  r^{3} (\hodgeCFT \form{\xi})\, . 
\end{eqnarray}

In the following, we first evaluate the contribution to the components $(\SpH)^{(\alpha\beta) r}$ 
from these three terms, $(\SpH)'^{ab}{}_c$, $(\SpH^{(1)})^{ab}{}_c$ and $(\SpH^{(2)})^{ab}{}_c$, 
separately. After this, we then combine them to obtain the term $T_3$ at the leading order. 

\begin{itemize}
\item \underline{\bf From $(\SpH)'$} \\
We notice that only non-zero components of $(\F{0})\wedge (\R{0})^{b}{}_c$
are $(b, c) = (\beta, \gamma)$ which are proportional to 
$dr\wedge \fu\wedge dx^\beta \wedge P_{\gamma\nu}dx^\nu$. 
It then follows that the only non-zero components of $(\SpH)'^{ab}{}_c$ are
$(\SpH)'^{\alpha\beta}{}_\gamma \propto \epsilon^{r \alpha \beta \mu \nu \ldots}  u_\mu P_{\gamma \nu}$
which thus vanishes after contracting with $u^\gamma$,  i.e. $(\SpH)'^{\alpha\beta}{}_\gamma u^\gamma=0$. 
By substituting this into \eqref{eq:sigmaalphabetarads5}, we see that 
$(\SpH)'^{(\alpha\beta) r}=0$
and thus $(\SpH)'$ term does not contribute to $T_3$\,.  

\item \underline{\bf From $(\SpH^{(1)})$} \\
In this case, using  $(\F{0}) \propto dr \wedge \fu\,$ and $(\R{1})^\beta{}_r\propto \fu$, 
we obtain
$(\SpH^{(1)})^{rb}{}_c=0$ and  
$(\SpH^{(1)})^{\alpha \beta}{}_r =0\,$. 
Therefore, we have 
$( \SpH^{(1)})^{(\alpha\beta) r} = 
-4c_{_M}r^{-4}
(\partial_r \Phi)(r \Phi_T) V^{(\alpha} u^{\beta)}\,$ .

\item \underline{\bf From $(\SpH^{(2)})$} \\
We first notice that $ ( \SpH^{(2)})^{\alpha \beta}{}_r =0$ which follows from the 
fact that $ (\R{0})^{\beta}{}_r$ is proportional to $ dx^\beta$ and $(\F{1})\propto \fomega$. 
Then the components $(\SpH^{(2)})^{(\alpha\beta) r}$ is computed as 
$ ( \SpH^{(2)})^{(\alpha \beta) r} = 
  -4c_{_M}r^{-4} 
\Phi \partial_r( r\Phi_T)(V^{(\alpha} u^{\beta)}
)\,$.
\end{itemize}

From the above calculations, non-trivial components of the spin Hall current 
relevant to the evaluation of $T_3$ at the leading order are $(\SpH)^{(\alpha\beta) r}$ 
only and this can be computed by summing the contribution from $(\SpH^{(1)})$
and $(\SpH^{(2)})$ terms as 
\be\label{eq:ads4symmetricalphabetar}
(\SpH)^{(\alpha\beta) r}
=- 
\frac{1}{r^{4}}\frac{d}{dr}\left[
    r\frac{\partial \GV}{\partial \Phi_T}
         \right] u^{(\beta} V^{\alpha)}   \,,
\ee 
where $\GV=c_{_M} \Phi (2\Phi_T^2)$. Then, 
by using $\ic_{\xi}  [(\delta \fu)\wedge \fu  \wedge (2\fomega)] =
V^\beta (\delta u_\beta) \hodgeCFT \form{\xi}$ 
we obtain  the following expression for $T_3$ for arbitrary fixed $r$ at the leading order of the 
derivative expansion:       
\bea \label{eq:T3ads5generalr}
T_3&=&
\frac{1}{2r}(2 \Psi -r^2)  
   \frac{d}{dr} \left[
    r\frac{\partial \GV}{\partial \Phi_T}
         \right]\ic_{\xi}
    \left[    (\delta \fu)\wedge \fu \wedge (2\fomega)\right]\,.
\eea 

Now we evaluate $T_3$ at the horizon and boundary. 
At the horizon, by substituting  $\xi^a \athor=\xiH^a$, $\Lambda\athor = \LambdaH$ as in Eq.~\eqref{eq:horsub} while at the boundary by choosing $\xi^a$ 
 as in Eq.~\eqref{eq:infsub} and using Eq.~\eqref{eq:fields_limitinfty}, we finally obtain 
\bea
T_3 \athor = -\frac{r_H  }{2 T}  
\left[\frac{\partial \GV}{\partial \Phi_T}
+ r\frac{d}{dr}\frac{\partial \GV}{\partial \Phi_T}\right]_{hor}
 (\delta \fu) \wedge (2\fomega) \, ,  \qquad 
T_3 \atinfty \to {\cal O}\left(\frac{1}{r^8}\right) \, . 
\eea

\subsection{Term \texorpdfstring{$T_4$}{T4}}
To evaluate the term $T_4$ at the leading order of the derivative expansion, we 
first evaluate $(\tilde{T}^{(a)}_4)_\mu$ and $(\tilde{T}^{(b)}_4)_\mu$ defined in  \eqref{eq:T4dividedads5} separately and then combine them. In the evaluation of these two contributions, 
we use some results on $(\SpH)'$, $(\SpH^{(1)})$ and $(\SpH^{(2)})$ that we obtained 
in the middle of the computation of $T_3$. 

\begin{itemize}
\item \underline{\bf Term $\tilde{T}^{(a)}_4$} \\
By using the fact $(\SpH)^{r\beta r}=0$ up to $\omega^1$ order 
and the expression of $(\SpH)^{(\alpha\beta) r}$ in \eqref{eq:ads4symmetricalphabetar} 
we obtained in the evaluation of $T_3$, the term $\tilde{T}^{(a)}_4$ is computed as  
\bea
(\tilde{T}_4^{(a)})_\mu
&=&
-\frac{1}{2r} 
\left[
2 \Psi u_\mu \hodgeCFT \fV 
+ r^2 V_\mu \hodgeCFT \fu
\right]\frac{d}{dr}\left[
r \frac{\partial \GV}{\partial \Phi_T} 
\right] \, , 
\eea 
where $\GV= c_{_M}\Phi (2\Phi_T^{2})$. 

\item \underline{\bf Term $\tilde{T}^{(b)}_4$} \\
As we have shown $(\SpH)'^{r\alpha}{}_\mu = (\SpH^{(1)})^{r\alpha}{}_\mu=0$, 
in the middle of the evaluation of $T_3$,  the non-trivial contribution to $\tilde{T}_4^{(b)}$
comes only from $(\SpH^{(2)})^{r\alpha}{}_\mu$ which is computed as 
\begin{eqnarray}
(\SpH^{(2)})^{r \alpha}{}_\mu=
\frac{1}{r^{4}}
\left[ 
 r^2  V_\mu u^\alpha-
 2 \Psi    
u_\mu V^\alpha
\right]\left[
\frac{\partial \GV}{\partial \Phi_T} 
+8r^{-1}\Psi \,\HVtemp 
\right]\,, 
\end{eqnarray}
where $\GV= c_{_M}\Phi (2\Phi_T^{2})$ and
  $\HVtemp = c_{_M}\Phi$. 
Here we have used the identity
\be\label{eq:epstouV}
    \varepsilon^{\mu\alpha \lambda \sigma }\omega_{\lambda \sigma}
     =u^\mu V^\alpha-u^\alpha V^\mu\, .
 \ee
Then, by noticing $\hodgeCFT \fV = \fu\wedge (2\fomega)$, 
we finally have the expression for $\tilde{T}_4^{(b)}$  as follows :
\begin{eqnarray}
 (\tilde{T}_4^{(b)})_\mu=
\frac{1}{2r} 
\left[
 r^2  V_\mu \hodgeCFT \fu
-2 \Psi    
u_\mu \fu\wedge (2\fomega)\right]
\left[
\frac{\partial \GV}{\partial \Phi_T} 
+8r^{-1}\Psi\, \HVtemp 
\right]  \, . 
\end{eqnarray}
\end{itemize}

Putting the results for $\tilde{T}^{(a)}_4$ and $\tilde{T}^{(b)}_4$ together, 
we obtain $T_4$ at the leading order of the derivative expansion valid at arbitrary fixed $r$ as follows :
\bea\label{eq:T4b}
T_4 &=&\xi^\mu\delta\Biggl\{
-\frac{ r^2}{2} \frac{d}{dr}
\left[ \frac{\partial \GV}{\partial \Phi_T} 
\right]V_\mu \hodgeCFT \fu
- u_\mu \frac{\Psi}{r^2}\frac{d}{dr}
\left[ r^2\frac{\partial \GV}{\partial \Phi_T} 
\right]\fu\wedge(2\fomega) \nonumber\\
&&\qquad\qquad
+ 4 
r^{-2}\Psi \,\HVtemp 
\left[
    r^2 V_\mu \hodgeCFT \fu
-2 \Psi u_\mu  \fu\wedge (2\fomega)\right]\Biggr\}\,, 
\eea
where $\GV= c_{_M}\Phi (2\Phi_T^{2})$ and $\HVtemp = c_{_M}\Phi$. 

Let us then evaluate the leading order term of $T_4$ at the horizon and boundary, 
by substituting Eq.~\eqref{eq:horsub} at the horizon and 
by using the fall-off behaviors given in Eqs.~\eqref{eq:infsub} and \eqref{eq:fields_limitinfty}
at the boundary respectively : 
\bea
T_4 \athor &=&
\frac{r^2_H}{2T}
 \frac{ d}{dr}
\left[ \frac{\partial \GV}{\partial \Phi_T} 
\right]_{hor}
(\delta \fu) \wedge (2\fomega)\,,  \qquad 
T_4  \atinfty \to 
2c_{_M}\form{\xi}\wedge\delta (2\fomega q)\, . 
\eea
Here we have also used $\hodgeCFT(   V_\mu u_\nu dx^\nu-   u_\mu   V_\nu dx^\nu)
=\eta_{\mu\nu}dx^\nu\wedge (2\fomega)$\,.

\subsection{\fixform{$(\fQNoether)_{H}$} at horizon and boundary}
Let us in the end summarize the above computation of $T_0$, $T_1$, $T_2$, $T_3$ and $T_4$ 
to evaluate $(\fQNoether)_{H}$ both at the horizon and boundary. 

For the anomaly polynomial $\fP_{CFT}= c_{_M} \fF \wedge \tr[\fR^2]$ in AdS$_5$, we find that $(T_0+T_1+T_2)$ at the horizon and boundary respectively
takes the following form :
 \bea
 (T_0+T_1 + T_2) \athor
&=&
\delta\left\{  (\JSanomCSV)\, [\fu
\wedge (2 \fomega)]\right\}
-\left(\frac{r_H}{4\pi T}\right)(\JSanomCSV)
\left[ (\delta \fu)
 \wedge (2 \fomega)\right]\, ,
 \nonumber \\
 (T_0+T_1 + T_2) \atinfty 
&=&-2c_{_M}\form{\xi} \wedge\delta(2\fomega q)\,, 
 \eea   
 where in the computation at the horizon, we have used Eq.~\eqref{eq:integrationbypart} and neglected the terms proportional to $dr$ as well as the total derivative term. 
On the other hand, the terms $T_3$ and $T_4$ at arbitrary $r$ are given in 
\eqref{eq:T3ads5generalr} and \eqref{eq:T4b}
with $\GV= c_{_M}\,\Phi (2\Phi_T^{2})$ and $\HVtemp = c_{_M}\, \Phi$. 
In particular, the sum of these two terms $(T_3+T_4)$ at the horizon and boundary
is respectively evaluated as 
\bea\label{eq:T3nT4horAdS5}
(T_3+T_4) \athor&=&
\left(\frac{r_H  }{4\pi T }\right)\,( \JSanomCSV)\,
        [  (\delta \fu) \wedge (2\fomega)]\,,\nonumber\\
(T_3+T_4)  \atinfty &= &
2c_{_M}\form{\xi}\wedge\delta (2\fomega q)\, , 
\eea
with $\JSanomCSV= - (2\pi)\times 2^2\times \,c_{_M} \times (2\pi T)\times \mu\,$. 

Combining all the contributions, we finally have the expression for $(\fQNoether)_{H}$ 
at the horizon and the boundary as follows :
 \bea
(\fQNoether)_{H} \athor&=&\delta\left[
 (\JSanomCS)_\mu \right] \hodgeCFT dx^\mu\,, \\
 (\fQNoether)_{H} \atinfty&=&- \delta\left[
\xi^\mu \TmnanomCS+(\Lambda+\ic_{\xi} \fA)|_{ \infty}\JnanomCS \right]\hodgeCFT dx^\nu
\,, 
 \eea
 where the anomaly-induced currents are given by 
 \be
 \TmnanomCS=\JmanomCS=0\,,\quad
 (\JSanomCS)_\mu =
    \JSanomCSV V_\mu\, , 
\ee with $\JSanomCSV= - (2\pi)\times 2^2\times \,c_{_M} \times (2\pi T)\times \mu\,$.
This result verifies Eqs.~\eqref{eq:TJanomCS0} and \eqref{eq:defJSanomCSV} 
for the anomaly polynomial $\fP_{CFT}= c_{_M} \fF \wedge \tr[\fR^2]$ in AdS$_5$.


\section{Replacement rule for \texorpdfstring{$T_3$}{T3} and \texorpdfstring{$T_4$}{T4}}
\label{app:replacementT3T4}
As in Eq.~(\ref{eq:deltaQcssum}), we divide 
$(\fQNoether)_{H}$ into 
$T_0$,  $T_1$, $T_2$, $T_3$ and $T_4$. In particular, the terms $T_3$ and $T_4$, \begin{eqnarray}
T_3 &\equiv & -\half \delta G_{ab} (\SpH)^{abc}\, i_\xi \hodge dx_c\, ,   \label{eq:t3_general}\\
T_4 &\equiv & -\frac{1}{4} \xi^a \delta\left\{
\left[(\SpH)_a{}^{bc} + (\SpH)^b{}_a{}^c + (\SpH)^{cb}{}_a\right] \hodge (dx_b \wedge dx_c)
\right\},  \label{eq:t4_general}
\end{eqnarray} 
are related to the spin Hall current $(\SpH)^{cb}{}_a \hodge dx_c\equiv -2 (\partial \fP_{CFT} /\partial \fR^a{}_b)$
and are relatively easy to deal with. The final expression for $T_3$ and $T_4$ at a fixed arbitrary $r$ is simple. 
Moreover, by using essentially the same argument as 
we used for the Einstein source in Appendix D.6 Ref.~\cite{Azeyanagi:2013xea}, 
one can prove the statement that the 2nd and higher order building blocks
do not contribute to $T_3$ and $T_4$ at the leading order of the derivative expansion. 

In the following, we will first begin with the general single trace case $\fP_{CFT}= c_{_M}\, \fF^{l} \wedge \tr[\fR^{2k}]$ and then move on to 
the purely gravitational multi-trace case
$\fP_{CFT}= c_{_g} \, \tr[\fR^{2k_1}]\wedge \tr[\fR^{2k_2}]\wedge \ldots\wedge \tr[\fR^{2k_p}]$.
 Finally, we will consider the most general anomaly polynomial  of the form 
 $\fP_{CFT}= c_{_M}\, \fF^{l} \wedge \tr[\fR^{2k_1}]\wedge \tr[\fR^{2k_2}]\wedge \ldots \wedge \tr[\fR^{2k_p}]$.

In the first part of this Appendix, we will only consider the zeroth and first order building blocks.
At the end, we briefly explain why the 2nd and higher order terms do not contribute 
to $T_3$ and $T_4$ at the leading order based on the argument for the Einstein source 
in Appendix D.6 of Ref.~\cite{Azeyanagi:2013xea}. We stress that the computation here is valid for any fixed $r$. 
In some places, we will specifically evaluate them
at infinity $r\to\infty$ and at the horizon $r = r_H$ to explicitly display the expressions of 
$T_3$ and $T_4$. 

Before computing $T_3$ and $T_4$ for a specific $\fP_{CFT}$,  similar to the case in AdS$_5$ (see around Eq.~\eqref{eq:t3t4simplifiedads5}),
we massage the expression of $T_3$ and $T_4$ into the following forms for later use :
\bea
T_3 &=& -\half (\delta G_{a\beta}) (\SpH)^{a\beta r}
r^{d-1} \, \hodgeCFT \form{\xi}
\, , \qquad
T_4 \equiv \xi^\mu \delta( \tilde{T}_4)_\mu\, , \label{eq:t3t4simplified}
\eea
where 
\bea
(\SpH)^{a\beta r} &=& (\SpH)^{a\beta }{}_r (2\Psi) + (\SpH)^{a\beta}{}_\gamma u^\gamma\, , 
\label{eq:sigmaalphabetar}
 \\
(\tilde{T}_4)_\mu
&=&
\left[
G_{\mu a}(\SpH)^{(a \gamma)r}
+\frac{1}{2}
(\SpH)^{ r \gamma}{}_\mu
\right]
\times  (r^{d-1} \hodgeCFT \eta_{\gamma\rho}dx^\rho)
\equiv (\tilde{T}_4^{(a)} )_\mu+ (\tilde{T}_4^{(b)})_\mu
\, .  \label{eq:T4divided}
\eea
To derive \eqref{eq:t3t4simplified} and \eqref{eq:sigmaalphabetar}, we have used 
$\xi^r=0$ and the identities 
\be
\hodge dx_r=r^{d-1} \hodgeCFT 1\, , \quad
\ic_{\xi} \hodgeCFT \form{C} = \hodgeCFT(\form{C} \wedge \form{\xi})\, , 
\ee
for any boundary form $\form{C}$
as well as the anti-symmetric property of the spin Hall current, 
$(\SpH)^{abc}=-(\SpH)^{acb}$.

\subsection{\fixform{$\fP_{CFT}=  c_{_M}\, \fF^{l} \wedge \tr[\fR^{2k}]$}}
\label{sec:gen1T3nT4}
Let us consider the general single-trace anomaly polynomial 
$\fP_{CFT}=  c_{_M}\, \fF^{l} \wedge \tr[\fR^{2k}]$ admitted by 
AdS$_{d+1}$ with $d=2n=2l+4k-2$ for $d\ge6$. The spin Hall current in this case is 
given by 
\bea
(\SpH)^{cb}{}_a&=&-2c_{_M}\,(2k)\,\varepsilon^{c p_1 q_1\ldots p_l q_l r_1 s_1 \ldots r_{2k-1} s_{2k-1}}
\left(\half\right)^{l}\left( F_{p_1 q_1 } \ldots F_{p_l q_l}\right)  \\
&&\quad\times \left(\half\right)^{2k-1} R^{b}{}_{c_1 r_1 s_1} R^{c_1}{}_{c_2 r_2 s_2}\ldots
R^{c_{2k-3}}{}_{c_{2k-2} r_{2k-2} s_{2k-2}}
R^{c_{2k-2}}{}_{a r_{2k-1} s_{2k-1}}\, .\nonumber
\eea

As is known from the Einstein source computation carried out in Appendix D.6 of Ref.~\cite{Azeyanagi:2013xea},
there are three types of terms in $(\SpH)^{ab}{}_c$
that can contribute up to $\omega^{n-1}$ order,  depending on how $(\F{0})$ and $(\R{0})$ are
distributed. For $k=1$ (thus $n=l+1$),  
\bea
{\cal O}(\fomega^{n-2}) &:& 
(\SpH)'^{ab}{}_c\propto \epsilon^{a p_1q_1\cdots p_l q_l r_1s_1
 } (\F{0}\F{1}^{l-1})_{p_1q_1\cdots p_l q_l} (\R{0})^b{}_{c r_1 s_1} \, , \nonumber\\
{\cal O}(\fomega^{n-1}) &:&
(\SpH^{(1)})^{ab}{}_c\propto  
\epsilon^{a  p_1q_1\cdots p_l q_l r_1s_1 }
(\F{0}\F{1}^{l-1})_{p_1q_1\cdots p_l q_l}(\R{1})^b{}_{cr_1s_1} \, ,  \\
&& (\SpH^{(2)})^{ab}{}_c \propto \epsilon^{a 
p_1q_1\cdots p_l q_l r_1 s_1 } (\F{1}^l)_{p_1q_1\cdots p_l q_l} (\R{0})^b{}_{c r_1 s_1},  \nonumber 
\eea which is just a generalization of Eq.\eqref{eq:ads5SpH} in the case of AdS$_5$.
For $k>1$, 
\bea
&&{\cal O}(\omega^{n-2}) :\nonumber\\
&& (\SpH)'^{ab}{}_c
 \propto 
\epsilon^{ap_1q_1\cdots p_l q_l r_1s_1\cdots r_{2k-1}s_{2k-1}
 } (\F{1}^{l})_{p_1q_1\cdots p_l q_l} (\R{0}\R{1}\R{0})^b{}_{c r_1s_1r_2s_2r_3s_3}
(\fomega^{2k-4})_{r_4s_4\cdots r_{2k-1}s_{2k-1}} \, ,\nonumber \\
\nonumber\\
&&{\cal O}(\omega^{n-1}) :\nonumber\\
&&(\SpH^{(1)})^{ab}{}_c\propto  
\epsilon^{a p_1q_1\cdots p_l q_l r_1s_1\cdots r_{2k-1}s_{2k-1} }
(\F{0}\F{1}^{l-1})_{p_1q_1\cdots p_l q_l}(\R{1}^{2k-1})^b{}_{cr_1s_1r_2\cdots r_{2k-1}s_{2k-1}}\, , \nonumber \\
&& (\SpH^{(2)})^{ab}{}_c \propto\epsilon^{a  p_1q_1\cdots p_l q_l r_1s_1\cdots r_{2k-1}s_{2k-1}} (\F{1}^l)_{p_1q_1\cdots p_l q_l} (\R{0}\R{1}\R{1})^b{}_{cr_1s_1r_2s_2r_3s_3}(\fomega^{2k-4})_{r_4s_4\cdots r_{2k-1}s_{2k-1}}\, ,\nonumber \\
&&\qquad{\rm or} \qquad\epsilon^{a  p_1q_1\cdots p_l q_l r_1s_1\cdots r_{2k-1}s_{2k-1} } (\F{1}^l)_{p_1q_1\cdots p_l q_l} (\R{1}\R{0}\R{1})^b{}_{cr_1s_1r_2s_2r_3s_3}(\fomega^{2k-4})_{r_4s_4\cdots r_{2k-1}s_{2k-1}} \, , \nonumber\\
&&\qquad  {\rm or}\qquad \epsilon^{a p_1q_1\cdots p_l q_l r_1s_1\cdots r_{2k-1}s_{2k-1}} (\F{1}^l)_{p_1q_1\cdots p_l q_l} (\R{1}\R{1}\R{0})^b{}_{cr_1s_1r_2s_2r_3s_3}(\fomega^{2k-4})_{r_4s_4\cdots r_{2k-1}s_{2k-1}}\, . \nonumber
\\
\eea

Here we have used the fact that $(\F{0})\propto dr\wedge \fu$ and $(\R{0}^2)=0$. 
We note that, for $k>1$, the wedge product of odd numbers of the curvature 2-forms
$(\fR^{2k-1})$ consisting of all $(\R{1})$'s except exactly one $(\R{0})$
reduces to $(\R{1}\R{1}\R{0})^b{}_c$, $(\R{1}\R{0}\R{1})^b{}_c$ or $(\R{0}\R{1}\R{1})^b{}_c$ 
wedged by $(2\Phi_T\fomega)^{2k-4}$.  
We also notice that $(\F{0})\wedge (\R{0}\R{1})=(\F{0})\wedge (\R{1}\R{0})=0$,  
because $(\F{0}) \propto dr\wedge \fu$ and all the terms in $(\R{0}\R{1})$ and $(\R{1}\R{0})$ 
are either proportional to $dr$ or $\fu$ .

\subsubsection{Term $T_3$} 
From the expression of \eqref{eq:t3t4simplified}, to evaluate $T_3$, 
we only need to evaluate some specific components of the spin Hall current $(\SpH)^{a\beta r}$ 
in \eqref{eq:sigmaalphabetar}.  We evaluate contribution to these components 
from $(\SpH)'^{ab}{}_c$, $(\SpH^{(1)})^{ab}{}_c$ and $(\SpH^{(2)})^{ab}{}_c$ separately.
As we will see, these computations follow in the same vein as the AdS$_5$ computations in Appendix~\ref{sec:T3ads5}. \\
\vspace{0.01cm} \\
\noindent
{\bf \underline{$(\SpH)'$ term:}} \\
The same argument as the AdS$_5$ case in Appendix~\ref{sec:T3ads5} holds by replacing $(\R{0})$ with $(\R{0}\R{1}\R{0})$ for $k>1$. 
Therefore, $(\SpH)'$ term does not contribute to the third term $T_3$\,.
\vspace{0.01cm} \\
\noindent
{\bf \underline{$(\SpH^{(1)})$ term:}} \\
By noticing that $(\F{0}) \propto dr \wedge \fu$ and $(\R{1}^{2k-1})^\beta{}_r\propto \fu$, 
we have
\begin{eqnarray}
(\SpH^{(1)})^{rb}{}_c=0\, , \qquad 
(\SpH^{(1)})^{\alpha \beta}{}_r =0\, , 
\end{eqnarray}
(and thus $(\SpH^{(1)})^{r\beta r}=0$).
Hence, the non-trivial components of Eq. \eqref{eq:sigmaalphabetar} are calculated as 
\be
 ( \SpH^{(1)})^{\alpha\beta r} = 
 ( \SpH^{(1)})^{\alpha\beta}{}_\gamma u^\gamma=-c_{_M}
\frac{2(2k)}{r^{d-1}}
\partial_r(  \Phi^{l}) \Phi_T^{2k-1} V^\alpha u^\beta \, . 
 \ee 
\noindent
{\bf \underline{$(\SpH^{(2)})$ term:}}\\
We first notice that 
$(\R{0})^{\beta r}$, $(\R{0}\R{1}\R{1})^{\beta r}$, $(\R{1}\R{0}\R{1})^{\beta r}$, $(\R{1}\R{1}\R{0})^{\beta r}$
are all proportional to $dr$. This leads to $( \SpH^{(2)})^{r\beta r}=0$ 
and thus we can rewrite \eqref{eq:t3t4simplified} for $( \SpH^{(2)})$ as 
\begin{eqnarray}
T_3 |_{\SpH^{(2)}} &=&
  -\half (\delta G_{\alpha\beta}) (\SpH^{(2)})^{(\alpha\beta) r} r^{d-1}( \hodgeCFT \form{\xi})
\, , 
\end{eqnarray}
where 
\bea
 ( \SpH^{(2)})^{(\alpha \beta) r} 
  =
   ( \SpH^{(2)})^{(\alpha \beta)}{}_\gamma u^\gamma=
  -c_{_M}\frac{2(2k)}{r^{d}} 
\Phi^{l} \partial_r( r\Phi_T^{2k-1})(V^{(\alpha} u^{\beta)}
)\, . 
\eea
In the first equality, we have used the fact that  the objects
$ (\R{0})^{\beta}{}_r$,  $(\R{0}\R{1}\R{1})^{\beta}{}_{ r}$, $(\R{1}\R{0}\R{1})^{\beta}{}_{ r}$ and 
$ (\R{1}\R{1}\R{0})^{\beta}{}_{ r}$ are either zero or proportional to $ dx^\beta$, 
which leads to $ ( \SpH^{(2)})^{(\alpha \beta) }{}_r =0$. 

From the above computation, non-trivial components of
the spin Hall current relevant to the computation of $T_3$
is given up to $\omega^{n-1}$ order by
\bea
( \SpH)^{(\alpha \beta) r}
=-\frac{1}{r^d} \frac{d}{dr}\left[
r \frac{\partial \GV}{\partial \Phi_T} 
\right] V^{(\alpha} u^{\beta)},  \label{eq:sigmaalphabetasymr}
\eea where $\GV= c_{_M}\,\Phi^l (2\Phi_T^{2k})$. 
Therefore, the leading order expression of $T_3$   
for $\fP_{CFT}= c_{_M}\, \fF^{l} \wedge \tr[{\fR^{2k}}]$
turns out to be
\bea
T_3 = \frac{1}{2r} 
(2\Psi- r^2 )
 \frac{d}{dr}\left[
r \frac{\partial \GV}{\partial \Phi_T} 
\right] \ic_{\xi}\left[
(\delta \fu)\wedge \fu \wedge (2\fomega)^{n-1}\right] \,.
\eea 

When we evaluate at $r=r_H$ (at the horizon) or $r\to \infty$ (at the boundary), substituting $\xi^a \athor=\xiH^a$, $\Lambda\athor = \LambdaH$ as in Eq.~\eqref{eq:horsub} while at the boundary choosing $\xi^a$ 
 as in Eq.~\eqref{eq:infsub}, we have the following expressions : 
\bea
T_3 \athor&=&
-\frac{r_H}{2 T} 
  \left[
 \frac{\partial \GV}{\partial \Phi_T} 
+ r\frac{d}{dr}
 \frac{\partial \GV}{\partial \Phi_T} 
 \right]_{hor} 
(\delta \fu) \wedge (2\fomega)^{n-1} \, , \\
T_3 \atinfty &\to& {\cal O}\left(
\frac{1}{r^{d-1}}\frac{1}{r^{l(d-2)}} \frac{1}{r^{(d-1)(2k-1)}} \right)\,,
\eea where we have used Eq.~\eqref{eq:fields_limitinfty}. Hence, $T_3 \atinfty =0$ since we are considering $n\ge 2$ (where $2k+l\ge3$ and $k\ge1$). 

\subsubsection{Term $T_4$}
To calculate $T_4$ at the leading order of the derivative expansion, 
we start with the evaluation of $(\tilde{T}_4^{(a)})_\mu$ and $(\tilde{T}_4^{(b)})_\mu$ 
as defined in \eqref{eq:T4divided}. \\
\vspace{0.01cm} \\
\noindent
\underline{\bf $(\tilde{T}_4^{(a)})$:} \\ 
By using the results $(\SpH)^{r\beta r}=0$ up to $\omega^{n-1}$ order 
and the expression of $( \SpH)^{(\alpha \beta) r}
$ from Eq.\eqref{eq:sigmaalphabetasymr}, first term $(\tilde{T}_4^{(a)})_\mu$ is easily calculated as 
\bea
(\tilde{T}_4^{(a)})_\mu
&=&
-\frac{1}{2r} 
\left[
2 \Psi u_\mu \hodgeCFT \fV 
+ r^2 V_\mu \hodgeCFT \fu
\right]\frac{d}{dr}\left[
r \frac{\partial \GV}{\partial \Phi_T} 
\right] \, .
\eea
\vspace{0.01cm} \\
\noindent
\underline{\bf $(\tilde{T}_4^{(b)})$:} \\ 
As we have shown in the case of $T_3$ above, 
$(\SpH)'^{r\alpha}{}_\mu = (\SpH^{(1)})^{r\alpha}{}_\mu=0$. 
Therefore only $(\SpH^{(2)})^{r\alpha}{}_\mu$ can contribute to 
the second term $(\tilde{T}_4^{(b)})$. Direct computations show
that $(\SpH^{(2)})^{r\alpha}{}_\mu$ is 
\begin{eqnarray}
{\rm for } \,\,k=1 &:&  \quad 
(\SpH^{(2)})^{r \alpha}{}_\mu=
\frac{1}{r^{d}}
\left[ 
 r^2  V_\mu u^\alpha-
 2 \Psi    
u_\mu V^\alpha
\right]\left[
\frac{\partial \GV}{\partial \Phi_T} 
+8r^{-1}\Psi \,\HVtemp 
\right]\,, \\
{\rm for }\,\, k>1 &:&  \quad 
(\Sigma^{(2)})^{r \alpha}{}_\mu
=
\frac{1}{r^{d}}
\left[
r^2 V_\mu u^\alpha 
-2 \Psi  u_\mu V^\alpha
\right]\frac{\partial \GV}{\partial \Phi_T}\, , 
\end{eqnarray}
where $\GV= c_{_M}\,\Phi^l (2\Phi_T^{2k})$ and $\HVtemp = c_{_M}\, \Phi^l$. 
In deriving the results above, it is helpful to remember 
\be
    \varepsilon^{\mu\alpha \lambda_1 \sigma_1 \ldots \lambda_{n-1} \sigma_{n-1}}\omega_{\lambda_1 \sigma_1}\ldots \omega_{\lambda_{n-1} \sigma_{n-1}}
     =u^\mu V^\alpha-u^\alpha V^\mu\, ,
\ee which is just a generalization of Eq.~\eqref{eq:epstouV} (in the case of AdS$_5$).
We note that in the expression for $(\Sigma^{(2)})^{r \alpha}{}_\mu$, the only difference between $k=1$ case and $k\ge1$ case 
comes from $\HVtemp$.
Finally, we have the expression for $(\tilde{T}_4^{(b)})_\mu$  as follows
(note that $\hodgeCFT \fV = \fu\wedge (2\fomega)^{n-1}$) :
\begin{eqnarray}
{\rm for }\,\, k=1 &:&
 (\tilde{T}_4^{(b)})_\mu=
\frac{1}{2r} 
\left[
 r^2  V_\mu \hodgeCFT \fu
-2 \Psi    
u_\mu \fu\wedge (2\fomega)^{n-1}\right]
\left[
\frac{\partial \GV}{\partial \Phi_T} 
+8r^{-1}\Psi \,\HVtemp 
\right], \\
{\rm for }\,\, k>1 &:&
 (\tilde{T}_4^{(b)})_\mu= 
 \frac{1}{2r}
\left[
r^2 V_\mu\hodgeCFT \fu
-2  \Psi  u_\mu \fu\wedge (2\fomega)^{n-1}
\right]\frac{\partial \GV}{\partial \Phi_T}  . 
\end{eqnarray}

Combining the results for $ (\tilde{T}_4^{(a)})$ and $(\tilde{T}_4^{(b)})$ above, 
we obtain $T_4$ in the leading order of the derivative expansion as 
\begin{eqnarray}
{\rm for }\,\, k=1 &:&
T_4 = 
\xi^\mu \delta\left\{
-\frac{r^2}{2} 
\frac{d}{dr}\left[
 \frac{\partial \GV}{\partial \Phi_T} 
\right]  V_\mu \hodgeCFT \fu
-  u_\mu \frac{\Psi}{r^2}\frac{d}{dr}\left[r^2
 \frac{\partial \GV}{\partial \Phi_T} 
\right]\fu\wedge(2\fomega)^{n-1}
\right.\nonumber\\
&&\qquad\qquad\qquad\qquad \left.
+4r^{-2}\Psi \,\HVtemp 
\left[
r^2V_\mu \hodgeCFT\fu
-2 \Psi u_\mu  \fu\wedge (2\fomega)^{n-1}
\right]  \right\}\, , \\
{\rm for }\,\, k>1 &:&
 T_4=\xi^\mu\delta
 \left\{- 
\frac{r^2}{2}
\frac{d}{dr} \left[\frac{\partial \GV}{\partial \Phi_T} 
\right]  V_\mu \hodgeCFT \fu
 -u_\mu \frac{\Psi}{r^2}   
\frac{d}{dr}\left[r^2 \frac{\partial \GV}{\partial \Phi_T} \right]\fu\wedge(2\fomega)^{n-1} \right\}\,. \nonumber\\
\end{eqnarray}

At the horizon, the term related to ${\mathbb H}^{(V)}$ vanishes and upon substituting  $\xi^a \athor=\xiH^a$, $\Lambda\athor = \LambdaH$ as in Eq.~\eqref{eq:horsub} , $T_4$ at the horizon is given by 
\begin{eqnarray}
T_4 \athor&=&
\frac{r_H^2}{2T}
\frac{d}{dr} \left[\frac{\partial \GV}{\partial \Phi_T} 
\right] _{hor}
(\delta \fu)\wedge(2\fomega)^{n-1} \,,
\end{eqnarray}  for all $k \ge 1$.

On the other hand, at $r\to \infty$, using $\xi^a$ as in Eq.~\eqref{eq:infsub}, depending on the value of $k$, we have 
\begin{eqnarray}
{\rm for }\,\, k=1 &:&  
T_4 \atinfty \to {\cal O}\left(
 \frac{q^l}{r^{l(d-2)-2}}\right)\, ,
\\
{\rm for }\,\, k>1 &:&  
T_4\atinfty \to {\cal O}\left(
\frac{1}{r^{d(d-1)/2-(l+1)}}\right)\, .
\end{eqnarray}
For $k>1$, which exists for $d\ge6$, the term $T_4$ obviously vanishes at $r\rightarrow \infty$.
On the other hand, in $d\ge6$ with $k=1$, we have $l\ge 2$ and thus $T_4$ vanishes at $r\rightarrow \infty$.
Thus, $T_4$ vanishes at infinity for $d\ge 6$. 

 \subsubsection{$T_3+T_4$}
 To summarize, for  $\fP_{CFT}= 
c_{_M}\, \fF^l \wedge \tr[\fR^{2k}]$, we obtained the following results at any fixed $r$ :
\bea
T_3 
&=&
\frac{1}{2r} 
(2\Psi- r^2 )
 \frac{d}{dr}\left[
r \frac{\partial \GV}{\partial \Phi_T} 
\right] \ic_{\xi}\left[
(\delta \fu)\wedge \fu \wedge (2\fomega)^{n-1}\right]\, , \nonumber\\
T_4 &= & 
\xi^\mu \delta\left\{
-\frac{r^2}{2} 
\frac{d}{dr}\left[
 \frac{\partial \GV}{\partial \Phi_T} 
\right]  V_\mu \hodgeCFT \fu
-  u_\mu \frac{\Psi}{r^2}\frac{d}{dr}\left[r^2
 \frac{\partial \GV}{\partial \Phi_T} 
\right]\fu\wedge(2\fomega)^{n-1}
\right.\nonumber\\
&&\left.\qquad\qquad
+4r^{-2}\Psi \,\HVtemp 
\left[
r^2V_\mu \hodgeCFT\fu
-2 \Psi u_\mu  \fu\wedge (2\fomega)^{n-1}
\right]  \right\}\, , 
\eea  where $\GV=c_{_M}\, \Phi^l (2\Phi_T^{2k})$. 
For $\HVtemp$, it takes nonzero value  $\HVtemp = c_{_M}\, \Phi^l $ for $k=1$, 
while it vanishes for $k>1$.  
In particular, $T_3+T_4$ at infinity $r\to\infty$ vanishes for $d\ge6$, while 
at the horizon $r=r_H$ we have 
\begin{eqnarray}
(T_3 +T_4) \athor
&=&
\left(\frac{r_H  }{4\pi T }\right)\, \JSanomCSV\,
          (\delta \fu) \wedge (2\fomega)^{n-1}\,, 
\end{eqnarray}  where in this case $\JSanomCSV=-2\pi  c_{_M}\times \mu^l \times2\times (2k)\times(2\pi T)^{2k-1}$\,.
 
\subsection{\fixform{$\fP_{CFT}= c_{_g} \, \tr[\fR^{2k_1}]\wedge \tr[\fR^{2k_2}]\wedge \ldots \wedge \tr[\fR^{2k_p}]$}}
\label{sec:T3nT4purelyGrav}
Here we consider the case with the purely gravitational anomaly polynomial consisting of multiple traces,   
$\fP_{CFT}=c_{_g} \, \tr[\fR^{2k_1}]\wedge \tr[\fR^{2k_2}]\wedge \ldots \wedge \tr[\fR^{2k_p}]$ in AdS$_{d+1}$ 
where $d=2n=4(\sum_{i=1}^p k_i)-2\equiv 4k_{tot}-2$.
Without loss of generality, we can assume that $k_i=1$ for $i\le p_0$, i.e. the first $p_0$ $\tr[\fR^{2k_i}]$'s are $\tr[\fR^2]$'s
(in particular $p_0=0$ means that the anomaly polynomial $\fP_{CFT}$ does not contain any $\tr[\fR^2]$).
As in the previous case, here we first consider zeroth and first order building blocks
and will show at the end of this Appendix that 2nd and higher order building blocks do not contribute to 
$T_3$ and $T_4$ at the leading order. 

We first recall that 
\be
\tr[(\R{1}^{2k_i})]=2(2\Phi_T\fomega)^{2k_i}, \quad
\tr[(\R{0}\R{1}^{2k_i-1})]=2\Phi_T' dr\wedge \fu\wedge (2 \Phi_T \fomega)^{2k_i-1} \, , 
\ee  while
\be
(\F{1}^{2l})=(2\Phi \fomega)^{2l},\quad
(\F{0}\F{1}^{2l-1}) = \Phi' dr\wedge \fu \wedge (2\Phi \fomega)^{2l-1}.
\ee Because of these, by replacing the contributions of 
$\fF^l$ consisting of $(\F{1})$'s and at most one $(\F{0})$ by products of traces of $(\fR^{2k_i})$ consisting of all $(\R{1})$'s and at most one $(\R{0})$ (upon sending $\Phi$ to $\Phi_T$ with some appropriate factors of $2$'s), we can carry out the same
classification of the spin Hall current $\SpH$ as we did for $\fP_{CFT}= c_{_M} \,\fF^l\wedge \tr[\fR^{2k}]$ 
and define $(\SpH)'$, $(\SpH^{(1)})$ and $(\SpH^{(2)})$. As in the previous case, the first term $(\SpH)'$ 
does not contribute to both $T_3$ and $T_4$. We thus restrict ourselves to the order $\omega^{n-1}$ 
contribution $(\SpH^{(1)})$ and $(\SpH^{(2)})$ only. 

Since $\tr[\fchi_2]=0$, the derivative of the anomaly polynomial $(\partial \fP_{CFT}/\partial \fR^a{}_b)$
up to order $\omega^{n-1}$ is given by 
\bea
\frac{\partial \fP_{CFT}}{\partial \fR^a{}_b}
&= &
c_{_g} \sum_{i=1}^{p_0}(2 k_i)2^{p-1}
(\R{0})^b{}_a \wedge  (2\Phi_T\fomega)^{2k_{tot}-2}
\nonumber\\
&&+c_{_g} \sum_{i=p_0+1}^{p}(2 k_i)2^{p-1}
[(\R{0}\R{1}\R{1})+(\R{1}\R{1}\R{0})+(2k_i-3)(\R{1}\R{0}\R{1})]^b{}_a\wedge (2\Phi_T\fomega)^{2k_{tot}-4}
\nonumber\\
& &
+c_{_g} \sum_{i,j; i\neq j}^{p}(2 k_i)(2k_j)2^{p-2}(\R{1})^b{}_a\wedge\tr[(\R{0}\R{1})]
\wedge (2\Phi_T\fomega)^{2k_{tot}-4}\, ,  \label{eq:twoterms_case3} 
\eea where $2k_{tot}=n+1$.

The first two terms on the right hand side are essentially in the same form as $(\SpH^{(2)})$ 
appearing in Appendix~\ref{sec:gen1T3nT4}, while the third term is the same as $(\SpH^{(1)})$ there 
under the replacement of $\fF^l$ by the traces of the curvature two-forms explained above. 
Therefore we can straightforwardly compute the nontrivial components of
$(\SpH^{(1)})$ and $(\SpH^{(2)})$ related to $T_3$ as 
\bea
 ( \SpH^{(1)})^{\alpha\beta r} &= &
-c_{_g}\frac{2^{p}}{r^{d}} \left[(2k_{tot})^2-\sum_{i=1}^p (2k_i)^2\right]
 (r\Phi_T' )
 \Phi_T^{2k_{tot}-2}V^\alpha u^\beta \, , \nonumber\\
  ( \SpH^{(2)})^{(\alpha\beta) r} &= &
-c_{_g}\frac{2^{p}}{r^d} \sum_{i=1}^{p}(2 k_i)
\partial_r(r \Phi_T^{2k_{i}-1})
\Phi_T^{2k_{tot}-2k_i}V^{(\alpha} u^{\beta)}
\nonumber\\
&= &
-c_{_g} \frac{2^{p}}{r^d}
\left[
(2 k_{tot})\partial_r(r \Phi_T)
+\left(\sum_i^p (2k_i)^2-4 k_{tot}\right)(r  \Phi_T')
\right]
\Phi_T^{2k_{tot}-2}V^{(\alpha} u^{\beta)}~,\nonumber\\
\eea and thus nontrivial component of the spin Hall current relevant to $T_3$
is at the leading order given by 
\be
( \SpH)^{(\alpha\beta) r} = 
- \frac{1}{r^d}
\frac{d}{dr}\left[r \frac{\partial \GV}{\partial \Phi_T}\right]
V^{(\alpha} u^{\beta)},
\ee where $\GV=c_{_g}\, 2^p\, \Phi_T^{2k_{tot}}$. Therefore we finally have 
the expression of $T_3$ at the leading order as follows :
\bea
T_3 &=&\frac{1}{2r} 
(2\Psi- r^2 )
 \frac{d}{dr}\left[
r \frac{\partial \GV}{\partial \Phi_T} 
\right] \ic_{\xi}\left[
(\delta \fu)\wedge \fu \wedge (2\fomega)^{n-1}\right] .\eea

In a similar manner, the computation of $T_4$ also follows from the computation in 
Appendix \ref{sec:gen1T3nT4}. That is, as a result of $(\SpH)^{r\beta r}=0$ and 
$(\SpH^{(1)})^{r\alpha}{}_\mu=0$, the
nontrivial contribution to $T_4$ comes only from $(\SpH^{(2)})^{r\alpha}{}_\mu$ given by 
\bea
(\SpH^{(2)})^{r \alpha}{}_\mu&=&
\frac{1}{r^{d}}
\left[ 
 r^2  V_\mu u^\alpha-
 2 \Psi    
u_\mu V^\alpha
\right]\left[
\frac{\partial \GV}{\partial \Phi_T} 
+8r^{-1}\Psi \,\HVtemp 
\right],
  \eea where $\HVtemp=p_0 c_{_g}  (2^{p-1}\Phi_T^{2k_{tot}-2}) $.
Therefore, $T_4$ at the leading order is evaluated as 
\bea
T_4 &= & 
\xi^\mu \delta\left\{
-\frac{r^2}{2} 
\frac{d}{dr}\left[
 \frac{\partial \GV}{\partial \Phi_T} 
\right]  V_\mu \hodgeCFT \fu
-  u_\mu \frac{\Psi}{r^2}\frac{d}{dr}\left[r^2
 \frac{\partial \GV}{\partial \Phi_T} 
\right]\fu\wedge(2\fomega)^{n-1}
\right.\nonumber\\
&&\qquad\qquad\left.
+4r^{-2}\Psi \,\HVtemp 
\left[
r^2V_\mu \hodgeCFT\fu
-2 \Psi u_\mu  \fu\wedge (2\fomega)^{n-1}
\right]  \right\}.
  \eea

Let us now summarize the expression of $T_3+T_4$ at $r=r_H$ and  $r\to\infty$.  
At the horizon, substituting $\xi^a \athor=\xiH^a$, $\Lambda\athor = \LambdaH$ as in Eq.~\eqref{eq:horsub}, we obtain 
\be
\label{eq:T3nT4atrH}
(T_3+T_4) \athor=
\left(\frac{r_H  }{4\pi T }\right)\, \JSanomCSV\,
          (\delta \fu) \wedge (2\fomega)^{n-1}\, ,
 \ee  where in this case $\JSanomCSV=
 -2\pi  c_{_g} \times 2^p \times (2k_{tot})(2\pi T)^{2k_{tot}-1}
 $.
 
 On the other hand, at  $r\to \infty$, we have 
\bea
T_3 \atinfty&\to& {\cal O}\left(
 \frac{1}{r^{(d-1)(2k_{tot})}}\right)\, , \nonumber\\
T_4 \atinfty&\to& {\cal O}\left(
\frac{1}{r^{(d-1)(2k_{tot}-2)-2}}\right) \, ,
\eea  both of which vanish since $k_{tot}\ge2$ and $d\ge 6$.
 
\subsection{\fixform{$\fP_{CFT}= c_{_M}\, \fF^{l} 
\wedge \tr[\fR^{2k_1}]\wedge \tr[\fR^{2k_2}]\wedge \ldots  \wedge \tr[\fR^{2k_p}]$ }}
\label{eq:t3pt4general}
As a final case, we consider a general mixed anomaly polynomial consisting of multiple traces
$\fP_{CFT}= c_{_M}\, \fF^{l} \wedge \tr[\fR^{2k_1}]\wedge \tr[\fR^{2k_2}]\wedge \ldots  \wedge \tr[\fR^{2k_p}]$ 
admitted by AdS$_{d+1}$ with $d=2n=2l+4k_{tot}-2$.
Again, without loss of generality, we assume that $k_i=1$ for $i\le p_0$ ($p_0\ge 0$). 

In the current case, by treating $\fF^l$ and traces of the curvature two-forms 
as we did in Appendix~\ref{sec:gen1T3nT4}. We can classify contribution to
the spin Hall current up to $\omega^{n-1}$ order into three cases,  
$(\SpH)'$, $(\SpH^{(1)})$, $(\SpH^{(2)})$. 
Then it is straightforward to show that $(\SpH)'$ does not contribute to $T_3$ and $T_4$,  
and we only need to consider the order $\omega^{n-1}$ contributions which are from $(\SpH^{(1)})$ and $(\SpH^{(2)})$.  

The expression of $(\partial \fP_{CFT}/\partial \fR^a{}_b)$,  up to order $\omega^{n-1}$, is
\bea \label{eq:derivative01only_case4}
\frac{\partial \fP_{CFT}}{\partial \fR^a{}_b}=
& &c_{_M} \sum_{i=1}^{p}(2 k_i)2^{p-1}
\form{S}_{(i)} ^b{}_a
\wedge (2\Phi_T\fomega)^{2k_{tot}-4}\wedge (2\Phi\fomega)^{l}
\nonumber\\
& &+
c_{_M} \sum_{i,j; i\neq j}^{p}(2 k_i)(2k_j)2^{p-2}(\R{1})^b{}_a\wedge\tr[(\R{0}\R{1})]
\wedge (2\Phi_T\fomega)^{2k_{tot}-4}\wedge (2\Phi\fomega)^{l} \nonumber \\
&& +c_{_M}\,l\,(\F{0})\wedge\sum_{i=1}^{p} (2k_i)2^{p-1}(\R{1})^b{}_a\wedge (2\Phi_T\fomega)^{2k_{tot}-2}
\wedge(2\Phi\fomega)^{l-1}\, ,
  \label{eq:threeterms}
\eea
with
\bea
\form{S}_{(i)}^b{}_a&=& [(\R{0}\R{1}\R{1})+(\R{1}\R{1}\R{0})+(2k_i-3)(\R{1}\R{0}\R{1})]^b{}_a\, ,
\,\,\,\, {\rm for}\,\,\,\, i\ge p_0+1\, ,   \nonumber \\
\form{S}_{(i)}^b{}_a&=& (\R{0})^b{}_a\, ,
\,\,\,\, {\rm for}\,\,\,\,1\le i \le p_0\,.
\eea
We can see that the first term on the right hand side of Eq.~(\ref{eq:threeterms}) contributes only to $(\SpH^{(2)})$ 
while the rest of the terms contribute only to $(\SpH^{(1)})$, and they can be evaluated essentially in the same way 
as in Appendix~\ref{sec:gen1T3nT4}. More practically, 
the contribution to the spin Hall current from the first two terms
can be obtained by wedging the result for $l=0$ (see Appendix \ref{sec:T3nT4purelyGrav}) with $(2 \Phi \fomega)^l$, 
while the one from the third term 
can be computed essentially in the same way as $(\SpH^{(1)})$ for the case of $p=1$ (see Appendix \ref{sec:gen1T3nT4}) because the former just contains extra powers of $(2 \Phi \fomega)$ and $(2 \Phi_T \fomega)$ 
compared to the latter.  In the end, we finally have the expression for $T_3$ and $T_4$ at any fixed $r$ as  
\bea
T_3 &=&\frac{1}{2r} 
(2\Psi- r^2 )
 \frac{d}{dr}\left[
r \frac{\partial \GV}{\partial \Phi_T} 
\right] \ic_{\xi}\left[
(\delta \fu)\wedge \fu \wedge (2\fomega)^{n-1}\right] \,,\nonumber\\
T_4 &= & 
\xi^\mu \delta\left\{
-\frac{r^2}{2} 
\frac{d}{dr}\left[
 \frac{\partial \GV}{\partial \Phi_T} 
\right]  V_\mu \hodgeCFT \fu
-  u_\mu \frac{\Psi}{r^2}\frac{d}{dr}\left[r^2
 \frac{\partial \GV}{\partial \Phi_T} 
\right]\fu\wedge(2\fomega)^{n-1}
\right.\nonumber\\
&&\left. \qquad\qquad
+4r^{-2}\Psi \,\HVtemp 
\left[
r^2V_\mu \hodgeCFT\fu
-2 \Psi u_\mu  \fu\wedge (2\fomega)^{n-1}
\right]  \right\}\, ,
\eea where
\be\label{eq:GVHVgeneral}
\GV=c_{_M}\, \mu^l \, 2^p\, \Phi_T^{2k_{tot}};
\quad \HVtemp= p_0 c_{_M}  (2^{p-1}\Phi^l \Phi_T^{2k_{tot}-2})\,.
\ee

Let us finally evaluate $T_3+T_4$ at the horizon and infinity. 
The expression at the horizon (upon substituting Eq.~\eqref{eq:horsub}) is the same as Eq.~(\ref{eq:T3nT4atrH}) 
where $\JSanomCSV$ is replaced by the expressions
$\JSanomCSV=-2\pi  c_{_M}\times  \mu^l \times 2^p\,\times (2k_{tot}) (2\pi T)^{2k_{tot}-1}.$

On the other hand,  at infinity $r\to\infty$, we have 
\bea
T_3 \atinfty&\rightarrow& {\cal O}\left(
 \frac{1}{r^{(d-2)l+(d-1)(2k_{tot})}}\right)\,, \nonumber\\
T_4 \atinfty &\rightarrow& {\cal O} \left(
\frac{1}{r^{(d-2)l+(d-1)(2k_{tot}-2)-2}} \right)\,,
\eea  both of which vanish since $d\ge 6$ and $k_{tot}\ge 2$\,.

\subsection{On 2nd and higher order building blocks}
\label{sec:2ndorderT3andT4}
Up to this point, we only considered the zeroth and first order building blocks. Here we briefly explain why 
2nd and higher order building blocks do not contribute to $T_3$ and $T_4$ up to $\omega^{n-1}$ order. 

In Appendix D.6 of Ref.~\cite{Azeyanagi:2013xea}, we have proved that these 2nd order and higher order building blocks do not 
contribute to the Einstein sources. One of the key points of the proof is that
the 2nd and higher order building blocks do not contribute to 
$(\partial \fP_{CFT}/ \partial \fR^a{}_b)$ (and thus to the spin Hall current)
up to $\omega^{n-1}$ order. 
Since both $T_3$ and $T_4$ are linear in the spin Hall current, this result 
is sufficient to prove that 2nd and higher order building blocks do not contribute to $T_3$ and $T_4$ 
up to $\omega^{n-1}$ order.


\section{Contributions to \texorpdfstring{$T_0,T_1$ and $T_2$}{T0,T1 and T2} at horizon:  without higher order}
\label{sec:T0T1T2hor0th1st}
In this Appendix, we evaluate $T_0$, $T_1$ and $T_2$ terms 
for a general anomaly polynomials admitted on AdS$_{2n+1}$ with $n\ge3$. 
We also combine with the results on $T_3$ and $T_4$ we obtained in 
Appendix~\ref{app:replacementT3T4}
to calculate $(\fQNoether)_{H}$ at the horizon. 
We start with some specific warm-up examples and then consider the general cases.    

For the curvature and gauge field strength 2-form  at the zeroth and first orders and its products, drastic simplifications occur at the horizon.
As we will explain, most of terms appearing in the products vanish or is proportional to $dr$
while the remaining nontrivial terms have simple structures. In particular, for the $\fF^l$ part, 
since $(\F{0})\propto dr\wedge \fu$, we can replace $\fF^l$ by $(\F{1}^l)$.  
Therefore, we essentially only need to deal with purely gravitational 
anomaly polynomials. 

Furthermore, due to Eq.~\eqref{eq:horsub}, $(\Lambda + \ic_{\xi} \fA) \athor=\LambdaH+\mu/T=0 $ (up to 1st order) and hence $T_0 \athor=0$ for a general anomaly polynomial $\fP_{CFT}$. Thus, we only need to evaluate $T_1$ and $T_2$ at the horizon. 

For the rest of this Appendix, we shall first evaluate $(\fQNoether)_{H} \athor$ using zeroth and first order 
building blocks. Similar to the case of the Einstein sources in Ref.~\cite{Azeyanagi:2013xea}, 2nd and higher order building blocks do not contribute to $T_1$ and $T_2$ at the horizon 
up to $\omega^{n-1}$ order of the derivative expansion. 
As we will show in Appendix~\ref{sec:entropyhighercontribution}, the restriction at the horizon makes this proof simpler too.

\subsection{Warm-up example 1 : \fixform{$\fP_{CFT}=c_{_M} \fF^l\wedge  \tr[\fR^2]$}}
As a first warm-up example, we consider $\fP_{CFT}=c_{_M} \fF^l\wedge  \tr[\fR^2]$ in AdS$_{2n+1}$ with $n=l+1$. 
We notice that all the $\fF$'s in $T_1$ and $T_2$
can be set to be $(\F{1})=(2\Phi \fomega)$,  since $(\F{0})\propto dr\wedge \fu$.  
Then the evaluation of $T_1$ and $T_2$ is essentially the same as the case of $\fP_{CFT}=c_{_M} \fF \wedge \tr[\fR^2]$ in Appendix~\ref{sec:deltaQcsads5eq0} (with extra $(2\Phi \fomega)$'s). Therefore, using Eq.~(\ref{eq:integrationbypart}) and 
substituting $\xi^a \athor=\xiH^a$, $\Lambda\athor = \LambdaH$ as in Eq.~\eqref{eq:horsub},
we obtain
\bea
T_1 \athor&=&
\delta\left[
 (\JSanomCSV)_l \,\fu\right] \wedge(2 \mu \fomega)^{n-1}]
-\left(\frac{r_H}{4\pi T}\right)\JSanomCSV[
 (\delta \fu)\wedge(2\fomega)^{n-1}]\, , \nonumber\\
T_2  \athor&=& (\JSanomCSV)_l \,\fu\wedge \delta [(2 \mu \fomega)^{n-1}]\, , 
\eea where $\JSanomCSV=-2\pi  c_{_M}\times  \mu^l \times 2^2\times (2\pi T)$ and $
(\JSanomCSV)_l=-2\pi  c_{_M}   \times 2^2\times (2\pi T)$.
This leads to
\be
(T_0+T_1 +T_2) \athor=
\delta\left\{
 (\JSanomCSV) \,[\fu\wedge (2 \fomega)^{n-1}]\right\}
-\left(\frac{r_H  }{4\pi T }\right)\, \JSanomCSV\,
         [ (\delta \fu) \wedge (2\fomega)^{n-1}]\,.
\ee
Combining with the results of $T_3$ and $T_4$ in Appendix~\ref{app:replacementT3T4},  
\be
(T_3+T_4) \athor=
\left(\frac{r_H  }{4\pi T }\right)\, \JSanomCSV\,
          (\delta \fu) \wedge (2\fomega)^{n-1}\,,
\ee we finally have the full expression for $(\fQNoether)_{H} \athor$
\bea
(\fQNoether)_{H} \athor
&=& \delta \bigbr{ (J_{S}^{\text{\tiny{CFT,anom}}})_\mu \hodgeCFT dx^\mu }\,,
\eea where the entropy current is
\be
(J_{S}^{\text{\tiny{CFT,anom}}})_\mu =
\JSanomCSV \, V_\mu\,.
\ee

\subsection{Warm-up example 2 :  \fixform{$\fP_{CFT}=c_{_g}\tr[\fR^2]\wedge \tr[\fR^2]$}}
As a second warmup example, we consider $\fP_{CFT}=c_{_g}\tr[\fR^2]\wedge \tr[\fR^2]$ in AdS$_{7}$ where $n=3$.
Using Eq.~(\ref{eq:hordeltaGgradxi})-(\ref{eq:hordeltaGR}) and Eq.~(\ref{eq:integrationbypart}) while
substituting $\xi^a \athor=\xiH^a$, $\Lambda\athor = \LambdaH$ as in Eq.~\eqref{eq:horsub}, we have
\bea
T_1 \athor
&=&4c_{_g}\left.\left\{
\tr[\nabla \xi( \delta \fGamma) ]  \wedge\tr[\fR^2] 
+2\tr[( \delta \fGamma)\wedge \fR]\wedge\tr[\nabla \xi \fR] \right\} \right |_{r=r_H}
\nonumber\\
&=&
\delta\left\{
 (\JSanomCSV) \,[\fu\wedge (2 \fomega)^{2}]\right\}
-\left(\frac{r_H  }{4\pi T }\right)\, \JSanomCSV\,
          (\delta \fu) \wedge (2\fomega)^{2}\,,
\eea where
$\JSanomCSV=-2\pi  c_{_g}\,    2^2 \times 4 \times (2\pi T)^3$.
Note that $T_2$ is zero since $\fP_{CFT}$ does not contain any $\fF$'s. 

Combining with the result of $T_3+T_4$ in Appendix~\ref{app:replacementT3T4}
evaluated at the horizon, 
\be
(T_3+T_4) \athor=
\left(\frac{r_H  }{4\pi T }\right)\, \JSanomCSV\,
          (\delta \fu) \wedge (2\fomega)^{2}\,,
\ee 
we finally have the expression for $(\fQNoether)_{H} \athor$ : 
\be
(\fQNoether)_{H} \athor
= \delta \bigbr{ (J_{S}^{\text{\tiny{CFT,anom}}})_\mu \hodgeCFT dx^\mu }\quad , \quad (J_{S}^{\text{\tiny{CFT,anom}}})_\mu =
\JSanomCSV \, V_\mu\,.
\ee

\subsection{Warm-up example 3 :  \fixform{$\fP_{CFT}=c_{_g}  \tr[\fR^4]$}}
As a final warm-up example, we consider $\fP_{CFT}=c_{_g}  \tr[\fR^4]$ in AdS$_{7}$ where $n=3$.
In this case, the term $T_1$ is computed as  
\bea
T_1 \athor
&=&4 c_{_g}\,\left. \left\{
\tr[(\delta \fGamma)(\fR^2)(\nabla \xi)]
+\tr[ (\delta \fGamma) \fR ( \nabla \xi)  \fR]
+ \tr[(\delta \fGamma) (\nabla \xi)  \fR^2]\right\} \right|_{r=r_H}\nonumber\\
&=&
\delta\left\{
 (\JSanomCSV) \,[\fu\wedge (2 \fomega)^{2}]\right\}
-\left(\frac{r_H  }{4\pi T }\right)\, \JSanomCSV\,
          (\delta \fu) \wedge (2\fomega)^{2}\,,
  \eea  where $\JSanomCSV=-2\pi  c_{_g} \times 2\times 4 (2\pi T)^3.$
 Here we have used
  \be\label{eq:useful1}
  \delta\left[\fu\wedge (2\fomega)^{n-1}\right]
  =
  n(\delta \fu) \wedge  (2\fomega)^{n-1}  \, , 
  \ee 
  which comes from Eq.~(\ref{eq:integrationbypart}).
As in the previous example, $T_2$ is obviously zero for purely gravitational anomaly polynomials.

For $T_3+T_4$, by setting $r=r_H$ in the result 
obtained in  Appendix~\ref{app:replacementT3T4}, we have 
\be
(T_3+T_4) \athor=\left(\frac{r_H  }{4\pi T }\right)\, \JSanomCSV\,
          (\delta \fu) \wedge (2\fomega)^{2}\,.
\ee  
Finally, we obtain the following result for $(\fQNoether)_{H}$ evaluated at the horizon :
\be
(\fQNoether)_{H} \athor
= \delta \bigbr{ (J_{S}^{\text{\tiny{CFT,anom}}})_\mu \hodgeCFT dx^\mu }\quad , \quad (J_{S}^{\text{\tiny{CFT,anom}}})_\mu =
\JSanomCSV \, V_\mu\,.
\ee

\subsection{\fixform{$\fP_{CFT}=c_{_M}\,\fF^l\wedge \tr[\fR^{2k}]$}}
Let us start with the single trace case,
$\fP_{CFT}=c_{_M}\, \fF^l \wedge \tr[\fR^{2k}]$ in AdS$_{2n+1}$ with $n=2k+l-1$.
Derivatives of the anomaly polynomial with respect to the curvature two-form and 
the $U(1)$ field strength are given respectively by \bea
\frac{\partial^2 \fP_{CFT}}{\partial \fF \partial \fR^a{}_b} &=& c_{_M}\,(2 k l ) \fF^{l-1}\wedge (\fR^{2k-1})^b{}_a\, , \nonumber\\
\frac{\partial^2 \fP_{CFT}}{\partial \fR^a{}_b\partial \fR^c{}_d} &=&c_{_M}\,( 2  k) \fF^l\wedge
\sum_{m=0}^{2k-2} (\fR^{m})^b{}_c (\fR^{2k-2-m})^d{}_a\, ,\label{eq:ddPdRdR}
\eea
where $(\fR^0)^b{}_c\equiv \delta^b{}_c$.

As for the $U(1)$ field strength, since $(\F{0})\propto dr\wedge\fu$ and
$(\F{1}) =(2\Phi \fomega)$,
in the evaluation of $T_1$ and $T_2$ at the leading order 
under the assumption that 2nd and higher order terms of $\fF$, $\fR$ etc. do not contribute,   
we can replace  $\fF$ by $(2\Phi \fomega)$. 
Thus, in the evaluation of $T_1$, 
we can just concentrate on the case of the anomaly polynomial of the form
\be
\fP_{CFT}=c_{_g}\, \tr[\fR^{2k}] \, , 
\ee in AdS$_{2n+1}$ with $n=2k-1$, 
and the result for more general single trace anomaly polynomial $\fP_{CFT}=c_{_M}\,\fF^l\wedge \tr[\fR^{2k}]$ 
follows from this straightforwardly. 

\subsubsection{$T_1$ at horizon}
In the evaluation of $T_1$, we encounter three kinds of terms which can contribute nontrivially:
\bea
&& \delta \fGamma^a{}_b(\fR\wedge \ldots \wedge \fR)^b{}_c\nabla_d\xi^c
 \wedge (\fR\wedge \ldots \wedge \fR)^{d} {}_a\,,  \label{eq:1st_possibility1} \\
&& \delta \fGamma^a{}_b(\fR\wedge \ldots \wedge \fR)^b{}_c\nabla_a\xi^c\,,
\quad \delta \fGamma^a{}_b\nabla_c\xi^b
 \wedge (\fR\wedge \ldots \wedge \fR)^{c} {}_a\,. \label{eq:1st_possibility2}
\eea
We first note that $(\fR\wedge \ldots \wedge \fR)$ with more than one  $(\R{0})$ vanishes or
is proportional to $dr$. Thus, we only have to consider the products with one or no $(\R{0})$.
Then for the first kind of possibility,  \eqref{eq:1st_possibility1}, we can classify the potential
nontrivial contribution as summarized in Table \ref{table:structure1}.
The entries in the second and third column (i.e. the $0$'s and $1$'s) under $\nabla \xi$ and $\delta \fGamma$ indicate which order of derivative expansion in $\nabla \xi$ and $\delta \fGamma$ are appearing in that structure. For example, when it is $0$ under $\nabla \xi$ and $1$ under $\delta \fGamma$, it means that we are considering the structure containing $({}^{(0)} \nabla \xi)$ and $({}^{(1)} \delta \fGamma)$.
Since we are considering contributions
composed of 0th or 1st order terms of $\nabla \xi, \delta\fGamma$ and $\fR$,
we can classify the possible nontrivial contributions by the numbers of the 0th order terms.
In Table \ref{table:structure1}, we list up all the possibilities containing two or more zeroth order terms.
We note that when the number of the 0th order terms are more than four, then
at least one of the two $(\fR\wedge \ldots \wedge \fR)$'s contains two or more $(\R{0})$'s and thus this kind of
contribution vanishes.
If the number of the zeroth order terms are less than two, then the contributions are of higher order.
Note that in the table, we did not specify the explicit contractions of the indices. 
They should be contracted according to Eq.~\eqref{eq:1st_possibility1}. 
For the possibilities listed in the table, there are usually two ways that the $(\fR\wedge \ldots \fR)$ can be contracted, depending if it is next to $\delta \fGamma$ or $\nabla \xi$. For example, in the second row, we see that we could have $\fchi_1$ in the 3rd column and $\fchi_0$ in the 4th column. This means that we can have the following two structures
\be
({}^{(0)} \delta \fGamma)^a{}_b(\fchi_1)^b{}_c ({}^{(0)} \nabla_d\xi^c)
  (\fchi_0)^{d} {}_a,\quad
({}^{(0)} \delta \fGamma)^a{}_b(\fchi_0)^b{}_c ({}^{(0)} \nabla_d\xi^c)
  (\fchi_1)^{d} {}_a\,.
\ee

We also note that when both of $(\fR\wedge \ldots \wedge \fR)$'s are elements of
$\fchi_1$ (the first and third case in Table  \ref{table:structure1}), then such a kind of contribution is
proportional to $\fu\wedge\fu$ and thus vanish (we again stress that we neglect the terms
proportional to $dr$ during the evaluation of the first term).
As we have shown in Eq.~(\ref{eq:chi1horizon}), the nontrivial elements
in $\fchi_1$ are $(\R{0}\R{1}^m)^{\mu}{}_r$ and
$(\R{1}^m\R{0})^{\mu}{}_\rho$ only, since $(\R{1}\R{0}\R{1})^a{}_b$ is proportional to 
$dr$ or zero. 
\begin{table}[ht]
\caption{Single trace anomaly polynomial : contributions to $T_1$ from terms with $\delta \fGamma$ and $\nabla \xi$ not next to each other.} 
\centering 
\begin{tabular}{c c c c c c} 
\hline\hline 
\# of 0th order terms & $\nabla \xi$ & $\delta \fGamma{}$ &
$ (\fR\wedge\ldots \wedge \fR)$ &$ (\fR\wedge\ldots \wedge \fR)$& Results \\ [0.5ex]
\hline
4 & 0 & 0 & $\fchi_1$ $ \propto \fu$& $\fchi_1$$\propto \fu $& zero\\
3  & 0 & 0 & $\fchi_1$&$\fchi_0$&Case A (vanish in the end) \\
3 & 0 & 1 & $\fchi_1$ $\propto \fu$&$\fchi_1$ $\propto \fu$&zero \\
2  & 0 & 0 & $\fchi_0$ &$\fchi_0$ &Case B\\
2& 0 & 1 & $\fchi_1$ &$\fchi_0$ &Case C\\
\hline 
\end{tabular}
\label{table:structure1}
\end{table}

In a similar way, in the case of the two possibilities given in \eqref{eq:1st_possibility2},
we can summarize potential nontrivial contributions as in Table  \ref{table:structure2}.
\begin{table}[ht]
\caption{Single trace anomaly polynomial : contributions to $T_1$ from terms with $\delta \fGamma$ and $\nabla \xi$ next to each other.} 
\centering 
\begin{tabular}{c c c c c} 
\hline\hline 
\# of  0th order terms& $\nabla \xi$ & $\delta \fGamma$ &$ \fR\wedge\ldots \wedge \fR$ & Results \\ [0.5ex]
\hline
3 & 0 & 0 & $\fchi_1$ &Case D  (vanish in the end)\\
2 & 0 & 0 &$\fchi_0$ &Case E\\
2 & 0 & 1 & $\fchi_1$&Case F \\
\hline 
\end{tabular}
\label{table:structure2}
\end{table}

In the rest of this subsection, we directly evaluate the possibilities Case A, B, C, D, E and F in the above tables to calculate the leading order contribution to the first term $T_1$.

\noindent
\underline{\bf Case A (vanish in the end)}\\
The followings are four potential nontrivial terms  in Case A, but all of them vanish 
as a result of explicit computations (for $1\le m\le 2k-3$) :
\begin{eqnarray}
&&^{(0)}\delta \fGamma^a{}_b\wedge (\R{0}\R{1}^{m-1})^b{}_c\wedge\nabla_d \xi^c\wedge
(\R{1}^{2k-2-m})^d{}_a = 0 \, , \nonumber \\
&&^{(0)}\delta \fGamma^a{}_b\wedge (\R{1}^{m-1}\R{0})^b{}_c \wedge \nabla_d \xi^c\wedge
(\R{1}^{2k-2-m})^d{}_a = 0 \, , \\
& &{}^{(0)}\delta \fGamma^a{}_b\wedge (\R{1}^m)^b{}_c \wedge \nabla_d \xi^c\wedge
(\R{0}\R{1}^{2k-3-m})^d{}_a=0\, ,  \nonumber \\
& &{}^{(0)}\delta \fGamma^a{}_b\wedge (\R{1}^m)^b{}_c \wedge \nabla_d \xi^c\wedge
(\R{1}^{2k-3-m}\R{0})^d{}_a=0\, .  \nonumber 
\end{eqnarray} 
Therefore, there is no contribution to $T_1$ from this type of terms. \\

\noindent
\underline{\bf Case B}\\
There is only one type of contribution in Case B 
which can be computed as follows (for $1\le m\le 2k-3$) :
\bea
& &{}^{(0)}\delta \fGamma^a{}_b\wedge (\R{1}^m)^b{}_c \wedge
(\nabla_d \xi^c) \wedge(\R{1}^{2k-2-m})^d{}_a \nonumber
 \\
  &&\qquad\qquad
  =
  -(2\pi)\left[2\delta (2\pi T\fu)
  -r_H \delta \fu\right]
\wedge
  (2(2\pi T)\fomega)^{2k-2}\, . 
  \eea 

\noindent
\underline{\bf Case C}\\
As in Case A, there are four potential nontrivial terms in Case C 
and only one out of these terms contribute nontrivially (for $1\le m\le 2k-3$) :
\begin{eqnarray}
&&{}^{(1)}\delta \fGamma^a{}_b\wedge (\R{0}\R{1}^{m-1})^b{}_c \wedge \nabla_d \xi^c
\wedge(\R{1}^{2k-2-m})^d{}_a=0\, , \nonumber \\
&&{}^{(1)}\delta \fGamma^a{}_b\wedge (\R{1}^{m-1}\R{0})^b{}_c \wedge \nabla_d \xi^c
\wedge(\R{1}^{2k-2-m})^d{}_a=0\, , \nonumber \\
&&{}^{(1)}\delta \fGamma^a{}_b\wedge (\R{1}^m)^b{}_c \wedge \nabla_d \xi^c \wedge(\R{0}\R{1}^{2k-3-m})^d{}_a
= 0 \, , \\
&&{}^{(1)}\delta \fGamma^a{}_b\wedge (\R{1}^m)^b{}_c \wedge \nabla_d \xi^c \wedge(\R{1}^{2k-3-m}\R{0})^d{}_a
= 
  (-2\pi )  r_H(\delta \fu)\wedge 
  (2(2\pi T)\fomega)^{2k-2},\nonumber 
\end{eqnarray}  where we have used Eq.~(\ref{eq:integrationbypart}).

\noindent
\underline{\bf Case D (vanish in the end)}\\
The following four terms are potentially nontrivial but the direct computation 
shows that all of them vanish :
\bea
&&{}^{(0)}\delta \fGamma^a{}_b(\R{0}\R{1}^{2k-3})^b{}_c \nabla_a \xi^c = 0\, , \qquad
{}^{(0)}\delta \fGamma^a{}_b(\R{1}^{2k-3}\R{0})^b{}_c \nabla_a \xi^c = 0\, , \\
&&{}^{(0)}\delta \fGamma^a{}_b \nabla_c \xi^b(\R{0}\R{1}^{2k-3})^c{}_a = 0\, , \qquad
{}^{(0)}\delta \fGamma^a{}_b \nabla_c \xi^b(\R{1}^{2k-3}\R{0})^c{}_a = 0\, . \nonumber 
\eea
Therefore there is no nontrivial contribution from Case D. 

\noindent
\underline{\bf Case E}\\
There are two non-trivial structures in Case E :
\bea
& &{}^{(0)}\delta \fGamma^a{}_b (\R{1}^{2k-2})^b{}_c \nabla_a \xi^c
= -2\pi
\left[2
\delta (2\pi T\fu )-r_H \delta \fu\right]\wedge(2 (2\pi T)\fomega)^{2k-2}\,,  \\
& &{}^{(0)}\delta \fGamma^a{}_b \nabla_c \xi^b (\R{1}^{2k-2})^c{}_a
=
-2\pi \left[2\delta(2\pi T \fu)
-  r_H  \delta \fu\right]\wedge
(2 (2\pi T)\fomega)^{2k-2} .
\eea

\noindent
\underline{\bf Case F}\\
There are two potential nontrivial cases one of which vanishes as a result of direct computation~\,:
 \bea
&& {}^{(1)}\delta \fGamma^a{}_b \nabla_c \xi^b (\R{1}^{2k-3}\R{0})^c{}_a 
= (-2\pi) r_H  (\delta \fu)\wedge(2(2\pi T)\fomega)^{2k-2}\, ,  \\
&& {}^{(1)}\delta \fGamma^a{}_b \nabla_c \xi^b (\R{0}\R{1}^{2k-3})^c{}_a 
= 0\, . 
\eea
Here we have used Eq.~(\ref{eq:integrationbypart}).

\noindent
\underline{\bf Summary for $T_1$}\\
Now that we have calculated all the contributions in the Table \ref{table:structure1} and  \ref{table:structure2}, we sum them up to calculate
the leading order contribution to the first term $T_1$. The above calculation
shows that all the contributions containing more than two 0th order terms vanish.
Then the leading order contribution to $T_1$ contains two 0th order terms and is calculated by
summing up the results in Case B, C, E and F.  For $\fP_{CFT} = c_{_g}\, \tr[\fR^{2k}]$ with $n=2k-1$,
we have
\bea
T_1 \athor&=& (-2\pi)\,c_{_g}\,(2k)\Bigl[
2(2k-1)\delta(2\pi T\fu)
 -r_H(\delta \fu)
\Bigr ]\wedge (2(2\pi T)\fomega)^{2k-2} \nonumber \\
&=&
\delta\left\{
(\JSanomCSV)\, [\fu 
\wedge (2 \fomega)^{n-1}]\right\}
-\left(\frac{r_H}{4\pi T}\right)
\JSanomCSV\,[(\delta \fu)\wedge(2 \fomega)^{n-1}]\,, 
\eea
where we have used  Eq.~(\ref{eq:useful1}) and $\JSanomCSV=-2\pi \times c_{_g} 2\times (2k) \times (2\pi T)^{n}.$

As explained before, $T_1$ in the case of $\fP_{CFT}=c_{_M}\, \fF^{l}\wedge \tr[\fR^{2k}]$ (with $n=2k+l-1$) is easily obtained by multiplying $(2\Phi\fomega)^l$ to the result for $l=0$, which yields
\bea
T_1&=&
\delta\left\{
(\JSanomCSV)_l\,[ \fu\wedge (2 \fomega)^{n-l-1}]\right\}\wedge (2\mu \fomega)^{l}
-\left(\frac{r_H}{4\pi T}\right)\JSanomCSV\,[(\delta \fu)\wedge(2 \fomega)^{n-1}]\, ,
\eea where in this case
$\JSanomCSV=-2\pi  c_{_M}\times \mu^l\times 2\times (2k) \times (2\pi T)^{2k-1}$ and
$(\JSanomCSV)_l=-2\pi  c_{_M} \times2\times (2k) \times (2\pi T)^{2k-1}.$

\subsubsection{$T_2$ at horizon}
Here we evaluate the second term $T_2$. For $\fP_{CFT}=c_{_g} \,\tr[\fR^{2k}]$, it is obviously 
zero and thus we consider $\fP_{CFT}=c_{_M} \fF^{l}\wedge \tr[\fR^{2k}]$ for which $T_2$ is of the form
\be
T_2 
 =  c_{_M} (2k)l \delta \fA\wedge
 \fF^{l-1} \wedge [\nabla_b \xi^a (\fR^{2k-1})^b{}_a]\, .
\ee
We note that $ \nabla_b \xi^a (\fR^{2k-1})^b{}_a$ is either zero or proportional to
$dr$ when it contains at least one $(\R{0})$,
the $U(1)$ gauge field $\fA$ at the leading order is $\Phi\fu$,
and $\fF$ can be replaced by the  $(2\mu\fomega)$.
Then the leading order contribution to the second term $T_2$ is given by
\bea
T_2 \athor&=&
(\JSanomCSV)_l\,[ \fu\wedge (2 \fomega)^{n-l-1}]
    \wedge
\delta[ (  2\mu \fomega)^{l}]\, ,
\eea where $(\JSanomCSV)_l=-2\pi  c_{_M} \times2\times (2k) \times (2\pi T)^{2k-1}$.

\subsubsection{\fixform{$(\fQNoether)_{H}$} at horizon}

We recall that for $\fP_{CFT}=c_{_M}\, \fF^{l}\wedge \tr[\fR^{2k}]$,
$T_3+T_4$ evaluated at the horizon is given by (see Appendix~\ref{app:replacementT3T4}) 
\be
(T_3+T_4) \athor=
\left(\frac{r_H  }{4\pi T }\right)\, \JSanomCSV\,
          [(\delta \fu) \wedge (2\fomega)^{n-1}]\, .
\ee 
Collecting all these terms, we finally obtain
\be
(\fQNoether)_{H} \athor
= \delta \bigbr{ (J_{S}^{\text{\tiny{CFT,anom}}})_\mu \hodgeCFT dx^\mu }\quad , \quad (J_{S}^{\text{\tiny{CFT,anom}}})_\mu =
\JSanomCSV \, V_\mu\,.
\ee

\subsection{\fixform{$\fP_{CFT}=c_{_M}\, \fF^l\wedge \tr[\fR^{2k_1}] \wedge\ldots \wedge \tr[\fR^{2 k_p}]$}}
As in the previous case, $\fF$ can be replaced by $(2\Phi\fomega)$ to evaluate the leading order contribution.
We thus concentrate on the anomaly polynomial of the form $\fP_{CFT}=c_{_g}\,\tr[\fR^{2k_1}]\wedge \tr[\fR^{2k_2}]\wedge\ldots\wedge \tr[\fR^{2k_p}]$ from which the result for the case with $\fF^l$ is derived straightforwardly.

The crucial difference from the single trace case is that when we consider two derivatives of $\fP_{CFT}$
with respect to $\fR$, there are two kinds of terms:
the two $\fR$-derivatives can act on the same trace or on different traces.
As we know that $\tr[\fchi_1]\propto dr\wedge\fu$, for the trace $\tr[\fR^{2k_i}]$
on which the $\fR$-derivatives do not act, we can replace it by $\tr[(\R{1}^{2k_i})] = 2(2\Phi_T\fomega)^{2k_i}$. 
Therefore,  we only need to consider the two-trace case, $\fP_{CFT}=c_{_g}\,\tr[\fR^{2k_1}]\wedge \tr[\fR^{2k_2}]$  
with $k_1,k_2 \ge 1$,  
and the more general case follows from this straightforwardly. 

In this two-trace case, we have 
\bea \label{eq:twotracehorizon}
\frac{\partial^2 \fP_{CFT}}{\partial \fR^a{}_b \fR^c{}_d}
&=&c_{_g}\,(2k_1)(2k_2)\left[
 (\fR^{2k_1-1})^b{}_a (\fR^{2k_2-1})^d{}_c
+(\fR^{2k_1-1})^d{}_c (\fR^{2k_2-1})^b{}_a\right]  \label{eq:ddPdRdR_multi}\\
& &
+c_{_g}\,(2k_1) \tr[\fR^{2k_2}] \sum_{m=0}^{2k_1-2}(\fR^{m})^b{}_c (\fR^{2k_1-2-m})^d{}_a\nonumber\\
&&
+c_{_g}\,(2k_2)  \tr[\fR^{2k_1}]  \sum_{m=0}^{2k_2-2}(\fR^{m})^b{}_c (\fR^{2k_2-2-m})^d{}_a\,.\nonumber
\eea The results for the terms in the second and third lines can be obtained from the single trace answer by multiplying
$\tr[(\R{1}^{2k_1})]=2(2\Phi_T\fomega)^{2k_1}$ (or, the one with $k_1$ replaced by $k_2$).
Thus, we will concentrate on the first term.
We also restrict ourselves to the
case in which the derivatives with respect to $\fR^{a}{}_b$ and $\fR^{c}{}_d$
act respectively on $\tr[\fR^{2k_1}]$ and $\tr[\fR^{2k_2}]$ (the other one can be derived
by just interchanging $k_1$ and $k_2$).

\subsubsection{$T_1$ at horizon}
In the case of $\fP_{CFT}$ with a single trace, the possible nontrivial contributions are classified
as in Table \ref{table:structure1} and \ref{table:structure2}. 
In the double trace case, there is the following extra contribution 
related to the first line of \eqref{eq:ddPdRdR_multi} :
\be
[\delta \fGamma^{a}{}_{b} \wedge (\fR\wedge\ldots\wedge \fR)^{b}{}_a]
\wedge[ \nabla_{d}\xi^c (\fR\wedge\ldots\wedge \fR)^{d}{}_c],
\ee
whose potential nontrivial contribution at low orders are classified as in Table  \ref{table:structure1},
though the contraction structures are different. For clarify, we shall present such classification in Table ~\ref{table:structure3}.
\begin{table}[ht]
\caption{Multi-trace anomaly polynomial : contributions to $T_1$ from terms with $\delta \fGamma$ and $\nabla \xi$ in different traces.} 
\centering 
\begin{tabular}{c c c c c c} 
\hline\hline 
\# of 0th order terms & $\nabla \xi$ & $\delta \fGamma{}$ &
$ (\fR\wedge\ldots \wedge \fR)$ &$ (\fR\wedge\ldots \wedge \fR)$& Results \\ [0.5ex]
\hline
4 & 0 & 0 & $\fchi_1$ $ \propto \fu$& $\fchi_1$$\propto \fu $& zero\\
3  & 0 & 0 & $\fchi_1$&$\fchi_0$&Case A (vanish in the end) \\
3 & 0 & 1 & $\fchi_1$ $\propto \fu$&$\fchi_1$ $\propto \fu$&zero \\
2  & 0 & 0 & $\fchi_0$ &$\fchi_0$ &Case B\\
2& 0 & 1 & $\fchi_1$ &$\fchi_0$ &Case C\\
\hline 
\end{tabular}
\label{table:structure3}
\end{table}

\noindent
\underline{\bf Case A}\\
For Case A, there are four potential nontrivial contributions but all of them vanish (or proportional to $dr$) as follows :
\bea
&&[{}^{(0)}\delta \fGamma^a{}_b \wedge (\R{0}\R{1}^{2k_1-2})^{b}{}_a]\wedge
[\nabla_d \xi^c(\R{1}^{2k_2-1})^d{}_c] =0\, , \nonumber \\
&&[{}^{(0)}\delta \fGamma^a{}_b \wedge (\R{1}^{2k_1-1})^b{}_a]\wedge
[\nabla_d \xi^c(\R{0}\R{1}^{2k_2-2})^{d}{}_c] = 0 \,,  \\
&&[{}^{(0)}\delta \fGamma^a{}_b \wedge (\R{1}^{2k_1-2}\R{0})^{b}{}_a]\wedge
[\nabla_d \xi^c(\R{1}^{2k_2-1})^d{}_c] =0\, , \nonumber \\
&&[{}^{(0)}\delta \fGamma^a{}_b \wedge (\R{1}^{2k_1-1})^b{}_a]\wedge
[\nabla_d \xi^c(\R{1}^{2k_2-2}\R{0})^{d}{}_c] = 0 \, , \nonumber 
\eea due to Eq.~\eqref{eq:usefulrelation105}-\eqref{eq:usefulrelation106}.
We note that, strictly speaking, there are other four cases where $k_1$ and $k_2$ are 
interchanged, but they are obviously zero (or proportional to $dr$) for the same reason.

\noindent
\underline{\bf Case B}\\
For Case B, we have only one possibility (and the case with $k_1$ and $k_2$ interchanged) :
\bea
& & [{}^{(0)}\delta \fGamma^a{}_{b} \wedge(\R{1}^{2k_1-1})^b{}_a]\wedge
[\nabla_d \xi^c (\R{1}^{2k_2-1})^d{}_c]
\\
&&\quad=
-(2\pi)2
 \left[2
\delta(2\pi T\fu)
-r_H 
 \delta \fu
\right] \wedge (2(2\pi T)\fomega)^{2(k_1+k_2)-2}
\,.\nonumber
\eea

\noindent
\underline{\bf Case C}\\
In this case, there are four potential non-trivial cases but one out of the four cases gives 
non-trivial contribution (as in the case of Case A and B, there is 
another nontrivial contribution from $k$ and $l$ interchanged) :
\bea
&&
 [{}^{(1)}\delta \fGamma^a{}_b \wedge(\R{1}^{2k_1-1})^b{}_a]
\wedge[ \nabla_d \xi^c (\R{0}\R{1}^{2k_2-2})^d{}_c] =0 
\, ,\nonumber \\
&&
 [{}^{(1)}\delta \fGamma^a{}_b \wedge(\R{1}^{2k_1-1})^b{}_a]
\wedge[ \nabla_d \xi^c (\R{1}^{2k_2-2}\R{0})^d{}_c] =0 
\, , \\
&&
 [{}^{(1)}\delta \fGamma^a{}_b \wedge(\R{0}\R{1}^{2k_1-2})^b{}_a]
\wedge[ \nabla_d \xi^c (\R{1}^{2k_2-1})^d{}_c] =0 
\, ,\nonumber \\
&&[{}^{(1)}\delta \fGamma^a{}_b \wedge(\R{1}^{2k_1-2}\R{0})^b{}_a]
\wedge[ \nabla_d \xi^c (\R{1}^{2k_2-1})^d{}_c]
=-
(2\pi)2 r_H  (\delta \fu)\wedge (2 (2\pi T)\fomega)^{2(k_1+k_2)-2}\,, \nonumber 
\eea where we have used Eq.~(\ref{eq:integrationbypart}).

\noindent
\underline{\bf Summary for $T_1$} \\
Now we collect all the terms to calculate the first term $T_1$.
For $\fP_{CFT}=c_{_g}\,\tr[\fR^{2k_1}]\wedge \tr[\fR^{2k_2}]$ (with $n=2k_1+2k_2-1$), by using Eq.~(\ref{eq:integrationbypart}),
the first term $T_1$ at the leading order is given by (do not forget the second and third terms in \eqref{eq:ddPdRdR_multi})
\bea
T_1
&=&2\,c_{_g}\,(8k_1 k_2)(-2\pi)\left[2
\delta\left(2\pi T \fu\right)\right]\wedge (2(2\pi T) \fomega)^{2k_1+2k_2-2}
\nonumber\\
& &
+2\,c_{_g}\,(2k_1)(-2\pi)\left[
(2k_1-1)2\delta\left(2\pi T\fu\right)-r_H \delta \fu
\right](2(2\pi T)\fomega)^{2k_1+2k_2-2}\nonumber\\
& &
+2\,c_{_g}\,(2k_2)(-2\pi)\left[
(2k_2-1)2\delta\left( 2\pi T\fu\right)-r_H \delta \fu
\right]\wedge (2(2\pi T)\fomega)^{2k_1+2k_2-2}\nonumber\\
&=&
\delta\left\{(\JSanomCSV )\,[\fu \wedge (2\fomega)^{n-1}]\right\}
-\left(\frac{r_H}{4\pi T}\right) \JSanomCSV\,[( \delta \fu)\wedge(2\fomega)^{n-1}]\, ,
\eea where we have used Eq.~(\ref{eq:useful1}) and the fact that in this case
$\JSanomCSV=-2\pi  c_{_g} \times    2^2\times(2k_1+2k_2)\times(2\pi T)^{k_1+k_2-1}$.

For more general multi-trace terms in the anomaly polynomial,
$\fP_{CFT}=c_{_M}\,\fF^l\wedge \tr[\fR^{2k_1}]\wedge \tr[\fR^{2k_2}]\wedge\ldots \tr[\fR^{2k_p}]$ (with $n=2k_{tot}+l-1$ and  $k_{tot}=\sum_{i=1}^p k_i$),
by recalling that $\fF$ can be replaced by $(2\Phi\fomega)$ and $\tr[\fR^{2k_i}]$ by $2(2\Phi_T\fomega)^{k_i}$
when the derivatives with respect to $\fR$ do not act there, the first term $T_1$ at the leading order is
given by
\bea
&&T_1  
=
\delta\left\{ (\JSanomCSV)_l\,[\fu \wedge (2\fomega)^{n-l-1}]\right\}\wedge (2\mu \fomega)^l -\left(\frac{r_H}{4\pi T}\right) 
 \JSanomCSV \,[( \delta \fu)\wedge (2\fomega)^{n-1}] \, , 
\eea where in this case
$\JSanomCSV=-2\pi  c_{_M} \times \mu^l \times   2^p \times(2k_{tot})\times(2\pi T)^{k_{tot}-1}$ and
$(\JSanomCSV)_l=-2\pi  c_{_M}  \times   2^p \times(2k_{tot})\times(2\pi T)^{k_{tot}-1}$.

\subsubsection{$T_2$ at horizon}
\label{sec:SecondTermWaldEntropy}
For the anomaly polynomial without $\fF$'s, this term is obviously zero and thus 
we consider more general cases including the multiple traces,  $\fP_{CFT}=c_{_M}\,\fF^l\wedge \tr[\fR^{2k_1}]\wedge \tr[\fR^{2k_2}]\wedge\ldots \wedge\tr[\fR^{2k_p}]$. 
The evaluation of $T_2$ at the leading order
is the same as the single trace case except that we need to replace the traces $\tr[\fR^{2k_i}]$
without the derivative acted on by $\tr[(\R{1}^k_i)]= 2(2\Phi_T\fomega)^{k_i}$. 
Then the final result is given by
\be
T_2
=\left\{ (\JSanomCSV)_l\,[\fu \wedge (2\fomega)^{n-l-1}]\right\}\wedge\delta[ (2\mu \fomega)^l]\, .
 \ee
 
\subsubsection{\fixform{$(\fQNoether)_{H}$} at horizon}
We recall that $T_3+T_4$ evaluated at the horizon is given by 
(see Appendix~\ref{app:replacementT3T4})
\be
(T_3+T_4) \athor =\left(\frac{r_H  }{4\pi T }\right)\, \JSanomCSV\,
          (\delta \fu) \wedge (2\fomega)^{2}\,.
\ee 
Therefore, collecting all the terms, we finally obtain 
\be
(\fQNoether)_{H} \athor
= \delta \bigbr{ (J_{S}^{\text{\tiny{CFT,anom}}})_\mu \hodgeCFT dx^\mu }\,,  \qquad (J_{S}^{\text{\tiny{CFT,anom}}})_\mu =
\JSanomCSV \, V_\mu\,.
\ee

\section{Contributions to \texorpdfstring{$T_0,T_1$ and $T_2$}{T0,T1 and T2} at horizon:  with higher order}
\label{sec:entropyhighercontribution}
In Appendix \ref{sec:T0T1T2hor0th1st}, we have assumed that the 2nd and higher order 
building blocks do not generate any non-trivial contribution to $(\fQNoether)_{H} \athor
$ up to $\omega^{n-1}$ order. We confirm this statement here.  
As we have proved this statement for $T_3$ and $T_4$ in Appendix \ref{sec:2ndorderT3andT4}, here we will prove 
that 2nd and higher order building blocks
do not contribute to $T_0,T_1$ and $T_2$ (at the horizon)
at the leading order. In the following, we check this statement term by term. 
Since $T_0 \athor=0$ due to $(\Lambda+\ic_{\xi} \fA) \athor =0$ to all order, 
we do not need to deal with $T_0 \athor$.

Here is one important remark that will be useful : As we have seen,  when 2nd and higher order 
building blocks are neglected, the leading order contribution to $T_1$ and $T_2$ is of order $\omega^{n-1}$. 

We emphasize that in this Appendix, we are considering all contributions which contain at least one 2nd or higher order building block.

\subsection{\texorpdfstring{$T_1$}{T1} at horizon}
Let us consider the anomaly polynomial of the form
$\fP_{CFT} = c_{_g} \,\tr[\fR^{2k_1}]\wedge\tr[\fR^{2k_2}]\wedge \ldots \wedge\tr[\fR^{2k_p}]$ with $n=2k_{tot}-1$.
We will take into account $\fF^l$ later. 
Without loss of generality, we assume $k_i=1$ for $i \le p_0$ ($p_0\ge0$). 
Furthermore, let us denote  the total number of the derivative in $\nabla \xi$ and $\delta \fGamma$ by $\tilde{N}$.
One important remark is that at the horizon, $(\R{0})$ is either proportional to $dr$ or $\fu$ and 
thus the number of $(\R{0})$'s needs to be less than two to have nontrivial contribution to $T_1$.

When we evaluate the derivative $(\partial^2 \fP_{CFT} /\partial \fR^a{}_{b}\partial \fR^c{}_{d})$, 
there are four possibilities :
\begin{itemize}
\item Case 1 : $\tilde{N}\ge 1$.
\item Case 2 : $\tilde{N}= 0$ and both of the derivatives act on 
the same trace $\tr[\fR^{2k_i}]$ with  $i \le p_0$, i.e.
\be
T_1 \propto \tr[({}^{(0)}\delta\fGamma)({}^{(0)} \nabla \xi)] \wedge \tr[\fR^{2k_2}]\wedge\cdots\wedge\tr[\fR^{2k_p}]\, ,
\ee where without loss of generality, we have assumed that 
$i=1$.

\item Case 3 : $\tilde{N}= 0$ and both of the derivatives act on 
the same trace $\tr[\fR^{2k_i}]$ with  $i> p_0$, i.e.
\be
T_1 \propto\tr [({}^{(0)} \delta \fGamma) \fR^{2k_p-1} ({}^{(0)} \nabla \xi )\fR^{2k_q-1} ].
\ee

\item Case 4 : $\tilde{N}= 0$ and the two derivatives act on different traces
\be 
T_1 \propto\tr [({}^{(0)} \delta \fGamma )\fR^{2k_p-1} ] \wedge  \tr [ ({}^{(0)}\nabla \xi) \fR^{2k_q-1} ]. 
\ee

\end{itemize}
\subsubsection{Case 1} 
Since we are considering the contributions containing at least one 2nd or higher order term in $T_1$, 
at least two $\fR$'s must be $(\R{0})$ to contribute to $T_1$ up to $\omega^{n-1}$ order.  Therefore, this gives vanishing contribution. 
\subsubsection{Case 2} 
In this case, to circumvent the appearance of 
two or more $(\R{0})$'s,  the $\fR$'s need to contain one $(\R{0})$, one $(\R{2})$ and the rest are set 
to $(\R{1})$. This indicates that  $\tr[\fR^{2k_2}]\wedge\cdots\wedge\tr[\fR^{2k_p}]$ contains 
$\tr[\R{2}\fchi_1]$ or $\tr[\fchi_1]$, but both of them are zero or proportional to $dr$. 
We thus conclude that this case give vanishing contribution to $T_1$. 
\subsubsection{Case 3 and Case 4} 
For Case 3 and Case 4, we note that the derivative $(\partial^2 \fP_{CFT} /\partial \fR^a{}_{b}\partial \fR^c{}_{d})$
contains two $(\fR\wedge\fR\wedge\ldots\wedge\fR)$'s  
(we note that one of the $(\fR\wedge\fR\wedge\ldots\wedge\fR)$'s can be $\delta^a{}_b$ which means we are in Case 3)
and a product of traces of the form ${\cal T} = {\cal T}_1\wedge{\cal T}_2\wedge{\cal T}_3$, where ${\cal T}_1$ (${\cal T}_2$, ${\cal T}_3$, respectively) is the wedge product of the trace of the form
$TR_{(1)}$ ($TR_{(2)}$, $TR_{(3)}$, respectively) only, with the $TR_{(i)}$'s are defined Ref.~\cite{Azeyanagi:2013xea} as :
\bea
&&
TR_{(1)} \equiv  \tr[\fchi]\,,  \nonumber \\
&&TR_{(2)} \equiv
\tr[\form{\upsilon}\,\fchi\form{\upsilon}\,\fchi\cdots
\form{\upsilon}\,\fchi]\,,   \nonumber \\
&&TR_{(3)} \equiv  \tr[\form{\upsilon}]\,.
\eea
Here, the symbol $\fchi$ represents one of the elements in
 $\fchi_0\cup \fchi_1\cup\fchi_2$,
that is, it is a string made of $(\R{0})$'s and $(\R{1})$'s only. On the other hand, the symbol $\form{\upsilon}$ is defined to represent a string made of 2nd or higher order terms 
 $(\R{m})$ with $m\ge2$.
 We note that the existence of $(\R{0})$ in ${\cal T}_1$ is not allowed since
$\tr[\fchi_1]\propto dr$ and thus we set all the trace in ${\cal T}_1$ to be $\tr[\fchi_0]$.
Therefore, we might as well neglect the ${\cal T}_1$ in the following discussion.
 
For the two $(\fR\wedge\fR\wedge\ldots\fR)$'s, each of them is of the form
denoted in Case $\barA, \barB, \barC, \barD$ or $\barE$ (as was done in Ref.~\cite{Azeyanagi:2013xea} and reviewed in Appendix \ref{sec:classificationRpq}) :
 \bea\label{eq:casestructures2}
\mbox{Case } \barA&:& \left(\form{\upsilon}\,\fchi\,\form{\upsilon}\,\fchi
\ldots \form{\upsilon}\,\fchi\, \form{\upsilon}\right)\,, \nonumber \\
\mbox{Case }\barB&:& \left(\form{\upsilon}\,\fchi\form{\upsilon}\,\fchi
\ldots \form{\upsilon}\,\fchi\right)\,,\quad
 \left(\fchi\,\form{\upsilon}\,\fchi\,\form{\upsilon}
\ldots \fchi\,\form{\upsilon}\right)\, , 
\nonumber\\
\mbox{Case }\barC&:&  \left(\fchi\,\form{\upsilon}\,\fchi\,\form{\upsilon}
\ldots \fchi\,\form{\upsilon}\,\fchi\,\right)\,,   \nonumber \\
\mbox{Case }\barD&:&  \left(\form{\upsilon}\right)\,, \nonumber \\
\mbox{Case }\barE&:&  \left(\fchi\right)\,,
\eea where ($I\ge1$). We notice that we regard $\delta^a{}_b$ 
as an element of Case $\barE$ with $\fchi_0$.    
\vspace{0.2cm} \\
\noindent
\underline{\bf Case 3 :} \\
When we consider the contribution of order $\omega^{n-1}$ or lower,
to avoid appearance of more than one $(\R{0})$
in $(\partial^2 \fP_{CFT} /\partial \fR^a{}_{b}\partial \fR^c{}_{d})$,  we require that one of the $(\fR\wedge\fR \wedge\ldots \wedge \fR)$'s is of the form in Case $\barE$ while the other $(\fR\wedge\fR \wedge \ldots \wedge \fR)$'s is either of the form in Case $\barB, \barC$ or $\barE$ (see Eq.~(\ref{eq:casestructures2})) and that
there is no terms in ${\cal T}_3$. More precisely, there are three possible nontrivial cases
(we note that non-trivial terms in 
$\fchi_{1}$ are $(\R{1}^{m-1}\R{0})$ or $(\R{0}\R{1}^{m-1})$ with $m\ge1$) :

\begin{itemize}
\item
The first case is when both of the $(\fR\wedge\fR\wedge\ldots\wedge\fR)$'s are $\fchi_0$ and
${\cal T}_2$ contains only one trace of the form $\tr[\R{2}\fchi_1]$. In this case,
we know that $\tr[\R{2}\fchi_1]=0$
and thus this does not contribute.

\item The second case is when
one of the $(\fR\wedge\fR\wedge\ldots\wedge\fR)$'s is $\fchi_1$ and the other is
$\fchi_0$, while ${\cal T}_2$ contains only one trace of the form $\tr[\R{2}\fchi_0]$.
$T_1$ contains one of the following four structures 
but all of them vanish :
\bea
&&(\nabla_b\xi^a)(\fchi_0)^b{}_c({}^{(0)}\delta\fGamma^c{}_d)(\fchi_1)^d{}_a=0\,, \qquad({}^{(0)}\nabla_c\xi^a)({}^{(0)}\delta\fGamma^c{}_d)(\fchi_1)^d{}_a=0\,, \nonumber\\
&&
({}^{(0)}\nabla_b\xi^a)(\fchi_1)^b{}_c({}^{(0)}\delta\fGamma^c{}_d)(\fchi_0)^d{}_a=0 \,,\quad
({}^{(0)}\nabla_b\xi^a)(\fchi_1)^b{}_c {}^{(0)} (\delta\fGamma^c{}_a)=0\,,
\eea which follow from Eq.~(\ref{eq:someeq1})-(\ref{eq:anotheruseful1}).

\item
 Finally, the third case is when one of the $(\fR\wedge\fR\wedge\ldots\wedge\fR)$'s is of Case $\barB$ or $\barC$ (where there exists only one $\form{\upsilon}$ which is set to $(\R{2})$) while the other $(\fR\wedge\fR\wedge\ldots\wedge\fR)$ is of Case $\barE$.
 The trace part ${\cal T}$ is exactly ${\cal T}_1$ with all of them set to $\tr[\fchi_0]$.
In this case,  all possibilities vanish because they are either $(^{(0)}\delta \fGamma^b{}_a)$ contracted with
\bea
&&({}^{(0)}\nabla_c\xi^a)(\R{2}\R{0})^c{}_b 
=({}^{(0)}\nabla_c\xi^a)(\R{0}\R{2})^c{}_b 
=(\R{2}\R{0})^a{}_c({}^{(0)}\nabla_b\xi^c)
= (\R{0} \R{2})^a{}_c({}^{(0)}\nabla_b\xi^c)=0\, , \nonumber \\
\eea or they are proportional to :
\bea
&&
({}^{(0)}\delta\fGamma^c{}_d)(\fchi_1)^d{}_a({}^{(0)}\nabla_b\xi^a)=0\,,\nonumber\\
&&
({}^{(0)}\nabla_b\xi^a)(\fchi_1)^b{}_c({}^{(0)}\delta\fGamma^c{}_d)(\R{2})^d{}_a=0\, , 
\quad ({}^{(0)}\nabla_b\xi^a)(\fchi_1)^b{}_c({}^{(0)}\delta\fGamma^c{}_d)(\fchi_0\R{2}\fchi_0)^d{}_a=0\, ,
\nonumber\\
&&
({}^{(0)}\nabla_b\xi^a)(\fchi_1)^b{}_c({}^{(0)}\delta\fGamma^c{}_d)(\R{2}\fchi_0)^d{}_a=0\, , 
\qquad
({}^{(0)}\nabla_b\xi^a)(\fchi_1)^b{}_c({}^{(0)}\delta\fGamma^c{}_d)(\fchi_0 \R{2})^d{}_a=0\,.\nonumber\\
\eea 

Here, we have used Eqs.~(\ref{eq:someeq1}), (\ref{eq:anotheruseful1}) and (\ref{eq:anotheruseful2}).
\end{itemize} 
\noindent
\underline{\bf Case 4 :} \\
As in Case 1, up to $\omega^{n-1}$ order,
to avoid appearance of more than one $(\R{0})$'s
in two derivatives of the anomaly polynomial, i.e. $(\partial^2 \fP_{CFT} /\partial \fR^a{}_{b}\partial \fR^c{}_{d})$,  we need to require that one of the $(\fR\wedge\ldots \wedge \fR)$'s is of the form in Case $\barE$ while the other $(\fR\wedge \ldots \wedge \fR)$'s is either of the form in Case $\barB, \barC$ or $\barE$ and that there is no terms in ${\cal T}_3$. There are three possibilities as before :
\begin{itemize}
\item 
When both of $(\fR\wedge\fR\wedge\ldots\wedge\fR)$'s are $\fchi_0$ and
${\cal T}_2$ contains only one trace of the form $\tr[\R{2}\fchi_1]$, 
this type of terms vanishes since $\tr[\R{2}\fchi_1]=0$. 
\item Let us next consider the case where
one of the $(\fR\wedge\fR\wedge\ldots\wedge\fR)$'s is $\fchi_1$ and the other is
$\fchi_0$, while ${\cal T}_2$ contains only one trace of the form $\tr[2\fchi_0]$.
Then $T_1$ contains either $\tr[({}^{(0)} \delta\fGamma) \fchi_1]$ or $\tr[({}^{(0)}\nabla\xi) \fchi_1]$ 
but both of them are zero. 
\item
The final case is when one of the $(\fR\wedge\fR\wedge\ldots \wedge \fR)$'s is of Case $\barB$ or $\barC$ (where there exists only one $\form{\upsilon}$ and it is set to $(\R{2})$) while the other $(\fR\wedge\fR\wedge\ldots\wedge\fR)$ is in Case $\barE$.
 The trace part ${\cal T}$ is exactly ${\cal T}_1$ with all of them set to $\tr[\fchi_0]$.
By noticing that both $\tr[({}^{(0)} \delta\fGamma)\fchi_1]$ and $\tr[({}^{(0)}\nabla\xi )\fchi_1]$ vanish,  
$T_1$ needs to be proportional to $\tr[({}^{(0)} \delta\fGamma) \fchi_0]$ or $\tr[({}^{(0)}\nabla\xi) \fchi_0]$. 
Then, the only potential nontrivial cases are when $T_1$ is proportional to one of the following terms, 
but all of them vanish from explicit computations :  
\begin{eqnarray}
&&\tr[({}^{(0)}\delta \fGamma)  (\fchi_0 \R{2}\fchi_1)]=
 \tr[({}^{(0)}\delta \fGamma)  (\fchi_1 \R{2}\fchi_0)]=0\nonumber\\
 &&\tr[({}^{(0)}\nabla\xi )(\fchi_0 \R{2}\fchi_1)]=\tr[({}^{(0)}\nabla\xi ) (\fchi_1 \R{2}\fchi_0)]=0,\nonumber\\
&&\tr[({}^{(0)}\delta \fGamma)  ( \R{2}\fchi_1)]= \tr[({}^{(0)}\delta \fGamma)  ( \fchi_1 \R{2})]=0\,,
\nonumber \\ 
&&\tr[({}^{(0)}\nabla\xi )(\R{2}\fchi_1)]= \tr[({}^{(0)}\nabla\xi ) (\fchi_1 \R{2})]=0, 
\end{eqnarray} which follow from Eqs.~(\ref{eq:deltaGamma02}), (\ref{eq:fchi02fchi1}), (\ref{eq:anotheruseful2})
and the fact that the non-zero possibilities in $(\R{2}\fchi_1)$ and $(\fchi_1\R{2})$ are reducible to
$ (\R{2}\R{0})$ and $(\R{0}\R{2})$.

\end{itemize}

To summarize up to this point, for 
$\fP_{CFT} = c_{_g}\, \tr[\fR^{2k_1}] \wedge \tr[\fR^{2k_2}]\wedge \cdots\wedge \tr[\fR^{2k_p}]$, 
we have shown that 2nd and higher order building blocks 
do not give any nontrivial contribution to $T_1$ up to $\omega^{n-1}$ order. 

Let us next consider the case with 
$\fP_{CFT} = c_{_M}\, \fF^l \wedge \tr[\fR^{2k_1}] \wedge \tr[\fR^{2k_2}]\wedge \cdots \wedge\tr[\fR^{2k_p}]$. 
Since $(\F{0})\propto dr\wedge \fu$, $\fF^l$ starts at order $\omega^l$. As for the rest part of $T_1$, 
the argument goes in the same way as we did for 
$\fP_{CFT} = c_{_g}\, \tr[\fR^{2k_1}] \wedge \tr[\fR^{2k_2}]\wedge \cdots\wedge \tr[\fR^{2k_p}]$ 
(that is, we know that for this anomaly polynomial, $T_1$ at the leading order is 
$\omega^{n-1}$ and this does not contain 2nd and higher order terms). 

From the above arguments, we conclude that there is no contribution
to the first term $T_1$ up to $\omega^{n-1}$ order from second or higher order building blocks.

\subsection{\texorpdfstring{$T_2$}{T2} at horizon}
Let us consider the general term in the anomaly polynomial of the form
$\fP_{CFT} = c_{_M}\, \fF^l \wedge\tr[\fR^{2k_1}]\wedge\tr[\fR^{2k_2}]\wedge \ldots \wedge\tr[\fR^{2k_p}]$. 
The second term $T_2$ is of the form 
\bea 
T_2&=&(\delta \fA) \wedge (l\fF^{l-1} ) \wedge\sum_{i=1}^p\left\{(2k_i) \tr[(\nabla \xi)\fR^{2k_i-1}] \right.\nonumber\\
&&\left.
\times  \tr[\fR^{2k_1}]\wedge \ldots \wedge \tr[\fR^{2k_{i-1}}] \wedge \tr[\fR^{2k_{i+1}}] 
\wedge \ldots \wedge
\tr[\fR^{2k_p}] \right\}\, . \label{eq:T2higherhorizon}
\eea
Without loss of generality, we concentrate on the term with $i=1$ only. 

Now we denote the total number of derivatives in $\delta \fA$ and $\nabla \xi$ as $\tilde{N}$ 
and classify based on it. 
When $\tilde{N}\ge 1$, under the assumption that $T_2$ contains at least one 2nd or higher order terms, 
to contribute up to $\omega^{n-1}$ order, $T_2$ needs to have at least two $(\R{0})$'s or one $(\F{0})$. 
Therefore, this case does not contribute $T_2$. 

Let us next consider the case with $\tilde{N}= 0$. To circumvent 
$(\F{0})\propto dr\wedge \fu $ and the appearance of more than one $(\R{0})$, 
$T_2$ needs to contain one $(\R{0})$ as well as one $(\R{2})$, 
while the rest of $\fF$'s and $\fR$'s are set to be of 1st order. 
In addition, by noticing that $\tr[\R{2}\fchi_1]=0$,  
we can reduce the possibilities to potentially nontrivial cases for which the $\tr[(\nabla \xi)\fR^{2k_1-1}]$ part in $T_2$ is 
given by one of the following forms :
\begin{eqnarray}
&&({}^{(0)}\nabla_a \xi^b)(\R{2}\fchi_1)^a{}_b\, , \quad  ({}^{(0)}\nabla_a \xi^b)(\fchi_1\R{2})^a{}_b\, , \nonumber \\
&&({}^{(0)}\nabla_a \xi^b)(\fchi_0\R{2}\fchi_1)^a{}_b\, , \quad  ({}^{(0)}\nabla_a \xi^b)(\fchi_1\R{2}\fchi_0)^a{}_b\, , \nonumber \\
&& ({}^{(0)}\nabla_a \xi^b)(\fchi_1)^a{}_b\,.
\end{eqnarray} However, all of the above vanish due to Eq.~(\ref{eq:someeq111}), Eq.~(\ref{eq:fchi02fchi1})-(\ref{eq:anotheruseful2})
and the fact that the non-zero possibilities in $(\R{2}\fchi_1)$ and $(\fchi_1\R{2})$ are reducible to
$ (\R{2}\R{0})$ and $(\R{0}\R{2})$.


 \section{Why \fixform{$(\fQNoether)_{H}\atinfty = 0$}  in \texorpdfstring{AdS$_{d+1>7}$}{~} ? :  without higher order}
\label{sec:roughrefinedestimate}
\subsection{Rough and refined estimates }
The goal of this Appendix is to prove that
 $(\fQNoether)_{H}\atinfty=0$.
For AdS$_5$, we have already shown explicitly that
$(\fQNoether)_{H} \atinfty=0$ in Appendix \ref{sec:chargeshighercontribution}.
Thus, here we will concentrate on AdS$_{2n+1}$ for $n\ge3$.
Here we ignore terms proportional to $dr$.

Since we have already shown 
that both $T_3$  and $T_4$ vanish at infinity in Appendix \ref{eq:t3pt4general},  here we prove that $T_0$, $T_1$ and $T_2$ vanish at infinity for AdS$_{2n+1}$ with $n\ge 3$. 
We first neglect the 2nd and higher order building blocks. Eventually, in Appendix \ref{sec:chargeshighercontribution} we will take them 
into account and show that their contributions vanish at the leading order of the derivative expansion.

Our strategy goes as follows. Based on the asymptotic behavior of 
the building blocks of $T_0$, $T_1$ and $T_2$, 
we first carry out the rough estimate for the asymptotic behaviors of these terms, 
without contracting their indices. 
This rough estimate is enough to confirm that each term vanishes at infinity for sufficiently 
high spacetime dimensions (more precisely, the rough estimate is enough for AdS$_{11}$ and higher),  
while more detailed analysis is still required
for some low-dimensional examples (that is, AdS$_7$ and AdS$_9$). 
We thus analyze the latter cases 
more carefully after the rough estimate. 

Here, we define more precisely what we mean by the rough estimate and 
introduce some notation we will employ in this Appendix. For the building blocks,  we define their fall-offs by the slowest damping component and denote it by `$\sim$' :
\begin{eqnarray}
&&\delta \fGamma \sim r\, ,\quad 
\nabla \xi \sim r^3 \,,\quad
 (\R{0})\sim r^2 \,,\quad 
(\R{1})\sim r^{-(d-3)} \,,   \nonumber \\ 
&& (\F{0})=(\ldots)dr,\quad
(\F{1})\sim r^{-d+2}\, . 
\end{eqnarray}
For example, for $\nabla \xi$, the component with the slowest fall-off is $\nabla_\mu \xi^r$ 
which behaves as $r^3$ at the boundary, and thus in the rough estimate we have $\nabla \xi \sim r^3$.   
Here we remind the readers that we are ignoring terms proportional to $dr$ in this Appendix.

The rough estimate of the fall-off behavior for the products of $\fR$'s goes as follows : 
we just forget all their index-contractions and replace {\it each} building block by its fall-off from the rough estimate. For example,
\bea
(\R{1}^{a_1}{}_{b_1}\wedge \ldots \wedge\R{1}^{a_{2k-2}}{}_{b_{2k-2}}) &\sim&  r^{-(d-3)(2k-2)}\,,\nonumber\\
(\R{0}^p{}_q\wedge 
\R{1}^{a_1}{}_{b_1}\wedge \ldots \wedge\R{1}^{a_{2k-3}}{}_{b_{2k-3}}
) &\sim& r^{-(d-3)(2k-3)+2}\,,\nonumber\\
(\R{0}^{p_1}{}_{q_1}\wedge \R{0}^{p_2}{}_{q_2}\wedge 
\R{1}^{a_1}{}_{b_1}\wedge \ldots \wedge\R{1}^{a_{2k-4}}{}_{b_{2k-4}}
) &\sim&  r^{-(d-3)(2k-4)+4}\, .  
\eea
Of course, we can consider the case with more $(\R{0})$'s, but the above possibilities are enough 
for our purpose, as we will see later.

The refined estimate takes into account the contraction structure of each building block. To distinguish the refined estimate from the rough estimate, we will denote the refined estimate with `$\leadsto$'.
Here is a simple example:  for the rough estimate, we have 
\be
 (\R{1}\R{1})
\sim r^{-2(d-3)} \,,\quad
(\R{0}\R{1}) \sim r^{-(d-5)} \,,\quad
(\R{1}\R{0}) \sim r^{-(d-5)}\,,\quad
(\R{0}\R{0}) \sim r^{4}\,, 
\ee  while for the refined estimate, the fall-offs are much faster (due to explicit contractions of the indices) :\footnote{In the refined estimate, we only aim to improve upon the rough estimate by explicitly contracting the indices. We do not attempt to explicitly compute the exact fall-off, i.e. figuring out which is the first non-zero leading behavior, but merely want to bound how slow the exact fall-off could be.
}
\be
(\R{1}\R{1})^{a_1}{}_{b_2} \leadsto r^{-2(d-1)} \,,\quad
(\R{0}\R{1}) \leadsto r^{-2(d-1)-1} \,,\quad
(\R{1}\R{0}) \leadsto r^{-2(d-1)+3}\,, \quad
(\R{0}\R{0}) = 0\,.
\ee

In the course of the rough estimate, since 
we do not take into account the index contraction, there is no 
distinction between the anomaly polynomial with a single trace or multiple traces. 
For example, for the rough estimate, we do not distinguish $\fP_{CFT} = c_{_g} \,\tr[\fR^{2(k_1+k_2)}]$ 
and $\fP_{CFT} = c_{_g}\,\tr[\fR^{2k_1}]\wedge\tr[\fR^{2k_2}]$.  In the following part, we therefore carry out
the rough estimate for the following two cases: 
(1) purely gravitational anomaly polynomial (including both single- and multi-trace cases)
(2) mixed anomaly polynomial (including both single- and multi-trace cases). 

\subsection{Rough estimate for purely gravitational anomaly polynomials}
Let us consider a general purely gravitational 
anomaly polynomial containing $2k$ curvature two-forms $\fR$ ($k\ge2$) on AdS$_{2n+1}$ with $n=2k-1$ 
(note that $n\ge3$ and $n$ is an odd number in this case).  
In this case, since both $T_0$ and $T_2$ are zero from the definition of these terms, 
we only need to consider the first term $T_1$. 

Before the estimate, we remind the readers that when there are three or more $(\R{0})$'s in $T_1$, 
then $T_1$ contains at least one $\fchi_2$ or $\tr[\fchi_1]$ both of which 
are proportional to $dr\wedge\fu$. 
The symbol $\fchi_m$ is defined as products of $(\fR)$'s consist of $m$ number of $(\R{0})$'s with the remaining $(\fR)$'s being $(\R{1})$'s
(see Appendix \ref{sec:app_useful_previouspaper01only} for useful facts about $\fchi_m$).
Therefore, we only need to consider
the cases with two or less $(\R{0})$'s.  

Through the rough estimate, the first term $T_1$ behaves as 
\be
T_1 \sim ( \delta \fGamma)  \nabla \xi \wedge (\fR^{a_1}{}_{b_1}\wedge \ldots \wedge\fR^{a_{2k-2}}{}_{b_{2k-2}})
\sim r^4  (\fR^{a_1}{}_{b_1}\wedge \ldots \wedge\fR^{a_{2k-2}}{}_{b_{2k-2}}) \, . 
\ee Thus, depending on the number of $(\R{0})$'s in $T_1$,  there are three possibilities 
with the following fall-off behaviors: 
\bea\label{eq:T1roughestsingletrace}
r^4 (\R{1}^{a_1}{}_{b_1}\wedge \ldots \wedge\R{1}^{a_{2k-2}}{}_{b_{2k-2}}) &\sim & r^{-(2n-3)(n-1)+4}\, ,  
\nonumber\\ r^4 (\R{0}^p{}_q\wedge 
\R{1}^{a_1}{}_{b_1}\wedge \ldots \wedge\R{1}^{a_{2k-3}}{}_{b_{2k-3}}
)  &\sim & r^{-(2n-3)(n-2)+6}\, , \nonumber \\
 r^4  (\R{0}^{p_1}{}_{q_1}\wedge \R{0}^{p_2}{}_{q_2}\wedge 
\R{1}^{a_1}{}_{b_1}\wedge \ldots \wedge\R{1}^{a_{2k-4}}{}_{b_{2k-4}}
) &\sim&  r^{-(2n-3)(n-3)+8} \, .
\eea
Since $n\ge 3$, the first case vanishes at infinity while 
the second and third cases vanish for $n\ge 5$ (note that $n$ is an odd integer). Therefore, for AdS$_{11}$ and higher, the rough estimate is sufficient to imply that $T_1 |_{r\rightarrow \infty} \sim 0$.
Thus, the only nontrivial case is 
when $n=3$ which corresponds to the anomaly polynomial $\fP_{CFT} = c_{_g}\, \tr[\fR^{4}]$ and $\fP_{CFT} = c_{_g}\, \tr[\fR^{2}] \wedge \tr[\fR^2]$ on AdS$_7$. 
We will carry out the refined estimate for these cases later in Appendix~\ref{sec:T1T2AdS7}.

\subsection{Rough estimate for mixed anomaly polynomials}
Now we consider the anomaly polynomial containing $\fF^l$ and 
$2k$ curvature two-forms (where $l\ge 1$ and $k\ge 1$). This anomaly polynomial is admitted on 
AdS$_{2n+1}$ with $n=2k+l-1$. Here we consider $n\ge 3$ only and thus 
$2k+l\ge 4$ needs to be satisfied. In the following, 
we evaluate the three terms $T_0$, $T_1$, $T_2$ one by one. 

\subsubsection{Zeroth term $T_0$}
There are two terms in $T_0$. From Eq.~(\ref{eq:T0toT4sec2}), we see that the first and the second term each contains $(\partial^2 \fP_{CFT}/\partial \fF\partial \fF)$ 
and $(\partial^2 \fP_{CFT}/\partial \fR^{a}{}_b\partial \fF)$ respectively.
Here we assume $(\Lambda + \ic_{\xi}\fA )$ fall-offs as $r^0$ at most. 

For the first term, since $(\F{0})\propto dr\wedge \fu$, all the $\fF$'s need to 
be replaced by $(\F{1})$.  We also note that all the $\fR$'s need to 
be replaced by $(\R{1})$ since $\tr[\fchi_1]\propto dr\wedge \fu$ and 
$\fR$'s only appear in the form of $\tr[\fR^{2k_i}]$ (since there is no derivative with respect to $\fR^a{}_b$). 
Then the fall-off behavior is roughly estimated as 
\begin{eqnarray}
(\delta \fA)\wedge (\Lambda + i_{\xi}\fA )\, \wedge \fF^{l-2} 
\wedge (\fR^{a_1}{}_{b_1}\wedge\cdots \wedge \fR^{a_{2k}}{}_{b_{2k}})
\sim \Phi^{l-1} \Phi_T^{2k} \sim 
r^{-(d-1)d/2+(l-1)} 
\, ,  
\end{eqnarray}  
which vanishes at infinity.

Similarly, for the second term in $T_0$,
we can replace all the $\fF$'s by $(\F{1})$'s, thus the second term in $T_0$ behaves as
\begin{eqnarray}
&&\delta \fGamma^a{}_b (\Lambda + \ic_\xi \fA)\wedge \fF^{l-1} 
\wedge (\fR^{a_1}{}_{b_1}\wedge\cdots \wedge \fR^{a_{2k-1}}{}_{b_{2k-1}})\nonumber\\
&&\quad\sim r\Phi^{l-1} (\fR^{a_1}{}_{b_1}\wedge\cdots \wedge \fR^{a_{2k-1}}{}_{b_{2k-1}})
\sim
 r \times r^{-2(n-1)(l-1)} (\fR^{a_1}{}_{b_1}\wedge\cdots \wedge \fR^{a_{2k-1}}{}_{b_{2k-1}})\,. \nonumber\\
\end{eqnarray}  
Since the second term in $T_0$ contains only one derivative with respect to $\fR^{a}{}_b$, there is only one 
$(\fR\wedge \cdots \wedge\fR)$ (without trace) wedged by $\tr[\fR^{2k_i}]$'s. 
Since $\fchi_2\propto dr\wedge \fu$ and $\tr[\fchi_1]\propto dr\wedge \fu$, 
we only need to consider the case with one or less $(\R{0})$.
In particular, the $(\R{0})$ must be located in the 
$(\fR\wedge \cdots \wedge\fR)$ part, while all the  
$\tr[\fR^{2k_i}]$'s must be replaced by $\tr[\R{1}^{2k_i}]$. Therefore,
there are 
two potential nontrivial terms which fall-off as 
\begin{eqnarray}
r^{1-2(n-1)(l-1)}(\R{1}^{a_1}{}_{b_1}\wedge \ldots \wedge\R{1}^{a_{2k-1}}{}_{b_{2k-1}})
&\sim & r^{1-2(n-1)(l-1)} \times r^{-(2n-3)(2k-1)}\, , \nonumber\\ 
r^{1-2(n-1)(l-1)}
(\R{0}^p{}_q\wedge 
\R{1}^{a_1}{}_{b_1}\wedge \ldots \wedge\R{1}^{a_{2k-2}}{}_{b_{2k-2}} )
&\sim& r^{1-2(n-1)(l-1)}  \times r^{2-(2n-3)(2k-2)}, \nonumber\\
\end{eqnarray}
both of which vanish at infinity (note that $2k+l\ge4$ for $n\ge 3$). 

To summarize, the zeroth term $T_0$ vanishes at infinity for $n\ge 3$. 

\subsubsection{First term $T_1$}
\label{eq:roughtesimateT1}
Similar to $T_0$, all the $\fF$'s need to be replaced by $(\F{1})$ to have 
a non-trivial result. Then the fall-off behavior of this term is 
\begin{eqnarray}
T_1 \sim \Phi^l ( \delta \fGamma) \nabla \xi \wedge(\fR^{a_1}{}_{b_1}\wedge \ldots \wedge\fR^{a_{2k-2}}{}_{b_{2k-2}})
\sim r^{-2(n-1)l+4} (\fR^{a_1}{}_{b_1}\wedge \ldots \wedge\fR^{a_{2k-2}}{}_{b_{2k-2}})\, .\nonumber\\
\end{eqnarray} 
Then, as in the case of a purely gravitational anomaly polynomial, there are three possibilities 
depending on the number of $(\R{0})$'s :
\begin{eqnarray} \label{eq:roughT1mixed}
r^{-(2n-2)l+4} (\R{1}^{a_1}{}_{b_1}\wedge \ldots \wedge\R{1}^{a_{2k-2}}{}_{b_{2k-2}})&\sim &
r^{-(2n-3)(n-1)+4-l}\, , 
 \nonumber\\
r^{-(2n-2)l+4} (\R{0}^p{}_q\wedge 
\R{1}^{a_1}{}_{b_1}\wedge \ldots \wedge\R{1}^{a_{2k-3}}{}_{b_{2k-3}}
)  &\sim &
r^{-(2n-3)(n-2)+6-l} \, , \nonumber\\
r^{-(2n-2)l+4} (\R{0}^{p_1}{}_{q_1}\wedge \R{0}^{p_2}{}_{q_2}\wedge 
\R{1}^{a_1}{}_{b_1}\wedge \ldots \wedge\R{1}^{a_{2k-4}}{}_{b_{2k-4}}
)&\sim &
r^{-(2n-3)(n-3)+8-l} \, . 
\end{eqnarray} 
These three terms vanish when $n\ge 3$, $n\ge 4$ and $n\ge 5$, respectively. 
We thus need to carry out the refined estimate for AdS$_7$ ($n=3$) and AdS$_9$ ($n=4$). 

\subsubsection{Second term $T_2$}
As a final part for the rough estimate, we evaluate $T_2$. Again, by noting that 
$(\F{0})\propto dr\wedge \fu$, we replace all the $\fF$'s by $(\F{1})$. 
Then the fall-off behavior of this term is 
\be
T_2\sim (\delta \Phi) \nabla \xi 
\Phi^{l-1} (\fR^{a_1}{}_{b_1}\wedge \ldots \wedge\fR^{a_{2k-1}}{}_{b_{2k-1}})
\sim r^{-2n+5}\times  r^{-(2n-2)(l-1)}(\fR^{a_1}{}_{b_1}\wedge \ldots \wedge\fR^{a_{2k-1}}{}_{b_{2k-1}})
\, .
\ee
As in the case of the second term in $T_0$, this term contains only one derivative with respect to 
the curvature two-form and thus at most one $(\R{0})$ is allowed to have a nontrivial contribution. 
Therefore, the followings are two non-trivial possibilities : 
\begin{eqnarray}
r^{-2n+5-(2n-2)(l-1)}(\R{1}^{a_1}{}_{b_1}\wedge \ldots \wedge \R{1}^{a_{2k-1}}{}_{b_{2k-1}})
&\sim& r^{-2n+5-(2n-2)(l-1)}\times r^{-(2n-3)(2k-1)}\, , 
\nonumber \\
r^{-2n+5-(2n-2)(l-1)}
(\R{0}^p{}_q\R{1}^{a_1}{}_{b_1}\wedge \ldots \wedge \R{1}^{a_{2k-2}}{}_{b_{2k-2}})
&\sim& 
r^{-2n+5-(2n-2)(l-1)}\times r^{-(2n-3)(2k-2)+2}\, , 
\nonumber \\
\end{eqnarray}
both of which vanish for $n\ge3$ (note again that $2k+l\ge4$ for $n\ge 3$). 

To summarize, we conclude that:
\begin{itemize}
\item $T_0$, $T_1$ and $T_2$ vanish at infinity for AdS$_{2n+1}$ with $n\ge5$.

\item For AdS$_7$ and AdS$_9$ the refined estimate needs to be done for $T_1$ only.

\item
In the case of AdS$_7$,  from \eqref{eq:roughT1mixed} (which is valid for $l=0$ case too), 
the subtle case is when one or two $\fR$'s are $(\R{0})$ 
and $T_1$ in the rough estimate behaves as $r^{3-l}$ and $r^{8-l}$, respectively.  

\item On the other hand, for AdS$_9$,  the only subtle case is when there exist two $(\R{0})$'s 
where $T_1$ in the rough estimate behaves as $r^{3-l}$.  
\end{itemize}

\subsection{Refined estimate of \texorpdfstring{$T_1$ for AdS$_7$}{~}}
\label{sec:T1T2AdS7}
In this part, we carry out the refined estimate of $T_1$ for the anomaly polynomials 
admitted on AdS$_7$. 
For AdS$_7$, there are three types of the anomaly polynomials,   
$\fP_{CFT}= c_{_M} \fF^2 \wedge \tr[\fR^2]$, $c_{_g} \tr[\fR^2] \wedge \tr[\fR^2]$ 
and $c_{_g} \tr[\fR^4]$. We will evaluate how they behave at infinity one by one. 
We stress again that here we still neglect 2nd and higher order building blocks, but we will take them into account at the end of this Appendix. 
We also recall that we have confirmed through the rough estimate that 
$T_1$ vanishes at infinity when it contains no $(\R{0})$'s. We thus consider 
the case with one or two $(\R{0})$'s. 

\subsubsection{\fixform{$\fP_{CFT}= c_{_M} \fF^2 \wedge \tr[\fR^2]$}}
In this case, the first term $T_1$ is given by 
\begin{eqnarray}
T_1 = 2 c_{_M} \fF^2 \wedge \delta \fGamma^a{}_b \nabla_a \xi^b\, . 
\end{eqnarray}
This case does not contain any $(\R{0})$ and thus we conclude that 
$T_1$ vanishes at infinity. 

\subsubsection{\fixform{$\fP_{CFT}= c_{_g} \tr[\fR^2] \wedge \tr[\fR^2]$}}

The first term $T_1$ for this anomaly polynomial is composed of two types of terms, 
depending on where the two derivatives with respect to $\fR$ act : 
\bea\label{eq:ads7trRsqT1}
T_1 &=& 4c_{_g} \delta \fGamma^a{}_b \nabla_a \xi^b  \wedge\tr[\fR^2] 
+8c_{_g}( \delta \fGamma^a{}_b\wedge \fR^b{}_a)\wedge(\nabla_d \xi^c \fR^c{}_d )\, .
\eea   
For the first term in \eqref{eq:ads7trRsqT1}, 
since $(\R{0}^2)=0$ and $\tr[(\R{0}\R{1})]\propto dr\wedge \fu$, 
this term does not contribute at infinity.  
For the second term in \eqref{eq:ads7trRsqT1}, direct computations show that 
\bea
\delta \fGamma^a{}_b\wedge ( \R{0})^b{}_a&\leadsto&
r^{-10}\, ,\qquad 
\delta \fGamma^a{}_b\wedge ( \R{1})^b{}_a\leadsto
r^{-4}\,,\nonumber\\
\nabla_d \xi^c (\R{0})^d{}_c  &\leadsto& r^2\,,\qquad
 \nabla_d \xi^c( \R{1})^c{}_d \leadsto 
r^{-4}\, , 
\eea
which lead to 
\begin{eqnarray}
8c_{_g}( \delta \fGamma^a{}_b\wedge \fR^b{}_a)\wedge(\nabla_d \xi^c \fR^c{}_d ) \leadsto r^{-2}\, .  
\end{eqnarray}

To summarize, we have shown that $T_1$ vanishes at infinity for the anomaly polynomial
$\fP_{CFT}= c_{_g} \tr[\fR^2] \wedge \tr[\fR^2]$. 

\subsubsection{\fixform{$\fP_{CFT}= {c}_g \tr[\fR^4]$}}
As a final example for AdS$_7$, we carry out the refined estimate of $T_1$ at infinity 
for the anomaly polynomial $\fP_{CFT}= {c}_g \tr[\fR^4]$.  The first term $T_1$ 
in this case is given by 
\bea\label{eq:ads7trR4T1}
T_1 &=&4 {c}_g\left[
\delta \fGamma^a{}_b\fR^b{}_g \fR^g{}_c \nabla_a \xi^c  
+ \delta \fGamma^a{}_b \nabla_d \xi^b  \fR^d{}_f \fR^f{}_a
+ \delta \fGamma^a{}_b \fR^b{}_c \nabla_d \xi^c  \fR^d{}_a
 \right]\, .
\eea  

The first and the second terms in \eqref{eq:ads7trR4T1} have three potential contributions, 
but all of them vanish at infinity in the following way :
\bea
\delta \fGamma^a{}_b (\R{0}\R{1})^b{}_c \nabla_a \xi^c  \leadsto r^{-7}\,,  \quad
\delta \fGamma^a{}_b  \nabla_d \xi^b  (\R{0}\R{1})^d{}_a   &\leadsto& r^{-7}\, , 
\nonumber\\
\delta \fGamma^a{}_b (\R{1}\R{0})^b{}_c \nabla_a \xi^c \leadsto r^{-3}\,,  \quad
\delta \fGamma^a{}_b  \nabla_d \xi^b  (\R{1}\R{0})^d{}_a   &\leadsto& r^{-3}\, ,
\nonumber\\
\delta \fGamma^a{}_b (\R{0}\R{0})^b{}_c \nabla_a \xi^c = 0, \quad
\delta \fGamma^a{}_b  \nabla_d \xi^b  (\R{0}\R{0})^d{}_a   &=& 0\, . 
\eea  

For the third term, there are three potential contributions: 
\begin{eqnarray} \label{eq:four-possibility}
&& \delta \fGamma^a{}_b (\R{0})^b{}_c \nabla_d \xi^c  (\R{1})^d{}_a 
\leadsto r^{-3}\, , \nonumber \\
&&
\delta \fGamma^a{}_b (\R{1})^b{}_c \nabla_d \xi^c  (\R{0})^d{}_a 
\sim r^{-1} \, , \nonumber \\
&&
\delta \fGamma^a{}_b (\R{0})^b{}_c \nabla_d \xi^c  (\R{0}) ^d{}_a
\leadsto r^{-8}\, . 
\end{eqnarray}
For the first line, we have used
\begin{eqnarray} \label{eq:deltagammazeroinfinity}
\delta \fGamma^a{}_{b} (\R{0})^b{}_c \leadsto r^{-(d-3)}\,,
\end{eqnarray} which behaves as  $r^{-3}$ for AdS$_7$.
For the other two terms, we evaluated them directly to show the fall-off behavior.
In the end, we confirmed that $T_1\to 0$ at infinity.

\subsection{Refined estimate of \texorpdfstring{$T_1$ for AdS$_9$}{~}}
\label{sec:T1T2AdS9}
Here we carry out the refined estimate for AdS$_9$. 
We recall that, as a result of the rough estimate, the only potential nontrivial case is 
when $T_1$ contains two $(\R{0})$'s. 
For this Appendix, we will focus on such possibilities.
In this case, $T_1$ behaves as 
$T_1\sim r^{3-l}$. 
We also notice the relation \eqref{eq:deltagammazeroinfinity}
in this case is $\delta\fGamma^a{}_b (\R{0})^b{}_c\leadsto r^{-(d-3)}=r^{-5}$, 
while the rough estimate gives  
$\delta\fGamma^a{}_b (\R{0})^b{}_c\sim r^3$.

\subsubsection{\fixform{$\fP_{CFT}= c_{_M} \fF^3 \wedge \tr[\fR^2]$}}
In this case, $T_1$ is given by 
\begin{eqnarray}
T_1 = 2c_{_M}\, \fF^3\wedge \delta\fGamma^{a}{}_b \nabla_a\xi^b\, , 
\end{eqnarray}
which does not contain any $\fR$. Therefore, this term vanishes at 
infinity as a result of the rough estimate. 

\subsubsection{\fixform{$\fP_{CFT}= c_{_M} \fF \wedge \tr[\fR^2] \wedge \tr[\fR^2]$}}
The first term $T_1$ for this anomaly polynomial is 
\begin{eqnarray}
T_1 = 4{c}_M \, \fF \wedge \delta\fGamma^{a}{}_b \nabla_a\xi^b\wedge \tr[\fR^2] + 
8{c}_M \,  \fF \wedge (\delta \fGamma^a{}_b \fR^b{}_a)\wedge (\nabla_d\xi^c \fR^c{}_d)\, ,
\end{eqnarray}
which vanishes since $(\R{0}^2)=0$. 

As for the second term, as a result of  \eqref{eq:deltagammazeroinfinity} (for AdS$_9$), 
the refined estimate changes the fall-off behavior of $T_1$ by at least a factor $r^{-8}$
compared to the rough estimate 
This is enough to confirm that $T_1$, which fall-off as $r^{3-l}=r^2$ under the rough estimate, 
vanishes at infinity. 

\subsubsection{\fixform{$\fP_{CFT}= c_{_M} \fF \wedge \tr[\fR^4]$}}
In this case, $T_1$ is given by 
\bea\label{eq:ads9FtrR4T1}
T_1 &=&4 c_{_M}\left[
\fF\wedge \delta \fGamma^a{}_b\fR^b{}_g \fR^g{}_c \nabla_a \xi^c  
+\fF\wedge \delta \fGamma^a{}_b \nabla_d \xi^b  \fR^d{}_f \fR^f{}_a
+ \fF\wedge\delta \fGamma^a{}_b \fR^b{}_c \nabla_d \xi^c  \fR^d{}_a
 \right]\, . \nonumber\\
\eea  
The first and second terms vanish as a result of $(\R{0}^2)=0$. 
On the other hand, for the third term,  
Eq.~\eqref{eq:deltagammazeroinfinity} (for AdS$_9$) improves the fall-off behavior 
at infinity by at least a factor $r^{-8}$ compared to the rough estimate. 
As in the previous case, this is enough to prove that $T_1$
in this case vanishes at infinity.  

 \section{Why \texorpdfstring{\fixform{$(\fQNoether)_{H}\atinfty = 0$}  in AdS$_{d+1>7}$}{~} ? : with higher order}
\label{sec:chargeshighercontribution}
In this Appendix, we will focus on contributions containing at least a 2nd or higher order building block and prove that the 2nd and higher order building blocks do not contribute to 
$(\fQNoether)_{H}$ at $r\to \infty$.  
Before we begin the general argument, it is instructive to study in details all cases in AdS$_7$ and a particular case in AdS$_{15}$.

\subsection{Example :  \texorpdfstring{AdS$_7$}{AdS7}}
\label{sec:deltaQinfinityAdS7}
In AdS$_7$, we consider contributions up to $\omega^2$ order.
Hence, we only need to consider the case containing exactly one 2nd order building block (which contributes at $\omega^2$) with the rest of the building blocks set to zeroth order. We will show that such contributions are  zero at any fixed $r$. 
In particular, they vanish at $r\rightarrow \infty$.
\subsubsection{\fixform{$\fP_{CFT}= c_{_M} \fF^2\wedge \tr[\fR^2]$}}

The terms $T_0$, $T_1$ and $T_2$ for this anomaly polynomial are given by 
\bea
T_0&= &2c_{_M}(\Lambda+ i_\xi \fA)\delta \fA  \wedge \tr[\fR^2]
+4c_{_M}(\Lambda+ \ic_\xi \fA)\fF \wedge \tr[\delta \fGamma \fR] \, , 
\nonumber\\
T_1 &=& 2 c_{_M} \fF^2 \wedge \tr[\delta \fGamma \nabla \xi]\,, \nonumber\\
T_2 &=&4c_{_M} ( \delta \fA) \wedge \fF\wedge \tr[ \nabla \xi \fR]\,.
\eea
The first term in $T_0$ does not contain $\omega^2$ order contribution
because of $\tr[(\R{0})^2]=0$ and $\tr[(\R{0}\R{2})]=0$.    
For $T_1$, $T_2$ and the second term in $T_0$,  since $(\F{0})\propto dr\wedge \fu$, 
one needs to set one $\fF$ to be $(\F{2})$. In that case,  
since $\fF$,  $\tr[\delta \fGamma \fR]$ and $\tr[\nabla \xi  \fR]$ start at $\omega^1$ order, 
these terms all become $\omega^3$ or higher order.  
\subsubsection{\fixform{$\fP_{CFT}= c_{_g} \tr[\fR^2]\wedge \tr[\fR^2]$}}

In this case, $T_0$ and $T_2$ are trivially zero, while $T_1$ is
\bea
T_1 &\equiv&
4c_{_g} \,
\tr[\delta \fGamma \nabla \xi]\wedge
 \tr[\fR^2]
+8c_{_g} \,
\tr[\delta \fGamma \fR]\wedge\tr[\nabla \xi \fR]\,. 
\eea

The first term does not contribute to $\omega$ order because of 
$\tr[(\R{0})^2]=0$ and $\tr[(\R{2}\R{0})]=0$. 
For the second term, let us first consider the case when the second order term is located in 
$\tr[\delta \fGamma \fR]$ . Then, since $\tr[\nabla \xi  \fR]$ starts at $\omega^1$ order, 
the second term in $T_1$ starts at $\omega^3$ or higher order. 
The argument is the same when $\tr[\delta \fGamma \fR]$ and $\tr[\nabla \xi  \fR]$ are interchanged. 
\subsubsection{\fixform{$\fP_{CFT}= c_{_g} \tr[\fR^4]$}}
This is the final case admitted in AdS$_7$, in which a little bit detailed analysis is 
required. For this anomaly polynomial,  $T_1$ is given by 
($T_0$ and $T_2$ are trivially zero as in the previous example)
\bea\label{eq:ads7trR4T1ver2}
T_1 &=&4 c_{_g}\left\{
\tr[\delta \fGamma \fR^2  (\nabla \xi)]
+\tr[ \delta \fGamma (\nabla \xi) \fR^2]
+ \tr[\delta \fGamma \fR  (\nabla \xi) \fR]
 \right\}.
\eea  
Similar to the previous examples, to have order $\omega^2$ contribution, 
exactly one of the building blocks
is of second order while the rest need to be of zeroth order.  Then, we have the following possibilities : 
\begin{eqnarray}
{\cal O}(\omega^2)&:&
({}^{(0)} \delta \fGamma^a{}_b)( {}^{(0)}\nabla_c \xi^b)(\R{0}\R{2})^c{}_a  \,,\quad
({}^{(0)} \delta \fGamma^a{}_b)( {}^{(0)}\nabla_c \xi^b)(\R{2}\R{0})^c{}_a    \,,\nonumber\\
 && 
({}^{(0)} \delta \fGamma^a{}_b)(\R{0}\R{2})^b{}_c( {}^{(0)}\nabla_a \xi^c)  \,,\quad
({}^{(0)} \delta \fGamma^a{}_b)(\R{2}\R{0})^b{}_c( {}^{(0)}\nabla_a \xi^c) \,,
\nonumber\\
&&
 ({}^{(0)}\delta \fGamma^a{}_b) (\R{0})^b{}_c  ({}^{(0)}\nabla_d \xi^c ) (\R{2})^d{}_a \,,\quad
 ({}^{(0)}\delta \fGamma^a{}_b) (\R{2})^b{}_c  ({}^{(0)}\nabla_d \xi^c ) (\R{0})^d{}_a\, .
 \end{eqnarray}
These terms all turn out to be zero due to Eq.~\eqref{eq:useful101}.
 
\subsection{Example : \texorpdfstring{AdS$_{15}$}{AdS15}} 
\label{sec:ads15deltaQ}
Now we consider one specific example in AdS$_{15}$ 
which will be useful when we deal with the general cases. 
\subsubsection{\fixform{$\fP_{CFT}= c_{_g} \tr[\fR^4]\wedge \tr[\fR^4]$}}
For this anomaly polynomial, $T_0$ and $T_2$ are trivially zero, while 
the first term $T_1$ is 
\bea
T_1&=& 
24\, c_{_g}\left\{
\tr[ \delta \fGamma  (\fR^2) \nabla \xi]
+  \tr[  \delta \fGamma \nabla \xi (\fR^2)]
+   \tr[ \delta \fGamma \fR \nabla \xi\fR] \right\}\wedge \tr[\fR^4]\nonumber\\
&&  + 32\, c_{_g}\,\tr[\delta \fGamma \fR^3]\wedge \tr[\nabla \xi (\fR^3)]\, .
\eea 
In AdS$_{15}$, we consider up to $\omega^{6}$ order terms.
Here we discuss the first and second line separately. 

The first line in $T_1$ is easy to deal with.  
There are two points to notice: first of all, 
$\tr[\fR^4]$ already starts at order $\omega^{4}$ and  
it is equal to $\tr[(\R{1}^4)]$.  
Secondly, the terms in the curly bracket starts at order $\omega^2$ 
and 2nd (and higher order) building blocks  do not contribute to the curly bracket terms
up to $\omega^{2}$ order.
Thus, the lowest order contributions therefore start at order $\omega^6$, 
and this order $\omega^6$ contribution does not contain any 2nd or higher order building blocks.

For the second line in $T_1$, as we have shown in Appendix~\ref{sec:trgradxiR}, 
both $\tr[\nabla \xi \fR^3]$ and $\tr[\delta \fGamma \fR^3]$ start at order $\omega^3$ 
and such contributions do not contain 2nd or higher order building blocks.

To summarize, we conclude that 2nd and higher order building blocks 
do not contributing to $(\fQNoether)_{H}$ at infinity up to order $\omega^{6}$.

\subsection{General arguments on 2nd and higher order terms}
In Appendix.~\ref{sec:roughrefinedestimate}, we used only zeroth and first order building blocks 
in the estimates of $(\fQNoether)_{H}$ at $r\rightarrow \infty$. 
Here we will deal with contributions containing at least one higher order building blocks, i.e. 
at least one ${}^{(m)}\delta \fGamma$, ${}^{(m)} \nabla \xi$, $(\F{m})$ or $ (\R{m})$ for $m\ge2$. 
We will show that all such contributions to $(\fQNoether)_{H}$ vanish up to $\omega^{n-1}$ order.

Before the proof, we here summarize some important facts. 
Let us consider $\tr[\fR^{2k}]$, $\tr[\delta \fGamma \fR^{2k-1}]$, 
and  $\tr[\nabla\xi \fR^{2k-1}]$. As we have seen in 
the computation for the Maxwell sources in Appendix C.3 of Ref.~\cite{Azeyanagi:2013xea} and Appendix \ref{sec:trgradxiR}, these traces start at $\omega^{2k}$, 
 $\omega^{2k-1}$ and  $\omega^{2k-1}$ order respectively, 
 and these leading order terms contain zeroth and first order building blocks only. 
 For the wedge product of the gauge field strength $\fF^l$, 
 the leading order contribution is of the form $(\F{1})^l$ and thus 
 is of order $\omega^l$. Second and higher order terms in 
 $\fF$ can only start contributing to $\fF^l$ at $\omega^{l+1}$-order or higher. 
 These results will become important in the later computations. 
 
 We stress that in this Appendix, by `all contributions', we mean all contributions containing at least a second or higher order building block.
\\
\vspace{0.1cm} \\
\noindent
\underline{\bf Strategy} \\
\noindent
Let us first explain our strategy. 
Our arguments rely on the following two facts (which will be shown next):
\begin{enumerate}
\item Fact 1: At $\omega^{n-2}$ order and lower, all contributions to $T_1$ vanish at any fixed $r$. \label{fact1}
\item Fact 2: 
 At $\omega^{n-1}$- order, all contributions vanish at any fixed $r$ with one exception :  terms having exactly one $(\R{2})$, $({}^{(0)} \delta \fGamma), ({}^{(0)} \nabla \xi)$ along with products of $(\R{0})$'s and $(\R{1})$'s.
\label{fact2}
\end{enumerate}

Once these two facts are confirmed, we only need to deal with the exceptional case in Fact \ref{fact2} which contains exactly one $(\R{2})$. 
Moreover, to realize $\omega^{n-1}$ order contribution,  the derivative
$(\partial^2 \fP_{CFT}/\partial \fR\partial \fR)$ needs to 
contain one $(\R{2})$, one $(\R{0})$ and the rest of $\fR$'s are $(\R{1})$. 
Now we consider $r\to\infty$. From Appendix \ref{sec:falloffR2} (in particular Eq.~(\ref{eq:R2falloff})),
we note that the fall-offs of $(\R{2})$ in the rough estimate are
\be
(\R{2})\sim r^2 \sim (\R{0})\,.
\ee Thus, we can replace $(\R{2})$ by $(\R{0})$ under the rough estimate. Therefore, what we need to evaluate is 
\be
(\R{0}^a{}_b \R{0}^c{}_d \R{1}^{p_1}{}_{q_1} 
\R{1}^{p_2}{}_{q_2}\ldots \R{1}^{p_{2k-4}}{}_{q_{2k-4}} )\, .
\ee 
This term, however, has been estimated in Appendix~\ref{eq:roughtesimateT1} to fall-off sufficiently fast and vanish in AdS$_{11}$ and higher. 
We note that, for AdS$_7$ ($n=3$), in Appendix.~\ref{sec:deltaQinfinityAdS7}, 
we have already shown explicitly that at any fixed $r$, 
there is no lower contributions than $\omega^{n-1}$ 
and that the nontrivial $\omega^{n-1}$ order contributions 
do not contain any 2nd and higher order building blocks. 

In the case of AdS$_9$ ($n=4$), we note that the anomaly polynomials are just the ones in AdS$_7$
wedged with an extra $\fF$. Since $\fF$ starts at $\omega^1$ (since $(\F{0})\propto dr\wedge \fu$), 
essentially the same argument as the AdS$_7$ case lead to the proof that 
there is no lower contributions than $\omega^{n-1}$ 
and that the nontrivial $\omega^{n-1}$ order contributions 
do not contain any 2nd and higher order building blocks.

In the rest of this part,  we will prove 
Fact \ref{fact1} and Fact \ref{fact2} case by case. 
Before doing so, we remind the readers that $\fchi_m$ denotes all possible
products of $(\R{0})$'s and $(\R{1})$'s  containing exactly $m$-number of $(\R{0})$'s wedged with an arbitrary number of $(\R{1})$'s (see Appendix \ref{sec:app_useful_previouspaper01only}).
We sometimes use $\fchi_m$ to simply denote an element in $\fchi_m$.
A useful symbol
\be
(\cR^q_{(p)})
\ee is defined to denote all possible structures (including those consist of zeroth and first order 
building blocks only) that can contribute to $(\fR^q)$ at $\omega^p$ order. 
Here is a summary of the results from Appendix~\ref{eq:singleproductstructure} regarding classifications of $(\cR^q_{(p)})$:
\bea\label{eq:classification}
(\cR^q_{(q-1)})&\equiv&
 (\fchi_1)\,, \nonumber\\
(\cR^q_{(q)})&\equiv&\left\{ 
(\fchi_0)\, ,\quad 
 (\R{2} \R{0})\,,\quad
 (\R{0} \R{2} )\,,\quad
 (\R{0} \R{3} \R{0})\right\} \,, \nonumber\\
(\cR^q_{(q+1)})&\equiv&
\left\{
 (\R{2}) \,,\quad
  (\R{2}\R{0}\R{2}) \,,\right.
   \nonumber\\
       &&
 ( \R{3}\fchi_1) \,,\quad
(\fchi_1 \R{3}) \,,\quad
\nonumber\\
&&
 (\R{2}\fchi_0) \,,\quad
      (\R{2}\R{2}\R{0}) \,,\quad
            (\R{0}\R{2}\R{2}) \,,\quad
  (\fchi_0 \R{2}) \,,\quad
\nonumber\\
&&
 (\R{0} \R{4}\R{0}) \,,
\nonumber\\
&&
 (\fchi_0 \R{3}\fchi_1)\,, \quad
(\fchi_1 \R{3}\fchi_0)\,,\nonumber\\
&&\left.
 (\fchi_0 \R{2}\fchi_0) \,,\quad
   (\fchi_0 \R{2}\R{2}\R{0}) \,,\quad
  (\R{0} \R{2}\R{2}\fchi_0) \right\}.
\eea 
We note that, when Einstein sources were evaluated in Appendix D.6 of Ref.~\cite{Azeyanagi:2013xea}, it was shown that that 2nd and higher order terms 
in $\fR$ do not contribute to $(\fR^q)$ at order $\omega^{q-1}$ (and lower). 
This is why 2nd and higher order terms in $\fR$ do not appear in $(\cR^q_{(q-1)})$.  
Furthermore, up to  $\omega^{q-2}$,  all contributions (including zeroth and first order building blocks) to 
$\fR^q$ vanish.

\subsubsection{Single-trace case 1: \fixform{$\fP_{CFT}=c_{_g}\, \tr[\fR^{2k}]$}}
We begin by first dealing with the case of $\fP_{CFT}=c_{_g}\, \tr[\fR^{2k}]$  for $k\ge2$ 
in AdS$_{2n+1}$ with $n=2k-1$.
In this case, $T_0$ and $T_2$ are trivially zero and thus we consider $T_1$ only.

When we compute $T_1$ for 
$\fP_{CFT}=c_{_g}\, \tr[ \fR^{2k}]$ (for $k\ge2$), 
we encounter two types of terms:
\bea
\mbox{single-product}\quad&:&\quad\tr[\delta \fGamma   (\fR^{2k-2}) \nabla \xi]\, , \quad
\tr[\delta \fGamma \nabla \xi  (\fR^{2k-2})] \,,\nonumber\\
\mbox{double-product}\quad&:&\quad
\tr[\delta \fGamma (\fR^q) \nabla \xi (\fR^{2k-2-q})]\,,
\eea
 where $1 \le q \le 2k-3$. 
We will first prove Fact \ref{fact1} and Fact \ref{fact2} for the single-product terms 
 and then for the double-product terms.
\\
\vspace{0.01cm} \\
\noindent
{\bf \underline{Single-product terms:}} \\
\noindent
For such terms, the contributions up to order $\omega^{n-1}$ are :
\bea\label{eq:singleproduct}
{\cal O }(\omega^{n-1})&:& 
\tr[
({}^{(0)}\delta \fGamma) (\cR^{(2k-2)}_{(2k-2)}) ({}^{(0)}\nabla\xi)]\, ,\quad
\tr[({}^{(0)}\delta \fGamma) ({}^{(0)}\nabla\xi) (\cR^{(2k-2)}_{(2k-2)} 
)]\,.
\eea
We note that all the structures above contain exactly one $(\R{2})$ except the following cases:
\bea
\label{eq:eq030v1}
&&
\tr[({}^{(0)}\delta \fGamma)  (\R{0} \R{3} \R{0})({}^{(0)}\nabla\xi)]\,,\quad
\tr[({}^{(0)}\delta \fGamma) ({}^{(0)}\nabla\xi)  (\R{0} \R{3} \R{0})]\,.
\eea These two terms are in fact zero due to
\be\label{eq:useful61}
 ({}^{(0)}\delta \fGamma)^a{}_b (\R{0})^b{}_c ( \R{m} \R{0})^d{}_f
=
 ( \R{m} \R{0})^f{}_a({}^{(0)}\delta \fGamma)^a{}_b ({}^{(0)}\nabla_c\xi^b)  (\R{0})^c{}_d 
 =0\, , 
\ee
for arbitrary $(\R{m})$  ($m\ge 0$). 
Thus, for the single-product, we proved Fact \ref{fact1} and Fact \ref{fact2}\,.
\\
\vspace{0.01cm} \\
\noindent
{\bf \underline{Double-product terms:}} \\
\noindent
The contributions of this type up to $\omega^{n-1}$ are :
\bea\label{eq:doubleproductclass}
{\cal O}(\omega^{n-2})&:&
\tr[({}^{(0)}\delta \fGamma)(\cR^q_{(q)}) ({}^{(0)}\nabla\xi
)(\fchi_1)] \,,\quad
\tr[({}^{(0)}\delta \fGamma)(\fchi_1) 
({}^{(0)}\nabla\xi)
(\cR^q_{(q)})]\,,
 \nonumber\\
 \nonumber\\
{\cal O}(\omega^{n-1})&:& 
\tr[({}^{(2)}\delta \fGamma)(\tilde{ \fchi_1}) ({}^{(0)}\nabla\xi)(\R{0})]\,,\quad
\tr[({}^{(0)}\delta \fGamma)(\tilde{ \fchi_1}) ({}^{(2)}\nabla\xi)(\R{0})]\,,\quad
\nonumber\\
&&
\tr[({}^{(1)}\delta \fGamma)(\cR^q_{(q)}) ({}^{(0)}\nabla\xi
)(\fchi_1)]\,,\quad
\tr[({}^{(0)}\delta \fGamma)(\cR^q_{(q)}) ({}^{(1)}\nabla\xi
)(\fchi_1)]\,,\nonumber\\
&&
\tr[
({}^{(1)}\delta \fGamma)(\fchi_1) 
({}^{(0)}\nabla\xi)(\cR^q_{(q)})]\,,\quad
\tr[
({}^{(0)}\delta \fGamma)(\fchi_1) 
({}^{(1)}\nabla\xi)(\cR^q_{(q)})]\,,
\nonumber\\
&&
\tr[
({}^{(0)}\delta \fGamma)(\cR^q_{(q)})({}^{(0)}\nabla\xi
)(\fchi_0)]\,,\quad
\tr[
({}^{(0)}\delta \fGamma)(\fchi_0) 
({}^{(0)}\nabla\xi)
(\cR^q_{(q)})]\,,
\nonumber\\
&&
\tr[({}^{(0)}\delta \fGamma)({\fchi_1})({}^{(0)}\nabla\xi)(\cR^q_{(q+1)})]\,,\quad
\tr[
({}^{(0)}\delta \fGamma)(\cR^q_{(q+1)})({}^{(0)}\nabla\xi)({\fchi_1} )]\,,
\nonumber\\
&&\tr[
({}^{(0)}\delta \fGamma)(\cR^q_{(q)})({}^{(0)}\nabla\xi)(\cR^{n-1-q}_{(n-1-q)})]\,,
\eea
where we have used
Eq.~(\ref{eq:fchi1fchi1}) and $\tilde{\fchi}_1$ is defined in Eq.~\eqref{eq:fchitilde}.
First, the $\omega^{n-2}$ order terms all vanish as a result of 
\bea&&
\label{eq:R0gradxiR0}
(\delta \fGamma)^a{}_b( \R{0})^b{}_c (\tilde{\fchi}_1)^d{}_e=
(\delta \fGamma)^a{}_b(\R{0}\R{1})^b{}_c=0\,, 
\nonumber\\
&&
(\R{0})^a{}_b( {}^{(0)}\delta \fGamma)^b{}_c(\fchi_1)^c{}_d=
(\R{0})^a{}_b( {}^{(0)}\delta \fGamma)^b{}_c(\R{1}\R{1} \R{0})^c{}_d=
(\R{0})^a{}_b( {}^{(0)}\delta \fGamma)^b{}_c(\R{2}\R{0})^c{}_d=0\,, 
\nonumber\\
&&
(\R{0})^a{}_b ({}^{(0)} \nabla_c \xi^b) (\R{0})^c{}_d
=(\R{0}\R{2})^a{}_b ({}^{(0)} \nabla_c \xi^b) (\R{0})^c{}_d
=0\,, 
\nonumber\\
&&
(\tilde{\fchi}_1)^a{}_b (\R{0}\R{2})^c{}_d=(\tilde{\fchi}_1)^a{}_b(\R{m} \R{0})^c{}_d=0\, , 
\eea for arbitrary $(\R{m})$ with $m\ge 0$. Thus, Fact \ref{fact1} is confirmed.

For the $\omega^{n-1}$ order contributions, one can show that all $\omega^{n-1}$ order contributions vanish except 
for the case containing $({}^{(0)} \delta \fGamma), ({}^{(0)} \nabla \xi)$, 
one $(\R{0})$,  one $(\R{2})$ (and the rest of $\fR$'s are all $(\R{1})$'s). 
To show that, we use Eq.~(\ref{eq:R0gradxiR0}) and the identities
\bea\label{eq:useful71}
&&
(\delta \fGamma)^a{}_b (\R{0}\R{1}^{p+1})^b{}_c =
(\delta \fGamma)^a{}_b (\R{0}\R{m} \R{0})^b{}_c =
\tr[(\R{1}^{2k+1}) ({}^{(0)} \nabla \xi)
(\R{0}\R{m}\R{0}) ({}^{(0)} \delta \fGamma)]  =
0\, , \nonumber\\
&&
(\R{m}\R{0}\R{1}^{p+1})
=(\R{1}^{p+1}\R{0}\R{m})
=
0\, ,
\eea for any $m\ge0$ and any $\delta \fGamma$ of any order. We could prove such identities by making use of the first Bianchi identity for $(\R{m})$ and the fact that $\delta \Gamma^a{}_{[bc]}=0$ together with the following :
\be
(\R{0})^r{}_b=(\R{0}\R{1}^{p+1})^r{}_b=0, \quad
(\R{0}\R{1}^{p+1})^\mu{}_b \propto (\ldots)_b dx^\mu,\quad
(\R{0})^\mu{}_b\propto (\ldots)^\mu{}_b \fu+(\ldots)_b dx^\mu\,
\ee
and
\be
(\R{1}^{2k+1})^a{}_b=(\R{1})^a{}_b\wedge (2\Phi_T\fomega)^{2k}\,,\quad
(\R{1})^{rr}=0\,,\quad
(\R{1})^{r\beta}\propto (\ldots) u^\beta+\fu \wedge \omega^{\beta}{}_\nu dx^\nu\,.
\ee 

Thus, we have proved Fact \ref{fact2}.


\subsubsection{Single-trace case 2: 
\fixform{$\fP_{CFT}=c_{_M}\,  \fF^l\wedge \tr[\fR^{2k}]$}}
Now we include a $U(1)$ gauge field and consider 
the most general anomaly polynomial in the single-trace form, 
$\fP_{CFT}=c_{_M}\, \fF^l\wedge \tr[\fR^{2k}]$, admitted in AdS$_{2n+1}$ 
with $n=2k+l-1$.  
In this case, the gauge field part $\fF^l$ at the lowest order is $(\F{1}^l)$ 
and is of order $\omega^{l}$. Combining with the result for the previous 
case, for $T_1$, we confirmed Fact \ref{fact1} and Fact \ref{fact2}. 

For $T_2$, we encounter the term of the form 
\begin{eqnarray}
\delta \fA \wedge \fF^{l-1} \wedge \tr[\delta\fGamma \fR^{2k-1}]\, . 
\end{eqnarray}
From Appendix.~\ref{sec:trgradxiR}, we know that $\tr[\delta\fGamma \fR^{2k-1}]$ starts at $\omega^{2k-1}$ and consists of purely zeroth and first order building blocks. Hence,
the lowest order contribution to this term is of order $\omega^{l-1+2k-1}=\omega^{n-1}$, 
which does not contain any 2nd and higher order terms in the building blocks 
 
Finally, let us consider $T_0$. It contains the following two types of terms :
\begin{eqnarray}
\delta \fA(\Lambda+\ic_\xi \fA) \wedge\fF^{l-2} \wedge\tr[\fR^{2k}]\,,\quad
(\Lambda+\ic_\xi \fA) \wedge\fF^{l-1} \wedge\tr[\delta \fGamma \fR^{2k-1}] \,. 
\end{eqnarray}
We note that the gauge field parts at the leading order 
are  ${}^{(0)}\delta \fA {}^{(0)}(\Lambda+\ic_\xi \fA) \wedge(\F{1}^{l-2})$  
(thus $\omega^{l-2}$-order)
and ${}^{(0)}(\Lambda+\ic_\xi \fA) \wedge(\F{1}^{l-1})$ 
(thus $\omega^{l-1}$-order) only. 
From the Maxwell sources computations (see Appendix C.3 of Ref.~\cite{Azeyanagi:2013xea}) and Appendix \ref{sec:trgradxiR}, the terms $\tr[\fR^{2k}]$ and $\tr[(\delta \fGamma)\fR^{2k-1}]$ starts at $\omega^{2k}$ and $\omega^{2k-1}$ order respectively, 
and these leading order terms do not contain 2nd and higher order building blocks.
Therefore, the leading order contribution to $T_0$ is of order $\omega^{n-1}$ 
and does not contain any 2nd and higher order terms of the building blocks. 

To summarize, we confirmed Fact \ref{fact1} and Fact \ref{fact2}.

\subsubsection{Multi-trace case 1: 
\fixform{$\fP_{CFT}=c_{_g}\,  \tr[\fR^{2k_1}]\wedge \ldots \wedge \tr[\fR^{2k_p}]$}}

Let us begin with the purely gravitational anomaly case. 
Since the  second order and higher terms in the building blocks 
do not contribute to $\tr[\fR^{2q}]$ at $\omega^{2q}$ order,    
without loss of generality, we only need to consider the two-trace case, 
$\fP_{CFT}=c_{_g}\,\tr[\fR^{2k_1}] \wedge \tr[\fR^{2k_2}]$  (in AdS$_{2n+1}$ where $n=2k_1+2k_2-1$
and $k_1,k_2\ge 1$). 
Depending on how the two $\fR$-derivatives act on this anomaly polynomial, 
there are two type of contribution: single-trace case and double-trace case.  
For the former, by noticing again the fact that the leading order 
contribution to $\tr[\fR^{2k}]$ is $\omega^{2k}$ and does not contain 
any 2nd order building blocks, we can deal with this type of term in 
the same way as the single-trace anomaly polynomial case. 
We therefore concentrate on the double-trace terms appearing 
in $T_1$ :
\be\label{eq:derivativeontracesv2}
\tr[\delta \fGamma \fR^{2k_1-1}]\wedge\tr[\nabla \xi \fR^{2k_2-1}]\, .
\ee
We note that there is another case where $k_1$ and $k_2$ are interchanged, 
but  since we can treat it in the same way,  we consider the above case only. 
The rest of the arguments follow along the same line of arguments as the AdS$_{15}$ case in 
Appendix.~{\ref{sec:ads15deltaQ}}. As we have summarized, 
the traces in this expression start at $\omega^{2k_1-1}$ and $\omega^{2k_2-1}$ order, 
respectively, and these leading order contributions do not contain any 
2nd order building blocks. More precisely, they are of the form 
 \bea
 {\cal O}( \omega^{2k_1-1}) :
 \quad \tr[(\delta \fGamma)  \fR^{2k_1-1}]
 &=&\{ 
 \tr[{}^{(0)} \delta \fGamma \fchi_0]\,, \quad
 \tr[ {}^{(1)} \delta \fGamma \fchi_1] 
 \}\,,  \nonumber\\
 {\cal O}( \omega^{2k_2-1}) :
\quad  \tr[(\nabla \xi) \fR^{2k_2-1}]&=&
  \{ 
 \tr[ {}^{(0)}\nabla \xi \fchi_0 ]\,,\quad
 \tr[ {}^{(1)} \nabla \xi \fchi_1 ]
 \}\, . 
\eea   

To summarize, we have proven even stronger statements than 
Fact \ref{fact1} and Fact \ref{fact2} that we used in the single-trace case for the terms in Eq.~(\ref{eq:derivativeontracesv2}). That is, here we 
have shown that for such type of terms,  the first non-trivial contributions start at $\omega^{n-1}$ and that 2nd (or higher) order building blocks 
do not contribute at order $\omega^{n-1}$ for any fixed $r$-surface.

\subsubsection{Multi-trace case 2: 
\fixform{$\fP_{CFT}=c_{_M}\,  \fF^l\wedge\tr[\fR^{2k_1}]\wedge \ldots \wedge \tr[\fR^{2k_p}]$}}
As in the previous case, we only need to consider the double-trace case, 
$\fP_{CFT}=c_{_M}\, \fF^l\wedge \tr[\fR^{2k_1}] \wedge \tr[\fR^{2k_2}]$  (in AdS$_{2n+1}$ where $n=2k_1+2k_2+l-1$, 
and $k_1,k_2,l\ge 1$), 
and more general cases follow from this immediately. 

Let us start with the first term $T_1$. We recall that 
the lowest order contribution from $\fF^l$ is $(\F{1})^l$ only and thus is at order $\omega^{l}$. 
Then the above results for the purely gravitational anomaly polynomial cases 
(double-trace and single-trace cases) are enough to prove Fact \ref{fact1} and Fact \ref{fact2}.  

For $T_0$ and $T_2$, since they contain only single $\fR$-derivative, 
the proof of Fact \ref{fact1} and Fact \ref{fact2} essentially follows from the one 
for the case with the mixed single-trace anomaly polynomial we have 
investigated above (by recalling that the leading order term of $\tr[\fR^{2k}]$ 
is $\tr[(\R{1}^{2k})]$ only and is of order $\omega^{2k}$). 

To summarize, we have confirmed Fact \ref{fact1} and Fact \ref{fact2} for general anomaly polynomials. 
We therefore see that in $T_0$ and $T_2$ the second and higher order building blocks
do not contribute up to order $\omega^{n-1}$ for any fixed $r$-surface. 
For $T_1$, we have a weaker statement that these contributions do not contribute at $r\rightarrow \infty$.


\subsection{Weyl covariance in \texorpdfstring{AdS$_{2n+1}$/CFT$_{2n}$}{}}
\label{sec:WeylWeight}
In this paper, since we are studying AdS$_{2n+1}$ with a CFT$_{2n}$ dual, it is useful to understand how fields in the boundary CFT$_{2n}$  transform under the Weyl rescaling.
We follow closely the discussions in Ref.~\cite{Kharzeev:2011ds,Baier:2007ix,Loganayagam:2008is}.
We first review the Weyl scaling of boundary fields and then extend it to bulk fields. 
We then apply this analysis to estimate the fall-off of the higher-order terms in the 
curvature two-forms at the boundary.

\subsubsection{Weyl scaling at boundary }

First we recall that the Weyl weight $\WeylW$ of a boundary object 
${\cal O}^{\,\mu_1\mu_2\cdots \mu_p}_{\,\nu_1\nu_2\cdots \nu_q}$ 
is defined by
\be
\WeylW ({\cal O}^{\,\mu_1\mu_2\cdots \mu_p}_{\,\nu_1\nu_2\cdots \nu_q})
=(\mbox{mass dimension\,\,\,of}\,\,\, {{\cal O}^{\,\mu_1\mu_2\cdots \mu_p}_{\,\nu_1\nu_2\cdots \nu_q}})
+p-q\,.
\ee 
We note that every  time we act an extra $\partial_\mu$, we get no net change in the Weyl weight.
Similarly, if we wedge with an extra $dx^\mu$, we also get no net change in the Weyl weight. From these rules, it is straightforward to figure out the Weyl weight for the following boundary objects :
\bea
\WeylW(x^\mu ) &=&0\,,\quad
\WeylW(x_\mu ) =-2\,,\nonumber\\
\WeylW(\eta_{\mu\nu} )& =&\WeylW(P_{\mu\nu} ) =-2 \,,\quad
\WeylW(\eta^{\mu\nu} ) =\WeylW(P^{\mu\nu} ) =+2\, ,\quad
\nonumber\\
\WeylW(u^\mu)
&=&+1\,,\quad
\WeylW(u_\mu)= -1\,,\quad
\WeylW(\fu)=-1\,, \nonumber\\
\WeylW(\omega_{\mu\nu})&=&-1\,,\quad
\WeylW(\fomega)=-1\, ,\quad
\WeylW\left(\partial_{\mu_1}\ldots \partial_{\mu_n} u_\nu\right)
=-1\,, 
 \nonumber\\
\WeylW(A_\mu)&=&\WeylW(\fA)=
\WeylW(F_{\mu\nu})=\WeylW(\fF)=0\,,
\eea
 for the boundary gauge field $\fA=A_\mu\, dx^\mu$ and field strength 
$\fF = (1/2!)F_{\mu\nu}\,dx^\mu\wedge dx^\nu= 
(1/2!)(\partial_\mu A_\nu - \partial_\nu A_\mu)\,dx^\mu\wedge dx^\nu$.
In our convention, the gauge field $A_{\mu}$ has mass dimension $1$ and thus 
$F_{\mu\nu}$ does mass dimension $2$. 

For a general curved boundary metric, we could deduce the following Weyl weights
\be \label{eq:weylweighgammarboundary}
\WeylW(\fGamma^\mu{}_{\nu})=0\,, \quad
\WeylW(\fR^\mu{}_{\nu})=0\, .
\ee since $\Gamma^\mu{}_{\alpha\beta}$ has mass dimension $1$ and $R^\mu{}_{\alpha\beta\gamma}$ has mass dimension $2$.

\subsubsection{Weyl scaling at bulk}
Now we extent the Weyl scaling to the bulk fields. 
We first need to assign the Weyl weight to the radial coordinate $r$ in a consistent way. To do this, we recall that the Weyl scaling is generated by a particular bulk asymptotic Killing vector by the standard AdS/CFT dictionary. More concretely,  the Weyl scaling preserves the combination $r^2 \eta_{\mu\nu}$ 
asymptotically and hence $r^2 \eta_{\mu\nu}$ has Weyl weight zero. Combining with the fact 
$\WeylW(\eta_{\mu\nu} )=-2$, we conclude that the Weyl weight of $r$ is one,  
\begin{eqnarray}
\WeylW(r)=+1\,.
\end{eqnarray}

As a second step, we identify the Weyl weight of objects defined at the bulk. 
Let us consider the connection 1-form and curvature 2-form (we note that 
$\fR^a{}_b$ has the same Weyl weight as $\fGamma^a{}_b$ since the former 
contains a term $d\fGamma^a{}_b$). 
We first notice 
$\fGamma^\mu{}_{\nu}=\Gamma^\mu{}_{\nu r}dr+\Gamma^\mu{}_{\nu \gamma}dx^\gamma$ 
which has the Weyl weight 0 as can be seen from the second term. Therefore, 
from the first term we conclude $\WeylW(\Gamma^\mu{}_{ r \nu} )=-1$. 
Let us next consider $\fGamma^\mu{}_{r} $ and $\fR^\mu{}_{r} $. Since $\fGamma^\mu{}_{r}$ contains  
the terms $\Gamma^\mu{}_{r\nu}\, dx^\nu$,
we can identify the Weyl weight for  $\fGamma^\mu{}_{r} $ and $\fR^\mu{}_{r} $ as 
\begin{eqnarray} \label{eq:murgammaweyl}
\WeylW(\fGamma^\mu{}_{r} )=\WeylW(\fR^\mu{}_{r} )=-1\,.
\end{eqnarray}
We then note that $\fR^\mu{}_{r} $ contains $ \fGamma^\mu{}_r \wedge \fGamma^r{}_r $ as well as 
$\fR^r{}_r$ does $\fGamma^r{}_\nu \wedge \fGamma^\nu{}_r$.  From these, we have the following assignment 
of the Weyl weight on the connection 1-form and curvature 2-form 
(here we list up the results in \eqref{eq:weylweighgammarboundary} 
and \eqref{eq:murgammaweyl} again) :
\begin{eqnarray} \label{eq:WeylR}
&&\WeylW(\fGamma^r{}_r)=0\, ,  \qquad
\WeylW(\fGamma^\mu{}_{r} )=-1\,, \qquad
\WeylW(\fGamma^r{}_\nu)=+1\, , \qquad
\WeylW(\fGamma^\mu{}_{\nu})=0\,, \\
&&\WeylW(\fR^r{}_r)=0\, , \qquad 
\WeylW(\fR^\mu{}_{r} )=-1\, , \qquad 
\WeylW(\fR^r{}_\nu)=+1\,, \qquad 
\WeylW(\fR^\mu{}_{\nu})=0\, .
\end{eqnarray}
 
For the fields $\Phi$ and $\Phi_T$, since  ${}^{(0)} \fA= \Phi \fu$ and $\Phi_T$ has mass dimension $+1$, 
we can assign the Weyl weight for any $r$ as follows :  
\be
\WeylW(\Phi)=\WeylW(\Phi_T)= 1\, .
\ee 
 
Before we apply for this Weyl scaling in the next part, we provide an alternative argument 
for the second step above: in the following, we determine 
the Weyl weight of  $\fGamma^a{}_b$ (and hence $\fR^a{}_b$) by starting with 
the identification of the Weyl weights for the bulk metric components. 

For asymptotically AdS spacetimes, the components $G_{\mu\nu}$ contain a term  of the form 
$r^2 \eta_{\mu\nu}$ and thus, by using $\WeylW(r)=+1$, the Weyl weight of these components is 
\begin{eqnarray}
\WeylW(G_{\mu\nu})=0\, . 
\end{eqnarray}
Then the line element 
$ds^2 = G_{\mu\nu} dx^\mu dx^\nu+2 G_{\mu r} dx^\mu dr  + G_{rr} dr^2$ has 
the Weyl weight $0$ and thus we can identify the Weyl weights of 
$G_{\mu r}$ and $G_{rr}$ as 
\begin{eqnarray}
\WeylW(G_{\mu r}) = -1\,,\qquad
\WeylW(G_{r r}) = -2\,.
\end{eqnarray}

Let us next determine the Weyl weights for the inverse metric. 
By using the relation $G^{\alpha c} G_{c\beta}=\delta^\alpha_\beta$ with $\WeylW(\delta^\alpha_\beta)=0$
and  $G^{ab}\partial_a \partial_b=G^{rr}\partial_r \partial_r + G^{\mu r} \partial_\mu \partial_r+\ldots$, 
we can have 
\begin{eqnarray}
\WeylW(G^{\mu r})=+1\,,\qquad
\WeylW(G^{\mu\nu})=0\,, \qquad  \WeylW(G^{rr})=+2\, . 
\end{eqnarray}
Then from the definition of the Christoffel symbols, we obtain the Weyl weights of 
the connection 1-form as well as the curvature 2-form as follows :
\bea
\fGamma^r{}_r
&=&G^{r\mu} \partial_r G_{\mu r} \,  dr+\ldots
\, \Rightarrow\quad
\WeylW(\fR^r{}_r)=\WeylW(\fGamma^r{}_r)=0\,,\nonumber\\
\fGamma^\mu{}_r
&=&G^{\mu \beta} \partial_r G_{\beta r}\, dr+\ldots
\, \Rightarrow\quad
\WeylW(\fR^\mu{}_r)=\WeylW(\fGamma^\mu{}_r)=-1\, ,\nonumber\\
\fGamma^r{}_\nu
&=&G^{r\beta } \partial_{\beta} G_{\nu r}\, dr+\ldots
\, \Rightarrow\quad
\WeylW(\fR^r{}_\nu)=\WeylW(\fGamma^r{}_\nu)=+1\, ,\nonumber\\
\fGamma^\mu{}_\nu
&=&G^{\mu r} \partial_{r} G_{\nu r}\, dr+\ldots
\, \Rightarrow\quad
\WeylW(\fR^\mu{}_\nu)=\WeylW(\fGamma^\mu{}_\nu)=0\, .
\eea These results are consistent with Eq.~\eqref{eq:WeylR}.

\subsubsection{Fall-off of higher order terms in curvature two-form}
Now we are ready to use Weyl covariance to estimate fall-offs of $(\R{m})$ for any positive integer $m$.
We can easily deduce these fall-offs based on the following facts. 
We consider the two-form at the $\omega^m$ order which is made of  
$\partial_\mu$, $dx^\nu$ and $u_{\rho}$  only. We call them the boundary two-form here. 
We first neglect the contraction structure. 
There are many possibilities for this type of the two-from. 
For example, at order $\omega^2$ we can have the following possibilities :
\be
(\partial_{\alpha_1} \partial_{\alpha_2} u_\beta)\, dx^\kappa\wedge dx^\delta\,,\quad
(\partial_{\alpha_1} u_{\beta_1})(\partial_{\alpha_2} u_{\beta_2}) \,dx^\kappa\wedge dx^\delta\, , 
\ee and etc..  However, the important point is that  their  mass dimension is all $m-2$,
no matter how the structures look like. We notice that even when we multiply or contract appropriately by using 
\be\label{eq:stuff1}
u_\mu\,, \quad P_{\mu\nu}\,, \quad\eta_{\mu\nu}\, , 
\ee (or those with some indices up), the mass dimensions are not changed at all
since the objects in Eq.~\eqref{eq:stuff1} all have the mass dimension $0$. 
Therefore, the two-forms at $\omega^{m}$ order constructed in this manner 
all have the mass dimension $m-2$. Now we take into account the index structure 
of the two-form to evaluate the Weyl weight. When there is one free upper and one free lower boundary indices 
or no boundary indices, then the Weyl weight of this type of the two-form turns out to be $m-2$. 
On the other hand, when there is one free upper (resp. lower) boundary indices, 
then the Weyl weight turns out to be $m-1$ (resp. $m-3$).  

Now we consider the curvature two-form. Asymptotically, the $\omega^m$ order terms 
in the curvature two-form are written as the boundary two-form classified above 
with appropriate power of $r$ multiplied (we notice that the curvature two-form also contains 
terms proportional to $dr$ in general, but they do not contribute to the Noether charge. 
We thus neglect these terms). 
Combining the above result with the Weyl weight of the curvature two-form summarized in \eqref{eq:WeylR}, 
we can identify what power of $r$ needs to be multiplied. Then the fall-off behaviors of 
the curvature two-form at $\omega^m$ order is estimated as   
\bea
&&(\R{m})^\mu{}_r \atinfty \rightarrow {\cal O}(r^{1-m})\, ,\quad
(\R{m})^r{}_r \atinfty \rightarrow {\cal O}(r^{2-m})\, ,\nonumber\\
&&(\R{m})^\mu{}_\rho \atinfty \rightarrow {\cal O}(r^{2-m})\,,\quad
(\R{m})^r{}_\rho \atinfty \rightarrow {\cal O}(r^{4-m})\,. 
\eea
In particular, for $m=2$, this gives the estimate
\be
(\R{2})^r{}_r \atinfty \rightarrow {\cal O}(r^{0})\, ,\quad
(\R{2})^\mu{}_\rho\atinfty \rightarrow {\cal O}(r^{0})\,,\quad
(\R{2})^\mu{}_r \atinfty\rightarrow {\cal O}(r^{-1})\,,\quad
(\R{2})^r{}_\rho \atinfty\rightarrow {\cal O}(r^{2})\,.
\ee We see that this reproduces all the fall-offs in Eq.~\eqref{eq:falloffR2} except that we can improve 
the estimate of the fall-off for $(\R{2})^\mu{}_r$ by one power of $1/r$. 

We note that in the above estimates, when we construct the two-forms appearing in 
a particular component of $(\R{m})$, we have started with the boundary two-form discussed above 
and then compensated the Weyl weight (difference between the Weyl weights of 
a particular component of $(\R{m})$ and  the boundary two-form) by just 
multiplying an appropriate power of $r$. 
This provides the estimates when $q=m=0$ only. With $q, m \ne 0$, we are also allowed to use $\Phi_T$ 
and $\Phi$ (both are with the Weyl weight $+1$) to compensate the Weyl weight. 
Since both $\Phi_T$  and $\Phi$ have faster fall-offs than $r$, 
they will not affect the modest fall-off estimates above.



\section{Entropy : Einstein-Maxwell contribution}

This Appendix is devoted to the evaluation of the Einstein-Maxwell contribution to the black hole entropy. 
The Einstein-Maxwell contribution to the differential Noether charge is given in Eq.~\eqref{eq:einmaxnoether} of \S\S\S\ref{ssec:TachiForm}. 
We first evaluate the Einstein part (the first line of Eq.~\eqref{eq:einmaxnoether}) to confirm that this part reproduces 
the non-anomalous CFT result. After this, we will confirm that the Maxwell part (the second line of Eq.~\eqref{eq:einmaxnoether})
does not give any nontrivial contribution at the horizon. 
In the evaluation, we use the two prescriptions for the differential Noether charge introduced in \S\S\ref{app:entropyBTZwhole}
and confirm the matching of the final result explicitly.

\label{app:entropyEM}
\subsection{Entropy current from Komar charge}
We first evaluate the Komar charge at the horizon by taking $\xi^\mu = u^\mu/T$ , the 
appropriately normalized Horizon generator. We note that this is the Killing vector which preserves the 
horizon. To the order we are working in, the horizon 
is given by the surface $r=\rH$ with $f(\rH)=0$ and $4\pi T=\rH^2 f'(\rH)$. We obtain from the Komar part that
\begin{equation}
\begin{split}
\brk{ \nabla_a\xi^b \frac{\hodge (dx^a\wedge dx_b)}{16\pi\GN}
}_{r=\rH,\ \xi^\mu = \frac{u^\mu}{T}} = \prn{ J_S^{\text{\tiny{CFT}}}}_\mu \ \hodgeCFT dx^\mu\, , 
\end{split}
\end{equation}
with
\begin{equation}
\begin{split}
\prn{ J_S^{\text{\tiny{CFT}}}}_\mu  &= \frac{\rH^{d-1}}{4\GN}\ u_\mu +\ldots\, . 
\end{split}
\end{equation}
We note that this is essentially the prescription to compute CFT entropy current given in Eq.~(\ref{eq:Sentanom0}).

\subsection{Entropy current from differential Noether charge : prescription I}
Alternately, let us first compute the variation of the Komar charge at a fixed $r$
\begin{equation}
\begin{split}
&\delta\brk{\nabla_a\xi^b \frac{\hodge (dx^a\wedge dx_b)}{16\pi\GN}}\\
&\ = \xi^\mu  \frac{r^{d-1}}{16\pi\GN}
\brk{2r\delta f \eta_{\mu\nu} -r^2\frac{d\delta f}{dr} u_\mu u_\nu
-r^2\frac{df}{dr} \prn{\delta u_\mu u_\nu+ u_\mu \delta u_\nu}
}\ \hodgeCFT dx^\nu +\ldots\\
&\quad +\xi^\mu 
\brk{\frac{r^{d+1}}{16\pi\GN} \frac{d}{dr}\prn{\frac{g_{_V}}{r^2}}\delta V_\mu u_\nu
+ \frac{r^{d+1}f^2}{16\pi\GN}\frac{d}{dr}\prn{\frac{g_{_V}}{r^2f}}u_\mu \delta V_\nu }
\ \hodgeCFT dx^\nu \\
&\quad +\xi^\mu 
\brk{\frac{r^{d+1}}{16\pi\GN} \frac{d}{dr}\prn{\frac{g_{_V}}{r^2}}V_\mu \delta u_\nu
+ \frac{r^{d+1}f^2}{16\pi\GN}\frac{d}{dr}\prn{\frac{g_{_V}}{r^2f}}\delta u_\mu V_\nu }
\ \hodgeCFT dx^\nu \\
&\quad +\xi^\mu 
\brk{\frac{r^{d+1}}{16\pi\GN} \frac{d}{dr}\delta\prn{\frac{g_{_V}}{r^2}}V_\mu u_\nu
+ \frac{r^{d+1}f^2}{16\pi\GN}\frac{d}{dr}\delta\prn{\frac{g_{_V}}{r^2f}}u_\mu V_\nu }
\ \hodgeCFT dx^\nu \\
&\quad +\xi^\mu 
\brk{ \frac{2r^{d+1}f\delta f}{16\pi\GN}\frac{d}{dr}\prn{\frac{g_{_V}}{r^2f}}\delta u_\mu V_\nu }
\ \hodgeCFT dx^\nu \\
&\quad+\ldots \, . 
\end{split}
\end{equation}
We will now pull-back this Komar variation onto the horizon at $r=\rH$ and 
set $\xi^\mu = u^\mu/T$ to get
\begin{equation}
\begin{split}
&\lim_{\{r=\rH,\ \xi^\mu = \frac{u^\mu}{T}\}} \delta\brk{\nabla_a\xi^b \frac{\hodge (dx^a\wedge dx_b)}{16\pi\GN}}\\
&\ = 
\bigbr{-2  \frac{\rH^{d-2}}{4\GN}\delta \rH u_\nu+\frac{\rH^{d-1}}{4\GN} \delta u_\nu }\ \hodgeCFT dx^\nu \\
&\quad
+ \bigbr{
 \frac{\rH^{d+1}}{16\pi\GN}\prn{\frac{d\delta f}{dr}}_{r=\rH}  
 -V^\mu\delta u_\mu  
\brk{\frac{r^{d+1}}{16\pi\GN} \frac{d}{dr}\prn{\frac{g_{_V}}{r^2}}}_{r=\rH}
  }\frac{u_\nu}{T}\ \hodgeCFT dx^\nu +\ldots \, , \\
\end{split}
\end{equation}
where we have used 
\begin{equation}
\begin{split}
(f)_{r=\rH} &= 0\ ,\quad 
\prn{\frac{df}{dr}}_{r=\rH} = 4\pi T \rH^{-2} \, , \\ 
(\delta f)_{r=\rH} &= -\prn{\frac{df}{dr}}_{r=\rH}\delta \rH =-4\pi T\ \rH^{-2}\delta \rH \, . \\
\end{split}
\end{equation}
We add to this the non-Komar variation evaluated at the horizon. 
By subtracting from the above expression for the Komar part the following expression 
\begin{equation}
\begin{split}
\ic_\xi&\brk{\delta\fGamma^b{}_a\wedge \frac{\hodge(dx^a \wedge dx_b)}{16\pi\GN}
}_{\{r=\rH,\ \xi^\mu = \frac{u^\mu}{T}\}}\\
&= \bigbr{-\frac{(d+1)\rH^{d-2}}{4\GN} \delta \rH }
u_\nu\ \hodgeCFT dx^\nu  \\
&\quad
+ \bigbr{
 \frac{\rH^{d+1}}{16\pi\GN}\prn{\frac{d\delta f}{dr}}_{r=\rH}  
 -V^\mu\delta u_\mu  
\brk{\frac{r^{d+1}}{16\pi\GN} \frac{d}{dr}\prn{\frac{g_{_V}}{r^2}}}_{r=\rH}
  }\frac{u_\nu}{T}\ \hodgeCFT dx^\nu +\ldots \, , \\
\end{split}
\end{equation}
we obtain  the differential Noether charge at the horizon as 
\begin{eqnarray}
\label{eq:deltaQhorEM}
\lim_{\{r=\rH,\ \xi^\mu = \frac{u^\mu}{T}\}} (\fQNoether)_{_\text{Einstein}}
&=&\left\{(d-1) \frac{\rH^{d-2}}{4\GN}\delta \rH u_\nu 
+\frac{\rH^{d-1}}{4\GN} \delta u_\nu \right\} \hodgeCFT dx^\nu \nonumber \\
&=& \delta\left\{ \prn{J_S^{\text{\tiny{CFT}}}}_\mu \ \hodgeCFT dx^\mu \right\} \, . 
\end{eqnarray}

\subsection{Entropy current from differential Noether charge : prescription II}
Here we illustrate how to use prescription II to obtain the entropy current from evaluating the differential Noether charge at the horizon :
\begin{equation}
\begin{split}
(&\fQNoether)_{_\text{Ein}} \athor \\
&\ =
 \ \half\  \nabla_a \xi^b\ \delta\left.\brk{ \frac{\hodge(dx^a\wedge dx_b)}{8\pi \GN} } \right\athor
+ \half\  \delta\fGamma^b{}_a\ \left. \frac{\ic_\xi\hodge(dx^a\wedge dx_b)}{8\pi \GN} \right\athor  \, .
\end{split}
\end{equation}

Using Eq.~\eqref{eq:2formsUpDownhor} and Eq.~\eqref{eq:gradxihorV}, we obtain for the Komar part, 
\begin{eqnarray}
 \ \half\  \nabla_a \xi^b\ \delta\left.\brk{ \frac{\hodge(dx^a\wedge dx_b)}{8\pi \GN} } \right\athor&=&
\half \frac{1}{4 \GN}
 \delta \brk{r_H^{d-1} \hodgeCFT \fu}
+\half \frac{1}{4 \GN}
   \delta (r_H^{d-1})  \hodgeCFT \fu   \nonumber\\
&&
+ \frac{r_H^{d-1}}{32\pi \GN} T^{-1}\left. \left[\frac{d g_{_V}}{dr}\right]\right \athor  V_\mu
(\delta u^\mu) \hodgeCFT \fu+\ldots, \nonumber\\
\end{eqnarray}  
while using Eq.~\eqref{eq:2formsUpDownhorixi} and Eq.~\eqref{eq:flugravGammahorform}, we obtain for the non-Komar part
\begin{eqnarray}
   \delta\fGamma^b{}_a \left. \frac{\ic_\xi\hodge(dx^a\wedge dx_b)}{2\times 8\pi \GN} \right\athor &=&
   \half\frac{1}{4 \GN}  
   r_H^{d-1}
\hodgeCFT (\delta\fu )-
   \frac{    r_H^{d-1}
}{32\pi \GN} T^{-1} 
  \left. \left[\frac{dg_{_V}}{dr} \right]\right\athor  
  V_\mu (\delta u^\mu)\hodgeCFT \fu
   +\ldots . \nonumber\\
\end{eqnarray}
Summing up both contributions, we finally obtain the Einstein contribution to the differential Noether charge at the horizon as 
\begin{equation}
(\fQNoether)_{_\text{Ein}} \athor= 
\delta\left[
\frac{1}{4 \GN}  
   r_H^{d-1}\hodgeCFT \fu\right]
=
\delta\left[ \prn{J_S^{\text{\tiny{CFT}}}}_\mu \ \hodgeCFT dx^\mu \right]\, . 
\end{equation} This agrees with the result in Eq.~\eqref{eq:deltaQhorEM} from prescription I above.

\subsection{Gauge field and Maxwell contribution}

Now we move to the Maxwell part and confirm that the this part vanishes when evaluated at the horizon. 

The Komar charge for the Maxwell part is given by
\begin{equation}
\prn{\Lambda+\ic_\xi \fA}  \cdot  \frac{\hodge \fF}{\gYM^2}\, , 
\end{equation}
and thus vanishes at the horizon due to  $\prn{\Lambda+\ic_\xi \fA}\athor=0$\,.

On the other hand, if we evaluate the differential Noether charge at the horizon, then we need to evaluate the following at the horizon :
\begin{equation}
\begin{split}
 (\fQNoether)_{\rm Max}\athor 
= \prn{\Lambda+\ic_\xi \fA}  \cdot  \delta\left.\brk{
\frac{\hodge \fF}{\gYM^2} 
 }\right\athor
+\left.
 \delta \fA\cdot \frac{\ic_\xi \hodge \fF}{\gYM^2}
\right \athor\, .
\end{split}
\end{equation} Again, the first term is zero at the horizon due to $\prn{\Lambda+\ic_\xi \fA}\athor=0$.
 For the second term, from Eq.~\eqref{eq:hodgeFconstr} and the fact that $\mathcal{Q}^\tV$ vanishes at the horizon, we obtain
 \begin{equation}
 \hodge\fF  \athor =
  -(d-2)q\ \hodgeCFT \fu + \ldots  \, . 
 \end{equation} However, in evaluating $\ic_\xi \hodge \fF$ at the horizon, using $\xi^\mu \athor =  u^\mu/T$ and the fact that $\ic_{u} \hodge \fu=\hodge(\fu\wedge\fu)=0$, we finally conclude that 
\begin{equation}
 \left.(\fQNoether)_{\rm Max}\right|\athor=0\,.
\end{equation}
We note that the above computations for the Maxwell part at the horizon apply to both of the two prescriptions.


 \bibliographystyle{utphys}
\bibliography{CS-entropy-bib}

\end{document}